\documentclass[a4paper,11pt]{article}
\usepackage[a4paper, hmargin=1.25in,vmargin=1.75in]{geometry}
\usepackage{amsmath,amssymb,amsthm,mathrsfs,amsfonts,mathtools,calculator,dsfont,bm} 
\usepackage[dvipsnames,svgnames, x11names]{xcolor}
\usepackage{natbib,bibentry,bibunits,lmodern}
\usepackage{grffile} 
\usepackage[bottom,hang,flushmargin]{footmisc} 
\usepackage{enumitem} 
\newlist{problems}{enumerate}{1}
\setlist[problems]{label={\arabic*.}, ref={\arabic{part}.\thechapter.\arabic*}}
\usepackage{pdfpages,selectp} 
\usepackage[hyphens]{url}
\usepackage[labelsep=period]{caption} 
\usepackage{subcaption}
\usepackage{fancyvrb} 
\usepackage{environ,trimspaces,xspace}
\usepackage[titletoc]{appendix}
\usepackage{fancyhdr,setspace,titlesec,titling} 
\usepackage{siunitx,booktabs,longtable,threeparttable,multirow,makecell,float,rotating,caption} 
\usepackage{tikz,pgfplots} 
\pgfplotsset{compat=1.11}
\usetikzlibrary{arrows,intersections,plotmarks}
\usetikzlibrary{backgrounds,positioning,fit,petri}
\usetikzlibrary{mindmap,trees,calendar,external,shapes}
\usetikzlibrary{decorations.fractals,decorations.pathmorphing,shadows,shadings,patterns}
\usetikzlibrary{spy,calc} 
\usepgfplotslibrary{groupplots}
\usepackage[bookmarksnumbered,bookmarksopen,psdextra,unicode,hyperindex,breaklinks,hypertexnames=false]{hyperref}
\usepackage[open,openlevel=2,numbered,atend]{bookmark} 
\usepackage{tabularx}
\usepackage{cleveref}
\renewcommand{\thetable}{\Roman{table}}
\usepackage{pdflscape}
\usepackage{dcolumn,adjustbox}
\newcolumntype{L}{D{.}{.}{2,7}}

\definecolor{Ocean}{rgb}{0,0,0.75}
\makeatletter
\renewcommand\paragraph{\@startsection{paragraph}{4}{\z@}%
            {-2.5ex\@plus -1ex \@minus -.25ex}%
            {1.25ex \@plus .25ex}%
            {\normalfont\normalsize\bfseries}}
\makeatother
\hypersetup{colorlinks, linkcolor = {NavyBlue}, citecolor = {NavyBlue}, urlcolor ={NavyBlue}}

\newtheorem{thm}{Theorem}[section]

\newtheorem{lem}{Lemma}[section]
\newtheorem{pro}{Proposition}[section]
\newtheorem{ass}{Assumption}[section]
\theoremstyle{definition}

\theoremstyle{definition}

\newtheorem*{ex*}{Example}

\newtheoremstyle{exctd}
{\topsep} {\topsep}%
{\upshape}
{}
{\bfseries\scshape}
{.}
{1em}
{\thmname{#1} \thmnumber{ #2}\thmnote{#3} (cont.)}
\theoremstyle{exctd}

\newcommand{\amax}{\operatornamewithlimits{arg\,max}}

\newcommand{\convp}{\xrightarrow{p}}

\pgfmathdeclarefunction{npdf}{3}{%
\pgfmathparse{1/(#3*sqrt(2*pi))*exp(-((#1-#2)^2)/(2*#3^2))}%
}
\tikzset{
    declare function={
        ncdf(\x,\m,\s)=1/(1 + exp(-0.07056*((\x-\m)/\s)^3 - 1.5976*(\x-\m)/\s));
    }
}


\defaultbibliographystyle{ecta}
\defaultbibliography{Mybibliography_Infinity}

\allowdisplaybreaks 
\usepackage{afterpage}

\begin{document}
\begin{bibunit}
\pdfbookmark[1]{Title}{title}
\title{\vspace{-2.5cm}Semiparametric Conditional Factor Models in Asset Pricing\thanks{We thank Torben Andersen, Bin Chen, Xiaohong Chen, Xu Cheng, Greg Duffee, Robert Korajczyk (discussant), Yuan Liao, Ye Luo, Hao Ma (discussant), Seth Pruitt (discussant), Andres Santos, Frank Schorfheide, Liangjun Su, Dacheng Xiu, Chu Zhang, and Linyan Zhu (discussant) as well as conference audiences and seminar participants at Nanyang Technological University, Peking University, Shanghai University of Finance and Economics, UC Riverside, University of Lausanne (Swiss Finance Institute), University of Pennsylvania, Xiamen University, 2021 YEAP, 2022 SFS Cavalcade, 2022 WFA, 2022 CICF, 2023 AFA, Spring 2023 Rochester Conference in Econometrics, 2024 HKUST IAS-SBM Financial Econometrics conference, the Inaugural Meeting of the Greater Bay Econometrics Study Group for helpful discussions and comments. Shengqi Yang provided excellent research assistance.}
}
\author{Qihui Chen\thanks{School of Management and Economics and Shenzhen Finance Institute, The Chinese University of Hong Kong, Shenzhen (CUHK-Shenzhen); qihuichen@cuhk.edu.cn}\\CUHK-Shenzhen
\and
Nikolai Roussanov\thanks{The Wharton School, University of Pennsylvania and NBER; nroussan@wharton.upenn.edu} \\The Wharton School
\and
Xiaoliang Wang\thanks{Department of Finance, HKUST Business School; xlwangfin@ust.hk} \\HKUST}
\date{\today}
\maketitle


\begin{abstract}
We introduce a simple and tractable methodology for estimating semiparametric conditional latent factor models. Our approach disentangles the roles of characteristics in capturing factor betas of asset returns from ``alpha.''  We construct factors by extracting principal components from Fama-MacBeth managed portfolios. Applying this methodology to the cross-section of U.S. individual stock returns, we find compelling evidence of substantial nonzero pricing errors, even though our factors demonstrate superior performance in standard asset pricing tests. Unexplained ``arbitrage'' portfolios earn high Sharpe ratios, which decline over time. Combining factors with these orthogonal portfolios produces out-of-sample Sharpe ratios exceeding 4.

\end{abstract}

\begin{center}

\end{center}
\newpage
%
%
%
%
%
%
%

\section{Introduction}\label{Sec: 1}

A central question in empirical asset pricing is why different assets earn different average returns. While asset pricing theory attributes cross-sectional differences in returns to variations in risk exposures, considerable evidence suggests that mispricing\textemdash captured by the dependence of returns on asset characteristics\textemdash also plays a significant role, suggesting potential market inefficiencies. Much of the debate centers on multi-factor models that aim to link average returns to factor loadings, building on the influential framework of \citet{FamaFrench_Commonrisk_1993}, who introduced a portfolio-sorting approach to constructing asset pricing factors. Following their seminal work, researchers have proposed hundreds of factors, leading to what \citet{Cochrane_Presidential_2011} memorably termed the ``\textit{factor zoo},'' a concept further explored by \citet{Harveyetal_FactorZoo_2016}. While some factor models have an explicit justification based on economic theory, many implicitly rely on the idea that factors capture common variation in returns, thus appealing to arbitrage pricing theory and its extensions \citep{Ross_APT_1976,ChamberlainRothschild_FactorStuctures_1982,ConnorKorajczyk_Performance_1986,ConnorKorajczyk_RiskReturn_1988,Reisman/IAPT:1992}. Since implementing the latter requires estimating the conditional covariance matrix of returns, which becomes impractical when the number of assets ($N$) exceeds the number of time periods ($T$), most studies rely on asset characteristics to proxy for (imperfectly measured) risk exposures, often employing the portfolio-sorting approach. However, this makes distinguishing between risk-based explanations and those rooted in mispricing virtually impossible, as exemplified by the ``characteristics versus covariances'' debate \citep{DanielTitman_Characteristics_1997}.


In our analysis we consider a canonical conditional factor model:
\begin{align}\label{Eqn: Model}
y_{it} = \alpha(z_{it}) + \beta(z_{it})^{\prime}f_{t} + \varepsilon_{it}, i=1,\ldots,N,t=1,\ldots,T.
\end{align}
Here, $y_{it}$ is the excess return of asset $i$ at time period $t$, $z_{it}$ is an $M\times 1$ vector of pre-specified asset characteristics (which may include a constant term) that is observed at the beginning of time period $t$,\footnote{In asset pricing,  $z_{i,t-1}$, the characteristics observed at time period $t-1$,  is usually used in \eqref{Eqn: Model}. For notational simplicity, here we use $z_{it}$ rather than $z_{i,t-1}$.} $f_t$ is a $K\times 1$ vector of unobserved {\em latent} factors, $\beta(\cdot)$ is a $K\times 1$ vector of unknown factor loading functions, $\alpha(\cdot)$ is an unknown intercept function, $\varepsilon_{it}$ is the idiosyncratic component that is orthogonal to the common factors $f_t$.\footnote{While our main focus is on cross-sectional asset pricing, the model has other potential applications, which include modelling the implied volatility of options \citep{Parketal_FactorDynamics_2009} and describing consumer demand system \citep{Lewbel_DemandSystems_1991}, among others.} The model describes a {\em conditional} factor model, in the sense that it captures time-variation in asset return exposures to the common factors (i.e., $\beta(z_{it})$) as well as the pricing errors (i.e., $\alpha(z_{it})$), which are both functions of characteristics. This model is well suited for resolving the ``characteristics versus covariances'' debate, since it potentially allows for distinguishing between the risk and mispricing explanations of the role of characteristics in predicting asset returns.\footnote{While useful, it might not be sufficient to resolve the debate, since distinguishing between the different explanations requires understanding the economic nature of the latent factors - e.g., see \citet{Kozaketal_Interpreting_2018}.} Meanwhile, the model allows for pooling the information in a multitude of characteristics and summarizing the common variation using a small number of factors, thereby helping to ``tame the \textit {factor zoo}.''  The challenge of using the model is threefold: first, the identities of the common factors $f_t$ are unknown since the factors are latent; second,  the functional forms of the $alpha$ and $beta$ functions are also generally unknown; finally, the cross-sectional dimension $N$ is typically much larger than the sample time-series length $T$, which renders standard tools of factor analysis inapplicable, especially when conditional covariances are time-varying.

We introduce a simple and tractable estimation method to recover both the latent factors and the functional parameters of the model, alongside formal inference procedures. First, we develop an easy-to-compute estimator for $\alpha(\cdot)$, $\beta(\cdot)$, and $f_t$ based on a sieve approximation to the nonparametric functions $\alpha(\cdot)$ and $\beta(\cdot)$. The estimation involves two steps: (i) regressing $y_{it}$ on sieve functions of $z_{it}$ for each $t$; and (ii) applying principal component analysis (PCA) to the estimated coefficients obtained in step (i). We refer to this approach as the \textit{regressed-PCA}. The first step aligns with the cross-sectional regressions of \cite{FamaMacBeth_RiskReturn_1973}, where the estimated coefficients at each point in time represent returns of ``pure play'' characteristic-managed portfolios. The second step is effectively a standard PCA on a relatively small set of characteristic-managed portfolios constructed via Fama-MacBeth regressions. Second, we develop a bootstrap inference framework to assess the significance of $\alpha(\cdot)$ as well as test the linearity of $\alpha(\cdot)$ and $\beta(\cdot)$.
Third, we establish large-sample properties of the estimators under mild conditions, including consistency, rate of convergence, and asymptotic normality, as well as the validity of the proposed tests. Notably, the asymptotic results possess several advantages: (i) they do not require large $T$; (ii) they accommodate time-varying and potentially nonstationary $z_{it}$; (iii) they apply to unbalanced panels, which is particularly beneficial for securities with varying lifespans. Our Monte Carlo simulations demonstrate that the proposed estimators and tests exhibit satisfactory finite-sample performance and remain robust even when $T$ is small, provided $N$ is large. In addition to offering formal inference procedures and well-founded asymptotic properties, regressed-PCA presents several advantages over existing methods such as instrumented PCA (IPCA) \citep{Kellyetal_Characteristics_2019} and projected-PCA \citep{Fanetal_ProjectedPCA_2016,Kimetal_Arbitrage_2019}. Specifically, regressed-PCA is computationally efficient and accommodates nonzero alphas, time-varying characteristics, unbalanced panels, and short samples, making it particularly well-suited for empirical asset pricing applications.

We apply our new methodology to analyzing the cross-section of individual stock returns. Our analysis uses the same dataset as \citet{Kellyetal_Characteristics_2019}, the study most closely aligned with ours in terms of empirical aims. However, our econometric approach and empirical findings differ significantly from theirs. First, unlike \citet{Kellyetal_Characteristics_2019, Kellyetal_IPCA_2017}, our method does not aim to simultaneously maximize the ``fit'' of the factor model to individual asset returns in both the time-series and cross-section. Instead, we extract factors that capture the most time-series comovement within a set of portfolios, which, in turn, reflect the most cross-sectional variation in individual asset returns. Second, our approach allows the $\alpha(\cdot)$ and $\beta(\cdot)$ functions to be nonlinear.  We test\textemdash and reject\textemdash the validity of linear specifications empirically, which reveals the strong evidence of nonlinearity in factor loadings and pricing errors.  Third, our inference procedure also enables us to test the significance of pricing errors.  Our empirical results reveal that the pricing errors associated with many characteristics are statistically significant, leading to the rejection of the risk-based model.  Lastly, our methodology facilitates rolling sub-sample analyses to accommodate evolving factor dynamics, as it does not rely on large $T$. We find that both in-sample and out-of-sample goodness-of-fit measures for all factor models decline from 1970 until roughly 2000 but improve thereafter. This pattern aligns with the findings in \citet{Campbell2001have} and \citet{Campbell2022idiosyncratic} on the time-variation in the amount of idiosyncratic volatility in the U.S. stock market. We also document a significant decline in pricing errors in more recent years, particularly since 2000. This decline may reflect the growing prevalence of quantitative investing, which reduces mispricing by exploiting characteristic-related anomalies, as suggested by \citet{McLeanPontiff_DoesAcademic_2016} and \citet{Greenetal_Characteristics_2017}.


Based on these findings, we construct trading strategies. The  pure-$alpha$ portfolios constructed based on nonzero pricing errors are associated  with annualized Sharpe ratios typically above 3 (as is common in the literature, we refer to these as ``arbitrage'' portfolios, even though their returns are far from riskless). Meanwhile, the  mean-variance efficient (MVE) portfolios constructed from the corresponding factors deliver substantially lower Sharpe ratios. This is different from the case in IPCA, where the Sharpe ratios of pure-\textit{alpha} and MVE factor portfolios are comparable. Moreover, we approximate the stock market's MVE portfolio with the combined MVE portfolios of the pure-\textit{alpha} portfolios and factors. Regressed-PCA consistently yields higher Sharpe ratios than IPCA, highlighting the advantages of our method.  The higher Sharpe ratios in regressed-PCA primarily stem from the pure-\textit{alpha} portfolios, in contrast to the case in IPCA, where both the pure-\textit{alpha} portfolios and factors contribute comparably.  Furthermore, we document that the nonlinear specifications consistently produce MVE factor and combined MVE portfolios with higher Sharpe ratios than the linear specification, underscoring the significance of incorporating nonlinearity. Our results indicate that low-dimensional factors are unlikely to span the conditional efficient frontier. At the same time they demonstrate that imposing a factor structure on the conditional covariance matrix of returns yields robust estimates of the stochastic discount factor, as evidenced by the high Sharpe ratios of the out-of-sample MVE portfolios that we obtain using our approach.

In order to further validate our factors, we evaluate their performance in standard asset pricing tests. Our empirical results demonstrate that the regressed-PCA factors consistently outperform IPCA's factors in pricing a large set of testing portfolios, as evidenced by smaller pricing errors, $t$-statistics, and $GRS$ statistics. IPCA factors' inferior performance primarily arises from its much larger regression $R^2$'s, which indicates that these factors capture more time-series variation in returns well but less cross-sectional variation. Moreover, our factors from  the nonlinear specifications also outperform \cite{FamaFrench_FiveFactor_2015}'s factors, which justifies the advantages of regressed-PCA over the traditional portfolio-sorting approach.


Our paper contributes to several strands of the literature. A number of studies have estimated models similar to \eqref{Eqn: Model} under the assumption that $z_{it}$ are time-invariant, at least within subsamples. These include \citet{ConnorLinton_Semparametric_2007}, \citet{Connoretal_EfficientFFFactor_2012}, \citet{Fanetal_ProjectedPCA_2016}, \citet{Kimetal_Arbitrage_2019}, \citet{LiGeLinton_Dynamic_2020}, and \citet{Fanetal_StructuralDeep_2022}.
 \citet{GagliardiniMa_Extracting_2019} and \citet{Guetal_Autoencoder_2021} explore conditional latent factor models that impose the absence of arbitrage, i.e., $\alpha(\cdot)=0$. There are numerous studies of conditional models with observed factors; see \citet{Gagliardinietal_Timevarying_2016} and \citet{Gagliardinietal_EstimationConditionalFactor_2019} for a comprehensive review. Another strand of literature studies time-varying factor models in which factor loadings evolve smoothly as functions of $t/T$ or aggregate variables;\footnote{The broader literature on conditional models with observable factors has extensively explored time-varying factor loadings that depend on aggregate variables rather than firm-specific characteristics. For example, \citet{FersonHarvey_conditioning_1999} use a linear specification, while \citet{Roussanov_Composition_2014} employs nonparametric kernel-based approaches. Building on our methodology, \citet{Chen_UnifiedFramework_2022} extends the estimation of conditional latent factor models to include heterogeneous \textit{alpha} and \textit{beta} functions, accommodating aggregate variables within $z_{it}$.} see, for example, \citet{Mottaetal_LocallyStationary_2011}, \citet{SuWang_TimeVarying_2017}, and \citet{PelgerXiong_State-varying_2019}.

The literature on the cross-section of asset returns is vast; here we focus on multi-factor models motivated by the arbitrage pricing theory. Empirical analysis that exploits the ability of characteristics to predict asset returns typically follows either the portfolio-sorting approach \citep{FamaFrench_Commonrisk_1993,FamaFrench_FiveFactor_2015,DanielTitman_Characteristics_1997} or the characteristic-based approach \citep{RosenbergMcKibben_ThePrediction_1973, Jacobs/Levy:88, Lewellen_Crosssection_2015, Greenetal_Characteristics_2017,Freybergeretal_Dissecting_2017, Kirby_FirmChar_2020, GiglioXiu_Asset_2019,KozakNagel_Whydospan_2022}.  The significance of nonlinear relationships in asset pricing has been underscored by several empirical studies \citep{Connoretal_EfficientFFFactor_2012, Kirby_FirmChar_2020} and more recently explored through machine learning methods \citep{Guetal_Autoencoder_2021, Chenetal_DeepLearning_2020}.

The remainder of the paper is organized as follows. Section \ref{Sec: 3} introduces the estimation method\textemdash regressed-PCA\textemdash along with its key properties and advantages. Section \ref{SubSec:AssetPricing} interprets the method in the context of asset pricing. Section \ref{Sec: 4} establishes large-sample properties of the estimators and develops bootstrap inference procedures. Section \ref{Sec: 8} applies the new methodology to analyze the cross-section of individual stock returns in the U.S. market. Finally, Section \ref{Sec: 9} briefly concludes. The Online Appendix includes estimators for the number of factors, assumptions, proofs of theoretical results, additional discussions, simulation results, and additional empirical findings.

\section{Estimation Method}\label{Sec: 3}
In this section, we introduce a method for estimating the model in \eqref{Eqn: Model}, which we term regressed principal component analysis or regressed-PCA, along with its key properties and advantages.

To illustrate the underlying idea of our regressed-PCA approach, we begin by assuming that $\alpha(\cdot)$ is null and $\beta(\cdot)$ is linear, i.e., $\alpha(\cdot)=0$ and $\beta(z_{it}) = \Gamma^{\prime}z_{it}$ for some $M\times K$ matrix $\Gamma$. Let $Y_t \equiv (y_{1t},\ldots, y_{Nt})^{\prime}$, $Z_t \equiv (z_{1t},\ldots, z_{Nt})^{\prime}$, and $\varepsilon_t \equiv (\varepsilon_{1t},\ldots, \varepsilon_{Nt})^{\prime}$. The model in \eqref{Eqn: Model} can then be written in matrix form as:
\begin{align}\label{Eqn: Model: Vector}
Y_t = Z_t\Gamma f_{t} + \varepsilon_{t}.
\end{align}
A key challenge in applying PCA to estimate $\Gamma$ and $f_t$ is the presence of $Z_t$ in the first term on the right-hand side of \eqref{Eqn: Model: Vector}. To address this, we first regress $Y_t$ on $Z_t$, yielding:
\begin{align}\label{Eqn: Model: Vector: Regressed}
(Z_t^{\prime}Z_t)^{-1}Z_t^{\prime}Y_t = \Gamma f_{t} + (Z_t^{\prime}Z_t)^{-1}Z_t^{\prime}\varepsilon_{t}.
\end{align}
Heuristically, variation in the common component $Z_t\Gamma f_{t}$ over $t$ comes from two sources: $Z_t$ and $f_t$, and regressing $Y_t$ on $Z_t$ disentangles these sources by isolating $Z_t$ from the common component. Given the factor structure on the right-hand side of \eqref{Eqn: Model: Vector: Regressed}, we can apply PCA to the series $\{(Z_t^{\prime}Z_t)^{-1}Z_t^{\prime}Y_t\}_{t\leq T}$ to obtain estimators for $\Gamma$ and $f_t$.

Alternatively, the  model in \eqref{Eqn: Model: Vector} can be viewed as a panel data model with time-varying slope coefficients $\Gamma f_{t} $, which exhibit a factor structure. Essentially, regressed-PCA first estimates the time-varying slope coefficients by period-by-period cross-sectional regressions and then exploits the underlying factor structure by using PCA.

\subsection{Regressed-PCA}
Now, we consider the general case where $\alpha(\cdot)$ is nonzero and show how to estimate $\alpha(\cdot)$ and $\beta(\cdot) = (\beta_{1}(\cdot),\ldots,\beta_{K}(\cdot))^{\prime}$ nonparametrically. To avoid the curse of dimensionality when $z_{it}$ is multivariate, we assume $\alpha(\cdot)$ and $\beta_k(\cdot)$ are separable. Specifically, we assume there exist functions $\{\alpha_{m}(\cdot)\}_{m\leq M}$ and $\{\beta_{km}(\cdot)\}_{m\leq M}$ such that:
\begin{align}\label{Eqn: Sep}
\alpha(z_{it}) = \sum_{m=1}^{M}\alpha_{m}(z_{it,m}) \text{ and } \beta_{k}(z_{it}) = \sum_{m=1}^{M}\beta_{km}(z_{it,m}),
\end{align}
where $z_{it,m}$ is the $m$th entry of $z_{it}$. We adopt the sieve method to estimate $\alpha_{m}(\cdot)$ and $\beta_{km}(\cdot)$. Let $\{\phi_{j}(\cdot)\}_{j\geq 1}$ be a set of basis functions (e.g., B-splines, Fourier series, polynomials) that span a dense linear space of the functional space for $\alpha_{m}(\cdot)$ and $\beta_{km}(\cdot)$. Then, we can express:
\begin{align}\label{Eqn: Sieve}
\alpha_{m}(z_{it,m}) &= \sum_{j=1}^{J}a_{m,j} \phi_{j}(z_{it,m}) + r_{m,J}(z_{it,m}),\\
\beta_{km}(z_{it,m}) &= \sum_{j=1}^{J}b_{km,j} \phi_{j}(z_{it,m}) + \delta_{km,J}(z_{it,m}).
\end{align}
Here, $\{a_{m,j}\}_{j\leq J}$ and $\{b_{km,j}\}_{j\leq J}$ are the sieve coefficients; $r_{m,J}(\cdot)$ and $\delta_{km,J}(\cdot)$ are ``remaining functions'' representing the approximation errors; $J$ denotes the sieve size.\footnote{For notational simplicity, we use the same basis functions in \eqref{Eqn: Sieve} for different $m$'s and the same sieve size. Our results remain valid if different basis functions and different sieve sizes are used for each $m$.} The basic assumption for the sieve method is that $\sup_{z}|r_{m,J}(z)|\to 0$ and $\sup_{z}|\delta_{km,J}(z)|\to 0$ as $J\to\infty$. Let $\bar{\phi}(z_{it,m}) \equiv (\phi_{1}(z_{it,m}),\ldots,\phi_{J}(z_{it,m}))^{\prime}$, ${\phi}(z_{it}) \equiv (\bar{\phi}(z_{it,1})^{\prime},\ldots, \bar{\phi}(z_{it,M})^{\prime})^{\prime}$, and define the vectors $a\equiv (a_{1,1},\ldots,a_{1,J},\ldots, a_{M,1},\ldots,a_{M,J})^{\prime}$,  $b_{k}\equiv (b_{k1,1},\ldots,b_{k1,J},\ldots, $ $b_{kM,1},\ldots,b_{kM,J})^{\prime}$, and the matrix $B\equiv(b_1,\cdots, b_{K})$. Let $r(z_{it}) \equiv  \sum_{m=1}^{M}r_{m,J}(z_{it,m})$ and $\delta(z_{it}) \equiv (\sum_{m=1}^{M}\delta_{1m,J}(z_{it,m}),\ldots, \sum_{m=1}^{M}\delta_{Km,J}(z_{it,m}))^{\prime}$. Thus, we have:
\begin{align}\label{Eqn: Sieve: together}
\alpha(z_{it}) = a^{\prime}\phi(z_{it}) + r(z_{it}) \text{ and } \beta(z_{it}) = B^{\prime}\phi(z_{it}) + \delta(z_{it}).
\end{align}
This shows that $\alpha(z_{it})$ and $\beta(z_{it})$ can be approximated by $a^{\prime}\phi(z_{it})$ and $B^{\prime}\phi(z_{it})$, respectively, and estimating $\alpha(\cdot)$ and $\beta(\cdot)$ reduces to estimating $a$ and $B$.

Next, we adapt the regressed-PCA to estimate $a$, $B$, and $f_t$ using the sieve approximation in \eqref{Eqn: Sieve: together}. Let $\Phi(Z_t) \equiv ({\phi}(z_{1t}),\ldots, {\phi}(z_{Nt}))^{\prime}$, $R(Z_t)\equiv (r(z_{1t}),\ldots,r(z_{Nt}))^{\prime}$, and $\Delta(Z_{t}) \equiv (\delta(z_{1t}),\ldots, \delta(z_{Nt}))^{\prime}$.  Using the sieve approximation, we write the model in \eqref{Eqn: Model} in matrix form as:
\begin{align}\label{Eqn: Model: Sieve: Vectors}
Y_t = \Phi(Z_t)a + \Phi(Z_t)Bf_{t} + R(Z_t) + \Delta(Z_{t})f_t + \varepsilon_{t}.
\end{align}
Under the basic sieve assumption, the term $ R(Z_t) + \Delta(Z_{t})f_t$ is negligible. The main challenge in applying PCA to estimate $a$, $B$, and $f_t$ lies in the presence of $\Phi(Z_t)$ in the first two terms on the right-hand side of \eqref{Eqn: Model: Sieve: Vectors}. To address this, we regress $Y_t$ on $\Phi(Z_t)$, yielding:
\begin{align}\label{Eqn: Model: Sieve: Vectors: Regressed}
\tilde{Y}_{t} = a+ Bf_{t} +(\Phi(Z_t)^{\prime}\Phi(Z_t))^{-1}\Phi(Z_t)^{\prime}(R(Z_t) + {\Delta}(Z_{t})f_t + {\varepsilon}_{t}),
\end{align}
where $\tilde{Y}_{t} = (\Phi(Z_t)^{\prime}\Phi(Z_t))^{-1}\Phi(Z_t)^{\prime}Y_t$. Thus, we estimate $a$, $B$, and $f_t$ as follows. First, since $\tilde{Y}_{t}\approx a +Bf_t$, we remove $a$ by subtracting $\bar{\tilde{Y}} = \sum_{t=1}^{T}\tilde{Y}_{t}/T$ from $\tilde{Y}_{t}$ and estimate $B$ by applying PCA to the demeaned series $\{\tilde{Y}_{t}-\bar{\tilde{Y}}\}_{t\leq T}$. Second, for identifying $a$ (and thus $\alpha(\cdot)$), we impose the condition $a^{\prime}B = 0$. Since $\bar{\tilde{Y}}\approx a +B\bar{f}$ (with $\bar{f}=\sum_{t=1}^{T}f_t/T$), we estimate $a$ as $a\approx [I_{JM} - B(B^{\prime}B)^{-1}B]\bar{\tilde{Y}}$. Finally, $f_t$ is estimated as $f_t\approx (B^{\prime}B)^{-1}B^{\prime}\tilde{Y}_{t}$.

The formal estimators for $a$, $B$, $\alpha(\cdot)$, $\beta(\cdot)$, and $F=(f_1,\ldots,f_{T})^{\prime}$ are defined as follows. Let $\hat{a}$, $\hat{B}$, $\hat{\alpha}(\cdot)$, $\hat{\beta}(\cdot)$, and $\hat{F}$ denote the respective estimators. Let $\tilde{Y} \equiv (\tilde{Y}_1,\ldots, \tilde{Y}_T)$ and $M_T\equiv I_{T} - 1_T1_T^{\prime}/T$, where $1_{T}$ is a $T\times 1$ vector of ones. Using the normalization $B^{\prime}B=I_{K}$ and $F^{\prime}M_T F/T$ being diagonal with descending diagonal entries, the columns of $\hat{B}$ are the eigenvectors corresponding to the largest $K$ eigenvalues of $\tilde{Y}M_T\tilde{Y}^{\prime}/T$. We then have: $\hat{a} = (I_{JM}-\hat{B}\hat{B}^{\prime})\bar{\tilde{Y}}$,
\begin{align}\label{Eqn: Estimators}
\hat{\alpha}(z) = \hat{a}^{\prime}\phi(z), \hat{\beta}(z) = \hat{B}^{\prime}\phi(z), \text{ and }\hat{F} = (\hat{f}_1,\ldots,\hat{f}_T)^{\prime} = \tilde{Y}^{\prime}\hat{B}.
\end{align}
We assume that the number of factors, $K$, is fixed and known. In Section \ref{Sec: 4}, we establish asymptotic properties of the estimators and develop inference methods. In Appendix \ref{Sec: 6}, we propose two consistent estimators for $K$, ensuring that our results extend to the case of an unknown $K$ by using a conditioning argument.

\subsection{Key Properties}
Our regressed-PCA method has several appealing properties and is straightforward to implement. First, as discussed in Section \ref{Sec: 41}, it accommodates time-varying $z_{it}$ and does not require a large time dimension $T$. This flexibility allows us to examine the evolving relationship between risk and return using both full-sample and sub-sample analyses.

Second, the estimation procedure is well-suited for unbalanced panels, which is particularly relevant in cross-sectional asset pricing applications. The key step of regressed-PCA is to compute $\tilde{Y}_t$. Specifically, we can express $\tilde{Y}_t$ as:
\begin{align}
\tilde{Y}_t=\left(\sum_{i=1}^{N}\phi(z_{it})\phi(z_{it})^{\prime}\right)^{-1}\sum_{i=1}^{N}\phi(z_{it})y_{it}.
\end{align}
For unbalanced panels, $\tilde{Y}_t$ can still be calculated by summing only over the $i$'s for which both $z_{it}$ and $y_{it}$ are observed at time period $t$. This approach is equivalent to treating missing data as zeros, allowing us to proceed as if working with a balanced panel. The asymptotic results we derive remain valid as long as $\min_{t\leq T} N_t \to\infty$, where $N_t$ is the sample size at time period $t$.

Third, our method continues to be effective even when the pricing errors and risk exposures are not fully explained by $z_{it}$. Let $e_{\alpha,it}$ and $e_{\beta,it}$ be the error terms in the pricing errors and the risk exposures, respectively, which are orthogonal to $z_{it}$. In this case, the model becomes:
\begin{align}
y_{it} = [\alpha(z_{it})+e_{\alpha,it}] + [\beta(z_{it})+e_{\beta,it}]^{\prime}f_{t} + \varepsilon_{it} = \alpha(z_{it}) + \beta(z_{it})^{\prime}f_{t} + \varepsilon^{\ast}_{it},
\end{align}
where $\varepsilon^{\ast}_{it} = \varepsilon_{it} + e_{\alpha,it} + e_{\beta,it}^{\prime}f_{t}$. Since we are not interested in estimating $e_{\alpha,it}$ and $e_{\beta,it}$, our asymptotic results remain valid if we replace $\varepsilon_{it}$ in the original model with $\varepsilon^{\ast}_{it}$.

Finally, efficiency of our estimation procedure could be improved by using generalized least squares in its first step. Our asymptotic results continue to hold if we replace $\Phi(Z_t)$ and $\varepsilon_t$ with their transformed counterparts $V_{t}^{-1/2}\Phi(Z_t)$ and $V_{t}^{-1/2}\varepsilon_t$, where $V_t$ is the conditional covariance matrix of $Y_t$ at each time period $t$. While $V_t$ is usually unknown, \cite{Hoberg_Welch_Optimized_2009} suggest some practical guidance to account for the cross-correlation and heteroskedasticity of the idiosyncratic noise $\varepsilon_{it}$.

\subsection{Comparing Methods}\label{Subsubsec:Compare}
How does our regressed-PCA compare with existing methods that have been proposed in the literature? What are the advantages of our regressed-PCA?

First, the projected-PCA proposed by \citet{Fanetal_ProjectedPCA_2016} applies PCA to the series $\{\Phi(Z_t)(\Phi(Z_t)^{\prime}\Phi(Z_t))^{-1}\Phi(Z_t)^{\prime}Y_t\}_{t\leq T}$. In contrast, our regressed-PCA applies PCA to the series $\{(\Phi(Z_t)^{\prime}\Phi(Z_t))^{-1}\Phi(Z_t)^{\prime}Y_t\}_{t\leq T}$. The two methods are fundamentally different: regressed-PCA applies PCA to estimated coefficients, while projected-PCA focuses on fitted values. The regression step in regressed-PCA is designed to extract $Z_t$ from the common component for consistent estimation, whereas projected-PCA aims to remove noise in non-time-varying factor loadings to achieve efficiency in estimation. Consequently, projected-PCA may yield inconsistent estimates when $Z_t$ is time-varying.\footnote{For instance, as illustrated in \eqref{Eqn: Model: Vector}, we have $Z_t(Z_t^{\prime}Z_t)^{-1}Z_t^{\prime}Y_t \approx Z_t\Gamma f_{t}$, which does not conform to a factor structure unless $Z_t$ remains constant over $t$. Therefore, applying PCA to $\{Z_t(Z_t^{\prime}Z_t)^{-1}Z_t^{\prime}Y_t\}_{t=1}^{T}$ can result in failure to estimate $f_t$.} As noted by \citet{Fanetal_ProjectedPCASupp_2016} and further investigated by \citet{Chengetal_UniformPredictive_2020}, ensuring the consistency of projected-PCA may necessitate imposing smoothness conditions on how $Z_t$ varies with $t$, which our regressed-PCA does not require. Additionally, projected-PCA often requires dropping certain observations to maintain a balanced panel, whereas regressed-PCA is applicable to unbalanced panels.


While \citet{Kimetal_Arbitrage_2019} extend projected-PCA to accommodate nonzero $\alpha(\cdot)$, they do not develop an inference procedure. Similarly, \citet{Fanetal_StructuralDeep_2022} extend projected-PCA by employing deep neural networks and propose a local version to capture slowly changing alphas and betas. However, while neural networks can alleviate the curse of dimensionality for prediction tasks, they are not typically well-suited for inference\textemdash a key focus of our paper. Our regressed-PCA not only allows for nonzero $\alpha(\cdot)$ but also provides formal testing procedures, which are crucial for evaluating and comparing factor models. Moreover, by incorporating rapidly varying $Z_t$, our regressed-PCA is able to capture abrupt changes in both alphas and betas effectively.


Second, consider the least squares estimation approach introduced by \citet{Parketal_FactorDynamics_2009}, which is at the core of the IPCA of \citet{Kellyetal_Characteristics_2019}. The least squares method minimizes the following objective function:
\begin{align}\label{Eqn: objls}
	 \sum_{t=1}^{T}({Y}_t - \Phi(Z_t)a -\Phi(Z_t)Bf_{t})^{\prime}  ({Y}_t - \Phi(Z_t)a -\Phi(Z_t) Bf_{t}),
\end{align}
while regressed-PCA minimizes:\footnote{The equality in \eqref{Eqn: objregpca} follows because $\tilde{Y}_{t} = (\Phi(Z_t)^{\prime}\Phi(Z_t))^{-1}\Phi(Z_t)^{\prime}Y_t$.
}
\begin{align}\label{Eqn: objregpca}
&\sum_{t=1}^{T}(\tilde{Y}_t - a -Bf_{t})^{\prime}  (\tilde{Y}_t - a - Bf_{t})\notag\\
	=&\sum_{t=1}^{T}({Y}_t - \Phi(Z_t)a -\Phi(Z_t)Bf_{t})^{\prime} S_{t} ({Y}_t - \Phi(Z_t)a -\Phi(Z_t) Bf_{t}),
\end{align}
where $S_{t} = \Phi(Z_t)(\Phi(Z_t)'\Phi(Z_t))^{-1}(\Phi(Z_t)'\Phi(Z_t))^{-1}\Phi(Z_t)^{\prime}$. The objective functions of these two approaches differ, except when $\Phi(Z_t)'\Phi(Z_t)/N=I_{JM}$.\footnote{See Appendix \ref{App: Sec: C1} for discussion.}
Essentially, the least squares approach in \eqref{Eqn: objls} maximizes in-sample $R^2$, while regressed-PCA in \eqref{Eqn: objregpca} optimizes time-series comovement of $\tilde Y_{t}$, which captures the most cross-sectional variations of individual asset returns. A key challenge with the least squares approach is that its minimization problem is nonconvex and cannot be solved explicitly. While \citet{Parketal_FactorDynamics_2009} develop a numerical algorithm to find the estimators, and \citet{Kellyetal_Characteristics_2019} propose an alternating least squares procedure, both methods may require careful selection of initial values to ensure convergence to the correct solution. Furthermore, the asymptotic properties of these algorithms are not well understood. In addition to the asymptotic properties that we derive, our regressed-PCA provides estimators that can always be explicitly solved. Unlike IPCA, which relies on a long time series of returns, our regressed-PCA does not require a large time dimension $T$, enabling sub-sample analyses and capturing potential time variation in the coefficients ($a$ and $B$).


Another advantage of our regressed-PCA approach lies in factor construction. As the number of factors, $K$, increases by one, regressed-PCA simply constructs an additional factor on top of existing ones through PCA. In contrast, IPCA requires reconstructing all factors by recomputing alternating least squares,  making it sensitive to the number of factors. For example, the first factor under $K=2$ may differ significantly from the first factor under $K=1$, and the factor space under $K=2$ does not necessarily nest the factor space under $K=1$. Moreover, IPCA's factors can be correlated. In contrast, regressed-PCA produces factors that are uncorrelated by construction and remain stable, regardless of the number of factors chosen.

Overall, in addition to its formal inference procedures and well-established asymptotic properties, our regressed-PCA offers significant computational simplicity. It accommodates nonzero alphas, time-varying characteristics,  unbalanced panels, and short samples, making it particularly suitable for empirical asset pricing. Moreover, it provides stable and reliable factor construction.

\section{Asset Pricing Interpretation}\label{SubSec:AssetPricing}

Regressed-PCA has deep roots in asset pricing. In a typical asset pricing application, $y_{it}$ represents the realized returns on asset $i$ at the end of time period $t$, while $z_{it,m}$ represents the $m$'th attribute or characteristic of asset $i$ that is known at the {\em beginning} of time period $t$ (or, alternatively, at the ``end'' of time period $t-1$). The regressed-PCA first estimates the time-varying slope coefficients by period-by-period cross-sectional regressions of returns on (functions of) characteristics, and then exploits the factor structure by using PCA. These period-by-period cross-sectional regressions are known as Fama-MacBeth regressions \citep{FamaMacBeth_RiskReturn_1973},  which help transform a large unbalanced panel of noisy individual asset returns into a lower-dimensional balanced panel of portfolio returns that are largely free of idiosyncratic noise, $\tilde{Y}_{t}$. Furthermore, $\tilde{Y}_{t}$ can be interpreted as the time $t$ realization of returns on a set of $JM$ characteristic-managed portfolios, sometimes referred to as ``optimized portfolios,'' ``characteristic pure plays,'' or ``cross-section factors'' (e.g., as in \citet{Hoberg_Welch_Optimized_2009}, \cite{BackKapadiaOstdiek_Testing_2015}, and \cite{FamaFrench_CSTS_2020}). We also refer to them as Fama-MacBeth managed portfolios.

In particular, if the basis functions $\phi(z_{it})$ include a constant term (e.g., as the first element in $\phi(z_{it})$) and are standardized to have a zero mean in each cross-section, then the intercept in the Fama-MacBeth regressions (the first element in $\tilde{Y}_{t}$) represents a ``level'' return. This is essentially the equal-weighted average excess return across all individual assets, with weights summing up to unity or costing \$1, and it has no ex-ante loadings on any characteristics.\footnote{\label{Footport}This is due to two properties. First, the intercept in $\tilde{Y}_{t}$ is equal to $\sum_{i=1}^{N} y_{it}/N$ when the nonconstant regressors have a zero mean. Second, the weights in $\tilde{Y}_{t}$,  given by $W_t = (\Phi(Z_t)^{\prime}\Phi(Z_t))^{-1}\Phi(Z_t)^{\prime}$, satisfy the property that $W_t\Phi(Z_t) = I_{JM}$. The second property also clarifies the property of the weights for other portfolios in $\tilde{Y}_{t}$.} This is sometimes referred to as a ``naively diversified'' or $1/N$  portfolio. As shown by \cite{Fama_76}, the period-by-period slope coefficients corresponding to the time-varying basis functions (all elements of $\tilde{Y}_{t}$ starting from the second) are excess returns on zero-cost portfolios, as long as a constant term is included in the basis functions. These portfolios have weights on individual assets that set the weighted average value of the relevant basis function to one and those of all the remaining basis functions to zeros;\footnote{See Footnote \ref{Footport}.} see \cite{Hoberg_Welch_Optimized_2009} and \cite{Kirby_FirmChar_2020} for other attractive properties. In simpler terms, each portfolio has spread in only one basis function, or a pure play on a particular basis function. Moreover, adding one more basis function in the Fama-MacBeth regressions to a benchmark corresponds to introducing one additional zero-cost portfolio, which raises average exposure to that specific basis function by one unit while maintaining average exposures to other basis functions unchanged. Therefore, these portfolios can also be interpreted as ``slope'' returns with respect to the time-varying basis functions.

Moreover, \cite{Fama_76} demonstrates that the portfolios in $\tilde{Y}_{t}$ exhibit minimum variance and often low correlations under OLS-like i.i.d. assumptions, rendering them maximally diversified.\footnote{See Appendix \ref{App: Sec: C11} for discussion.} Due to their low noise and correlations, several studies have explored the advantages of these portfolios in asset pricing tests. For instance, \citet{Hoberg_Welch_Optimized_2009}, \cite{BackKapadiaOstdiek_Testing_2015}, and \cite{Kirby_FirmChar_2020} highlight their utility as test assets (dependent variables), while \cite{BackKapadiaOstdiek_Slope_2013} and \cite{FamaFrench_CSTS_2020} emphasize their effectiveness as pricing factors (independent variables), particularly when compared with the portfolio-sorting approach. However, these studies primarily  focus on linear specifications (i.e., $\phi(z_{it})= z_{it}$ which includes a constant term) with a small number of characteristics, and their findings may not generalize to more complex cases.

In the presence of a large number of characteristics or basis functions, the dimension of $\tilde{Y}_{t}$ can become substantial, and the portfolios in $\tilde{Y}_{t}$ may exhibit high correlations, even though each portfolio is maximally diversified. Understanding the factor structure of the portfolios in $\tilde{Y}_{t}$ as test assets is crucial to avoid spurious fit of misspecified asset pricing models that happen to be correlated with some of the latent factors, as emphasized by \citet{Lewellenetal_Skeptical_2010}. A straightforward solution is to apply PCA to extract uncorrelated principal components from $\tilde{Y}_{t}$, which is the essence of our approach. 

\subsection{Comparison with Portfolio Sorting} \label{SubSec:Comparefactor}
The extracted factors from regressed-PCA in (\ref{Eqn: Estimators}) are linear combinations of characteristic-managed portfolios $\tilde Y_{t}$, and thus are themselves tradable portfolios. As a way of factor construction, regressed-PCA shares a strong connection with the portfolio-sorting approach. 
Sorting assets into portfolios can be equivalently framed as running cross-sectional regressions of returns on dummies that represent groups sorted by characteristics at each time period, whether the sorting is independent or dependent. Specifically, the return of each sorted portfolio (i.e., the average return of individual assets within a group) is the coefficient of the corresponding group dummy in the regression. Using our notation, the returns of sorted portfolios can be represented by $\tilde{Y}_{t}$ when the basis functions $\phi(z_{it})$ are group dummies.\footnote{See Appendix \ref{App: Sec: C13} for discussion.} Thus, the key difference between the first steps of the regressed-PCA and portfolio-sorting approaches lies in their choice of basis functions. Sorting has been recognized as a nonparametric method for examining the relationship between average returns and characteristics, as highlighted by \citet{FamaFrench_Dissecting_2008}, \citet{Cochrane_Presidential_2011}, and \citet{CattaneoCrumpFarrellSchaumburg_2020_Characteristic}.

However, regressed-PCA offers several advantages over sorting. First, sorting quickly encounters the curse of dimensionality and rarely handles more than four characteristics simultaneously. When multiple characteristics are present, double sorting is typically used for each pair of characteristics, making it difficult to infer which characteristics uniquely affect average returns. Regressed-PCA addresses this limitation by employing a separable additive specification (see \eqref{Eqn: Sep}) and B-splines basis functions, which allow for a large number of characteristics and facilitate testing their significance. Second, sorting fails to fully exploit the variation in characteristics within each sorted group. In contrast, regressed-PCA takes advantage of the full variation in characteristics. Third, sorting struggles to effectively explore the nonlinear relationship between average returns and characteristics: sorting essentially uses step functions, which suffer from several well-known shortcomings, such as discontinuities at cutoffs, poor extrapolation, and unstable estimates that are highly sensitive to outlier assets \citep{HastieTibshiraniFriedman_StatisticalLearning_2011}. These issues are mitigated by the use of B-splines, which our regressed-PCA incorporates. 


Although long-short factors (such as high-minus-low and small-minus-big factors) are straightforward to interpret, regressed-PCA offers several advantages over them. First, the long-short approach focuses only on portfolios in extreme groups and ignores those in the middle. As a result, it may fail when factor loadings exhibit non-monotonicity with respect to characteristics, such as a ``tent'' shape. Regressed-PCA, on the other hand, uses all portfolios formed by Fama-MacBeth sieve regressions, making it more adaptable to capturing underlying nonlinearities. Second, long-short factors do not effectively distinguish between the risk and mispricing explanations of the role of characteristics in predicting asset returns, a distinction at the heart of the ``characteristics versus covariances'' debate. Regressed-PCA is based on a latent factor model for individual asset returns, which is ideally suited to resolve this debate. Third, when multiple characteristics are present, long-short methods can quickly lead to the ``\textit{factor zoo}'' problem as distentangling the roles of correlated characteristics can be difficult in multi-way sorts with potentially too few securities in each bin. Regressed-PCA mitigates this issue by utilizing Fama-Macbeth managed portfolios, which are maximally diversified ``pure plays'' on characteristics, together with PCA, which has also been proven effective in reducing the dimensionality of sorted portfolios \citep{Kozaketal_Interpreting_2018,Kozaketal_Shrinking_2020,LettauPelger_Factorstimeseries_2020}.

\section{Econometric Analysis}\label{Sec: 4}
In this section, we establish the asymptotic properties of our estimators, including consistency, the rate of convergence, and their asymptotic distribution. Additionally, we develop bootstrap inference procedures. We begin by defining some notation that will be used throughout the paper. For a symmetric matrix $A$, we denote its $k$th largest eigenvalue by $\lambda_{k}(A)$, and its smallest and largest eigenvalues by $\lambda_{\min}(A)$ and $\lambda_{\max}(A)$, respectively. The operator norm of a matrix $A$ is denoted by $\|A\|_2$, and its Frobenius norm by $\|A\|_{F}$. The vectorization of $A$ is written as $\mathrm{vec}(A)$. The Euclidian norm of a column vector $x$ is denoted by $\|x\|$. Finally, for matrices $A$ and $B$, we use $A\otimes B$ to denote their Kronecker product.

\subsection{Asymptotic Properties}\label{Sec: 41}
Before presenting formal theorems, we revisit \eqref{Eqn: Model: Vector} to briefly illustrate why a large $T$ is not required and $Z_t$ can be nonstationary over $t$. Consider the case when $T\geq K+1$ and $M\geq K$. Since the columns of $\hat{B}$ and $\Gamma$ are the eigenvectors of $\tilde{Y}M_{T}\tilde{Y}^{\prime}$ and $\Gamma F^{\prime} M_{T} F \Gamma^{\prime}$, respectively, corresponding to the first $K$ largest eigenvalues, by the matrix perturbation theorem (see, for example, \citet{Yuetal_Useful_2014}), the consistency of $\hat{B}$ to $\Gamma$ (up to a rotational transformation) can be established if we can show that:
\begin{align}\label{Eqn: IllFiniteT1}
\|\tilde{Y}M_{T} - \Gamma F^{\prime} M_{T}\|_{F}  = o_{p}(1) \text{ as } N\to\infty.
\end{align}
Since $\tilde{Y}= \Gamma F^{\prime} + ((Z_1^{\prime}Z_1)^{-1}Z_1^{\prime}\varepsilon_1,\ldots, (Z_T^{\prime}Z_T)^{-1}Z_T^{\prime}\varepsilon_T)$, \eqref{Eqn: IllFiniteT1} simplifies to:
\begin{align}\label{Eqn: IllFiniteT2}
\|((Z_{1}^{\prime}Z_1)^{-1}Z_{1}^{\prime}\varepsilon_1, \ldots, (Z_{T}^{\prime}Z_T)^{-1}Z_{T}^{\prime}\varepsilon_T)M_{T}\|_{F}= o_{p}(1) \text{ as } N\to\infty.
\end{align}
When $T$ is fixed, \eqref{Eqn: IllFiniteT2} is equivalent to $(Z_{t}^{\prime}Z_t)^{-1}Z_{t}^{\prime}\varepsilon_t = o_{p}(1)$ for each $t$. Therefore, only regularity conditions on $Z_t$ and $\varepsilon_t$ for each $t$ are needed to apply the law of large numbers. This result also implies that $Z_t$ can vary over $t$ in a nonstationary fashion.

Let $H \equiv (F^{\prime}M_T\hat{F})(\hat{F}^{\prime}M_T\hat{F})^{-1}$, which represents a rotational transformation matrix that governs the convergence limit of $\hat{B}$, $\hat{F}$, and $\hat{\beta}(\cdot)$. Define $\xi_{J}\equiv \sup_{z}\|\bar{\phi}(z)\|$, which scales as $O(\sqrt{J})$ for B-splines and Fourier series, and $O(J)$ for polynomials (see, for example, \citet{BelloniChernozhukovChetverikovKato_SeriesEstimator_2015}).

\begin{thm}\label{Thm: ImprovedRates}
Suppose Assumptions \ref{Ass: Basis}-\ref{Ass: Improvedrates} hold. Let $\hat{a}$, $\hat{B},\hat{F}$, $\hat{\alpha}(\cdot)$, and $\hat{\beta}(\cdot)$ be given in \eqref{Eqn: Estimators}. Assume (i) $N\to\infty$; (ii) $T\geq K+1$ ($T$ may stay fixed or grow simultaneously with $N$); (iii) $J\to\infty$ with $J^{2}\xi^{2}_{J}\log J=o(N)$. Then
\begin{align*}
\|\hat{a} - a\|^{2}&=O_{p}\left(\frac{1}{J^{2\kappa}}+\frac{J}{N^2}+\frac{{J}}{{NT}}\right),\notag\\
\|\hat{B} - B H\|^{2}_{F}&=O_{p}\left(\frac{1}{J^{2\kappa}}+\frac{J}{N^2}+\frac{{J}}{{NT}}\right),\notag\\
\frac{1}{T}\|\hat{F}-F(H^{\prime})^{-1}\|_{F}^{2}&=O_{p}\left(\frac{1}{J^{2\kappa}}+\frac{1}{{N}}\right),\notag\\
\sup_{z}|\hat{\alpha}(z)-\alpha(z)|^{2}&=O_{p}\left(\frac{1}{J^{2\kappa-1}}+\frac{J^2}{N^2}+\frac{{J^2}}{{NT}}\right)\max_{j\leq J}\sup_{z}|\phi_{j}(z)|^{2},\notag\\
\sup_{z}\|\hat{\beta}(z)-H^{\prime}\beta(z)\|^{2}&=O_{p}\left(\frac{1}{J^{2\kappa-1}}+\frac{J^2}{N^2}+\frac{{J^2}}{{NT}}\right)\max_{j\leq J}\sup_{z}|\phi_{j}(z)|^{2},
\end{align*}
where $\kappa>1/2$ is a constant representing the smoothness of $\alpha(\cdot)$ and $\beta(\cdot)$.
\end{thm}

All the assumptions are outlined in Appendix \ref{App: assumptions}. Theorem \ref{Thm: ImprovedRates} establishes that $a$ and $\alpha(\cdot)$ can be consistently estimated by $\hat{a}$ and $\hat\alpha(\cdot)$, while $B$, $F$, and $\beta(\cdot)$ can be consistently estimated by $\hat{B}$, $\hat{F}$, and $\hat\beta(\cdot)$, respectively, up to a rotational transformation. This consistency holds as long as $J\to\infty$ under both large $N$ and either fixed or large $T$. Notably, the large $J$ requirement differs from \citet{Fanetal_ProjectedPCA_2016}. This distinction highlights the importance of controlling sieve approximation errors in $\alpha(\cdot)$ and $\beta(\cdot)$ to ensure consistent estimation of $F$. Appendix \ref{App: Sec: C12} illustrates how misspecifications in $\alpha(\cdot)$ and $\beta(\cdot)$ can result in inconsistent estimation of $F$, motivating the need for a specification test, which is addressed in Section \ref{Sec: 52}.

Additionally, Theorem \ref{Thm: ImprovedRates} shows that the estimators achieve fast convergence rates. In particular, $\hat{F}$ attains the optimal rate $1/N$ when $\alpha(\cdot)$ and $\beta(\cdot)$ are sufficiently smooth (i.e., $\kappa$ is sufficiently large).  
This implies that the nonparametric modelling of $\alpha(\cdot)$ and $\beta(\cdot)$ do not degrade the convergence rate for estimating $F$, as long as the smoothness conditions are satisfied. This result is crucial for developing the specification test for $\alpha(\cdot)$ and $\beta(\cdot)$ in Section \ref{Sec: 52}, and also important for utilizing  $\hat{F}$ in subsequential asset pricing tests of Section \ref{Sec: 82}. Notably, if the functional forms of $\alpha(\cdot)$ and $\beta(\cdot)$ are known and correctly specified, sieve approximation errors can be avoided, and the asymptotic results hold for a fixed $J$. Theorem \ref{Thm: ImprovedRates} allows for weak dependence of the errors $\{\varepsilon_{it}\}_{i\leq N,t\leq T}$ over both $i$ and $t$, which is relevant in asset pricing.

Let $\Omega\equiv\sum_{i=1}^{N}\sum_{t=1}^{T}\sum_{s=1}^{T}f_{t}^{\dag}f_{s}^{\dag\prime}Q_{t}^{-1}E[\phi(z_{it})\phi(z_{is})^{\prime}]$ $\times Q_{s}^{-1}E[\varepsilon_{it}\varepsilon_{is}]/NT$, where $f^{\dag}_t=(1,(f_t-\bar{f})^{\prime})^{\prime}$ and $Q_{t} = \sum_{i=1}^{N}E[\phi(z_{it})\phi(z_{it})^{\prime}]/N$. This defines a variance-covariance matrix, which appears in the asymptotic distributions of $\hat{a}$ and $\hat{B}$.

\begin{thm}\label{Thm: AsymDis}
Suppose Assumptions \ref{Ass: Basis}-\ref{Ass: Asym} hold. Let $\hat{a}$ and $\hat{B}$ be given in \eqref{Eqn: Estimators}. Assume (i) $N\to\infty$; (ii) $T\geq K+1$; (iii) $J\to\infty$ with $J^{2}\xi^{2}_{J}\log J=o(N)$. Then there is a $JM\times (K+1)$ random matrix $\mathbb{N}$ with $\mathrm{vec}(\mathbb{N})\sim N(0,\Omega)$ such that:
\begin{align*}
\|\sqrt{{NT}}(\hat{a} - a)-\mathbb{G}_{a}\| = O_{p}\left(\frac{\sqrt{NT}}{J^{\kappa}}+\frac{\sqrt{TJ}}{\sqrt{N}}+\frac{J^{5/6}}{N^{1/6}}+\frac{\sqrt{J\xi_{J}}\log^{1/4}J}{N^{1/4}}\right)
\end{align*}
and
\begin{align*}
\|\sqrt{{NT}}(\hat{B} - B H)-\mathbb{G}_{B}\|_{F} = O_{p}\left(\frac{\sqrt{NT}}{J^{\kappa}}+\frac{\sqrt{TJ}}{\sqrt{N}}+\frac{J^{5/6}}{N^{1/6}}+\frac{\sqrt{J\xi_{J}}\log^{1/4}J}{N^{1/4}}\right),
\end{align*}
where $\kappa>1/2$ is a constant representing the smoothness of $\alpha(\cdot)$ and $\beta(\cdot)$, $\mathbb{G}_{a} =(I_{JM}-B\mathcal{H}\mathcal{H}^{\prime}B^{\prime})(\mathbb{N}_1-\mathbb{G}_{B}\mathcal{H}^{-1}\bar{f})-B\mathcal{H}\mathbb{G}_{B}^{\prime}a$, and $\mathbb{G}_{B} = \mathbb{N}_2 B^{\prime}B\mathcal{M}$. The matrices $\mathcal{H}$ and $\mathcal{M}$ are nonrandom, as given in Lemma \ref{Lem: TechC3}, while $\mathbb{N}_1$ and $\mathbb{N}_2$ are the first column and the last $K$ columns of $\mathbb{N}$, respectively.
\end{thm}

Theorem \ref{Thm: AsymDis} establishes a strong approximation, demonstrating that $(\sqrt{{NT}}(\hat{a} - a),\sqrt{{NT}}(\hat{B} - B H))$ can be well approximated by a normal random matrix $(\mathbb{G}_{a},\mathbb{G}_{B})$. Specifically, the difference between them converges in probability to zero under the conditions $T=o(N)$, $NT J^{-2\kappa}=o(1)$, and  $J=o(\min\{N^{1/5},N/T\})$. Since the dimensions of $\sqrt{{NT}}(\hat{a} - a)$ and $\sqrt{{NT}}(\hat{B} - B H)$ grow with $N$, rendering the classical central limit theorem inapplicable, we employ Yurinskii’s coupling to establish this strong approximation. This approach accommodates weak temporal dependence in the errors $\{\varepsilon_{it}\}_{i\leq N,t\leq T}$. Furthermore, the result can be readily extended to allow for cluster-type dependence across $i$ in the errors $\{\varepsilon_{it}\}_{i\leq N,t\leq T}$; see the discussion following Assumption \ref{Ass: Asym}. Notably, distributional results of this kind are not provided in \citet{Fanetal_ProjectedPCA_2016}.



\subsection{Weighted Bootstrap}\label{Sec: 51}

We develop a weighted bootstrap approach to estimating the distribution of $(\mathbb{G}_{a},\mathbb{G}_{B})$. Let $\{w_i\}_{i\leq N}$ be a sequence of i.i.d. positive random variables, with $E[w_i]=1$ and $var(w_i)=\omega_0>0$. For instance, the $w_i$'s can be drawn from a standard exponential distribution, where $\omega_0=1$. To preserve the time dependence, we assign the same weight $w_i$ to all observations over $t$. Define $\Phi(Z_t)^{\ast}\equiv ({\phi}(z_{1t})w_1,\ldots, {\phi}(z_{Nt})w_N)^{\prime}$ and $\tilde{Y}_{t}^{\ast}\equiv (\Phi(Z_t)^{\ast\prime}\Phi(Z_t))^{-1}\Phi(Z_t)^{\ast\prime}Y_t$, which is the bootstrap version of $\tilde{Y}_t$. To define the bootstrap estimators of $a$ and $B$, let $\tilde{Y}^{\ast} \equiv (\tilde{Y}^{\ast}_1,\ldots, \tilde{Y}^{\ast}_T)$ and $\bar{\tilde{Y}}^{\ast}\equiv \sum_{t=1}^{T}\tilde{Y}^{\ast}_t/T$. The bootstrap estimators are given by:
\begin{align}\label{Eqn: BootstrappedEstimators}
\hat{B}^{\ast} = \tilde{Y}^{\ast}M_T\hat{F}(\hat{F}^{\prime}M_T\hat{F})^{-1} \text{ and } \hat{a}^{\ast} = (I_{JM} - \hat{B}^{\ast}(\hat{B}^{\ast\prime}\hat{B}^{\ast})^{-1}\hat{B}^{\ast\prime})\bar{\tilde{Y}}^{\ast},
\end{align}
which mimic the original estimators: $\hat{B} = \tilde{Y}M_T\hat{F}(\hat{F}^{\prime}M_T\hat{F})^{-1}$ and $\hat{a} = (I_{JM}-\hat{B}\hat{B}^{\prime})\bar{\tilde{Y}} =(I_{JM}-\hat{B}(\hat{B}^{\prime}\hat{B})^{-1}\hat{B}^{\prime})\bar{\tilde{Y}} $. We propose estimating the distribution of $(\mathbb{G}_{a},\mathbb{G}_{B})$ by the distribution of $(\sqrt{{NT/\omega_0}}$ $(\hat{a}^{\ast}-\hat{a}),\sqrt{{NT/\omega_0}}(\hat{B}^{\ast}-\hat{B}))$ conditional on the data.\footnote{A more natural bootstrap estimator for $B$ is given by the eigenvectors of $\tilde{Y}^{\ast}M_T\tilde{Y}^{\ast\prime}/T$ corresponding to its first $K$ largest eigenvalues. However, the approach generally fails due to rational transformation matrices; see Appendix \ref{App: Sec: C2} for discussion.}

The bootstrap procedure can be easily adapted for unbalanced panels. The key step is obtaining $\tilde{Y}^{\ast}_t$. For balanced panels, we write:
\begin{align}
\tilde{Y}^{\ast}_t= \left(\sum_{i=1}^{N}\phi(z_{it})\phi(z_{it})^{\prime}w_i\right)^{-1}\sum_{i=1}^{N}\phi(z_{it})y_{it}w_i.
\end{align}
In unbalanced panels, we adjust by taking the sums over the $i$'s for which both $z_{it}$ and $y_{it}$ are observed at time period $t$. This effectively replaces missing data with zeros, making the procedure identical to the balanced case. The asymptotic results established below remain valid as long as as $\min_{t\leq T} N_t \to\infty$, where $N_t$ represents the sample size at time period $t$. Moreover, the bootstrap can easily accommodate cluster-type dependence across $i$ by assigning the same weight within each cluster.

\begin{thm}\label{Thm: Boot}
Suppose Assumptions \ref{Ass: Basis}-\ref{Ass: Boot} hold.  Let $\hat{a}$, $\hat{B}$, $\hat{a}^{\ast}$, and $\hat{B}^{\ast} $ be given in \eqref{Eqn: Estimators} and \eqref{Eqn: BootstrappedEstimators}. Assume (i) $N\to\infty$; (ii) $T\geq K+1$; (iii) $J\to\infty$ with $J^{2}\xi^{2}_{J}\log J=o(N)$. Then there is a $JM\times (K+1)$ random matrix $\mathbb{N}^{\ast}$ with $\mathrm{vec}(\mathbb{N}^{\ast})\sim N(0,\Omega)$ conditional on $\{Y_t, Z_t\}_{t\leq T}$ such that:
\begin{align*}
\|\sqrt{{NT/\omega_0}}(\hat{a}^{\ast} - \hat{a})-\mathbb{G}^{\ast}_{a}\| = O_{p^{\ast}}\left(\frac{\sqrt{NT}}{J^{\kappa}}+\frac{\sqrt{TJ}}{\sqrt{N}}+\frac{J^{5/6}}{N^{1/6}}+\frac{\sqrt{J\xi_{J}}\log^{1/4}J}{N^{1/4}}\right)
\end{align*}
and
\begin{align*}
\|\sqrt{{NT/\omega_0}}(\hat{B}^{\ast} - \hat{B})-\mathbb{G}_{B}^{\ast}\|_{F} = O_{p^{\ast}}\left(\frac{\sqrt{NT}}{J^{\kappa}}+\frac{\sqrt{TJ}}{\sqrt{N}}+\frac{J^{5/6}}{N^{1/6}}+\frac{\sqrt{J\xi_{J}}\log^{1/4}J}{N^{1/4}}\right),
\end{align*}
where $p^{\ast}$ is the probability measure with respect to $\{w_i\}_{i\leq N}$ conditional on $\{Y_t, Z_t\}_{t\leq T}$, $\kappa>1/2$ is a constant representing the smoothness of $\alpha(\cdot)$ and $\beta(\cdot)$, $\mathbb{G}^{\ast}_{a} = (I_{JM}-B\mathcal{H}\mathcal{H}^{\prime}B^{\prime})(\mathbb{N}^{\ast}_1-\mathbb{G}^{\ast}_{B}\mathcal{H}^{-1}\bar{f})-B\mathcal{H}\mathbb{G}_{B}^{\ast\prime}a$, and $\mathbb{G}^{\ast}_{B} = \mathbb{N}^{\ast}_2B^{\prime}B\mathcal{M}$. The matrices $\mathcal{H}$ and $\mathcal{M}$ are nonrandom, as given in Lemma \ref{Lem: TechC3}, while $\mathbb{N}^{\ast}_1$ and $\mathbb{N}^{\ast}_2$ are the first column and the last $K$ columns of $\mathbb{N}^{\ast}$, respectively.
\end{thm}

Theorem \ref{Thm: Boot} demonstrates that the distribution of $(\mathbb{G}_{a},\mathbb{G}_{B})$, which aligns with the distribution of $(\mathbb{G}^{\ast}_{a},\mathbb{G}^{\ast}_{B})$, can be approximated by the distribution of $(\sqrt{{NT/\omega_0}}(\hat{a}^{\ast} - \hat{a}),\sqrt{{NT/\omega_0}}(\hat{B}^{\ast}-\hat{B}))$ conditional on the data, under the conditions $T=o(N)$, $NTJ^{-2\kappa}$ $=o(1)$, and $J=o(\min\{N^{1/5},N/T\})$. Theorems \ref{Thm: AsymDis} and \ref{Thm: Boot} can then be directly applied to conduct significance tests. To test whether $\alpha(\cdot) = 0$, we compare $NT\hat{a}^{\prime}\hat{a}$ with the $1-\alpha$ quantile of $NT(\hat{a}^{\ast}-\hat{a})^{\prime}(\hat{a}^{\ast}-\hat{a})/\omega_0$ conditional on the data for $0<\alpha<1$. Similarly, we can test whether each component of $\phi(z_{it})$ is significant in $\alpha(z_{it})$, which is equivalent to testing whether the corresponding element of $a$ is zero. We can also test the joint significance of each component of $\phi(z_{it})$ in $\beta(z_{it})$, which corresponds to testing whether the associated row of $BH$ is zero.

\subsection{Specification Test}\label{Sec: 52}

To test for linearity of $\alpha(\cdot)$ and $\beta(\cdot)$, we consider the following hypotheses:
\begin{align}\label{Eqn: Hypo}
&\mathrm{H}_0: \alpha(z_{it})=\gamma^{\prime} z_{it} \text{ and }  \beta(z_{it})=\Gamma^{\prime} z_{it} \text{ for some } \gamma \text{ and } \Gamma \text{ versus }\notag\\
&\mathrm{H}_1: \inf_{\pi}E[|\alpha(z_{it})-\pi^{\prime} z_{it}|^{2}]>0 \text{ or } \inf_{\Pi}E[\|\beta(z_{it})-\Pi^{\prime} z_{it}\|^{2}]>0.
\end{align}
We develop a test by comparing the estimators under $\mathrm{H}_0$ and $\mathrm{H}_1$. The estimators of $\alpha(\cdot)$ and $\beta(\cdot)$ under $\mathrm{H}_1$ are given by $\hat{\alpha}(\cdot)$ and $\hat{\beta}(\cdot)$, as defined in \eqref{Eqn: Estimators}. Let $\vec{Y}_{t}\equiv (Z_{t}^{\prime}Z_{t})^{-1}Z_{t}^{\prime}Y_t$, $\vec{Y}\equiv (\vec{Y}_{1},\ldots, \vec{Y}_{T})$, and $\bar{\vec{Y}}\equiv \sum_{t=1}^{T}\vec{Y}_{t}/T$. The estimators of $\alpha(z_{it})$ and $\beta(z_{it})$ under $\mathrm{H}_0$ are given by $\hat{\gamma}^{\prime}z_{it}$ and $\hat{\Gamma}^{\prime}z_{it}$, where $\hat{\Gamma}=\vec{Y}M_T\hat{F}(\hat{F}^{\prime}M_T\hat{F})^{-1}$ and $\hat{\gamma}=\bar{\vec{Y}}-\hat{\Gamma}\sum_{t=1}^{T}\hat{f}_t/T$.\footnote{It is crucial to use the unrestricted estimator $\hat{F}$ in both $\hat{\Gamma}$ and $\hat{\gamma}$, rather than the restricted one under $\mathrm{H}_0$. This ensures that $\hat{\Gamma}^{\prime}z_{it}$ and $\hat{\beta}(z_{it})$ share a common rotational transformation matrix, justifying the validity of the test. It also avoids the full-rank requirement for $\Gamma$. Thanks to the optimal rate of $\hat{F}$ established in Theorem \ref{Thm: ImprovedRates}, this does not cause an issue.} Our test statistic is given by:
\begin{align}\label{Eqn: TestStatistics}
\mathcal{S} = \frac{1}{J}\sum_{i=1}^{N}\sum_{t=1}^{T}|\hat{\gamma}^{\prime}z_{it} - \hat{\alpha}(z_{it})|^{2}+ \frac{1}{J}\sum_{i=1}^{N}\sum_{t=1}^{T}\|\hat{\Gamma}^{\prime}z_{it} - \hat{\beta}(z_{it})\|^{2}.
\end{align}

To obtain critical values, we adopt a bootstrap method. Let $\vec{Y}^{\ast}_{t}\equiv (Z_{t}^{\ast\prime}Z_{t})^{-1}Z_{t}^{\ast\prime}Y_t$, $\vec{Y}^{\ast}\equiv (\vec{Y}^{\ast}_{1},\ldots, \vec{Y}^{\ast}_{T})$, and $\bar{\vec{Y}}^{\ast}\equiv \sum_{t=1}^{T}\vec{Y}^{\ast}_{t}/T$, where $Z^{\ast}_t = (z_{1t}w_1,$ $\ldots, z_{Nt}w_{N})^{\prime}$. It is shown in the proof of Theorem \ref{Thm: SpecTest} that under $\mathrm{H}_0$, $\mathcal{S} = \sum_{i=1}^{N}\sum_{t=1}^{T}|(\hat{\gamma}-\gamma)^{\prime}z_{it} - (\hat{a}-a)^{\prime}\phi(z_{it})|^2/J+\sum_{i=1}^{N}\sum_{t=1}^{T}\|(\hat{\Gamma}-\Gamma H)^{\prime}z_{it} - (\hat{B}-BH)^{\prime}\phi(z_{it})\|^2/J+o_{p}(J^{-1/2})$. Given this, we estimate the null distribution of $\mathcal{S}$ by the distribution of
\begin{align}\label{Eqn: BootStatistics}
\mathcal{S}^{\ast} &=\frac{1}{J\omega_0}\sum_{i=1}^{N}\sum_{t=1}^{T}|(\hat{\gamma}^{\ast}-\hat{\gamma})^{\prime}z_{it} - (\hat{a}^{\ast}-\hat{a})^{\prime}\phi(z_{it})|^2\notag\\
&\hspace{0.5cm}+\frac{1}{J\omega_0}\sum_{i=1}^{N}\sum_{t=1}^{T}\|(\hat{\Gamma}^{\ast}-\hat{\Gamma})^{\prime}z_{it} -(\hat{B}^{\ast}-\hat{B})^{\prime}\phi(z_{it})\|^2
\end{align}
conditional on the data. Here, $\hat{\Gamma}^{\ast}\hspace{-0.05cm}=\hspace{-0.05cm}\vec{Y}^{\ast}\hspace{-0.05cm}M_T\hat{F}(\hat{F}^{\prime}\hspace{-0.05cm}M_T\hspace{-0.05cm}\hat{F})^{-1}$ and $\hat{\gamma}^{\ast}\hspace{-0.05cm}=\hspace{-0.05cm}\bar{\vec{Y}}^{\ast}\hspace{-0.05cm}-\hat{\Gamma}^{\ast}\hspace{-0.05cm}(\hat{B}^{\ast\prime}\hat{B}^{\ast})^{-1}\hspace{-0.05cm}\hat{B}^{\ast\prime}\bar{\tilde{Y}}^{\ast}$. For $0<\alpha<1$, let $c_{1-\alpha}$ be the $1-\alpha$ quantile of $\mathcal{S}^{\ast}$ conditional on the data. Thus, we construct the test as follows: reject $\mathrm{H}_0$ if $\mathcal{S}>c_{1-\alpha}$.

\begin{thm}\label{Thm: SpecTest}
Suppose Assumptions \ref{Ass: Basis}-\ref{Ass: SpecTest} hold.  Let $\mathcal{S}$ be given in \eqref{Eqn: TestStatistics} and $c_{1-\alpha}$ be given after \eqref{Eqn: BootStatistics} for $0<\alpha<1$.  Assume (i) $N\to\infty$; (ii) $T\geq K+1$; (iii) $J\to\infty$ with $J^{2}\xi^{2}_{J}\log J=o(N)$. In addition, assume $T=o(N)$, $J=o(\min\{N^{1/5},N/T\})$, and $NTJ^{-2\kappa}=o(1)$, where $\kappa>1/2$ is a constant representing the smoothness of $\alpha(\cdot)$ and $\beta(\cdot)$. Then
\begin{align*}
P(\mathcal{S}>c_{1-\alpha})\to\alpha \text{ under } \mathrm{H}_0 \text{ and } P(\mathcal{S}>c_{1-\alpha})\to1 \text{ under } \mathrm{H}_1.
\end{align*}
\end{thm}



\section{Empirical Analysis}\label{Sec: 8}
Our empirical analysis is based on the model specified in \eqref{Eqn: Model}.  Following standard practice in asset pricing, we depart from the notation used in previous sections and denote characteristics observed at time $t-1$ as $z_{i,t-1}$ instead of $z_{it}$. We begin by estimating the model and evaluating its overall performance, and then assess the performance of the extracted factors. The primary objective of our analysis is to investigate whether pricing errors are associated with characteristics and to evaluate the performance of our factors in asset pricing tests.

\subsection{Data and Methodology}\label{Sec: 81}
We use the same dataset as \citet{Kellyetal_Characteristics_2019}, which is originally from \citet{Freybergeretal_Dissecting_2017}. The dataset contains monthly returns of $12,813$ individual stocks and $36$ time-varying characteristics, covering the sample period from July 1962 to May 2014. The data is in the form of an unbalanced panel, for which our method is applicable. For the detailed descriptions, refer to these papers. To ensure comparability, we use the same $36$ characteristics as those authors. Following the procedure in \citet{Kellyetal_Characteristics_2019}, we transform the values of the characteristics into relative rankings within the range $[-0.5, 0.5]$.  This transformation standardizes the contributions of characteristics to pricing errors and risk exposures such that the estimation only depends on the rankings of characteristics and
is robust to extreme values, sharing the similar logic with the sorting procedure as in \cite{FamaFrench_Commonrisk_1993, FamaFrench_FiveFactor_2015}. To meet the large $N$ requirement, we select a sample period during which at least $1,000$ individual stocks have observations for both returns and the $36$ characteristics. This results in a sample spanning from September 1968 to May 2014. Based on this dataset, we construct the market factor and five long-short factors following \cite{FamaFrench_FiveFactor_2015}. These factors exhibit close means and standard deviations and show high correlations with the corresponding factors from Kenneth R. French’s website, as shown in Table \ref{Tab: factor_compare_app}.

We implement regressed-PCA estimation by selecting the basis functions to be either linear (i.e., $\phi(z_{it})=z_{it}$ including a constant term) or non-linear (via linear B-splines of $z_{it}$).\footnote{{Our econometric theory accommodates a variety of basis functions, such as Fourier series, polynomials, splines, and wavelets. Following \citet{Guetal_EmpiricalAsset_2020} and \citet{Freybergeretal_Dissecting_2017}, we employ splines due to their flexibility, which arises from increasing the number of knots. Unlike polynomials that require a higher degree for flexibility, splines generally produce more stable estimates \citep{HastieTibshiraniFriedman_StatisticalLearning_2011}.}} Using $\phi(z_{it})=z_{it}$ leads to linear specifications of $\alpha(\cdot)$ and $\beta(\cdot)$, while setting $\phi(z_{it})$ as linear B-splines of $z_{it}$ results in nonlinear specifications of $\alpha(\cdot)$ and $\beta(\cdot)$, where $\alpha(\cdot)$ and $\beta(\cdot)$ are continuous, piecewise linear functions.\footnote{The one dimensional linear B-splines $\{\psi_{j}(z)\}^{J}_{j=1}$ are defined over a set of consecutive, equidistant knots: $\{z_{1},...,z_{J+1}\}$.  For $j< J$,  $\psi_{j}(z)=(z-z_{j})/(z_{j+1}-z_{j})$ on $(z_{j}, z_{j+1}]$, $\psi_{j}(z)=(z_{j+2}-z)/(z_{j+2}-z_{j+1})$ on $(z_{j+1}, z_{j+2}]$, and 0 elsewhere. For $j=J$, $\psi_{j}(z)=(z-z_{j})/(z_{j+1}-z_{j})$ on $(z_{j}, z_{j+1}]$ and 0 elsewhere.} For ease of comparison, we maintain the same parameter dimension across different specifications. Specifically, we consider 18 characteristics with one internal knot and 12 characteristics with two internal knots in the linear B-splines specifications. The most significant 18/12 characteristics are selected based on the linear specification, which are collected in Table \ref{Tab: chars}. To implement the weighted bootstrap, we let the bootstrap weights $w_{i}$'s be i.i.d. random variables following the standard exponential distribution. For testing $\alpha(\cdot) = 0$ and linearity of $\alpha(\cdot)$ and $\beta(\cdot)$, we set the number of bootstrap draws to 499.

In order to evaluate the performance of the models estimated via regressed-PCA, we compute several measures of fit and prediction. First, we calculate Fama-MacBeth cross-sectional regression $R^{2}$, denoted as $R^2_{\tilde Y}$, which captures the variation in individual stock returns explained by the Fama-MacBeth managed portfolios $\tilde{Y}_t$ constructed from $\phi(z_{i,t-1})$. Next, we report the variation in these managed portfolios explained by the extracted factors $\hat{f}_t$, denoted as $R^{2}_{K}$. We then consider the following three types of $R^2$ measures that directly speak to the ability of the factor models to explain the cross-section of individual stock returns. The first measure (denoted as $R^2$) is total $R^2$ as used in \citet{Kellyetal_Characteristics_2019}. The second measure ($R^2_{T,N}$) calculates the cross-sectional average of time series $R^2$ across all stocks, which reflects the ability of the factors to capture common variation in stock returns. The third measure ($R^2_{N,T}$) computes the time-series average of cross-sectional goodness-of-fit measures, approximating the Fama-MacBeth cross-sectional regression $R^2$. This measure is particularly relevant for evaluating the model's capacity to explain the cross-section of average returns. The measures are defined as follows:\footnote{The differences among the three $R^2$'s are provided in Appendix \ref{App: Sec: D6}.}
\begin{align}
R^2 & = 1-\frac{\sum_{i, t}[y_{it}- \hat{\alpha}(z_{i,t-1}) - \hat{\beta}(z_{i,t-1})^{\prime}\hat{f}_t]^2}{\sum_{i, t} y_{it}^{2}}, \label{Eqn: R21}\\
R^2_{T,N} & = 1 - \frac{1}{N} \sum_{i} \frac{\sum_{t}[y_{it}- \hat{\alpha}(z_{i,t-1}) - \hat{\beta}(z_{i,t-1})^{\prime}\hat{f}_t]^2}{\sum_{t}y_{it}^{2}},\label{Eqn: R22}\\
R^2_{N,T} & = 1 - \frac{1}{T} \sum_{t} \frac{\sum_{i}[y_{it}- \hat{\alpha}(z_{i,t-1}) - \hat{\beta}(z_{i,t-1})^{\prime}\hat{f}_t]^2}{\sum_{i}y_{it}^{2}}.\label{Eqn: R23}
\end{align}
We also report a version of these goodness-of-fit measures that zero in on the role of factors in explaining the time-series as well as the cross-section of stock returns, by excluding the conditional alphas $\hat{\alpha}(z_{i,t-1})$; see \eqref{Eqn: R24}-\eqref{Eqn: R26} for the corresponding formulas.

Finally, we assess the out-of-sample prediction and fit using expanding-window estimations. For $t\geq 120$, we use the data up to time period $t-1$ to implement the regressed-PCA and obtain estimates such as $\hat{a}_{t-1}$, $\hat{B}_{t-1}$, $\hat{\alpha}_{t-1}(\cdot)$, $\hat{\beta}_{t-1}(\cdot)$, and $\hat{F}_{t-1}\equiv (\hat{f}^{(t-1)}_{1},\ldots, \hat{f}^{(t-1)}_{t-1})^{\prime}$. Using these, we compute the out-of-sample prediction of $y_{it}$ as $\hat{\alpha}_{t-1}(z_{i,t-1})+\hat{\beta}_{t-1}(z_{i,t-1})^{\prime}\hat{\lambda}_{t}$, where $\hat{\lambda}_{t} = \sum_{s\leq t-1}\hat{f}^{(t-1)}_s/(t-1)$, which is the average of factor estimates through time period $t-1$. The out-of-sample predictive $R^2$ is:
\begin{align}
R^2_O & = 1-\frac{\sum_{i, t\geq 120}[y_{it}- \hat{\alpha}_{t-1}(z_{i,t-1}) - \hat{\beta}_{t-1}(z_{i,t-1})^{\prime}\hat{\lambda}_t]^2}{\sum_{i, t\geq 120} y_{it}^{2}}; \label{Eqn: R21Predictive_main}
\end{align}
see \eqref{Eqn: R21Predictive}-\eqref{Eqn: R23Predictive} for another two versions. We calculate the out-of-sample realized factor returns at time period \(t\) as: $\hat{f}_{t-1, t} = \hat B^{\prime}_{t-1}\tilde Y_{t} = \hat B^{\prime}_{t-1}(\Phi(Z_{t-1})^{\prime} \Phi(Z_{t-1}))^{-1}\Phi^{\prime}(Z_{t-1})Y_{t}$. Although the resulting factor returns are only known ex post at time period $t$, they represent returns on portfolios that are constructed ex ante, using weights based on estimates obtained at time period $t-1$. Using these, we can access how much of the cross-sectional variation of individual stock returns can be explained by the pre-estimated $\hat{\beta}_{t-1}(z_{i,t-1})$. We then define the out-of-sample fit $R^2$ as:
\begin{align}
R^2_{f,O} & = 1-\frac{\sum_{i, t\geq 120}[y_{it}- \hat{\beta}_{t-1}(z_{i,t-1})^{\prime}{\hat f_{t-1,t}}]^2}{\sum_{i, t\geq 120} y_{it}^{2}} \label{Eqn: R21OOS_main};
\end{align}
see \eqref{Eqn: R21OOS}-\eqref{Eqn: R23OOS} for another two versions.  

\subsection{Empirical Results}\label{Sec: 82}

\subsubsection{Model Estimation}\label{Sec: 821}
The main findings presented in Tables \ref{Tab: Emprical1}-\ref{Tab: Emprical3} can be summarized as follows. First, all of our measures of fit indicate that a low-dimensional factor model is unlikely to explain the time-series\textemdash or the cross-section\textemdash of individual stock returns (rather than the managed portfolios). In all specifications at least 5 or 6 factors are required for most of the in-sample R-squared to exceed $10\%$. That said, the total in-sample $R^2$'s in our model are smaller than those of \citet{Kellyetal_Characteristics_2019}. This is not surprising, since the objective of their IPCA estimation is maximizing the total in-sample $R^2$, as discussed in \eqref{Eqn: objls}. We extract factors that capture the most time-series comovement within a set of portfolios, which, in turn, reflect the most cross-sectional variation in individual asset returns. In contrast, our out-of-sample $R_{O}^{2}$'s are 0.54\% for the linear specification and 0.59\%  and 0.57\% for the two nonlinear specifications,\footnote{Our out-of-sample predictive $R^2$'s are invariant to the number of factors, because $\hat{\alpha}_{t-1}(z_{i,t-1})+ \hat{\beta}_{t-1}(z_{i,t-1})^{\prime}\hat{\lambda}_t = \phi(z_{i,t-1})^{\prime}[\hat{a}_{t-1} + \hat{B}_{t-1}\hat{F}^{\prime}_{t-1}1_{t-1}/(t-1)] = \phi(z_{i,t-1})^{\prime}\sum_{s=1}^{t-1}\tilde{Y}_t/{t-1}$, which does not depend on $K$.} which are comparable to the 0.60\% in \citet{Kellyetal_Characteristics_2019}'s linear specification with six factors. Similarly, the out-sample-sample fits are close to those of \citet{Kellyetal_Characteristics_2019}. With six factors, the out-sample-sample $R^{2}_{f,O}$'s are 15.38\%, 16.20\%, and 16.16\% for the three specifications, which are comparable to the 17.80\% in \citet{Kellyetal_Characteristics_2019}. Moreover, all in-sample and out-of-sample fits improve as the number of factors increases, as factor loadings soak up more of the variation in managed portfolio returns that is otherwise attributed to the alphas. 
In addition, compared to the linear specification, the nonlinear specifications based on linear B-splines significantly improve both in-sample and out-of-sample fits in most cases. They also slightly enhance the out-of-sample predictive $R^2$'s. Both the improved fits and robustness highlight the advantage of the nonlinear specifications.

Second, the results of testing the linearity of $\alpha(\cdot)$ and $\beta(\cdot)$ explain why we observe lower $R^2$'s in the linear specification. Linearity is strongly rejected at the $1\%$ level in all factor models estimated with one to ten factors (Tables \ref{Tab: Emprical1}-\ref{Tab: Emprical3} report the $p$-values concisely to save space, all of which indicate rejection at the $1\%$ level). Moreover, we also find robust evidence rejecting the null hypothesis $\alpha(\cdot) = 0$ across all cases. 
Additional empirical results are provided in Appendix \ref{App: Sec: F} (see Tables \ref{Tab: Emprical_apd1}-\ref{Tab: Emprical_apd3}).


Before analyzing the contribution of each characteristic to pricing errors and risk exposures, we first determine the signs of the extracted factors. Under the normalization $B^{\prime}B= I_K$ and $F^{\prime}M_TF/T$ being diagonal with descending diagonal entries, the signs of the factors remain undetermined. To address this, we set the sample means of the factors to be positive, ensuring the unconditional risk premium on each factor is positive. To interpret the factors, we examine their correlations with the market factor and five long-short factors from Kenneth R. French’s website  as discussed in Table \ref{Tab: factor_compare_app}, and conduct projection regressions, as detailed in Appendix \ref{App: Sec: F} (see Tables \ref{Tab: Emprical_apd10}-\ref{Tab: Emprical_apd13}). We find the substantial correlations between these factors and our factors, with both sets explaining significant variations in each other.

\begin{figure}[!htbp]
\centering
\includegraphics[width=14.5cm, height=16cm]{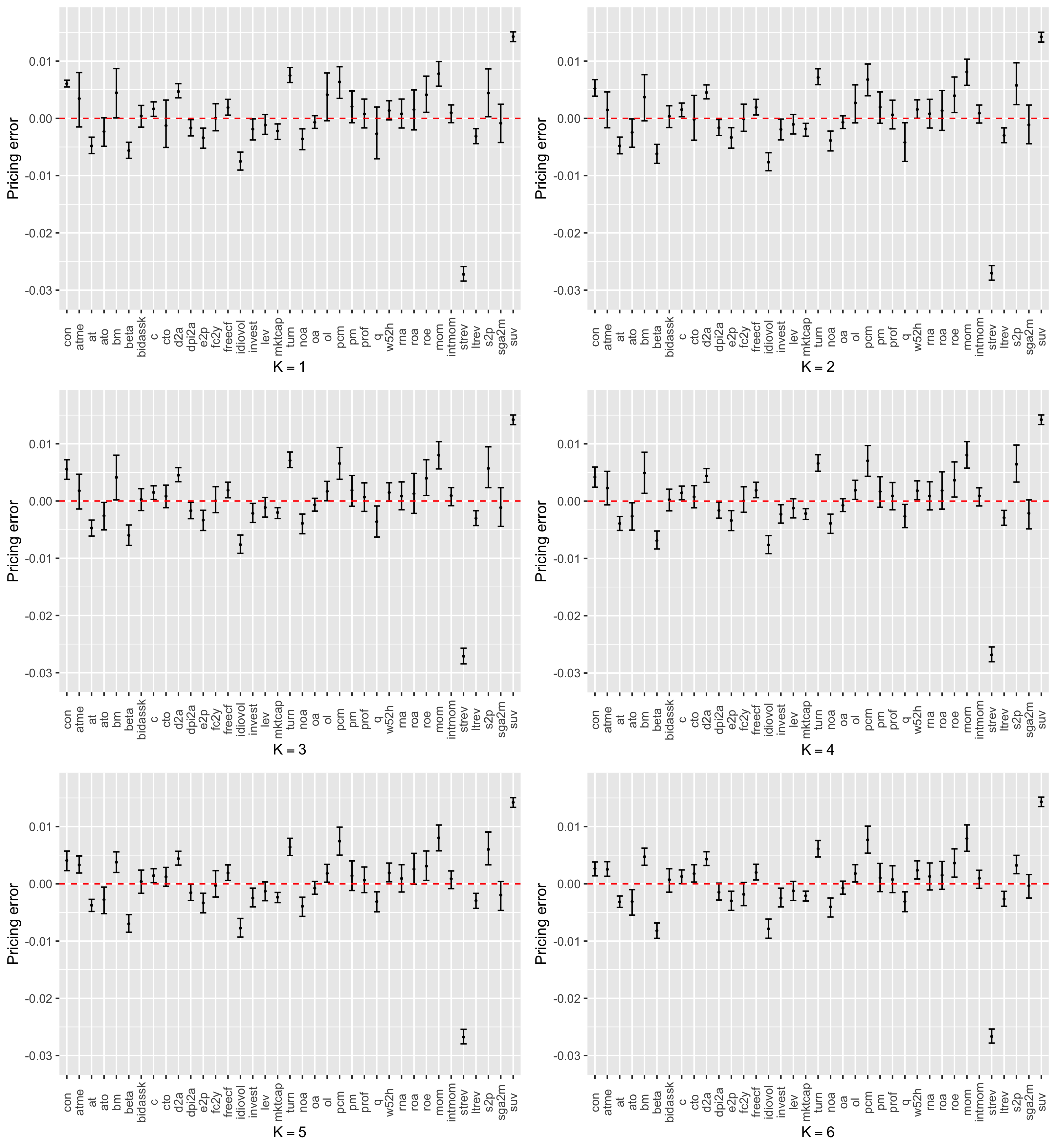}
\caption{$95\%$ confidence intervals for coefficients in $\alpha(\cdot)$ under linear specifications of $\alpha(\cdot)$ and $\beta(\cdot)$ with 36 characteristics}\label{Fig: Empirical4}
\end{figure}
\begin{figure}[!htbp]
	\centering
	\includegraphics[width=14.5cm, height=16cm]{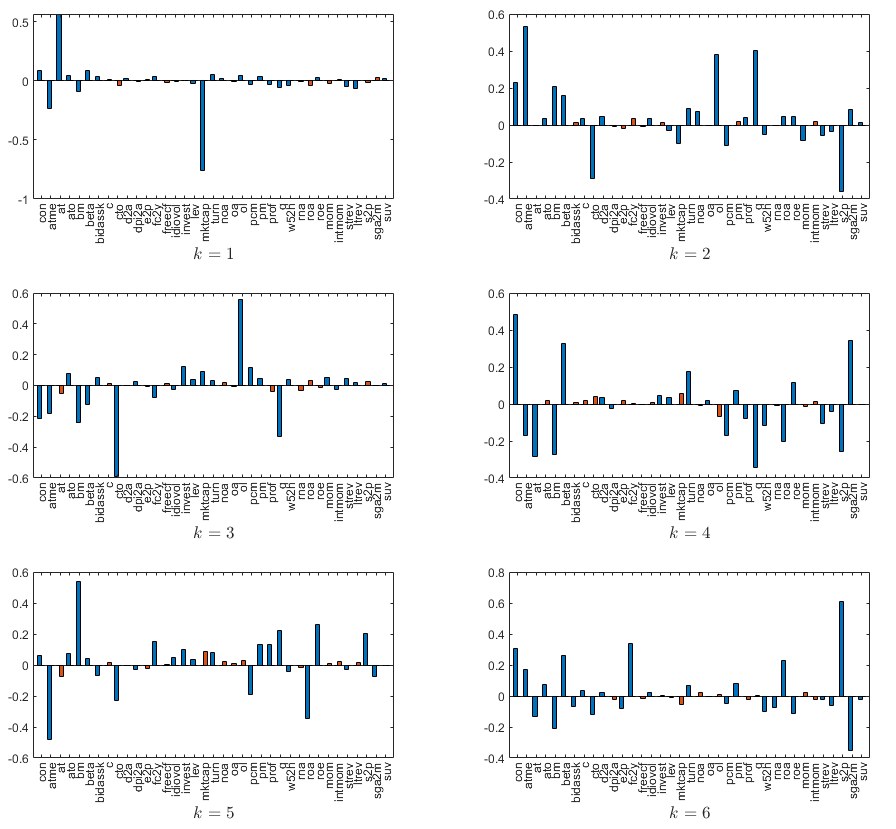}
	\caption{Estimates of coefficients in $\beta(\cdot)$ under linear specifications of $\alpha(\cdot)$ and $\beta(\cdot)$ with 36 characteristics (blue: significant at the $5\%$ level; red: insignificant)}\label{Fig: Empirical5}
\end{figure}

Figures \ref{Fig: Empirical4} and \ref{Fig: Empirical5} illustrate the contribution of each characteristic to pricing errors and risk exposures under the linear specification. Figure \ref{Fig: Empirical4} reveals that the $95\%$ confidence intervals of characteristic coefficients in pricing errors remain relatively stable as the number of factors increases from one to six. Notably, 22 out of 36 characteristics remain significant at the $5\%$ level for $K=6$, in stark contrast to the results reported by \citet{Kellyetal_Characteristics_2019}. Figure \ref{Fig: Empirical5} displays the characteristic coefficients in risk exposures for the first six factors. In contrast to \citet{Kellyetal_Characteristics_2019}'s finding that 13 out 36 characteristics are significant in driving risk exposures, we find 24 significant characteristics for $K=6$. Specifically, the coefficient of ``market cap" in the first factor is negative and large in magnitude; the fourth and sixth factors exhibit substantial positive loadings on ``market beta" (i.e., ``beta");  the second and fifth factors display significant positive loadings on  ``book-to-market ratio'' (i.e.,``bm"). These findings align with the traditional views of asset pricing anomalies as discussed in \citet{FamaFrench_Commonrisk_1993, FamaFrench_FiveFactor_2015}. More results for the two nonlinear specifications are provided in Appendix \ref{App: Sec: F} (see Figures \ref{Fig: Emprical_apd17}-\ref{Fig: Emprical_apd20}).

Taking advantage of the fact that our regressed-PCA method does not require a large time dimension ($T$), we perform subsample analyses using five-year intervals starting in January 1970. Figure \ref{Fig: Empirical6} presents the key results. The left panel shows that average pricing errors (measured by $\|a\|^2$) under the nonlinear specifications are significantly smaller than those under the linear specification. 
For the linear case, average pricing errors are the highest during the initial subsample period (1970-1974), decline over time, rise again during the equity market ``boom'' of the 1990s, peak in the early 2000s, and then drop sharply. Under the nonlinear specifications, the patterns differ somewhat, with average pricing errors spiking around 1990-1994 and subsequently decreasing to levels comparable to those in the linear case by the end of the sample period. This decline may reflect the growing prevalence of quantitative investing, which reduces mispricing by exploiting characteristic-related anomalies, as suggested by \citet{McLeanPontiff_DoesAcademic_2016} and \citet{Greenetal_Characteristics_2017}.

The right panel of Figure \ref{Fig: Empirical6} illustrates the proportion of time-series and cross-sectional variation in stock returns explained by common factors (measured by $R^{2}_{f,T,N}$ and $R^{2}_{f,N,T}$), which appears similar across model specifications. Notably, all the reported $R^2$ measures decline from 1970, reach a trough in the mid-1990s, and then steadily rise until the sample ends in 2014. This observation aligns with the empirical findings: \citet{Campbell2001have} document a noticeable increase in firm-level volatility between 1962 and 1997; while extending this analysis to 2021, \citet{Campbell2022idiosyncratic} find that idiosyncratic volatility declined after peaking in 1999-2000. Similar trends are evident in the out-of-sample fit measures shown in Figure \ref{Fig: Emprical_apd22}.

\subsubsection{Trading Strategies}\label{Sec: 822}
We construct trading strategies based on our estimated models. While constructing the MVE portfolio on individual stocks is usually infeasible due to the challenge in estimating a high-dimensional covariance matrix, the model in \eqref{Eqn: Model} enables us to devise trading strategies by leveraging the \textit{alpha} (i.e., mispricing)  and \textit{beta} (i.e., risk) roles of characteristics. Specifically, the initial step of the regressed-PCA method (i.e., Fama-MacBeth regressions) reduces a large number of individual stock returns into a smaller set of characteristic-managed portfolios. These portfolios exhibit a classical factor structure as outlined in \eqref{Eqn: Model: Sieve: Vectors: Regressed}, facilitating the construction of a pure-$alpha$ strategy and an MVE factor portfolio.

By \eqref{Eqn: Model: Sieve: Vectors: Regressed} and Theorem \ref{Thm: ImprovedRates}, $\hat{a}^{\prime}\tilde{Y}_{t}\convp \|a\|^{2}$ for each $t$ as $N\to\infty$, implying that $\hat{a}^{\prime}\tilde{Y}_{t}$ represents a portfolio with positive returns (if $a\neq0$) and no risk asymptotically. Using the expanding-window procedure in Section \ref{Sec: 81}, we construct a pure-$alpha$ ``arbitrage'' portfolio for $t\geq 120$ as follows:
\begin{align}\label{Eqn: purealpha}
  R_{\alpha,t} = \hat{a}_{t-1}^{\prime}\tilde{Y}_{t} = \hat{a}_{t-1}^{\prime} (\Phi(Z_{t-1})^{\prime} \Phi(Z_{t-1}))^{-1}\Phi(Z_{t-1})^{\prime} Y_{t}.
\end{align}
This portfolio can be equivalently constructed from individual stocks by using weights $\Phi(Z_{t-1}) (\Phi(Z_{t-1})^{\prime} \Phi(Z_{t-1}))^{-1}\hat{a}_{t-1}$. Since $R_{\alpha,t}$ relies on estimates obtained at time period $t-1$, it is tradable (ex ante), while $\hat{a}^{\prime}\tilde{Y}_{t}$ is based on full-sample estimates. We consider $R_{\alpha,t}$, the out-of-sample version of $\hat{a}^{\prime}\tilde{Y}_{t}$, and refer to their Sharpe ratios as the out-of-sample and in-sample Sharpe ratios of the pure-$alpha$ portfolio, respectively.

Similarly, we construct the out-of-sample version of $\hat{f}_t = \hat{B}^{\prime}\tilde{Y}_t$ as $\hat{f}_{t-1, t} = \hat B^{\prime}_{t-1}\tilde Y_{t}$, which has been introduced in \eqref{Eqn: R21OOS_main}. This enables the construction of an MVE factor portfolio based on the same expanding-window procedure: for $t\geq 120$,
\begin{align}\label{Eqn: purebeta}
R_{\beta,t} = \hat{\mu}_{t-1}^{\prime} \hat{\Sigma}^{-1}_{t-1} \hat{f}_{t-1, t} =  \hat{\mu}_{t-1}^{\prime} \hat{\Sigma}_{t-1}\hat B^{\prime}_{t-1}(\Phi(Z_{t-1})^{\prime} \Phi(Z_{t-1}))^{-1}\Phi^{\prime}(Z_{t-1})Y_{t},
\end{align}
where $\hat{\mu}_{t-1}$ and $\hat{\Sigma}_{t-1}$ are estimates of the (conditional) mean and covariance matrix of $f_t$ at time period $t-1$.\footnote{Specifically, $\hat{\mu}_{t-1}  = \sum_{s\leq t-1}\hat{f}^{(t-1)}_s/(t-1)$ and $\hat{\Sigma}_{t-1} =\sum_{s\leq t-1}(\hat{f}^{(t-1)}_s - \hat{\mu}_{t-1})(\hat{f}^{(t-1)}_s - \hat{\mu}_{t-1})^{\prime} /(t-2)$.}  Weights for $R_{\beta,t}$ can also be derived from and applied on individual stocks, making it tradable. The in-sample counterpart of $R_{\beta,t}$ is $\hat{\mu}^{\prime} \hat{\Sigma}^{-1} \hat{f}_{t}$, where $\hat{\mu}$ and $\hat{\Sigma}$ are the full-sample estimates of the (unconditional) mean and covariance matrix of $f_t$.\footnote{Specifically, $\hat{\mu}  = \sum_{s\leq T}\hat{f}_s/T$ and $\hat{\Sigma} =\sum_{s\leq T}(\hat{f}_s - \hat{\mu})(\hat{f}_s - \hat{\mu})^{\prime} /(T-1)$.} Their Sharpe ratios are referred to as the out-of-sample and in-sample Sharpe ratios of the MVE factor portfolio, respectively.

We further combine $R_{\alpha,t}$ and $\hat{f}_{t-1, t}$ to form a set of $K+1$ factor portfolios, constructing an MVE portfolio following the procedure outlined for $R_{\beta,t}$. The in-sample counterpart is derived from $\hat{a}^{\prime}\tilde{Y}_{t}$ and $\hat{f}_t$. The resulting Sharpe ratios are referred to as the out-of-sample and in-sample Sharpe ratios of the combined MVE portfolio, respectively. By imposing a factor structure on the conditional covariance of individual asset returns as in \eqref{Eqn: Model}, the combined MVE portfolio provides an approximation to the stock market's MVE portfolio. Tables \ref{Tab: in_SR} and \ref{Tab: OOS_SR} present the annualized in-sample and out-of-sample Sharpe ratios for the pure-$alpha$, MVE factor, and combined MVE portfolios. In all subsequent tables, ``Regressed-PCA'' denotes the results under linear specifications of $\alpha(\cdot)$ and $\beta(\cdot)$ with 36 characteristics, while ``Regressed-PCA S1'' and ``Regressed-PCA S2'' correspond to nonlinear specifications with 18 and 12 characteristics, respectively. ``IPCA'' represents the results based on the linear specification  using IPCA estimations with the same 36 characteristics.

Two remarks are essential for understanding Tables \ref{Tab: in_SR} and \ref{Tab: OOS_SR}. First, the in-sample Sharpe ratio of our MVE factor portfolio increases with $K$, whereas that of IPCA's MVE factor portfolio may not. This reflects the stability of our factor construction approach compared to IPCA, in particular the inherent orthogonality of regressed PCA factors (see Section \ref{Subsubsec:Compare}). We report the out-of sample Sharpe ratio of each incremental regressed PCA factor in Table \ref{Tab: OOS_SR} along with the pure-$alpha$ and MVE portfolios, but we do not report individual Sharpe ratios for IPCA factors as they are not incremental, since all of the factors are estimated jointly for each $K$.
%
 Second, for our regressed-PCA approach, the squared in-sample Sharpe ratio of the combined MVE portfolio equals the sum of those of the pure-\textit{alpha} and MVE factor portfolios as the two are orthogonal by construction.\footnote{This follows from $\sum_{t=1}^{T}\hat{a}^{\prime}(\tilde{Y}_{t}-\bar{\tilde{Y}})\hat{f}_t/T = \hat{B}^{\prime}\tilde{Y}M_T \tilde{Y}^{\prime}\hat{a}/T = \hat{\Lambda}\hat{B}^{\prime}\hat{a} = 0$, where $\hat{\Lambda}$ is a diagonal matrix with diagonal entries being the large $K$ eigenvalues of $\tilde{Y}M_T \tilde{Y}^{\prime}/T$.} As a result, the in-sample Sharpe ratios of the pure-\textit{alpha} and MVE factor portfolios are identical to their respective contributions to the combined MVE portfolio and are omitted in Table \ref{Tab: in_SR}. Similarly, the out-of-sample Sharpe ratio of our MVE factor portfolio equals its contribution to the combined MVE portfolio, which is also omitted in Table \ref{Tab: OOS_SR}. The orthogonality property does not hold for IPCA, highlighting another advantage of regressed-PCA.


The main findings are summarized as follows. First, under the linear specification, we compare the Sharpe ratios of the pure-\textit{alpha} and MVE factor portfolios constructed based on regressed-PCA and IPCA, separately. As $K$ increases, the Sharpe ratio of our pure-\textit{alpha} portfolio remains high (in-sample: 3.89 to 4.50; out-of-sample: 3.18 to 3.84), while that of the MVE factor portfolio is comparatively low (in-sample: 0.65 to 0.89; out-of-sample: 0.44 to 0.72). The observed increase in the Sharpe ratio of the pure-\textit{alpha} portfolio as $K$ grows suggests that factors play a crucial role in hedging common variation in stock returns, reducing the volatility of the pure-\textit{alpha} portfolio at a rate exceeding the decline in alphas. The findings also align with the testing evidence of nonzero pricing errors in the linear specification found in Section \ref{Sec: 821}, reinforcing our conclusions. While the Sharpe ratios of IPCA's pure-\textit{alpha} and MVE factor portfolios are more comparable (in-sample: 1.61 to 3.13 vs. 1.07 to 2.85; out-of-sample: 1.31 to 2.84 vs. 0.92 to 1.72), the former is higher than the latter.
Table \ref{Tab: OOS_SR} also shows that the higher Sharpe ratio of our pure-\textit{alpha} portfolio compared with IPCA's is due to its low volatility, consistent with the idea of no risk asymptotically. The lower volatility of our pure-\textit{alpha} portfolio underscores the superior hedging properties of our factors. Notably, regressed-PCA consistently yields higher in-sample and out-of-sample Sharpe ratios for the combined MVE portfolio than IPCA across all $K$. This indicates that the combined MVE portfolio constructed with regressed-PCA better approximates the stock market's MVE portfolio compared to IPCA. In particular, the high Sharpe ratios of the combined MVE portfolio primarily arise from the high Sharpe ratios of the pure-\textit{alpha} portfolio. The better approximation is also attributed to the stability and orthogonality properties of our regressed-PCA as discussed above.

Second, the nonlinear specifications yield higher in-sample and out-of-sample Sharpe ratios for the MVE factor and combined MVE portfolios than the linear specification across all values of $K$ (except $K=1$ in Regressed-PCA S2 of Table \ref{Tab: in_SR}). As $K$ increases, the Sharpe ratio of the pure-$alpha$ portfolio under the nonlinear specifications falls below that under the linear specification, indicating that nonlinear models yield smaller magnitudes of pricing errors. Meanwhile, the nonlinear specifications also yield better factors, reflected in higher Sharpe ratios of the MVE factor portfolio (in-sample: 0.61 to 3.96; out-of-sample: 0.51 to 3.33) than the linear specification (in-sample: 0.65 to 0.89; out-of-sample: 0.44 to 0.72). More importantly, the  Sharpe ratios of the combined MVE portfolio from the nonlinear specifications are substantially higher than those from the linear specification,  implying the potential nonlinearity of stochastic discount factor in the U.S. stock market.  Similarly,  the high Sharpe ratios of our combined MVE portfolio are primarily derived by the pure-\textit{alpha} portfolio for $K\leq 5$, the Sharpe ratios of the MVE factor portfolio catch up and become comparable for larger $K$.  Nevertheless, the Sharpe ratios of our pure-\textit{alpha} portfolio remain consistently high (in-sample: 3.47 to 4.80; out-of-sample: 3.09 to 4.26) compared with those of the MVE factor portfolio (in-sample: 0.61 to 3.96; out-of-sample: 0.51 to 3.33), providing strong evidence of nonzero pricing errors. This finding corroborates the evidence of nonzero pricing errors presented in Section \ref{Sec: 821}.   All these findings align with the strong evidence of nonlinearity found in Section \ref{Sec: 821} and underscore the advantage of the flexible nonlinear specifications.

\begin{landscape}
	\begin{figure}[!htbp]
		\centering
		\includegraphics[width=20cm, height=13cm]{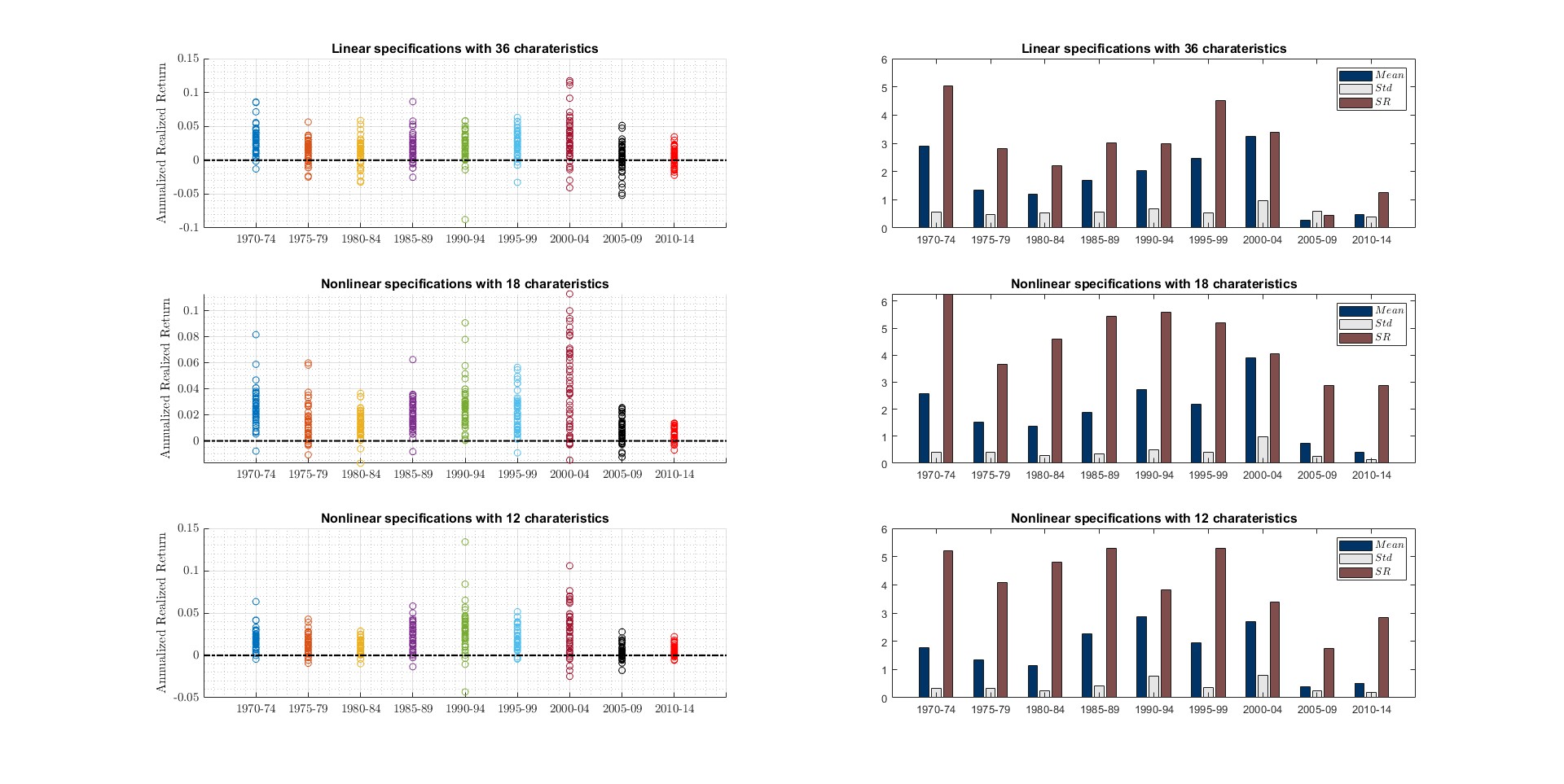}
	\caption{Annualized realized excess returns and Sharpe ratios of the pure-\textit{alpha} portfolio with $K=10$: subsample analysis}\label{Fig: Empirical7}
	\end{figure}
\end{landscape}

Lastly, we perform a subsample analysis of the pure-\textit{alpha} strategy, using five-year intervals starting in January 1970. Within each subsample, we construct the pure-\textit{alpha} ``arbitrage'' portfolio defined in \eqref{Eqn: purealpha}, employing expanding window estimation from the second year onward. The key results are presented in Figure \ref{Fig: Empirical7}. Notably, the left panel of Figure \ref{Fig: Empirical7} displays the decline in Sharpe ratios of the portfolio, and the right panel  indicates that the decline is primarily driven by a reduction in the portfolio’s average returns, rather than an increase in its standard deviations.  This is consistent with the findings in Figure \ref{Fig: Empirical6}, where we observe a significant decline in pricing errors since 2000.

In summary, we provide further evidence supporting the findings of nonlinearity and nonzero pricing errors, as well as their significant decline over time in Section \ref{Sec: 821}. This section also reinforces the advantages of regressed-PCA.  It is important to note that while nonzero pricing errors are observed, they do not necessarily imply market inefficiency unless the possibility of model misspecification is ruled out.


\subsubsection{Asset Pricing Tests}\label{Sec: 823}
We now evaluate the performance of our factors in asset pricing tests, considering a broad class of testing portfolios. Specifically, we examine three groups of Fama-MacBeth managed portfolios: Regressed-PCA, Regressed-PCA S1, and Regressed-PCA S2, as well as IPCA's managed portfolios. Additionally, we include two groups of single sorted portfolios based on 55 characteristics from \cite{Haddadetal_FactorTiming_2020} and our 36 characteristics, along with several groups of double sorted portfolios following \citet{FamaFrench_CSTS_2020}. Table \ref{Tab: corr_testasset_app} reports the bilateral correlations and standard deviations of these portfolios. The Fama-MacBeth managed portfolios exhibit lower bilateral correlations than others, and smaller standard deviations compared to the sorted portfolios. This supports the optimality of the Fama-MacBeth managed portfolios, which are maximally diversified, as discussed in Section \ref{SubSec:AssetPricing}.

We compare our factors with several existing sets: IPCA's factors, the five factors from \cite{FamaFrench_FiveFactor_2015} (denoted as FF5), and the factors constructed following \cite{Kozaketal_Interpreting_2018} (denoted as KNS). The comparison statistics from time series regressions are reported following the analysis in \citet{FamaFrench_CSTS_2020}. The results for five groups of testing portfolios with $K=5$ are presented in Tables \ref{Tab: factor_compare1_main} and \ref{Tab: factor_compare1_main_continue}, with additional results provided in Appendix \ref{App: Sec: F} (see Tables \ref{Tab: factor_compare3}-\ref{Tab: factor_compare_add_K6_5}).


The main findings are summarized as follows. First, for the Fama-MacBeth managed portfolios in Group I, both our factors and IPCA's factors outperform FF5 and KNS's factors in terms of average absolute intercepts ($A|a|$). They also achieve larger average regression $R^2$'s ($AR^2$), leading to smaller average standard errors ($As(a)$). The resulting average absolute $t$-statistics ($A|t(a)|$) and $GRS$ statistics ($GRS$) are comparable across all factors. Notably, our factors under the nonlinear specifications do not improve performance, as the testing portfolios are derived from the linear specification.

Second, our factors consistently outperform IPCA's factors in pricing the sorted portfolios in Groups II, III, V, and VI, as evidenced by smaller average absolute intercepts, $t$-statistics, and $GRS$ statistics. The higher average absolute $t$-statistics and $GRS$ statistics for IPCA's factors arise from their larger average regression $R^2$'s (or smaller average residual standard deviations ($As(e)$) or standard errors). To investigate this, we project IPCA's factors onto our factors under the linear specification (without a constant term) and treat the resulting residuals as new factors (denoted as IPCA$\setminus$Regressed-PCA). These residuals yield even larger average absolute intercepts and substantial average regression $R^2$'s, indicating that IPCA's factors capture more time-series variation in returns but less cross-sectional variation, likely due to overfitting idiosyncratic noise rather than extracting true signals. This overfitting is much less pronounced in Group I, as the Fama-MacBeth managed portfolios are maximally diversified, unlike the sorted portfolios, as discussed in Section \ref{SubSec:AssetPricing} and shown in Table \ref{Tab: corr_testasset_app}.

Third, our factors under the nonlinear specifications significantly reduce average absolute intercepts and $t$-statistics for sorted portfolios. This improvement stems from the nonparametric nature of sorting, as discussed in Section \ref{SubSec:Comparefactor}. However, due to the high correlations among sorted portfolios, as shown in Table \ref{Tab: corr_testasset_app}, no noticeable improvement in $GRS$ statistics is observed. Moreover, factors under the nonlinear specifications with 12 characteristics outperform FF5, yielding smaller average absolute $t$-statistics and $GRS$ statistics, with comparable average absolute intercepts. The higher $t$-statistics and $GRS$ statistics for FF5 also arise from their larger regression $R^2$'s, suggesting the presence of unpriced components similar to IPCA's factors \citep{Danieletal_CrossSection_2020,KozakNagel_Whydospan_2022}. Moreover, our factors also outperform KNS's factors, achieving smaller average absolute intercepts and $t$-statistics.

Lastly, for IPCA's managed portfolios in Group IV, IPCA's factors exhibit inferior performance compared to other factors, with larger average absolute intercepts, $t$-statistics, and $GRS$ statistics. This finding is surprising given that IPCA factors are derived from its managed portfolios.  As the testing portfolios are derived under the linear specification, our factors under the linear specification outperform those under the nonlinear specifications. The performance of our factors under the linear specification is comparable to FF5 and KNS's factors.
In addition, the performance comparisons based on relative metrics (e.g., $Aa^2/V\overline{r}$ or $A\lambda^2/V\overline{r}$) align with those based on absolute metrics ($A|a|$), reinforcing the robustness of our results.

In summary, our factors demonstrate superior performance compared to IPCA's factors and long-short factors in asset pricing tests, with robustness across a wide range of testing portfolios.

\section{Conclusion}\label{Sec: 9}
In this paper, we considered semiparametric conditional latent factor models to address the ``characteristics versus covariances'' debate and the ``\textit{factor zoo}'' problem in cross-sectional asset pricing. We proposed a simple and tractable sieve estimation approach combined with a weighted-bootstrap procedure for conducting inference on the \textit{alpha} and \textit{beta} functions. We established large-sample properties of the estimators and validity of the tests under large $N$, even when $T$ is small. In addition to offering formal inference procedures and well-founded asymptotic properties, our approach presents several advantages over existing methods such as IPCA and projected-PCA. Specifically, it is computationally efficient and accommodates nonzero alphas, time-varying characteristics, unbalanced panels, and short samples, making it particularly suitable for empirical asset pricing applications. These results enable the estimation of conditional factor structures for a large set of individual assets by incorporating numerous characteristics, accounting for nonlinearity without requiring pre-specified factors. Moreover, our approach disentangles the role of risk from the purely predictive power of return characteristics that is unrelated to common risk exposures.

We applied this method to analyze the cross-sectional differences in individual stock returns in the U.S. market. The findings provide robust evidence of large nonzero pricing errors and nonlinearity in both \textit{alpha} and \textit{beta} functions, leading to the formation of ``arbitrage'' portfolios with exceptionally high Sharpe ratios (exceeding 3). Additionally, we documented a significant decline in pricing errors since 2000. Our method  delivers stable and reliable factor construction without the risk of overfitting, yielding out-of-sample mean-variance efficient portfolios with Sharpe ratios in excess of 4. We also demonstrated that our factors outperform existing alternatives in explaining the cross-section of U.S. stock returns.



	\setlength{\tabcolsep}{3.3pt}
	\begin{table}[!htbp]
		\centering
		\begin{threeparttable}
			\renewcommand{\arraystretch}{1.6}
			\caption{Results under linear specifications of $\alpha(\cdot)$ and $\beta(\cdot)$ with 36 characteristics\tnote{\dag}}\label{Tab: Emprical1}
			\begin{tabular}{clccccccccccc}
				\hline\hline
				&\multicolumn{11}{c}{Unrestricted ($\alpha(\cdot) \neq 0$)}&\\
				\cline{2-12}
				&$K$&$R^{2}_{K}$&$R^{2}$&$R^{2}_{T,N}$&$R^{2}_{N,T}$&$R^{2}_{f}$&$R^{2}_{f,T,N}$&$R^{2}_{f,N,T}$&$R^{2}_{f,O}$&$R^{2}_{f,T,N,O}$&$R^{2}_{f,N,T,O}$\\
				\cline{2-12}
				&1&26.55& 2.54&1.37&0.36& 2.07&0.59&0.11& 6.23	&3.79&5.65\\
				&2&36.42&4.52&2.43&1.76&4.08&1.75&1.37&13.59&10.63&11.28\\
				&3&45.03&5.70&3.70&2.70&5.24&2.95&2.31&14.09&11.10&11.67\\
				&4&52.55&11.69&8.55&9.27&11.28&7.92&8.69&14.74&12.15&12.11\\
				&5&58.65&11.90&8.73&9.48&11.49&7.99&8.90&15.17&12.90&12.42\\
				&6&64.20&13.90&10.30&11.80&13.53&9.79&11.24&15.38&13.19&12.63\\
				&7&69.15&15.59&12.23&13.76&15.23&11.71&13.23&15.62&13.32&12.87\\
				&8&72.84&15.93&12.59&13.98&15.56&12.00&13.44&15.90&13.58&13.12\\
				&9&76.26&16.08&12.67&14.19&15.72&12.15&13.64&16.13&13.83&13.33 \\
				&10&79.15&16.23&12.82&14.35&15.87&12.34&13.80&16.29&14.06&13.47\\
				\cline{2-12}
				&$K$&$R^{2}_{\tilde Y}$&$R^{2}_{O}$&$R^{2}_{T,N,O}$&$R^{2}_{N,T,O}$&$p_{\alpha}$&$p_{\text{lin}}$\\
				\cline{2-12}
				&1-10&20.89& 0.54&0.64&0.21&$<1\%$&$<1\%$&\\
				\hline\hline
			\end{tabular}
			\begin{tablenotes}
				\small
				\item[\dag] $K$: the number of factors specified; $R^{2}_{\tilde Y}$: Fama-MacBeth cross-sectional regression $R^2$ ($\%$); $R^2_{K}$: the variation of the Fama-MacBeth managed portfolios $\tilde{Y}_t$ captured by the extracted factors $\hat{f}_t$ ($\%$);  $R^{2}$, $R^{2}_{T,N}$, $R^{2}_{N,T}$: various in-sample $R^2$'s ($\%$), see \eqref{Eqn: R21}-\eqref{Eqn: R23}; $R^{2}_{f}$, $R^{2}_{f,T,N}$, $R^{2}_{f,N,T}$: various in-sample $R^2$'s without $\alpha(\cdot)$ ($\%$), see \eqref{Eqn: R24}-\eqref{Eqn: R26};   $R^{2}_{f,O}$, $R^{2}_{f,T,N,O}$, $R^{2}_{f,N,T,O}$: various out-of-sample fit $R^2$'s ($\%$), see \eqref{Eqn: R21OOS}-\eqref{Eqn: R23OOS};  $R^{2}_O$, $R^{2}_{T,N,O}$, $R^{2}_{N,T,O}$: various out-of-sample predictive $R^2$'s ($\%$), see \eqref{Eqn: R21Predictive}-\eqref{Eqn: R23Predictive}; $p_{\alpha}$ and $p_{\text{lin}}$: the $p$-values of \textit{alpha} test ($\alpha(\cdot)=0$) and model specification test (joint linearity of $\alpha(\cdot)$ and $\beta(\cdot)$), respectively.
			\end{tablenotes}
		\end{threeparttable}
	\end{table}

	\setlength{\tabcolsep}{3.3pt}
	\begin{table}[!htbp]
		\centering
		\begin{threeparttable}
			\renewcommand{\arraystretch}{1.6}
			\caption{Results under nonlinear specifications of $\alpha(\cdot)$ and $\beta(\cdot)$ with 18 characteristics\tnote{\dag}}\label{Tab: Emprical2}
			\begin{tabular}{clccccccccccc}
				\hline\hline
				&\multicolumn{11}{c}{Unrestricted ($\alpha(\cdot) \neq 0$)}&\\
				\cline{2-12}
				&$K$&$R^{2}_{K}$&$R^{2}$&$R^{2}_{T,N}$&$R^{2}_{N,T}$&$R^{2}_{f}$&$R^{2}_{f,T,N}$&$R^{2}_{f,N,T}$&$R^{2}_{f,O}$&$R^{2}_{f,T,N,O}$&$R^{2}_{f,N,T,O}$\\
				\cline{2-12}
				&1 & 41.61& 5.94 &3.47  & 3.60 & 5.52  & 2.99  & 3.11  &11.27&7.81&8.93  & 	\\
				&2 & 59.05&  9.56  & 6.17  & 6.91  & 9.18  & 5.67  & 6.33 &14.04&11.31&11.29 &\\
				&3  & 64.47  & 10.42 & 6.78  & 7.96 & 10.03 & 6.27  & 7.38  &14.64&11.93&11.95  &	\\
				&4   & 68.99& 13.83 & 10.26 & 11.52  & 13.40 & 9.80  & 10.90 &15.44&12.98&12.54&\\
				&5 &72.33&   14.32 & 10.73 & 11.98  & 13.91 & 10.29 & 11.38 &15.78&13.43&12.89  & \\
				&6 & 75.35&  14.71 & 10.97 & 12.40 & 14.29 & 10.55 & 11.86 &16.20&14.16	&13.18 &\\
				&7  &77.63& 15.28 & 11.78 & 12.99 & 14.84 & 11.27 & 12.42 &16.45&14.34&13.37  &	\\
				&8 &80.83&  15.44 & 11.98 & 13.16  & 15.10 & 11.59 & 12.73  &16.59&14.50&13.52 & \\
				&9 &82.88& 15.84 & 12.33 & 13.49 & 15.48 & 11.87 & 13.05 &16.86&14.69&13.81 &\\
				&10 &85.61 &16.39 & 12.89 & 13.93  & 15.71 & 11.80 & 13.14 &16.98&14.72	&13.86  & 	\\
				\cline{2-12}
				&$K$&$R^{2}_{\tilde Y}$&$R^{2}_{O}$&$R^{2}_{T,N,O}$&$R^{2}_{N,T,O}$&$p_{\alpha}$&$p_{\text{lin}}$\\
				\cline{2-12}
				&1-10&21.11& 0.59&0.64&0.28&$<1\%$&$<1\%$&\\
				\hline\hline
			\end{tabular}
			\begin{tablenotes}
				\small
				\item[\dag] $K$: the number of factors specified; $R^{2}_{\tilde Y}$: Fama-MacBeth cross-sectional regression $R^2$ ($\%$); $R^2_{K}$: the variation of the Fama-MacBeth managed portfolios $\tilde{Y}_t$ captured by the extracted factors $\hat{f}_t$ ($\%$);  $R^{2}$, $R^{2}_{T,N}$, $R^{2}_{N,T}$: various in-sample $R^2$'s ($\%$), see \eqref{Eqn: R21}-\eqref{Eqn: R23}; $R^{2}_{f}$, $R^{2}_{f,T,N}$, $R^{2}_{f,N,T}$: various in-sample $R^2$'s without $\alpha(\cdot)$ ($\%$), see \eqref{Eqn: R24}-\eqref{Eqn: R26};   $R^{2}_{f,O}$, $R^{2}_{f,T,N,O}$, $R^{2}_{f,N,T,O}$: various out-of-sample fit $R^2$'s ($\%$), see \eqref{Eqn: R21OOS}-\eqref{Eqn: R23OOS};  $R^{2}_O$, $R^{2}_{T,N,O}$, $R^{2}_{N,T,O}$: various out-of-sample predictive $R^2$'s ($\%$), see \eqref{Eqn: R21Predictive}-\eqref{Eqn: R23Predictive}; $p_{\alpha}$ and $p_{\text{lin}}$: the $p$-values of \textit{alpha} test ($\alpha(\cdot)=0$) and model specification test (joint linearity of $\alpha(\cdot)$ and $\beta(\cdot)$), respectively.
			\end{tablenotes}
		\end{threeparttable}
	\end{table}%

	\setlength{\tabcolsep}{3.3pt}
	\begin{table}[!htbp]
		\centering
		\begin{threeparttable}
			\renewcommand{\arraystretch}{1.6}
			\caption{Results under nonlinear specifications of $\alpha(\cdot)$ and $\beta(\cdot)$ with 12 characteristics\tnote{\dag}}\label{Tab: Emprical3}
			\begin{tabular}{clccccccccccccc}
				\hline\hline
				&\multicolumn{11}{c}{Unrestricted ($\alpha(\cdot) \neq 0$)}&\\
				\cline{2-12}
				&$K$&$R^{2}_{K}$&$R^{2}$&$R^{2}_{T,N}$&$R^{2}_{N,T}$&$R^{2}_{f}$&$R^{2}_{f,T,N}$&$R^{2}_{f,N,T}$&$R^{2}_{f,O}$&$R^{2}_{f,T,N,O}$&$R^{2}_{f,N,T,O}$\\
				\cline{2-12}
				&1  & 42.78    & 5.57  & 2.98  & 3.32  & 5.19  & 2.54  & 2.83 &11.08&7.57&8.77 \\
				&2  & 61.36    & 9.56  & 5.97  & 6.87   & 9.18  & 5.51  & 6.26&  13.85	&11.12&	10.99 \\
				&3  & 67.77   & 10.59 & 6.65  & 7.88    & 10.20 & 6.15  & 7.29 &14.66&12.25	&11.83 \\
				&4  & 72.86   & 13.62 & 10.09 & 11.35   & 13.17 & 9.64  & 10.67 &15.39&13.53&12.53  \\
				&5  & 76.92   & 14.14 & 10.43 & 12.01   & 13.73 & 10.01 & 11.48&15.82&13.94	&12.90 \\
				&6  & 80.63   & 14.94 & 11.45 & 12.75  & 14.42 & 10.51 & 12.05 &16.16&14.20	&13.22\\
				&7   & 84.29  & 15.17 & 11.59 & 12.94  & 14.76 & 10.77 & 12.45 &16.57&14.59&13.57\\
				&8   & 87.42   & 15.45 & 11.87 & 13.23  & 15.26 & 11.47 & 12.98&16.94&14.83	&13.88 \\
				&9    & 89.11  & 16.33 & 12.68 & 13.94 & 16.16 & 12.31 & 13.72 &17.12&15.00	&14.09 \\
				&10 & 90.72    & 16.54 & 12.91 & 14.17  & 16.38 & 12.54 & 13.95 &17.30&15.19&14.29\\
				\cline{2-12}
				&$K$&$R^{2}_{\tilde Y}$&$R^{2}_{O}$&$R^{2}_{T,N,O}$&$R^{2}_{N,T,O}$&$p_{\alpha}$&$p_{\text{lin}}$\\
				\cline{2-12}
				&1-10 &20.72& 0.57  & 0.57  & 0.27 &$<1\%$&$<1\%$&\\
				\hline\hline
			\end{tabular}
			\begin{tablenotes}
				\small
				\item[\dag] $K$: the number of factors specified; $R^{2}_{\tilde Y}$: Fama-MacBeth cross-sectional regression $R^2$ ($\%$); $R^2_{K}$: the variation of the Fama-MacBeth managed portfolios $\tilde{Y}_t$ captured by the extracted factors $\hat{f}_t$ ($\%$);  $R^{2}$, $R^{2}_{T,N}$, $R^{2}_{N,T}$: various in-sample $R^2$'s ($\%$), see \eqref{Eqn: R21}-\eqref{Eqn: R23}; $R^{2}_{f}$, $R^{2}_{f,T,N}$, $R^{2}_{f,N,T}$: various in-sample $R^2$'s without $\alpha(\cdot)$ ($\%$), see \eqref{Eqn: R24}-\eqref{Eqn: R26};   $R^{2}_{f,O}$, $R^{2}_{f,T,N,O}$, $R^{2}_{f,N,T,O}$: various out-of-sample fit $R^2$'s ($\%$), see \eqref{Eqn: R21OOS}-\eqref{Eqn: R23OOS};  $R^{2}_O$, $R^{2}_{T,N,O}$, $R^{2}_{N,T,O}$: various out-of-sample predictive $R^2$'s ($\%$), see \eqref{Eqn: R21Predictive}-\eqref{Eqn: R23Predictive}; $p_{\alpha}$ and $p_{\text{lin}}$: the $p$-values of \textit{alpha} test ($\alpha(\cdot)=0$) and model specification test (joint linearity of $\alpha(\cdot)$ and $\beta(\cdot)$), respectively.
			\end{tablenotes}
		\end{threeparttable}
	\end{table}%

\begin{landscape}
\begin{figure}[!htbp]
	\centering
	\includegraphics[width=20cm, height=13cm]{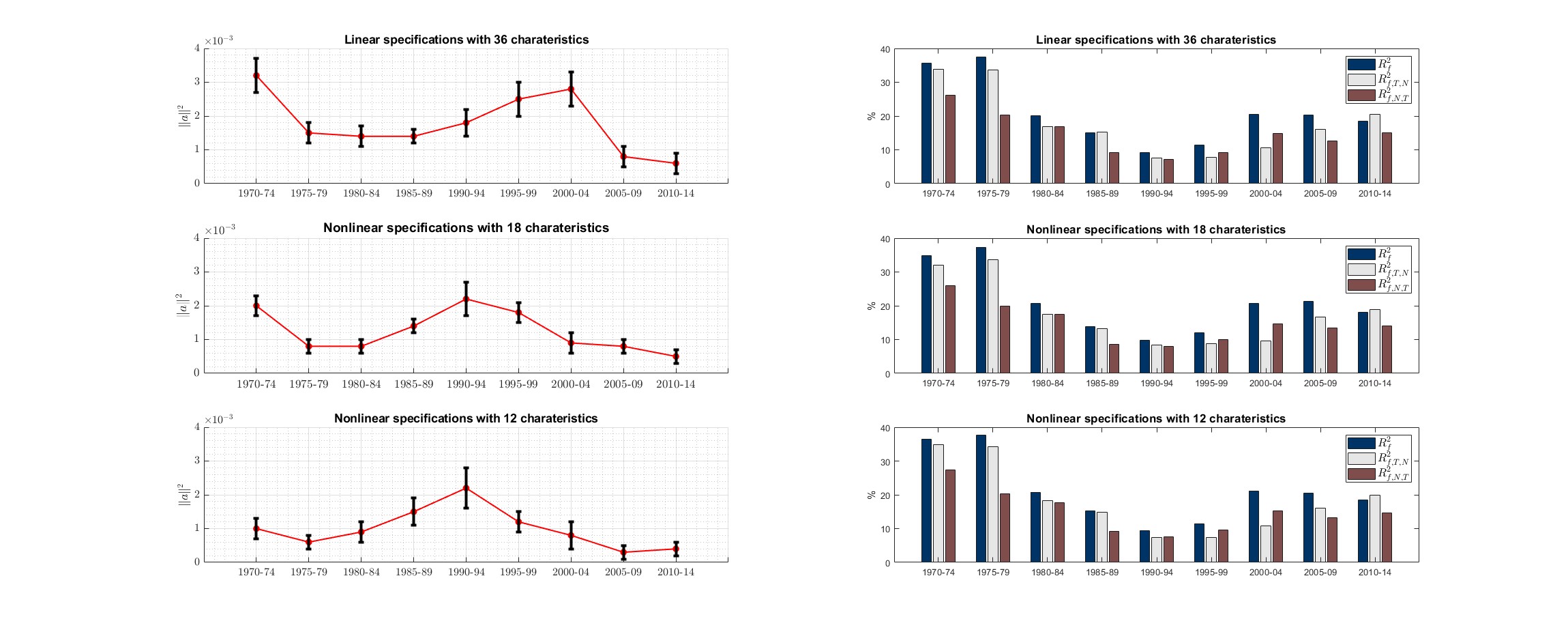}
    \caption{$95\%$ confidence intervals for $\|a\|^2$, $R^{2}_{f}$, $R^{2}_{f,T,N}$, and $R^{2}_{f,N,T}$ (\eqref{Eqn: R24}-\eqref{Eqn: R26}) with $K=10$: subsample analysis}\label{Fig: Empirical6}
\end{figure}
\end{landscape}

\begin{landscape}
\setlength{\tabcolsep}{5pt}
\begin{table}[!htbp]
	\centering
	\begin{threeparttable}
		\renewcommand{\arraystretch}{1.6}
		\caption{In-sample Sharpe ratios \tnote{\dag}}\label{Tab: in_SR}
		\begin{tabular}{ccccc|ccc|ccc|cccccc}
			\hline\hline
			&  &\multicolumn{3}{c}{Regressed-PCA} &\multicolumn{3}{c}{Regressed-PCA S1} & \multicolumn{3}{c}{Regressed-PCA S2}& \multicolumn{5}{c}{IPCA}&\\
			\cline{2-16}
			& $K$ & $SR_{\alpha}$ &$SR_{f}$& $SR_{M}$ & $SR_{\alpha}$ &$SR_{f}$& $SR_{M}$& $SR_{\alpha}$ &$SR_{f}$& $SR_{M}$&$SR_{\alpha}$ &$SR_{f}$& $SR_{M}$ & $SR_{M,\alpha}$& $SR_{M,f}$&  \\
			\cline{2-16}
			& 1 & 3.89 & 0.65 & 3.94 & 4.31 & 0.61 & 4.35  & 3.67 & 0.66 & 3.73  & 1.61 & 1.07 & 2.10 & 1.61 & 1.07& \\
			& 2 & 4.00 & 0.67 & 4.06  & 4.76 & 0.68 & 4.81  & 4.31 & 0.79 & 4.39  & 1.98 & 1.36 & 2.22 & 1.98 & 1.36& \\
			& 3 & 4.02 & 0.67 & 4.08 & 4.78 & 0.73 & 4.84  & 4.32 & 0.80 & 4.39  & 2.64 & 1.07 & 2.85 & 2.64 & 1.05& \\
			& 4 & 4.07 & 0.68 & 4.13& 4.80 & 0.87 & 4.88  & 4.32 & 0.89 & 4.41  & 3.13 & 1.07 & 3.32 & 3.13 & 1.03& \\
			& 5 & 4.10 & 0.69 & 4.15 & 4.75 & 1.38 & 4.95  & 4.28 & 1.24 & 4.45 & 3.00 & 1.11 & 3.21 & 3.00 & 1.07& \\
			& 6 & 4.19 & 0.72 & 4.25 & 4.71 & 1.63 & 4.98  & 3.90 & 2.25 & 4.51 & 2.57 & 1.99 & 3.20 & 2.57 & 1.97& \\
			& 7 & 4.37 & 0.80 & 4.44 & 4.69 & 1.69 & 4.99  & 3.47 & 3.07 & 4.63 & 2.74 & 2.09 & 3.41 & 2.74 & 2.08& \\
			& 8 & 4.48 & 0.88 & 4.56& 4.16 & 2.95 & 5.10  & 3.72 & 3.74 & 5.28   & 2.50 & 2.85 & 3.69 & 2.50 & 2.84& \\
			& 9 & 4.49 & 0.89 & 4.58& 4.07 & 3.11 & 5.12 & 3.74 & 3.75 & 5.29  & 2.41 & 2.77 & 3.56 & 2.41 & 2.76& \\
			& 10 & 4.50 & 0.89 & 4.58& 3.78 & 3.96 & 5.48  & 3.78 & 3.76 & 5.34  & 2.40 & 2.85 & 3.62 & 2.40 & 2.84& \\
			\hline\hline
		\end{tabular}
		\begin{tablenotes}
			\small
			\item[\dag] $K$: the number of factors specified; $SR_{\alpha}$: annualized Sharpe ratios of $\hat{a}^{\prime}\tilde{Y}_t$; $SR_{f}$: annualized Sharpe ratios of $\hat{\mu}^{\prime}\hat{\Sigma}\hat{f}_t$; $SR_{M}$: annualized Sharpe ratios of the combined MVE portfolios on $\hat{a}^{\prime}\tilde{Y}_t$ and $\hat{f}_t$;  $SR_{M,\alpha}$: annualized Sharpe ratios of the component from $\hat{a}^{\prime}\tilde{Y}_t$ in the combined MVE portfolios; $SR_{M,f}$: annualized Sharpe ratios of the component from $\hat{f}_t$ in the combined MVE portfolios.
		\end{tablenotes}
	\end{threeparttable}
\end{table}%
\end{landscape}

\begin{landscape}
\setlength{\tabcolsep}{6pt}
\begin{table}[!htbp]
	\centering
	\begin{threeparttable}
		\renewcommand{\arraystretch}{0.98}
		\caption{Out-of-sample Sharpe ratios \tnote{\dag}}\label{Tab: OOS_SR}
		\begin{tabular}{ccccccccc|ccccccccc}
			\hline\hline
			&  &\multicolumn{7}{c}{Regressed-PCA} & \multicolumn{7}{c}{IPCA}&\\
			\cline{2-16}
			& $K$ & Mean&Std&$SR_{\alpha}$ & $SR_{f,K}$ & $SR_{f}$  & $SR_{M}$ & $SR_{M,\alpha}$ & Mean&Std&$SR_{\alpha}$  & $SR_{f}$  & $SR_{M}$ & $SR_{M,\alpha}$& $SR_{M,f}$ \\
			\cline{2-16}
			& 1 & 1.72&0.54& 3.18 & 0.61&0.61 & 3.25&3.24 & 2.70&2.07&1.31&1.26&1.80&1.33&1.26\\
			& 2 & 1.74&0.52&3.36&-0.12&0.55 & 3.39	&3.38 & 2.38&1.41&1.69&1.44	&2.20&1.87&1.54 \\
			& 3 & 1.77&0.50&3.56&-0.34&0.46 & 3.55	&3.56& 2.01	&0.93&2.16&1.14	&2.33&2.20&1.17 \\
			& 4 & 1.77&0.47&3.74&0.02&0.44 & 3.67&3.68&1.85&0.68&2.70&0.92&2.71&2.74&0.93 \\
			& 5 &1.70&0.44&3.84&0.42&0.53 &3.81	&3.78&1.75&0.62	&2.84&0.98&2.79	&2.84&0.99 \\
			& 6 &1.68&0.44&3.78&0.23&0.57 & 3.80&3.76 &1.39	&0.52&2.68	&1.34&2.70&	2.66&1.31 \\
			& 7 & 1.63&0.44&3.73&0.60&0.68& 3.76&3.69&1.33&	0.52&2.57&	1.42&2.61&	2.52&	1.34 \\
			& 8 &1.61&0.42&3.79&0.24&0.72 & 3.80&3.74&1.22&	0.50&2.44&1.49&	2.67&2.44&	1.43\\
			& 9 &1.61&0.42&3.80&-0.06&0.69 & 3.83&3.76 & 1.23&	0.49&2.51&1.53&2.73&2.52&1.47\\
			& 10 &1.60&0.42&3.82&0.11&0.67& 3.87&3.79 & 1.19&0.48&	2.49&1.72&2.74	&2.45&	1.66 \\
			\cline{2-16}
			&  &\multicolumn{7}{c}{Regressed-PCA S1} & \multicolumn{7}{c}{Regressed-PCA S2}&\\
			\cline{2-16}
            & $K$ &  Mean&Std&$SR_{\alpha}$ & $SR_{f,K}$ & $SR_{f}$  & $SR_{M}$ & $SR_{M,\alpha}$ & Mean&Std&$SR_{\alpha}$ & $SR_{f,K}$ & $SR_{f}$  & $SR_{M}$ & $SR_{M,\alpha}$ \\
            \cline{2-16}
            & 1 &2.46  & 0.69  & 3.54 &0.51&	0.51 & 3.57&3.52 & 3.29  & 0.99  & 3.33 &0.54&	0.54 & 3.27&3.24 \\
            & 2 & 2.39& 0.57 & 4.22 &0.18	&0.53 & 4.16 &4.13 & 3.01  & 0.80  & 3.78&0.47&	0.70 & 3.77&3.72  \\
            & 3 & 2.36  & 0.57  & 4.17 &0.45&	0.64 & 4.13	&4.08& 2.94  & 0.80  & 3.69 &0.51&	0.78 & 3.75	&3.64  \\
            & 4 & 2.19  & 0.53  & 4.12& 0.85&	1.04 & 4.21	&4.08 &2.97  & 0.78  & 3.81&-0.18&	0.59 &3.84&	3.77 \\
            & 5 & 2.19  & 0.51  & 4.26 &-0.03&	0.93 & 4.29	&4.21 & 2.98  & 0.76  & 3.91& -0.04& 0.56 & 3.90&3.87\\
            & 6 & 1.95  & 0.49  & 3.96 &1.23&	1.62 & 4.34	&4.10 & 1.51  & 0.41  & 3.73 &2.47&	2.55& 4.08&3.74  \\
            & 7 &  1.90  & 0.48  & 3.93 &0.45&	1.66 & 4.36	&4.11 & 1.01  & 0.33  & 3.09 &1.87&	3.20 & 4.02&3.26\\
            & 8 &1.73  & 0.47  & 3.66& 1.11&	1.99 & 4.37	&3.92 & 0.73  & 0.22  & 3.36 &1.25&	3.29 & 4.41&3.60  \\
            & 9 &  1.31  & 0.40  & 3.26 &1.81&	2.80& 4.30	&3.49 & 0.71  & 0.20  & 3.62 &0.28&	3.24 & 4.54&3.74\\
            & 10 & 0.88  & 0.28  & 3.14& 1.72&	3.33 & 4.47	&3.50 & 0.69  & 0.18  & 3.90 &0.23&	3.19 & 4.64&3.90 \\
			\hline\hline
		\end{tabular}
		\begin{tablenotes}
			\small
			\item[\dag] $K$: the number of factors specified; Mean: annualized means of the pure-\textit{alpha} portfolios $R_{\alpha,t}$ in \eqref{Eqn: purealpha} ($\%$);  Std: annualized standard deviations of $R_{\alpha,t}$ ($\%$); $SR_{\alpha}$: annualized Sharpe ratios of $R_{\alpha,t}$; $SR_{f,K}$: annualized Sharpe ratios of the $K$th component in $\hat{f}_{t-1,t}$; $SR_{f}$: annualized Sharpe ratios of the MVE factor portfolios $R_{\beta,t}$ in \eqref{Eqn: purebeta};  $SR_{M}$: annualized Sharpe ratios of the combined MVE portfolios on $R_{\alpha,t}$ and $\hat{f}_{t-1,t}$; $SR_{M,\alpha}$: annualized Sharpe ratios of the component from $R_{\alpha,t}$ in the combined MVE portfolios;  $SR_{M,f}$: annualized Sharpe ratios of the component from $\hat{f}_{t-1,t}$ in the combined MVE portfolios.
		\end{tablenotes}
	\end{threeparttable}
\end{table}%
\end{landscape}

\begin{landscape}
	\setlength{\tabcolsep}{4.2pt}
	\begin{table}[!htbp]
		\centering
		\begin{threeparttable}
			\renewcommand{\arraystretch}{0.95}
			\caption{Comparing asset pricing tests: $K=5$\tnote{\dag}}\label{Tab: factor_compare1_main}
			\begin{tabular}{clcccccccccccccc}
				\hline\hline
				&Testing portfolios/Factors & $A|a|$ & $A|t(a)|$ & $Aa^2/V\overline{r}$ & $A\lambda^2/V\overline{r}$ & $As(a)$ & $As(e)$ & $Sh^2(a)$ & $Sh^2(f)$ & $AR^2$ & $GRS$ & $p(GRS)$&\\
				\cline{2-13}
				&\multicolumn{12}{l}{Group I: Regressed-PCA's 36 managed portfolios}\\
				\cline{2-13}
& Regressed-PCA & 0.40 & 3.40 & 0.64 & 0.61 & 0.13 & 2.93 & 2.57 & 0.04 & 28.93 & 34.89 & 0.00 \\
& Regressed-PCA S1 & 0.40 & 3.10 & 0.56 & 0.50 & 0.17 & 3.64 & 2.48 & 0.16 & 17.97 & 30.23 & 0.00 \\
& Regressed-PCA S2 & 0.43 & 3.28 & 0.59 & 0.53 & 0.17 & 3.73 & 2.71 & 0.13 & 16.08 & 33.86 & 0.00 \\
& IPCA & 0.40 & 3.15 & 0.64 & 0.58 & 0.17 & 3.73 & 3.59 & 0.10 & 17.76 & 45.92 & 0.00 \\
& IPCA$\setminus$Regressed-PCA  & 0.50 & 3.34 & 1.03 & 0.96 & 0.17 & 3.94 & 3.49 & 0.06 & 12.76 & 46.40 & 0.00 \\
& FF5 & 0.49 & 3.11 & 0.99 & 0.92 & 0.18 & 4.03 & 2.62 & 0.12 & 8.25 & 33.16 & 0.00 \\
& KNS & 0.50 & 3.34 & 1.07 & 1.01 & 0.17 & 3.99 & 2.56 & 0.04 & 9.92 & 34.89 & 0.00 \\

\cline{2-13}
&\multicolumn{12}{l}{Group II: 100 sorted portfolios (double sort on Size and BM, OP, INV, and MOM)}\\
\cline{2-13}
& Regressed-PCA & 0.85 & 5.04 & 14.14 & 13.60 & 0.17 & 3.97 & 1.00 & 0.04 & 50.81 & 4.26 & 0.00 \\
& Regressed-PCA S1 & 0.57 & 3.21 & 6.71 & 6.13 & 0.18 & 3.89 & 1.27 & 0.16 & 52.33 & 4.85 & 0.00 \\
& Regressed-PCA S2 & 0.44 & 2.42 & 4.18 & 3.59 & 0.18 & 4.00 & 1.35 & 0.13 & 49.67 & 5.32 & 0.00 \\
& IPCA & 0.90 & 11.07 & 16.08 & 15.93 & 0.09 & 1.96 & 4.56 & 0.10 & 86.75 & 18.38 & 0.00 \\
& IPCA$\setminus$Regressed-PCA  & 1.09 & 5.62 & 23.32 & 22.56 & 0.20 & 4.61 & 1.87 & 0.06 & 36.19 & 7.81 & 0.00 \\
& FF5 & 0.41 & 4.81 & 3.55 & 3.39 & 0.09 & 2.05 & 1.76 & 0.12 & 86.91 & 6.98 & 0.00 \\
& KNS & 0.96 & 6.25 & 17.25 & 16.82 & 0.16 & 3.55 & 0.92 & 0.04 & 59.83 & 3.96 & 0.00 \\
\cline{2-13}
&\multicolumn{12}{l}{Group III: 110 sorted portfolios (double sort on Size and Beta, Accruals, NI, and Variance)}\\
\cline{2-13}
& Regressed-PCA & 0.85 & 5.08 & 16.06 & 15.43 & 0.17 & 3.98 & 1.25 & 0.04 & 50.36 & 4.73 & 0.00 \\
& Regressed-PCA S1 & 0.55 & 3.09 & 7.42 & 6.74 & 0.18 & 3.94 & 1.55 & 0.16 & 50.83 & 5.29 & 0.00 \\
& Regressed-PCA S2 & 0.42 & 2.32 & 4.62 & 3.92 & 0.18 & 4.03 & 1.61 & 0.13 & 48.28 & 5.64 & 0.00 \\
& IPCA & 0.88 & 10.63 & 17.68 & 17.50 & 0.09 & 2.00 & 4.56 & 0.10 & 85.72 & 16.31 & 0.00 \\
& IPCA$\setminus$Regressed-PCA & 1.09 & 5.70 & 26.13 & 25.21 & 0.21 & 4.66 & 2.25 & 0.06 & 35.61 & 8.36 & 0.00 \\
& FF5 & 0.43 & 4.88 & 4.22 & 4.03 & 0.09 & 2.08 & 2.09 & 0.12 & 86.06 & 7.37 & 0.00 \\
& KNS & 0.95 & 6.15 & 19.41 & 18.90 & 0.16 & 3.59 & 1.11 & 0.04 & 58.32 & 4.22 & 0.00 \\
\hline\hline
\end{tabular}
\begin{tablenotes}
				\small
				\item[\dag]  $A|a|$: average absolute intercept;  $A|t(a)|$: average absolute $t$-statistic for the intercepts;  $Aa^2/V\overline{r}$: average squared intercept over the cross-section variance of $\overline{r}$, average returns of the testing portfolios; $A\lambda^2/V\overline{r}$: average difference between each squared intercept and its squared standard error divided by the variance of $\overline{r}$; $As(a)$: average standard error of the intercepts; $As(e)$: average residual standard deviation; $Sh^2(a)$: maximized squared Sharpe ratio for the intercepts; $Sh^2(f)$: maximized squared Sharpe ratio for the factors; $AR^2$: average regression $R^2$ (\%);  $GRS$: $GRS$ statistic of \citet{Gibbonsetal_efficiency_1989}; $p(GRS)$: $p$-value of $GRS$.
			\end{tablenotes}
		\end{threeparttable}
	\end{table}%
\end{landscape}

\begin{landscape}
	\setlength{\tabcolsep}{4.2pt}
	\begin{table}[!htbp]
		\centering
		\begin{threeparttable}
			\renewcommand{\arraystretch}{0.95}
			\caption{Comparing asset pricing tests: $K=5$ (continued)\tnote{\dag}}\label{Tab: factor_compare1_main_continue}
			\begin{tabular}{clcccccccccccccc}
				\hline\hline
				&Testing portfolios/Factors & $A|a|$ & $A|t(a)|$ & $Aa^2/V\overline{r}$ & $A\lambda^2/V\overline{r}$ & $As(a)$ & $As(e)$ & $Sh^2(a)$ & $Sh^2(f)$ & $AR^2$ & $GRS$ & $p(GRS)$&\\
				\cline{2-13}
				&\multicolumn{12}{l}{Group IV: IPCA's 36 managed portfolios}\\
				\cline{2-13}
& Regressed-PCA & 0.05 & 3.44 & 0.87 & 0.82 & 0.01 & 0.33 & 1.50 & 0.04 & 25.22 & 20.30 & 0.00 \\
& Regressed-PCA S1 & 0.06 & 5.12 & 1.41 & 1.38 & 0.01 & 0.27 & 1.85 & 0.16 & 47.24 & 22.50 & 0.00 \\
& Regressed-PCA S2 & 0.06 & 5.30 & 1.32 & 1.29 & 0.01 & 0.25 & 1.83 & 0.13 & 51.44 & 22.90 & 0.00 \\
& IPCA & 0.07 & 6.72 & 1.53 & 1.51 & 0.01 & 0.21 & 3.06 & 0.10 & 64.08 & 39.14 & 0.00 \\
& IPCA$\setminus$Regressed-PCA  & 0.06 & 4.53 & 1.22 & 1.17 & 0.01 & 0.31 & 2.20 & 0.06 & 39.29 & 29.28 & 0.00 \\
& FF5 & 0.04 & 3.32 & 0.71 & 0.67 & 0.01 & 0.27 & 1.42 & 0.12 & 50.00 & 17.96 & 0.00 \\
& KNS & 0.04 & 3.87 & 0.76 & 0.73 & 0.01 & 0.26 & 1.47 & 0.04 & 52.97 & 20.03 & 0.00 \\

\cline{2-13}
&\multicolumn{12}{l}{Group V: P1\&10 of sorted portfolios (single sort on 55 characteristics in \citet{Kozaketal_Interpreting_2018}, 110 portfolios)}\\
\cline{2-13}
& Regressed-PCA & 0.73 & 4.05 & 7.24 & 6.82 & 0.19 & 4.27 & 1.36 & 0.04 & 43.38 & 5.16 & 0.00 \\
& Regressed-PCA S1 & 0.47 & 2.61 & 3.78 & 3.37 & 0.19 & 4.01 & 1.42 & 0.16 & 49.01 & 4.82 & 0.00 \\
& Regressed-PCA S2 & 0.37 & 1.98 & 2.51 & 2.10 & 0.19 & 4.13 & 1.44 & 0.13 & 45.95 & 5.02 & 0.00 \\
& IPCA & 0.73 & 6.43 & 7.71 & 7.53 & 0.12 & 2.66 & 3.15 & 0.10 & 77.14 & 11.26 & 0.00 \\
& IPCA$\setminus$Regressed-PCA  & 0.90 & 4.68 & 10.96 & 10.45 & 0.21 & 4.67 & 1.85 & 0.06 & 34.10 & 6.87 & 0.00 \\
& FF5 & 0.40 & 4.50 & 2.40 & 2.25 & 0.11 & 2.34 & 2.54 & 0.12 & 82.37 & 8.97 & 0.00 \\
& KNS & 0.92 & 5.77 & 10.54 & 10.24 & 0.16 & 3.66 & 1.31 & 0.04 & 56.83 & 5.01 & 0.00 \\

\cline{2-13}
&\multicolumn{12}{l}{Group VI: P1\&10 of sorted portfolios (single sort on 36 characteristics, 72 portfolios)}\\
\cline{2-13}
& Regressed-PCA & 0.69 & 3.55 & 5.72 & 5.28 & 0.21 & 4.71 & 0.81 & 0.04 & 41.77 & 5.13 & 0.00 \\
& Regressed-PCA S1 & 0.49 & 2.52 & 3.25 & 2.83 & 0.20 & 4.36 & 0.96 & 0.16 & 48.78 & 5.40 & 0.00 \\
& Regressed-PCA S2 & 0.39 & 1.89 & 2.35 & 1.91 & 0.21 & 4.50 & 1.02 & 0.13 & 45.53 & 5.94 & 0.00 \\
& IPCA & 0.70 & 5.43 & 6.11 & 5.90 & 0.14 & 3.07 & 1.84 & 0.10 & 74.60 & 10.92 & 0.00 \\
& IPCA$\setminus$Regressed-PCA & 0.88 & 4.20 & 9.17 & 8.62 & 0.23 & 5.13 & 1.34 & 0.06 & 33.26 & 8.29 & 0.00 \\
& FF5 & 0.42 & 4.38 & 2.48 & 2.29 & 0.12 & 2.73 & 2.48 & 0.12 & 80.46 & 14.53 & 0.00 \\
& KNS & 0.91 & 5.61 & 9.35 & 9.07 & 0.17 & 3.79 & 0.88 & 0.04 & 59.94 & 5.54 & 0.00 \\
\hline\hline
			\end{tabular}
		\begin{tablenotes}
				\small
				\item[\dag]  $A|a|$: average absolute intercept;  $A|t(a)|$: average absolute $t$-statistic for the intercepts;  $Aa^2/V\overline{r}$: average squared intercept over the cross-section variance of  $\overline{r}$, average returns of the testing portfolios; $A\lambda^2/V\overline{r}$: average difference between each squared intercept and its squared standard error divided by the variance of $\overline{r}$; $As(a)$: average standard error of the intercepts; $As(e)$: average residual standard deviation; $Sh^2(a)$: maximized squared Sharpe ratio for the intercepts; $Sh^2(f)$: maximized squared Sharpe ratio for the factors; $AR^2$: average regression $R^2$ (\%);  $GRS$: $GRS$ statistic of \citet{Gibbonsetal_efficiency_1989}; $p(GRS)$: $p$-value of $GRS$.
			\end{tablenotes}
		\end{threeparttable}
	\end{table}%
\end{landscape}


\addcontentsline{toc}{section}{References}
\putbib
\end{bibunit}

\clearpage
\newpage

\begin{bibunit}
\begin{appendices} \sloppy
\bookmarksetup{open=false}
\allowdisplaybreaks 
\titleformat{\section}{\Large\center}{{\sc Appendix} \thesection}{0.25em}{- }

\renewcommand{\theass}{\Alph{section}.\arabic{ass}}
\setcounter{page}{1}
\setcounter{section}{0}
\setcounter{equation}{0}
\setcounter{table}{0}
\setcounter{figure}{0}
\setcounter{footnote}{1}
\numberwithin{equation}{section}
\numberwithin{table}{section}
\numberwithin{figure}{section}
\renewcommand{\thetable}{\thesection.\Roman{table}}
\renewcommand\thefigure{\thesection.\arabic{figure}}

\emptythanks
\phantomsection
\pdfbookmark[1]{Appendix Title}{title1}
\title{\vspace{-2.4cm} Online Appendix to ``Semiparametric Conditional Factor Models in Asset Pricing''}
\date{\today}
\maketitle
\vspace{-0.3in}

This online appendix is organized as follows. Appendix \ref{Sec: 6} introduces two estimators for the number of factors. Appendix \ref{App: assumptions} complies the assumptions. Appendix \ref{App: proofs of main} provides proofs of the theoretical results. Appendix \ref{App: Sec: C} presents additional discussions. Appendix \ref{App: Sec: E} shows simulation results, and Appendix \ref{App: Sec: F} collects additional empirical findings.

\section{Determining the Number of Factors}\label{Sec: 6}
In this appendix, we develop two estimators for the number of factors $K$: one based on maximizing the ratio of two adjacent eigenvalues \citep{AhnHorenstein_EigenvalueRatio_2013}, and another by counting the number of ``large'' eigenvalues \citep{BaiNg_NumberofFactors_2002}. To define the estimators, let $\lambda_{k}(\tilde{Y}M_T\tilde{Y}^{\prime}/T)$ denote the $k$th largest eigenvalue of the $JM\times JM$ matrix $\tilde{Y}M_T\tilde{Y}^{\prime}/T$. The first one is given by:
\begin{align}\label{Eqn: Kestiamtor}
\hat{K} = \amax_{1\leq k\leq JM/2}\frac{\lambda_{k}(\tilde{Y}M_T\tilde{Y}^{\prime}/T)}{\lambda_{k+1}(\tilde{Y}M_T\tilde{Y}^{\prime}/T)}.
\end{align}
Here, $\hat{K}$ is constrained to between $1$ and $JM/2$, which is not restrictive because we assume $K\geq 1$ is fixed and $J\to\infty$. The second one is defined as:
\begin{align}\label{Eqn: Kestiamtoralt}
\tilde{K}= \#\{1\leq k\leq JM: \lambda_{k}(\tilde{Y}M_T\tilde{Y}^{\prime}/T)\geq \lambda_{NT}\},
\end{align}
where $\# A$ denotes the cardinality of set $A$ and $0<\lambda_{NT}\to 0$ is a tuning parameter.

We differ from \citet{AhnHorenstein_EigenvalueRatio_2013} and \citet{BaiNg_NumberofFactors_2002} in two key aspects. First, in the presence of time-varying $Z_t$, methods based on the original data $\{Y_{t}\}_{t\leq T}$ or the projected data $\{\Phi(Z_t)(\Phi(Z_t)^{\prime}\Phi(Z_t))^{-1}\Phi(Z_t)^{\prime}Y_t\}_{t\leq T}$ may fail to estimate $K$. Instead, we work on the regressed data $\{\tilde{Y}_t\}_{t\leq T}$, where the regressed data matrix $\tilde{Y}M_T$ is approximately equal to $BF^{\prime}M_{T}$, whose rank is qual to the number of factors. Second, we allow $N/T\to\infty$.

\begin{thm}\label{Thm: Numberfactors}
(A) Suppose Assumptions \ref{Ass: Basis}-\ref{Ass: DGP}, \ref{Ass: Improvedrates}(i), and \ref{Ass: Numberfactors} hold. Let $\hat{K}$ be given in \eqref{Eqn: Kestiamtor}.  Assume (i) $N\to\infty$; (ii) $T\to\infty$; (iii) $J\to\infty$ with $J=o(\min\{\sqrt{N},\sqrt{T}\})$ and $NJ^{-2\kappa}=o(1)$. Then
\begin{align*}
P(\hat{K}=K)\to 1.
\end{align*}
(B) Suppose Assumptions \ref{Ass: Basis}-\ref{Ass: DGP} hold. Let $\tilde{K}$ be given in \eqref{Eqn: Kestiamtoralt}.  Assume (i) $N\to\infty$; (ii) $T\geq K+1$; (iii) $J\to\infty$ with $J=o(\sqrt{N})$; (iv) $0<\lambda_{NT}\to 0$ and $\lambda_{NT}\min\{N/J, J^{2\kappa}\}\to \infty$. Then
\begin{align*}
P(\tilde{K}=K)\to 1.
\end{align*}
\end{thm}

Theorem \ref{Thm: Numberfactors} shows that both $\hat{K}$ and $\tilde{K}$ are consistent estimators of $K$. The consistency of $\hat{K}$ requires $T\to\infty$, while the consistency of $\tilde{K}$ does not. The latter relies on the choice of $\lambda_{NT}$. In practice, $\hat{K}$ is recommended when $T$ is large, while $\tilde{K}$ is preferred when $T$ is small.

\section{Assumptions} \label{App: assumptions}

\begin{ass}[Basis functions]\label{Ass: Basis}
(i) There are positive constants $c_{\min}$ and $c_{\max}$ such that: with probability approaching one (as $N\to\infty$),
\[c_{\min}<\min_{t\leq T}\lambda_{\min}(\hat{Q}_t)\leq \max_{t\leq T}\lambda_{\max}(\hat{Q}_t)<c_{\max},\]
where $\hat{Q}_t=\Phi(Z_t)^{\prime}\Phi(Z_t)/N$; (ii) $\max_{m\leq M, j\leq J, i\leq N, t\leq T} E[\phi^{2}_{j}(z_{it,m})]<\infty$.
\end{ass}

Since $\hat{Q}_t = \sum_{i=1}^{N}\phi(z_{it})\phi(z_{it})^{\prime}/N$ is a $JM\times JM$ matrix with $JM$ much smaller than $N$, Assumption \ref{Ass: Basis}(i) can follow from the law of large numbers for finite $T$ and its uniform variant for $T\to\infty$; see Proposition \ref{Pro: JustNum1} for a set of sufficient conditions. The conditions can be easily verified for B-splines, Fourier series, and polynomials basis functions. In particular, we allow $Z_t$ to be nonstationary over $t$. When $Z_t$ is not changing over $t$, Assumption \ref{Ass: Basis} reduces to Assumptions 3.3 of \citet{Fanetal_ProjectedPCA_2016}.

\begin{ass}[Factor loading functions and factors]\label{Ass: LoadingsFactors}
There are positive constants $d_{\min}$ and $d_{\max}$ such that: (i) $d_{\min}<\lambda_{\min}(B'B)\leq \lambda_{\max}(B'B)<d_{\max}$; (ii) $\max_{t\leq T}\|f_{t}\|\hspace{-0.05cm}<\hspace{-0.05cm}d_{\max}$; (iii) $\lambda_{\min}(F^{\prime}M_TF/T)\hspace{-0.05cm}>\hspace{-0.05cm}d_{\min}$; (iv) $\max_{k\leq K,m\leq M}\sup_{z}|\delta_{km,J}(z)|$ $=O(J^{-\kappa})$ and $\max_{m\leq M}\sup_{z}|r_{m,J}(z)|=O(J^{-\kappa})$ for some constant $\kappa>1/2$.
\end{ass}

Assumption \ref{Ass: LoadingsFactors}(i) is similar to the \textit{pervasive} condition on the factor loadings in \citet{StockWatson_PCA_2002}. Similar assumptions also are imposed in Assumption B of \citet{Bai_Inferential_2003} and Assumption 4.1(ii) of \citet{Fanetal_ProjectedPCA_2016}. For simplicity of presentation, we assume $f_t$'s are nonrandom fixed parameters by following \citet{BaiLi_StatisticalAna_2012}. Our analysis holds if they are random variables. In this case, we assume $f_t$'s to be independent of all other variables, and all stochastic statements can then regarded as conditioning on $f_t$'s realizations.  Since the dimension of $B$ is $JM\times K$, Assumption \ref{Ass: LoadingsFactors}(i) requires $JM\geq K$. Since the rank of $M_T$ is $T-1$, Assumption \ref{Ass: LoadingsFactors}(iii) requires $T\geq K+1$, which implies $T\geq 2$. These two requirements are not restrictive, since we assume $K$ is fixed. Assumption \ref{Ass: LoadingsFactors}(iv) is standard in the sieve literature. It can be easily satisfied by using B-splines or polynomials basis functions under certain smoothness of $\alpha(\cdot)$ and $\beta(\cdot)$; see, for example, \citet{Lorentz_Approximation_1986} and \citet{Chen_Handbook_2007}.

\begin{ass}[Data generating process]\label{Ass: DGP}
\hspace{-0.2cm}(i)$\{\varepsilon_{t}\}_{t\leq T}$ is independent of $\{Z_{t}\}_{t\leq T}$; (ii) $E[\varepsilon_{it}]=0$ for all $i\leq N$ and $t\leq T$; (iii) there is $0<C_1<\infty$ such that
\[\max_{i\leq N,t\leq T}\sum_{j=1}^{N}|E[\varepsilon_{it}\varepsilon_{jt}]|<C_1 \text{ and } \frac{1}{NT}\sum_{i=1}^{N}\sum_{j=1}^{N}\sum_{t=1}^{T}\sum_{s=1}^{T}|E[\varepsilon_{it}\varepsilon_{js}]|<C_1.\]
\end{ass}

Assumption \ref{Ass: DGP}(iii) requires $\{\varepsilon_{it}\}_{i\leq N,t\leq T}$ to be weakly dependent over both $i$ and $t$, and is commonly imposed for high-dimensional factor analysis; see, for example, \citet{StockWatson_PCA_2002}, \citet{Bai_Inferential_2003}, and \citet{Fanetal_ProjectedPCA_2016}. When $Z_t$ is not changing over $t$, Assumption \ref{Ass: DGP} reduces to Assumptions 3.4 (i) and (iii) of \citet{Fanetal_ProjectedPCA_2016}.

\begin{ass}[Intercept function]\label{Ass: Intercept}
$a^{\prime}B=0$ and $\|a\|<C_0$ for some $0<C_0<\infty$.
\end{ass}

Assumption \ref{Ass: Intercept} is needed for the identification of $\alpha(\cdot)$. Similar assumption is imposed in \citet{Connoretal_EfficientFFFactor_2012} and Assumption 3.1(i) of \citet{Kimetal_Arbitrage_2019}.

\begin{ass}[Rate of convergence]\label{Ass: Improvedrates}
(i) $\max_{m\leq M, j\leq J, i\leq N, t\leq T}E[\phi^{4}_{j}(z_{it,m})]<\infty$; (ii) $0\hspace{-0.05cm}<\hspace{-0.05cm}\min_{i\leq N, t\leq T}\lambda_{\min}(Q_{it})\hspace{-0.05cm}\leq \hspace{-0.05cm}\max_{i\leq N, t\leq T}\lambda_{\max}(Q_{it})\hspace{-0.05cm}<\hspace{-0.05cm}\infty$, where $Q_{it}=E[\phi(z_{it})\phi(z_{it})^{\prime}]$; (iii) $\{z_{it}\}_{i\leq N, t\leq T}$ are independent across $i\leq N$; (iv) there is $0<C_2<\infty$ such that
\[\max_{t\leq T}\frac{1}{N^{2}}\sum_{i=1}^{N}\sum_{j=1}^{N}\sum_{k=1}^{N}\sum_{\ell=1}^{N}|E[\varepsilon_{it}\varepsilon_{jt}\varepsilon_{kt}\varepsilon_{\ell t}]|< C_2\]
and
\[\frac{1}{N^{2}T}\sum_{t=1}^{T}\left(\sum_{s=1}^{T}\sum_{i=1}^{N}\sum_{j=1}^{N}|E[\varepsilon_{it}\varepsilon_{js}]|\right)^2< C_2.\]
\end{ass}

Assumptions \ref{Ass: Basis}-\ref{Ass: Intercept} allow us to establish a preliminary rate of the estimators in Theorem \ref{Thm: Rate}. Assumption \ref{Ass: Improvedrates} is an additional assumption that we need to establish a fast rate in Theorem \ref{Thm: ImprovedRates}. Assumption \ref{Ass: Improvedrates}(i) strengthens Assumption \ref{Ass: Basis}(ii). Assumption \ref{Ass: Improvedrates}(ii) requires that the second moment matrix $E[\phi(z_{it})\phi(z_{it})^{\prime}]$ is bounded and nonsingular for all $i$ and $t$, which is widely used in the sieve literature; see, for example, \citet{Newey_SeriesEstimator_1997} and \citet{Huang_Series_1998}. Assumption \ref{Ass: Improvedrates}(iii) is commonly imposed in the sieve literature, which is used to justify the asymptotic convergence of $\hat{Q}_t$. Assumption \ref{Ass: Improvedrates}(iv) allows for weak dependence of $\{\varepsilon_{it}\}_{i\leq N,t\leq T}$ over both $i$ and $t$. The second condition is similar to the second condition in Assumption \ref{Ass: DGP}(iii); both are satisfied if $\max_{t\leq T}\sum_{s=1}^{T}\sum_{i=1}^{N}\sum_{j=1}^{N}$ $|E[\varepsilon_{it}\varepsilon_{js}]/N$ is bounded.

\begin{ass}[Asymptotic distribution]\label{Ass: Asym}
(i) $(F^{\prime}M_TF/T)B^{\prime}B$ has distinct eigenvalues; (ii) $\{\varepsilon_{it}\}_{i\leq N, t\leq T}$ are independent across $i\leq N$; (iii) there is $0<C_3<\infty$ such that
\[\max_{i\leq N}\frac{1}{T^{2}}\sum_{t=1}^{T}\sum_{s=1}^{T}\sum_{u=1}^{T}\sum_{v=1}^{T}|E[\varepsilon_{it}\varepsilon_{is}\varepsilon_{iu}\varepsilon_{iv}]|< C_3.\]
\end{ass}

Assumption \ref{Ass: Asym} is needed in Theorem \ref{Thm: AsymDis}. The distinct eigenvalue condition in Assumption \ref{Ass: Asym}(i) is necessary to establish the asymptotic normality, as known in the literature; see, for example, \citet{Bai_Inferential_2003} and \citet{ChenFang_ImprovedInference_2017}. Assumption \ref{Ass: Asym}(ii) imposes independence of $\{\varepsilon_{it}\}_{i\leq N,t\leq T}$ across $i$ for simplicity. Cross-sectional independence is also imposed in \citet{Fanetal_ProjectedPCA_2016} for studying specification test (Theorem 5.1). It is straightforward to modify the proof of Theorem \ref{Thm: AsymDis} to allow for clsuter-type dependence of $\{\varepsilon_{it}\}_{i\leq N,t\leq T}$ across $i$.\footnote{Assumption \ref{Ass: Asym}(ii) permits the use of Yurinskii’s coupling. Alternatively, one could apply the coupling method from \citet{LiLiao_NonparametricTimeSeries_2019}, which accommodates mixingale-type dependence. However, it remains unclear which dependence structure is more suitable for asset pricing models. For this reason, we adhere to Assumption \ref{Ass: Asym}(ii) in this context.} Assumption \ref{Ass: Asym}(iii) allows for weak dependence of $\{\varepsilon_{it}\}_{i\leq N,t\leq T}$ over $t$.
\begin{ass}[Bootstrap]\label{Ass: Boot}
(i) $\{w_i\}_{i\leq N}$ is a sequence of i.i.d. positive random variables with $E[w_i]=1$ and $var(w_i)=\omega_0>0$, and is independent of $\{Z_{t},\varepsilon_t\}_{t\leq T}$; (ii) there are positive constants $e_{\min}$ and $e_{\max}$ such that: with probability approaching one (as $N\to\infty$),
\[e_{\min}<\min_{t\leq T}\lambda_{\min}(\hat{Q}_t^{\ast})\leq \max_{t\leq T}\lambda_{\max}(\hat{Q}_t^{\ast})<e_{\max},\]
where $\hat{Q}_t^{\ast}=\Phi(Z_t)^{\ast\prime}\Phi(Z_t)/N$; (iii) $\lambda_{\min}(\Omega)>0$.
\end{ass}

Assumption \ref{Ass: Boot} is needed in Theorem \ref{Thm: Boot}. Assumption \ref{Ass: Boot}(i) defines the bootstrap weight $w_i$ for each $i$. It is straightforward to extend the bootstrap to accommodate clsuter-type dependence of $\{\varepsilon_{it}\}_{i\leq N,t\leq T}$ across $i$ by utilizing the same weight within each cluster. Since $\hat{Q}_t^{\ast} = \sum_{i=1}^{N}\phi(z_{it})\phi(z_{it})^{\prime}w_{i}/N$ is a $JM\times JM$ matrix with $JM$ much smaller than $N$, Assumption \ref{Ass: Boot}(ii) can follow from the law of large numbers for finite $T$ and its uniform variant for $T\to\infty$, similar to Assumption \ref{Ass: Basis}(i). Assumption \ref{Ass: Boot}(iii) requires nonsingularity of the variance-covariance matrix $\Omega$.

\begin{ass}[Specification test]\label{Ass: SpecTest}
(i) There are positive constants $g_{\min}$ and $g_{\max}$ such that: with probability approaching one (as $N\to\infty$),
\[g_{\min}<\min_{t\leq T}\lambda_{\min}(Z_{t}^{\prime}Z_t/N)\leq \max_{t\leq T}\lambda_{\max}(Z_{t}^{\prime}Z_t/N)<g_{\max},\]
(ii) $\max_{i\leq N,t\leq T}E[\|z_{it}\|^{4}]<\infty$; (iii) $\min_{i\leq N, t\leq T}\lambda_{\min}(E[z_{it}z_{it}^{\prime}])>0$; (iv) with probability approaching one (as $N\to\infty$),
\[g_{\min}<\min_{t\leq T}\lambda_{\min}(Z_{t}^{\ast\prime}Z_t/N)\leq \max_{t\leq T}\lambda_{\max}(Z_{t}^{\ast\prime}Z_t/N)<g_{\max};\]
(v) $\sup_{z}|\alpha(z)|<\infty$ and $\sup_{z}\|\beta(z)\|<\infty$.
\end{ass}

Assumption \ref{Ass: SpecTest} is needed in Theorem \ref{Thm: SpecTest}. Assumptions \ref{Ass: SpecTest}(i)-(iv) are analogous to Assumptions \ref{Ass: Basis}(i), \ref{Ass: Improvedrates}(i), (ii), and \ref{Ass: Boot}(ii), respectively. When $z_{it}$ is included as a part of $\phi(z_{it})$, which is true in the case of polynomial basis functions, the former are implied by the latter ones. In this case, Assumptions \ref{Ass: SpecTest}(i)-(iv) thus are redundant.

\begin{ass}[Determination of $K$]\label{Ass: Numberfactors}
(i) $0<\min_{t\leq T}\lambda_{\min}(E[\varepsilon_{t}\varepsilon_{t}^{\prime}])\leq \max_{t\leq T}$ $\lambda_{\max}(E[\varepsilon_{t}\varepsilon_{t}^{\prime}])<\infty$; ii) there is $0<C_4<\infty$ such that
\[\frac{1}{N^{2}T+T^{2}N}\sum_{t=1}^{T}\sum_{s=1}^{T}\sum_{i=1}^{N}\sum_{j= 1}^{N}\sum_{k=1}^{N}\sum_{\ell= 1}^{N}|cov(\varepsilon_{it}\varepsilon_{jt},\varepsilon_{ks}\varepsilon_{\ell s})|< C_4.\]
\end{ass}

Assumption \ref{Ass: Numberfactors} is needed in Theorem \ref{Thm: Numberfactors}(i). Assumption \ref{Ass: Numberfactors}(i) requires that the covariance matrix $E[\varepsilon_{t}\varepsilon_{t}^{\prime}]$ is bounded and nonsingular for all $t$. In particular, $\max_{t\leq T}\lambda_{\max}(E[\varepsilon_{t}\varepsilon_{t}^{\prime}])<\infty$ allows for weak dependence of $\{\varepsilon_{it}\}_{i\leq N,t\leq T}$ across $i$. When $\{\varepsilon_{it}\}_{i\leq N,t\leq T}$ are independent across $i$, the condition is satisfied when $\min_{i\leq N, t\leq T}E[\varepsilon_{it}^{2}]>0$ and $\max_{i\leq N, t\leq T}E[\varepsilon_{it}^{2}]<\infty$. Assumption \ref{Ass: Numberfactors}(ii) allows for weak dependence of $\{\varepsilon_{it}\}_{i\leq N,t\leq T}$ over both $i$ and $t$; see Proposition \ref{Pro: JustNum2} for a set of sufficient conditions.
Assumption \ref{Ass: Numberfactors} is distinct from Assumption 6.1 in \citet{Fanetal_ProjectedPCA_2016}, which may not be easy to verify.

\section{Proofs of Theoretical Results}\label{App: proofs of main}

\subsection{A Preliminary Rate of Convergence}
We first establish a preliminary convergence rate of $\hat{a}$, $\hat{B}$, $\hat{F}$, $\hat{\alpha}(\cdot)$, and $\hat{\beta}(\cdot)$, as an intermediate step toward proving Theorem \ref{Thm: ImprovedRates}.
\begin{thm}\label{Thm: Rate}
Suppose Assumptions \ref{Ass: Basis}-\ref{Ass: Intercept} hold. Let $\hat{a}$, $\hat{B},\hat{F}$, $\hat{\alpha}(\cdot)$, and $\hat{\beta}(\cdot)$ be given in \eqref{Eqn: Estimators}. Assume (i) $N\to\infty$; (ii) $T\geq K+1$; (iii) $J\to\infty$ with $J=o(\sqrt{N})$. Then
\begin{align*}
\|\hat{a} - a\|^{2}&=O_{p}\left(\frac{1}{J^{2\kappa}}+\frac{J^2}{N^2}+\frac{{J}}{{NT}}\right),\notag\\
\|\hat{B} - B H\|^{2}_{F}&=O_{p}\left(\frac{1}{J^{2\kappa}}+\frac{J^2}{N^2}+\frac{{J}}{{NT}}\right),\notag\\
\frac{1}{T}\|\hat{F}-F(H^{\prime})^{-1}\|_{F}^{2}&=O_{p}\left(\frac{1}{J^{2\kappa}}+\frac{{J}}{{N}}\right),\notag\\
\sup_{z}|\hat{\alpha}(z)-\alpha(z)|^{2}&=O_{p}\left(\frac{1}{J^{2\kappa-1}}+\frac{J^3}{N^2}+\frac{{J^2}}{{NT}}\right)\max_{j\leq J}\sup_{z}|\phi_{j}(z)|^{2},\notag\\
\sup_{z}\|\hat{\beta}(z)-H^{\prime}\beta(z)\|^{2}&=O_{p}\left(\frac{1}{J^{2\kappa-1}}+\frac{J^3}{N^2}+\frac{{J^2}}{{NT}}\right)\max_{j\leq J}\sup_{z}|\phi_{j}(z)|^{2}.
\end{align*}
\end{thm}

\noindent{\sc Proof:}  Let us begin by defining some notation. For $A_t = \Delta_t\equiv R(Z_t)+\Delta(Z_{t})f_{t}$ and $\varepsilon_t$, let $\tilde{A}_{t} \equiv (\Phi(Z_t)^{\prime}\Phi(Z_t))^{-1}\Phi(Z_t)^{\prime}A_t$ . Let $\tilde{\Delta}\equiv (\tilde{\Delta}_1,\ldots, \tilde{\Delta}_T)$ and $\tilde{E}\equiv (\tilde{\varepsilon}_1,\ldots, \tilde{\varepsilon}_T)$. Then \eqref{Eqn: Model: Sieve: Vectors: Regressed} can be written as
\begin{align}\label{Eqn: Thm: Rate: 1}
\tilde{Y} = a1_{T}^{\prime} + BF^{\prime} + \tilde{\Delta} + \tilde{E},
\end{align}
where $1_{T}$ denote a $T\times 1$ vector of ones. Recall $M_T= I_{T} - 1_T1_T^{\prime}/T$. Post-multiplying \eqref{Eqn: Thm: Rate: 1} by $M_T$ to remove $a$, we thus obtain
\begin{align}\label{Eqn: Thm: Rate: 2}
\tilde{Y}M_T = B(M_TF)^{\prime} + \tilde{\Delta}M_T + \tilde{E}M_T.
\end{align}
Let $V$ be a $K\times K$ diagonal matrix of the first $K$ largest eigenvalues of $\tilde{Y}M_T\tilde{Y}^{\prime}/T$. By the definitions of $\hat{B}$ and $\hat{F}$, $(\tilde{Y}M_T\tilde{Y}^{\prime}/T)\hat{B} = \hat{B}V$ and $M_T\hat{F} = M_T \tilde{Y}^{\prime}\hat{B}$. Thus, $\hat{F}^{\prime}M_T\hat{F}/T =\hat{B}^{\prime}(\tilde{Y}M_T\tilde{Y}^{\prime}/T)\hat{B}=V$ and $H = (F^{\prime} M_T \hat{F})(\hat{F}^{\prime} M_T \hat{F})^{-1}=(F^{\prime}M_T\tilde{Y}^{\prime}\hat{B}/T)V^{-1}$. We may substitute \eqref{Eqn: Thm: Rate: 2} to $(\tilde{Y}M_T\tilde{Y}^{\prime}/T)\hat{B} = \hat{B}V$ to obtain
\begin{align}\label{Eqn: Thm: Rate: 3}
\hat{B} - B H = [(\tilde{\Delta} + \tilde{E})M_{T}\tilde{Y}^{\prime}/T]\hat{B}V^{-1} = \sum_{j=1}^{6}D_{j}\hat{B}V^{-1},
\end{align}
where $D_1=\tilde{\Delta}M_TF B^{\prime}/T$, $D_2 = \tilde{\Delta}M_T\tilde{\Delta}^{\prime}/T$, $D_3 = D_{6}^{\prime}=\tilde{\Delta}M_T\tilde{E}^{\prime}/T$, $D_4 =\tilde{E}M_TFB^{\prime}/T$, and $D_5 = \tilde{E}M_T\tilde{E}^{\prime}/T$. By the Cauchy-Schwartz inequality and the facts that $\|C+D\|_{F}\leq \|C\|_{F} + \|D\|_{F}$ and $\|CD\|_{F}\leq \|C\|_{2}\|D\|_{F}$, \eqref{Eqn: Thm: Rate: 3} implies
\begin{align}\label{Eqn: Thm: Rate: 4}
\|\hat{B} - B H\|^{2}_{F} \leq 6\|\hat{B}\|^{2}_{2}\|V^{-1}\|^{2}_{2}\left(\sum_{j=1}^{6}\|D_{j}\|^{2}_{F}\right)=O_{p}\left(\frac{1}{J^{2\kappa}}+\frac{J^2}{N^2}+\frac{{J}}{{NT}}\right),
\end{align}
where the equality follows by Lemmas \ref{Lem: TechA1} and \ref{Lem: TechA2}(i), along with the fact that $\|D_3\|_{F}=\|D_6\|_{F}$. By the definition of $\hat{a}$,
\begin{align}\label{Eqn: Thm: Rate: 5}
\hat{a} - a &= -\hat{B}(\hat{B}-BH)^{\prime}a + (I_{JM} - \hat{B}\hat{B}^{\prime})(BH-\hat{B})H^{-1}\bar{f} \notag\\
&\hspace{0.5cm}+ (I_{JM} - \hat{B}\hat{B}^{\prime})\tilde{\Delta}1_T/T + (I_{JM} - \hat{B}\hat{B}^{\prime})\tilde{E}1_T/T.
\end{align}
where $H^{-1}$ is well defined with probability approaching one by \eqref{Eqn: Thm: Rate: 4} and Lemma \ref{Lem: TechA2}(ii), and we have used $a^{\prime}B=0$ and $(I_{JM} - \hat{B}\hat{B}^{\prime})\hat{B}=0$. By the Cauchy-Schwartz inequality and the facts that $\|x+y\|\leq \|x\| + \|y\|$ and $\|Ax\|\leq \|A\|_{2}\|x\|$, \eqref{Eqn: Thm: Rate: 5} implies
\begin{align}\label{Eqn: Thm: Rate: 6}
\|\hat{a} - a\|^2 &\leq 4\left(\|\hat{B}-BH\|^2_{F}\|a\|^2 +\|BH-\hat{B}\|^2_{F}\|H^{-1}\|^2_2\max_{t\leq T}\|f_t\|^2\right.\notag\\
&\hspace{0.5cm}\left.+ \frac{1}{T}\|\tilde{\Delta}\|^2_{F} +\frac{1}{T^2}\|\tilde{E}1_T\|^{2}\right)=O_{p}\left(\frac{1}{J^{2\kappa}}+\frac{J^2}{N^2}+\frac{{J}}{{NT}}\right),
\end{align}
where the equality follows by \eqref{Eqn: Thm: Rate: 4}, Assumptions \ref{Ass: LoadingsFactors}(ii) and \ref{Ass: Intercept}, as well as Lemmas \ref{Lem: TechA2}(ii), \ref{Lem: TechA3}(i), and \ref{Lem: TechA4}(ii). Noting $\hat{B}^{\prime}\hat{B}=I_{K}$, we may substitute \eqref{Eqn: Thm: Rate: 1} to $\hat{F} = \tilde{Y}^{\prime}\hat{B}$ to obtain
\begin{align}\label{Eqn: Thm: Rate: 7}
\hat{F} - F(H^{\prime})^{-1} = 1_{T}a^{\prime}(\hat{B}-BH) + F(H^{\prime})^{-1}(BH-\hat{B})^{\prime}\hat{B} + \tilde{\Delta}^{\prime}\hat{B} + \tilde{E}^{\prime}\hat{B}.
\end{align}
where $(H^{\prime})^{-1}$ is well defined with probability approaching one by \eqref{Eqn: Thm: Rate: 4} and Lemma \ref{Lem: TechA2}(ii), and we have used $a^{\prime}B=0$. By the Cauchy-Schwartz inequality and the facts that $\|C+D\|_{F}\leq \|C\|_{F} + \|D\|_{F}$ and $\|CD\|_{F}\leq \|C\|_{2}\|D\|_{F}$, \eqref{Eqn: Thm: Rate: 7} implies
\begin{align}\label{Eqn: Thm: Rate: 8}
\frac{1}{T}\|\hat{F} - F(H^{\prime})^{-1}\|^{2}_{F} &\leq \frac{4}{T}\left(\|{F}\|^{2}_{2}\|H^{-1}\|^{2}_{2}\|BH-\hat{B}\|^{2}_{F} + \|\tilde{\Delta}\|^{2}_{F} + \|\tilde{E}\|^{2}_{F}\right)\|\hat{B}\|^{2}_{2}\notag\\
                                     & \hspace{0.5cm}+\frac{4}{T}\|1_{T}\|^2\|BH-\hat{B}\|^{2}_{F}\|a\|^2=O_{p}\left(\frac{1}{J^{2\kappa}}+\frac{{J}}{{N}}\right),
\end{align}
where the equality follows from \eqref{Eqn: Thm: Rate: 4}, Assumptions \ref{Ass: LoadingsFactors}(ii) and \ref{Ass: Intercept}, as well as Lemmas \ref{Lem: TechA2}(ii) and \ref{Lem: TechA3}(i), (ii), by noting that $J=o(\sqrt{N})$. Since $\hat{\beta}(z) = \hat{B}^{\prime}\phi(z)$ and $\beta(z)=B^{\prime}\phi(z) + \delta(z)$,
\begin{align}\label{Eqn: Thm: Rate: 9}
\hat{\beta}(z)-H^{\prime}\beta(z)&= \hat{B}^{\prime}\phi(z)-(BH)^{\prime}\phi(z) + H^{\prime}\delta(z).
\end{align}
By the Cauchy-Schwartz inequality and the facts that $\|x+y\|\leq \|x\| + \|y\|$, $\|Ax\|\leq \|A\|_{2}\|x\|$ and $\|A\|_{2}\leq \|A\|_{F}$, \eqref{Eqn: Thm: Rate: 9} implies
\begin{align}\label{Eqn: Thm: Rate: 10}
\sup_{z}\|\hat{\beta}(z)-H^{\prime}\beta(z)\|^{2} &\leq2\|\hat{B}-BH\|^{2}_{F}\sup_{z}\|\phi(z)\|^{2} + 2\|H\|^{2}_{2}\sup_{z}\|\delta(z)\|^{2}\notag\\
&=O_{p}\left(\frac{1}{J^{2\kappa-1}}+\frac{J^3}{N^2}+\frac{{J^2}}{{NT}}\right)\max_{j\leq J}\sup_{z}|\phi_{j}(z)|^{2},
\end{align}
where the equality follows from \eqref{Eqn: Thm: Rate: 4} and Lemma \ref{Lem: TechA2}(i), noting that $\sup_{z}\|\phi(z)\|^{2}\leq JM\max_{j\leq J}\sup_{z}|\phi_{j}(z)|^{2}$ and $\sup_{z}\|\delta(z)\|^{2}\leq KM^{2}\max_{k\leq K,m\leq M}\sup_{z}|\delta_{km,J}(z)|^{2}=O(J^{-2\kappa})$ due to Assumption \ref{Ass: LoadingsFactors}(iv). The proof of the second last result is similar. This completes the proof of the theorem. \qed

\subsubsection{Technical Lemmas}
\begin{lem}\label{Lem: TechA1}
Let $D_1,D_2,D_3,D_4,D_5$ be given in the proof of Theorem \ref{Thm: Rate}.\\
(i) Under Assumptions \ref{Ass: Basis}(i), \ref{Ass: LoadingsFactors}(i), (ii), and (iv), $\|D_1\|^2_{F}=O_{p}(J^{-2\kappa})$.\\
(ii) Under Assumptions \ref{Ass: Basis}(i), \ref{Ass: LoadingsFactors}(ii), and (iv), $\|D_2\|^{2}_{F}=O_{p}(J^{-4\kappa})$.\\
(iii) Under Assumptions \ref{Ass: Basis}, \ref{Ass: LoadingsFactors}(ii), (iv), and \ref{Ass: DGP}, $\|D_3\|^{2}_{F}=O_{p}(J^{-2\kappa}J/N)$.\\
(iv) Under Assumptions \ref{Ass: Basis}, \ref{Ass: LoadingsFactors}(i), (ii), and \ref{Ass: DGP}, $\|D_4\|^2_{F}=O_{p}(J/NT)$.\\
(v) Under Assumptions \ref{Ass: Basis} and \ref{Ass: DGP}, $\|D_5\|^{2}_{F}=O_{p}(J^{2}/N^{2})$.
\end{lem}
\noindent{\sc Proof:} (i) Since $\|M_T\|_{2}=1$, $\|D_1\|_{F}\leq \|B\|_2\|F\|_{2}\|\tilde{\Delta}\|_{F}/{T}$. The result then immediately follows from Assumptions \ref{Ass: LoadingsFactors}(i) and (ii) as well as Lemma \ref{Lem: TechA3}(i).

(ii) Since $\|M_T\|_{2}=1$, $\|D_2\|_{F}\leq \|\tilde{\Delta}\|^{2}_{F}/T$. The result then immediately follows from Lemma \ref{Lem: TechA3}(i).

(iii) Since $\|M_T\|_{2}=1$, $\|D_3\|_{F} \leq\|\tilde{\Delta}\|_{F}\|\tilde{E}\|_{F}/T$. The result then immediately follows from Lemma \ref{Lem: TechA3}(i) and (ii).

(iv) Since $\|D_4\|_{F}\leq\|B\|_{2}\|\tilde{E}M_TF\|_{F}/T$, the result then immediately follows from Assumption \ref{Ass: LoadingsFactors}(i) and Lemma \ref{Lem: TechA3}(iii).

(v) Since $\|M_T\|_{2}=1$, $\|D_5\|_{F}\leq \|\tilde{E}\|^{2}_{F}/T$. The result then immediately follows from Lemma \ref{Lem: TechA3}(ii).\qed

\begin{lem}\label{Lem: TechA2}
Suppose Assumptions \ref{Ass: Basis}-\ref{Ass: DGP} hold. Let $V$ be given in the proof of Theorem \ref{Thm: Rate}. Assume (i) $N\to\infty$; (ii) $T\geq K+1$; (iii) $J\to\infty$ with $J=o(\sqrt{N})$. Then (i) $\|V\|_{2} = O_{p}(1)$, $\|V^{-1}\|_{2}=O_{p}(1)$, and $\|H\|_{2}=O_{p}(1)$; (ii) $\|H^{-1}\|_{2}=O_{p}(1)$, if $\|\hat{B} - B H\|_{F}=o_{p}(1)$.
\end{lem} 
\noindent{\sc Proof:} (i) Let $D_7\equiv D_1^{\prime}$ and $D_8\equiv D_4^{\prime}$. Then by \eqref{Eqn: Thm: Rate: 2}, $\tilde{Y}M_T\tilde{Y}^{\prime}/T=BF^{\prime}M_TFB^{\prime}/T + \sum_{j=1}^{8}D_j$, where $D_1,\ldots,D_6$ are given below \eqref{Eqn: Thm: Rate: 2}. By the fact that $\|C+D\|_{F}\leq \|C\|_{F} + \|D\|_{F}$,
\begin{align}\label{Eqn: Lem: TechA2: 1}
\|\tilde{Y}M_T\tilde{Y}^{\prime}/T-BF^{\prime}M_TFB^{\prime}/T\|_{F}\leq\sum_{j=1}^{8}\|D_j\|_{F} = O_{p}\left(\frac{1}{J^{\kappa}}+\frac{J}{N}+\frac{\sqrt{J}}{\sqrt{NT}}\right),
\end{align}
where the equality follows by Lemma \ref{Lem: TechA1} and the facts that $\|D_6\|_{F}=\|D_3\|_{F}$, $\|D_7\|_{F}=\|D_1\|_{F}$, and $\|D_8\|_{F}=\|D_4\|_{F}$. Let $\mathcal{V}$ be a $K\times K$ diagonal matrix of the eigenvalues of $(F^{\prime}M_TF/T)B^{\prime}B$, which are equal to the first $K$ largest eigenvalues of $BF^{\prime}M_TFB^{\prime}/T$. By the Weyl's inequality and the fact that $\|A\|_{2}\leq \|A\|_{F}$,
\begin{align}\label{Eqn: Lem: TechA2: 2}
\|V-\mathcal{V}\|_2 \leq \|{\tilde{Y}M_T\tilde{Y}^{\prime}}/{T}-{BF^{\prime}M_TFB^{\prime}}/{T}\|_{2} = O_{p}\left(\frac{1}{J^{\kappa}}+\frac{J}{N}+\frac{\sqrt{J}}{\sqrt{NT}}\right).
\end{align}
Thus, $\|V\|_{2} = O_{p}(1)$ and $\|V^{-1}\|_{2} = \lambda^{-1}_{\min}(V) =O_{p}(1)$ follow from \eqref{Eqn: Lem: TechA2: 2} and Assumptions \ref{Ass: LoadingsFactors}(i)-(iii). Let $H^{\diamond}\equiv(F^{\prime}M_TF/T)B^{\prime}\hat{B}V^{-1}$. Recall that $H = (F^{\prime}M_T\tilde{Y}^{\prime}\hat{B}/T)V^{-1}$. Then by the facts that $\|A\|_{2}\leq \|A\|_{F}$ and $\|M_T\|_2 = 1$,
\begin{align}\label{Eqn: Lem: TechA2: 3}
\hspace{-0.7cm}\|H - H^{\diamond}\|_{2}\leq\frac{1}{T}(\|F\|_{2}\|\tilde{\Delta}\|_{F}+\|\tilde{E}M_TF\|_{F})\|\hat{B}\|_2\|V^{-1}\|_2 = O_{p}\left(\frac{1}{J^{\kappa}}+\frac{\sqrt{J}}{\sqrt{NT}}\right),
\end{align}
where the equality follows from the second result in (i), Assumption \ref{Ass: LoadingsFactors}(ii), and Lemmas \ref{Lem: TechA3}(i) and (iii). Since $\|H^{\diamond}\|_{2} \leq \|F^{\prime}M_{T}F/T\|_{2}\|B\|_{2}\|\hat{B}\|_2\|V^{-1}\|_2$, the third result in (i) follows from \eqref{Eqn: Lem: TechA2: 3}, the second result in (i), and Assumptions \ref{Ass: LoadingsFactors}(i) and (ii).

(ii) By the facts that $\|C+D\|_{F}\leq \|C\|_{F}+\|D\|_{F}$ and $\|CD\|_{F}\leq \|C\|_{2}\|D\|_{F}$,
\begin{align}\label{Eqn: Lem: TechA2: 4}
\|\hat{B}^{\prime}\hat{B}-H^{\prime}B^{\prime}BH\|_{F}\leq\|\hat{B}\|_{2}\|\hat{B} - B H\|_{F} + \|\hat{B} - B H\|_{F}\|B\|_{2}\|H\|_{2}.
\end{align}
Thus, $I_{K}-H^{\prime}B^{\prime}BH=o_{p}(1)$ by Assumption \ref{Ass: LoadingsFactors}(i) and $\|H\|_2=O_{p}(1)$. It then follows that $I_{K}-\lambda_{\min}(B^{\prime}B)H^{\prime}H$ is negative semidefinite with probability approaching one, since  $H^{\prime}B^{\prime}BH-\lambda_{\min}(B^{\prime}B)H^{\prime}H$ is positive semidefinite. So, the eigenvalues of $H^{\prime}H$ are not smaller than $\lambda^{-1}_{\min}(B^{\prime}B)$ with probability approaching one. Thus, the result in (ii) follows from Assumption \ref{Ass: LoadingsFactors}(i).\qed

\begin{lem}\label{Lem: TechA3}
Let $\tilde{\Delta}$ and $\tilde{E}$ be given in the proof of Theorem \ref{Thm: Rate}. \\
(i) Under Assumptions \ref{Ass: Basis}(i), \ref{Ass: LoadingsFactors}(ii), and (iv), $\|\tilde{\Delta}\|_{F}^{2}/T = O_{p}(J^{-2\kappa})$.\\
(ii) Under Assumptions \ref{Ass: Basis} and \ref{Ass: DGP}, $\|\tilde{E}\|_{F}^{2}/T = O_{p}(J/N)$.\\
(iii) Under Assumptions \ref{Ass: Basis}, \ref{Ass: LoadingsFactors} (ii), and \ref{Ass: DGP}, $\|\tilde{E}M_TF\|^{2}_{F}/T=O_{p}(J/N)$.
\end{lem}
\noindent{\sc Proof:} (i) By the facts that $\|Ax\|\leq \|A\|_{2}\|x\|$ and $\|A\|_{2}\leq \|A\|_{F}$,
\begin{align}\label{Eqn: Lem: TechA3: 1}
\frac{1}{T}\|\tilde{\Delta}\|^{2}_{F}&= \frac{1}{T}\sum_{t=1}^{T}\|(\Phi(Z_t)^{\prime}\Phi(Z_t))^{-1}\Phi(Z_t)^{\prime}(R(Z_t)+\Delta(Z_t)f_{t})\|^{2}\notag\\
&\leq 2\max_{t\leq T}\|f_{t}\|^{2}\left(\min_{t\leq T}\lambda_{\min}(\hat{Q}_t)\right)^{-1} \frac{1}{NT}\sum_{t=1}^{T}\|\Delta(Z_t)\|_{F}^{2}\notag\\
&\hspace{0.5cm} + 2\left(\min_{t\leq T}\lambda_{\min}(\hat{Q}_t)\right)^{-1} \frac{1}{NT}\sum_{t=1}^{T}\|R(Z_t)\|^{2}=O_{p}\left(\frac{1}{J^{2\kappa}}\right),
\end{align}
where the last line follows from Assumptions \ref{Ass: Basis}(i) and \ref{Ass: LoadingsFactors}(ii), as well as Lemma \ref{Lem: TechA4}(iii).

(ii) By the fact that $\|Ax\|\leq \|A\|_{2}\|x\|$,
\begin{align}\label{Eqn: Lem: TechA3: 2}
\frac{1}{T}\|\tilde{E}\|^{2}_{F}&= \frac{1}{T}\sum_{t=1}^{T}\|(\Phi(Z_t)^{\prime}\Phi(Z_t))^{-1}\Phi(Z_t)^{\prime}\varepsilon_{t}\|^{2}\notag\\
&\leq \left(\min_{t\leq T}\lambda_{\min}(\hat{Q}_t)\right)^{-2} \frac{1}{N^{2}T}\sum_{t=1}^{T}\|\Phi(Z_t)^{\prime}\varepsilon_{t}\|^{2}= O_{p}\left(\frac{J}{N}\right),
\end{align}
where the last equality follows from Assumption \ref{Ass: Basis}(i) and Lemma \ref{Lem: TechA4}(i).

(iii) By the facts that $\|C+D\|_{F}\leq \|C\|_F +\|D\|_F$,
\begin{align}\label{Eqn: Lem: TechA3: 3}
\frac{1}{T}\|\tilde{E}M_TF\|^{2}_{F}&\leq \frac{2}{N^{2}T}\left\|\sum_{t=1}^{T}\hat{Q}_t^{-1}\Phi(Z_t)^{\prime}\varepsilon_t f_{t}^{\prime}\right\|^{2}_{F}\notag\\
&\hspace{0.5cm}+\frac{2\|\bar{f}\|^2}{N^{2}T}\left\|\sum_{t=1}^{T}\hat{Q}_t^{-1}\Phi(Z_t)^{\prime}\varepsilon_t\right\|^{2}=O_{p}\left(\frac{J}{N}\right),
\end{align}
where the equality follows from Assumption \ref{Ass: LoadingsFactors}(ii) and Lemma \ref{Lem: TechA4}(ii).\qed

\begin{lem}\label{Lem: TechA4}
(i) Under Assumptions \ref{Ass: Basis}(ii) and \ref{Ass: DGP},
\[\sum_{t=1}^{T}\|\Phi(Z_t)^{\prime}\varepsilon_{t}\|^{2}=O_{p}(NTJ).\]
(ii) Under Assumptions \ref{Ass: Basis}, \ref{Ass: LoadingsFactors}(ii), and \ref{Ass: DGP},
\[\left\|\sum_{t=1}^{T}\hat{Q}_{t}^{-1}\Phi(Z_t)^{\prime}\varepsilon_tf_{t}^{\prime}\right\|^{2}_{F}=O_{p}(NTJ) \text{ and } \left\|\sum_{t=1}^{T}\hat{Q}_{t}^{-1}\Phi(Z_t)^{\prime}\varepsilon_t\right\|^{2}=O_{p}(NTJ).\]
(iii) Under Assumption \ref{Ass: LoadingsFactors}(iv),
\[\sum_{t=1}^{T}\|\Delta(Z_t)\|_{F}^{2}=O_{p}(NTJ^{-2\kappa}) \text{ and } \sum_{t=1}^{T}\|R(Z_t)\|^{2}=O_{p}(NTJ^{-2\kappa}).\]
\end{lem}
\noindent{\sc Proof:} (i) The result follows by the Markov's inequality, since
\begin{align}\label{Eqn: Lem: TechA4: 1}
&\hspace{0.8cm}E\left[\sum_{t=1}^{T}\|\Phi(Z_t)^{\prime}\varepsilon_{t}\|^{2}\right]= E\left[\sum_{t=1}^{T}\sum_{i=1}^{N}\sum_{j=1}^{N}\phi(z_{it})^{\prime}\phi(z_{jt})\varepsilon_{it}\varepsilon_{jt}\right]\notag\\
&= \sum_{t=1}^{T}\sum_{i=1}^{N}\sum_{j=1}^{N}E[\phi(z_{it})^{\prime}\phi(z_{jt})]E[\varepsilon_{it}\varepsilon_{jt}]\notag\\
&\leq \max_{i\leq N,t\leq T}E[\|\phi(z_{it})\|^{2}]\sum_{t=1}^{T}\sum_{i=1}^{N}\sum_{j=1}^{N}|E[\varepsilon_{it}\varepsilon_{jt}]|\notag\\
&\leq TJM \max_{m\leq M, j\leq J, i\leq N, t\leq T} E[\phi^{2}_{j}(z_{it,m})]\max_{t\leq T}\sum_{i=1}^{N}\sum_{j=1}^{N}|E[\varepsilon_{it}\varepsilon_{jt}]|= O(NTJ),
\end{align}
where the second equality follows by the independence in Assumption \ref{Ass: DGP}(i), the first inequality is due to the Cauchy Schwartz inequality, the second inequality follows since $\max_{i\leq N, t\leq T}E[\|\phi(z_{it})\|^{2}]\leq JM\max_{m\leq M, j\leq J, i\leq N, t\leq T} E[\phi^{2}_{j}(z_{it,m})]$, and the last equality follows from Assumptions \ref{Ass: Basis}(ii) and \ref{Ass: DGP}(iii).


(ii) Let $E_{\varepsilon}$ be the expectation with respect to $\{\varepsilon_{t}\}_{t\leq T}$. Since $\|A\|^{2}_{F}\hspace{-0.05cm}=\hspace{-0.05cm}\mathrm{tr}(AA^{\prime})$,
\begin{align}\label{Eqn: Lem: TechA4: 2}
&E_{\varepsilon}\left[\left\|\sum_{t=1}^{T}\hat{Q}^{-1}_{t}\Phi(Z_t)^{\prime}\varepsilon_t f_{t}^{\prime}\right\|^{2}_{F}\right]\hspace{-0.1cm}=\hspace{-0.1cm} E_{\varepsilon}\left[\mathrm{tr} \hspace{-0.1cm}\left(\sum_{t=1}^{T}\sum_{s=1}^{T}\sum_{i=1}^{N}\sum_{j=1}^{N}\hat{Q}^{-1}_{t}\phi(z_{it})\varepsilon_{it}f_{t}^{\prime}f_{s}\varepsilon_{js}\phi(z_{js})^{\prime}\hat{Q}^{-1}_{s}\right)\right]\notag\\
&=\sum_{t=1}^{T}\sum_{s=1}^{T}\sum_{i=1}^{N}\sum_{j=1}^{N}\phi(z_{it})^{\prime}\hat{Q}_t^{-1}\hat{Q}_s^{-1}\phi(z_{js})f_t^{\prime}f_{s}E[\varepsilon_{it}\varepsilon_{js}]\notag\\
&\leq\max_{t\leq T}\|f_t\|^{2}\left(\min_{t\leq T}\lambda_{\min}(\hat{Q}_t)\right)^{-2}\sum_{t=1}^{T}\sum_{s=1}^{T}\sum_{i=1}^{N}\sum_{j=1}^{N}\|\phi(z_{it})\|\|\phi(z_{js})\||E[\varepsilon_{it}\varepsilon_{js}]|,
\end{align}
where the second equality follows from the independence in Assumption \ref{Ass: DGP}(i) and the linearity of both expectation and trace operators, and the inequality follows by  the fact that $\|Ax\|\leq \|A\|_{2}\|x\|$. Moreover,
\begin{align}\label{Eqn: Lem: TechA4: 3}
&\hspace{0.8cm}E\left[\sum_{t=1}^{T}\sum_{s=1}^{T}\sum_{i=1}^{N}\sum_{j=1}^{N}\|\phi(z_{it})\|\|\phi(z_{js})\||E[\varepsilon_{it}\varepsilon_{js}]|\right]\notag\\
&\leq \max_{i\leq N, t\leq T}E[\|\phi(z_{it})\|^{2}]\sum_{t=1}^{T}\sum_{s=1}^{T}\sum_{i=1}^{N}\sum_{j=1}^{N}|E[\varepsilon_{it}\varepsilon_{js}]|\notag\\
&\leq JM\max_{m\leq M, j\leq J, i\leq N, t\leq T} E[\phi^{2}_{j}(z_{it,m})]\sum_{i=1}^{N}\sum_{j=1}^{N}\sum_{t=1}^{T}\sum_{s=1}^{T}|E[\varepsilon_{it}\varepsilon_{js}]|,
\end{align}
where the first inequality is due to the Cauchy-Schwartz inequality, and the second one follows since $\max_{i\leq N, t\leq T}E[\|\phi(z_{it})\|^{2}]\leq JM\max_{m\leq M, j\leq J, i\leq N, t\leq T}E[\phi^{2}_{j}(z_{it,m})]$. Combining \eqref{Eqn: Lem: TechA4: 2} and \eqref{Eqn: Lem: TechA4: 3} implies that $E_{\varepsilon}[\|\sum_{t=1}^{T}\hat{Q}^{-1}_{t}\Phi(Z_t)^{\prime}\varepsilon_t f_{t}^{\prime}\|^{2}_{F}]=O_{p}(NTJ)$ by Assumptions \ref{Ass: Basis}, \ref{Ass: LoadingsFactors}(ii), and \ref{Ass: DGP}(iii). Thus, the first result of the lemma follows by the Markov's inequality and Lemma \ref{Lem: TechA5}. The proof of the second result is similar.

(iii) The first result follows since
\begin{align}\label{Eqn: Lem: TechA4: 4}
&\sum_{t=1}^{T}\|\Delta(Z_t)\|_{F}^{2}\leq NT K M^{2}\max_{k\leq K,m\leq M}\sup_{z}|\delta_{km,J}(z)|^{2}=O_{p}(NTJ^{-2\kappa}),
\end{align}
where the inequality follows since $\max_{i\leq N, t\leq T}\|\delta(z_{it})\|^2\leq M^{2}K\sup_{k\leq K,m\leq M}\sup_{z}$ $|\delta_{km,J}(z)|^{2}$, and the equality follows from Assumption \ref{Ass: LoadingsFactors}(iv). The proof of the second result is similar.\qed

\begin{lem}\label{Lem: TechA5}
Let $S_1, \ldots, S_N$ be a sequence of random variables and $\mathcal{D}_{1},\ldots, \mathcal{D}_{N}$ be a sequence of random vectors. Then $S_{N}=O_{p}(1)$ if and only if  $S_{N}=O_{p|\mathcal{D}_{N}}(1)$, where $p$ denotes the underlying probability measure and $p|\mathcal{D}_{N}$ denotes the probability measure conditional on $\mathcal{D}_{N}$.
\end{lem}
\noindent{\sc Proof:} By definition, $S_{N}=O_{p}(1)$ means that $P(|S_{N}|>\ell_{N})=o(1)$ for any $\ell_{N}\to\infty$, while $S_{N}=O_{p|\mathcal{D}_{N}}(1)$ means that $P(|S_{N}|>\ell_{N}|\mathcal{D}_{N})=o_{p}(1)$ for any $\ell_{N}\to\infty$. The second follows from the first by the Markov inequality because $E[P(|S_{N}|>\ell_{N}|\mathcal{D}_{N})]=P(|S_{N}|>\ell_{N})=o(1)$. Since $P(|S_{N}|>\ell_{N}|\mathcal{D}_{N})\leq 1$ for all $N$, $\{P(|S_{N}|>\ell_{N}|\mathcal{D}_{N})\}_{N\geq 1}$ are uniformly integrable. The first follows from the second by the fact that convergence in probability implies moments convergence for uniformly integrable sequences. \qed

\subsection{Proof of Theorem \ref{Thm: ImprovedRates}}
\noindent{\sc Proof of Theorem \ref{Thm: ImprovedRates}:} Theorem \ref{Thm: Rate} provides a preliminary rate of $\|\hat{a} - a\|^{2}$, $\|\hat{B} - B H\|^{2}_{F}$, and $\|\hat{F} - F(H^{\prime})^{-1}\|^2_{F}$ by using rough bounds based on \eqref{Eqn: Thm: Rate: 3}, \eqref{Eqn: Thm: Rate: 5}, and \eqref{Eqn: Thm: Rate: 7}.
To improve the rate of $\|\hat{B} - B H\|^{2}_{F}$, we need to treat $D_5\hat{B}$ in as \eqref{Eqn: Thm: Rate: 3} a whole to establish its rate. By the Cauchy-Schwartz inequality and the facts that $\|C+D\|_{F}\leq \|C\|_{F} + \|D\|_{F}$ and $\|CD\|_{F}\leq \|C\|_{2}\|D\|_{F}$, \eqref{Eqn: Thm: Rate: 3} implies
\begin{align}\label{Eqn: Thm: ImprovedRates: 1}
\|\hat{B} - B H\|^{2}_{F} &\leq 10\|\hat{B}\|^{2}_{2}\|V^{-1}\|^{2}_{2}\left(\sum_{j\neq 5}^{6}\|D_{j}\|^{2}_{F}\right) + 2\|V^{-1}\|^{2}_{2}\|D_5\hat{B}\|^{2}_{F},\notag\\
& = O_{p}\left(\frac{1}{J^{2\kappa}}+\frac{J}{N^2}+\frac{{J}}{{NT}}\right),
\end{align}
where the equality follows from $J=o(\sqrt{N})$, Lemmas \ref{Lem: TechA1}(i)-(iv), \ref{Lem: TechA2}(i), and \ref{Lem: TechB1}(ii), as well as the fact that $\|D_6\|_{F}=\|D_3\|_{F}$. Given the rate of $\|\hat{B} - B H\|^{2}_{F}$ in \eqref{Eqn: Thm: ImprovedRates: 1}, the rate of $\|\hat{a}-a\|^2$ immediately follows from the same argument in \eqref{Eqn: Thm: Rate: 6}. To improve the rate of $\|\hat{F} - F(H^{\prime})^{-1}\|^2_{F}$, we need to plug in the expansion of $\hat{B}-BH$ to \eqref{Eqn: Thm: Rate: 5}, and treat $a'D_{4}$, $D_4^{\prime}\hat{B}$, $D_5\hat{B}$, and $\tilde{E}^{\prime}\hat{B}$ as a whole to establish their rates.  By the facts that $\|C+D\|_{F}\leq \|C\|_{F} + \|D\|_{F}$ and $\|CD\|_{F}\leq \|C\|_{2}\|D\|_{F}$, combining \eqref{Eqn: Thm: Rate: 3} and \eqref{Eqn: Thm: Rate: 7} implies
\begin{align}\label{Eqn: Thm: ImprovedRates: 2}
\|\hat{F} - F(H^{\prime})^{-1}\|_{F}&= \left(\sum_{j\neq 4,5}^{6}\|D_{j}\|_{F}\|\hat{B}\|_{2}\|a\|+\|a^{\prime}D_4\|\|\hat{B}\|_{2}+ \|a\|\|D_5\hat{B}\|_{F}\right)\times \notag\\
&\hspace{0.8cm}\|V^{-1}\|_{2}\|1_{T}\|+\left(\sum_{j\neq 4,5}^{6}\|D_{j}\|_{F}\|\hat{B}\|_{2}+ \|D_4^{\prime}\hat{B}\|_{F}+ \|D_5\hat{B}\|_{F}\right)\notag\\
&\hspace{0.8cm}\times \|F\|_{2}\|H^{-1}\|_{2}\|V^{-1}\|_{2}\|\hat{B}\|_{2} + \|\tilde{\Delta}\|_{F}\|\hat{B}\|_{2} + \|\tilde{E}^{\prime}\hat{B}\|_{F}\notag\\
&=O_{p}\left(\frac{\sqrt{T}}{J^{\kappa}} + \sqrt{\frac{T}{N}}\right),
\end{align}
where the equality follows from $J=o(\sqrt{N})$, Assumptions \ref{Ass: LoadingsFactors}(ii) and \ref{Ass: Intercept}, Lemmas \ref{Lem: TechA1}(i)-(iii), \ref{Lem: TechA2}, \ref{Lem: TechA3}(i), \ref{Lem: TechB1}, and \ref{Lem: TechB2}(i), as well as the fact that $\|D_6\|_{F}=\|D_3\|_{F}$. Thus, the third result of the theorem follows from \eqref{Eqn: Thm: ImprovedRates: 2}. The proofs of the last two results of the theorem are similar to the proofs of the last two results of Theorem \ref{Thm: Rate}.\qed

\subsubsection{Technical Lemmas}
\begin{lem}\label{Lem: TechB1}
Let $D_4$ and $D_5$ be given in the proof of Theorem \ref{Thm: Rate}.  Assume (i) $N\to\infty$; (ii) $T\geq K+1$; (iii) $J\to\infty$ with $J^{2}\xi^{2}_{J}\log J=o(N)$.\\
(i) Under Assumptions \ref{Ass: Basis}-\ref{Ass: Improvedrates}, $\|D_4^{\prime}\hat{B}\|^{2}_{F}=O_{p}(1/NT)$.\\
(ii) Under Assumptions \ref{Ass: Basis}-\ref{Ass: Improvedrates}, $\|D_5\hat{B}\|^2_{F}=O_{p}(J/N^2)$.\\
(iii) Under Assumptions \ref{Ass: Basis}-\ref{Ass: Improvedrates}, $\|D_4^{\prime}a\|^{2}=O_{p}(1/NT)$.
\end{lem}
\noindent{\sc Proof:} (i) Since $\|D_4^{\prime}\hat{B}\|_{F}\leq \|B\|_2 \|\hat{B}^{\prime}\tilde{E}M_TF\|_{F}/T$, the result then immediately follows from Assumption \ref{Ass: LoadingsFactors}(i) and Lemma \ref{Lem: TechB2}(ii).

(ii) Since $\|M_T\|_2 =1$, $\|D_5\hat{B}\|_{F}\leq \|\tilde{E}\|_F \|\hat{B}^{\prime}\tilde{E}\|_{F}/T$. The result then immediately follows from Lemmas \ref{Lem: TechA3}(ii) and \ref{Lem: TechB2}(i).

(iii) Since $\|D_4^{\prime}a\|\leq \|B\|_2 \|a^{\prime}\tilde{E}M_TF\|/T$, the result then immediately follows from Assumption \ref{Ass: LoadingsFactors}(i) and Lemma \ref{Lem: TechB2}(iii).\qed

\begin{lem}\label{Lem: TechB2}
Let $\tilde{E}$ be given in the proof of Theorem \ref{Thm: Rate}. Assume (i) $N\to\infty$; (ii) $T\geq K+1$; (iii) $J\to\infty$ with $J^{2}\xi^{2}_{J}\log J=o(N)$.\\
(i) Under Assumptions \ref{Ass: Basis}-\ref{Ass: Improvedrates}, $\|\hat{B}^{\prime}\tilde{E}\|^2_{F}/T=O_{p}(1/N)$. \\
(ii) Under Assumptions \ref{Ass: Basis}-\ref{Ass: Improvedrates}, $\|\hat{B}^{\prime}\tilde{E}M_TF\|_{F}^{2}/T = O_{p}(1/N)$.\\
(iii) Under Assumptions \ref{Ass: Basis}-\ref{Ass: Improvedrates}, $\|a^{\prime}\tilde{E}M_TF\|^2/T=O_{p}(1/N)$.
\end{lem}
\noindent{\sc Proof:} By the facts that $\|C+D\|_{F}\leq \|C\|_{F}+\|D\|_{F}$ and $\|CD\|_{F}\leq \|C\|_{2}\|D\|_{F}$,
\begin{align}\label{Eqn: Lem: TechB2: 1}
\frac{1}{T}\|\hat{B}^{\prime}\tilde{E}\|^2_{F}&\leq \frac{2}{T}\|\tilde{E}\|_{F}^{2}\|\hat{B}-BH\|_{F}^{2} + \frac{2}{T}\|H\|_{2}^{2}\|B^{\prime}\tilde{E}\|^2_{F}\notag\\
&=\frac{2}{T}\|\tilde{E}\|_{F}^{2}\|\hat{B}-BH\|_{F}^{2} + \frac{2}{N^{2}T}\|H\|_{2}^{2}\left(\sum_{t=1}^{T}\|B^{\prime}\hat{Q}_{t}^{-1}\Phi(Z_t)^{\prime}\varepsilon_{t}\|^{2}\right)\notag\\
&=O_{p}\left(\frac{J}{N}\left(\frac{1}{J^{2\kappa}}+\frac{J^2}{N^2}+\frac{{J}}{{NT}}\right) + \frac{1}{N}\right) = O_{p}\left(\frac{1}{N}\right),
\end{align}
where the second equality follows from $J^{2}\xi^{2}_{J}\log J=o(N)$, Lemmas \ref{Lem: TechA2}(i), \ref{Lem: TechA3}(ii), and \ref{Lem: TechB3}(i), as well as Theorem \ref{Thm: Rate}, and the last line is due to $\kappa>1/2$ and $J=o(\sqrt{N})$.

(ii) By the facts that $\|C+D\|_{F}\leq \|C\|_{F}+\|D\|_{F}$ and $\|CD\|_{F}\leq \|C\|_{2}\|D\|_{F}$,
\begin{align}\label{Eqn: Lem: TechB2: 2}
\frac{1}{T}\|\hat{B}^{\prime}\tilde{E}M_TF\|^2_{F}&\leq \frac{2}{T}\|\tilde{E}M_TF\|_{F}^{2}\|\hat{B}-BH\|_{F}^{2} + \frac{2}{T}\|H\|_{2}^{2}\|B^{\prime}\tilde{E}M_TF\|^2_{F}\notag\\
&\leq \frac{2}{T}\|\tilde{E}M_TF\|_{F}^{2}\|\hat{B}-BH\|_{F}^{2} + \frac{4}{N^{2}T}\|H\|_{2}^{2}\left\|\sum_{t=1}^{T}B^{\prime}\hat{Q}_{t}^{-1}\Phi(Z_t)^{\prime}\varepsilon_{t}f_t^{\prime}\right\|_{F}^{2}\notag\\
&\hspace{0.5cm}+\frac{4\|\bar{f}\|^2}{N^{2}T}\|H\|_{2}^{2}\left\|\sum_{t=1}^{T}B^{\prime}\hat{Q}_{t}^{-1}\Phi(Z_t)^{\prime}\varepsilon_{t}\right\|^{2}\notag\\
&=O_{p}\left(\frac{J}{N}\left(\frac{1}{J^{2\kappa}}+\frac{J^2}{N^2}+\frac{{J}}{{NT}}\right) + \frac{1}{N}\right) = O_{p}\left(\frac{1}{N}\right),
\end{align}
where the first equality follows from $J^{2}\xi^{2}_{J}\log J=o(N)$, Assumption \ref{Ass: LoadingsFactors}(ii), Lemmas \ref{Lem: TechA2}(i), \ref{Lem: TechA3}(iii), and \ref{Lem: TechB3}(ii), as well as Theorem \ref{Thm: Rate}, and the last equality is due to $\kappa>1/2$ and $J=o(\sqrt{N})$.

(iii) By the fact that $\|x+y\|\leq \|x\|+\|y\|$,
\begin{align}\label{Eqn: Lem: TechB2: 3}
\frac{1}{T}\|a^{\prime}\tilde{E}M_TF\|^2&\leq \frac{2}{N^{2}T}\left\|\sum_{t=1}^{T}a^{\prime}\hat{Q}_{t}^{-1}\Phi(Z_t)^{\prime}\varepsilon_{t}f_t^{\prime}\right\|^{2}+\frac{2\|\bar{f}\|^2}{N^{2}T}\left|\sum_{t=1}^{T}a^{\prime}\hat{Q}_{t}^{-1}\Phi(Z_t)^{\prime}\varepsilon_{t}\right|^{2}\notag\\
&= O_{p}\left(\frac{1}{N}\right),
\end{align}
because $J^{2}\xi^{2}_{J}\log J=o(N)$, Assumption \ref{Ass: LoadingsFactors}(ii) and Lemma \ref{Lem: TechB3}(ii).\qed

\begin{lem}\label{Lem: TechB3}
Assume $J\geq 2$ and $\xi^{2}_{J}\log J=o(N)$.\\
(i) Under Assumptions \ref{Ass: Basis}(i), \ref{Ass: LoadingsFactors}(i), \ref{Ass: DGP}, and \ref{Ass: Improvedrates},
\[\sum_{t=1}^{T}\|B^{\prime}\hat{Q}_{t}^{-1}\Phi(Z_t)^{\prime}\varepsilon_{t}\|^{2}=O_{p}\left(NT\left(1+\frac{J\xi_{J}^2\log J}{N}\right)\right).\]
(ii) Under Assumptions \ref{Ass: Basis}(i), \ref{Ass: LoadingsFactors}(i), (ii), \ref{Ass: DGP}, \ref{Ass: Intercept}, and \ref{Ass: Improvedrates},
\begin{align*}
\left\|\sum_{t=1}^{T}B^{\prime}\hat{Q}_{t}^{-1}\Phi(Z_t)^{\prime}\varepsilon_tf_{t}^{\prime}\right\|^{2}_{F}&=O_{p}\left(NT\left(1+\frac{J\xi_{J}\sqrt{\log J}}{\sqrt{N}}\right)\right),\\
\left\|\sum_{t=1}^{T}B^{\prime}\hat{Q}_{t}^{-1}\Phi(Z_t)^{\prime}\varepsilon_t\right\|^{2}&=O_{p}\left(NT\left(1+\frac{J\xi_{J}\sqrt{\log J}}{\sqrt{N}}\right)\right),\\
\left\|\sum_{t=1}^{T}a^{\prime}\hat{Q}_{t}^{-1}\Phi(Z_t)^{\prime}\varepsilon_tf_{t}^{\prime}\right\|^{2}&=O_{p}\left(NT\left(1+\frac{J\xi_{J}\sqrt{\log J}}{\sqrt{N}}\right)\right),\\
\left|\sum_{t=1}^{T}a^{\prime}\hat{Q}_{t}^{-1}\Phi(Z_t)^{\prime}\varepsilon_t\right|^{2}&=O_{p}\left(NT\left(1+\frac{J\xi_{J}\sqrt{\log J}}{\sqrt{N}}\right)\right).
\end{align*}
\end{lem}
\noindent{\sc Proof:} (i) Let ${Q}_{t}\equiv E[\hat{Q}_{t}]$. By the fact that $\|x+y\|\leq \|x\|+\|y\|$,
\begin{align}\label{Eqn: Lem: TechB3: 1}
\sum_{t=1}^{T}\|B^{\prime}\hat{Q}_{t}^{-1}\Phi(Z_t)^{\prime}\varepsilon_{t}\|^{2}&\leq 2\sum_{t=1}^{T}\|B^{\prime}{Q}_{t}^{-1}\Phi(Z_t)^{\prime}\varepsilon_{t}\|^{2}\notag\\
&\hspace{-1.5cm}+ 2\sum_{t=1}^{T}\|B^{\prime}(\hat{Q}_{t}^{-1}-{Q}_{t}^{-1})\Phi(Z_t)^{\prime}\varepsilon_{t}\|^{2}\equiv 2\mathcal{T}_1 + 2\mathcal{T}_2.
\end{align}
Therefore, it suffices to show that $\mathcal{T}_1=O_{p}(NT)$ and $\mathcal{T}_2=O_{p}(TJ\xi_{J}^2\log J)$. The former holds by the Markov's inequality, since
\begin{align}\label{Eqn: Lem: TechB3: 2}
E[\mathcal{T}_1]&= E\left[\sum_{t=1}^{T}\sum_{i=1}^{N}\sum_{j=1}^{N}\phi(z_{it})^{\prime}Q_{t}^{-1}B B^{\prime}Q_{t}^{-1}\phi(z_{jt})\varepsilon_{it}\varepsilon_{jt}\right]\notag\\
&= \sum_{t=1}^{T}\sum_{i=1}^{N}\sum_{j=1}^{N}E[\phi(z_{it})^{\prime}Q_{t}^{-1}B B^{\prime}Q_{t}^{-1}\phi(z_{jt})]E[\varepsilon_{it}\varepsilon_{jt}]\notag\\
&\leq T \max_{i\leq N,t\leq T} E[\|B^{\prime}Q_{t}^{-1}\phi(z_{it})\|^2]\max_{t\leq T}\sum_{i=1}^{N}\sum_{j=1}^{N}|E[\varepsilon_{it}\varepsilon_{jt}]|= O(NT),
\end{align}
where the second equality follows by the independence in Assumption \ref{Ass: DGP}(i), the inequality is due to the Cauchy-Schwartz inequality, and the last equality follows from Assumption \ref{Ass: DGP}(iii) and Lemma \ref{Lem: TechB4}. The latter also holds, since
\begin{align}\label{Eqn: Lem: TechB3: 3}
\mathcal{T}_2 &\leq C_{NT}\sum_{t=1}^{T}\|\hat{Q}_{t}-{Q}_{t}\|_2^{2}\|\Phi(Z_t)^{\prime}\varepsilon_{t}\|^{2}\notag\\
&\leq C_{NT}\left(\sum_{t=1}^{T}\|\hat{Q}_{t}-{Q}_{t}\|_2^{4}\right)^{1/2}\left(\sum_{t=1}^{T} \|\Phi(Z_t)^{\prime}\varepsilon_{t}\|^{4}\right)^{1/2} = O_{p}(TJ\xi_{J}^2\log J),
\end{align}
where $C_{NT}= \|B\|_{2}^{2}(\min_{t\leq T}\lambda_{\min}(\hat{Q}_{t}))^{-2}(\min_{i\leq N, t\leq T}\lambda_{\min}({Q}_{it}))^{-2}$, the first inequality follows since $\min_{t\leq T}\lambda_{\min}(Q_{t})\geq \min_{i\leq N,t\leq T}\lambda_{\min}(Q_{it})$, the second inequality is due to the Cauchy-Schwartz inequality, and the equality follows from Assumptions \ref{Ass: Basis}(i), \ref{Ass: LoadingsFactors}(i), and \ref{Ass: Improvedrates}(ii), as well as Lemmas \ref{Lem: TechB5} and \ref{Lem: TechB6}.

(ii) Let ${Q}_{t}\equiv E[\hat{Q}_{t}]$. By the fact that $\|C+D\|_{F}\leq \|C\|_{F}+\|D\|_{F}$,
\begin{align}\label{Eqn: Lem: TechB3: 4}
\left\|\sum_{t=1}^{T}B^{\prime}\hat{Q}_{t}^{-1}\Phi(Z_t)^{\prime}\varepsilon_tf_{t}^{\prime}\right\|^{2}_{F}&\leq 2\left\|\sum_{t=1}^{T}B^{\prime}{Q}_{t}^{-1}\Phi(Z_t)^{\prime}\varepsilon_tf_{t}^{\prime}\right\|^{2}_{F}\notag\\
&\hspace{-1.5cm}+2\left\|\sum_{t=1}^{T}B^{\prime}(\hat{Q}_{t}^{-1}-{Q}_{t}^{-1})\Phi(Z_t)^{\prime}\varepsilon_tf_{t}^{\prime}\right\|^{2}_{F}\equiv 2\mathcal{T}_1 + 2\mathcal{T}_2.
\end{align}
Therefore, it suffices to show that $\mathcal{T}_1=O_{p}(NT)$ and $\mathcal{T}_2=O_{p}(\sqrt{N}TJ\xi_{J}\sqrt{\log J})$. Note that $\|A\|^{2}_{F}=\mathrm{tr}(AA^{\prime})$. The former holds by the Markov's inequality, since
\begin{align}\label{Eqn: Lem: TechB3: 5}
\hspace{-0.5cm}E[\mathcal{T}_1]&=E\left[\mathrm{tr}\left(\sum_{t=1}^{T}\sum_{s=1}^{T}\sum_{i=1}^{N}\sum_{j=1}^{N}B^{\prime}{Q}^{-1}_{t}\phi(z_{it})\varepsilon_{it}f_{t}^{\prime}f_{s}\varepsilon_{js}\phi(z_{js})^{\prime}{Q}^{-1}_{s}B\right)\right]\notag\\
&=\sum_{t=1}^{T}\sum_{s=1}^{T}\sum_{i=1}^{N}\sum_{j=1}^{N}E\left[\phi(z_{it})^{\prime}{Q}_t^{-1}BB^{\prime}{Q}_s^{-1}\phi(z_{js})\right]f_t^{\prime}f_{s}E[\varepsilon_{it}\varepsilon_{js}]\notag\\
&\leq C_{NT}\max_{i\leq N,t\leq T} E[\|B^{\prime}Q_{t}^{-1}\phi(z_{it})\|^2] \sum_{i=1}^{N}\sum_{j=1}^{N}\sum_{t=1}^{T}\sum_{s=1}^{T}|E[\varepsilon_{it}\varepsilon_{js}]|=O(NT),
\end{align}
where $C_{NT}=\max_{t\leq T}\|f_t\|^{2}$, the second equality follows from the independence in Assumption \ref{Ass: DGP}(i) and the linearity of both expectation and trace operators, the inequality is due to the Cauchy-Schwartz inequality, and the last equality follows from Assumptions \ref{Ass: LoadingsFactors}(ii) and \ref{Ass: DGP}(iii), as well as Lemma \ref{Lem: TechB4}. Let $E_{\varepsilon}$ denote the expectation with respect to $\{\varepsilon_{t}\}_{t\leq T}$. For the latter, we have
\begin{align}\label{Eqn: Lem: TechB3: 6}
E_{\varepsilon}[\mathcal{T}_2]&= E_{\varepsilon}\hspace{-0.1cm}\left[\mathrm{tr}\hspace{-0.1cm}\left(\sum_{t=1}^{T}\sum_{s=1}^{T}\sum_{i=1}^{N}\sum_{j=1}^{N}B^{\prime}(\hat{Q}_{t}^{-1}\hspace{-0.1cm}-\hspace{-0.1cm}{Q}_{t}^{-1})\phi(z_{it})\varepsilon_{it}f_{t}^{\prime}f_{s}\varepsilon_{js}\phi(z_{js})^{\prime}(\hat{Q}_{s}^{-1}\hspace{-0.1cm}-\hspace{-0.1cm}{Q}_{s}^{-1})B\right)\right]\notag\\
&=\sum_{t=1}^{T}\sum_{s=1}^{T}\sum_{i=1}^{N}\sum_{j=1}^{N}\phi(z_{it})^{\prime}(\hat{Q}_{t}^{-1}-{Q}_{t}^{-1})BB^{\prime}(\hat{Q}_{s}^{-1}-{Q}_{s}^{-1})\phi(z_{js})f_t^{\prime}f_{s}E[\varepsilon_{it}\varepsilon_{js}]\notag\\
&\leq C^{\ast}_{NT}\sum_{t=1}^{T}\sum_{s=1}^{T}\sum_{i=1}^{N}\sum_{j=1}^{N}\|\hat{Q}_{t}-{Q}_{t}\|_{2}\|\phi(z_{it})\|\|\phi(z_{js})\||E[\varepsilon_{it}\varepsilon_{js}]|\notag\\
&\leq C^{\ast\ast}_{NT}\left(\sum_{t=1}^{T}\left(\sum_{s=1}^{T}\sum_{i=1}^{N}\sum_{j=1}^{N}\|\phi(z_{it})\|\|\phi(z_{js})\||E[\varepsilon_{it}\varepsilon_{js}]|\right)^{2}\right)^{1/2},
\end{align}
where $C^{\ast}_{NT}= \|B\|_{2}^{2}\max_{t\leq T}\|f_t\|^{2}[(\min_{t\leq T}\lambda_{\min}(\hat{Q}_{t}))^{-1}+(\min_{i\leq N, t\leq T}\lambda_{\min}({Q}_{it}))^{-1}]$ $(\min_{t\leq T}\lambda_{\min}(\hat{Q}_{t}))^{-1}(\min_{i\leq N, t\leq T}\lambda_{\min}({Q}_{it}))^{-1}$ and $C^{\ast\ast}_{NT}=C^{\ast}_{NT}(\sum_{t=1}^{T}\|\hat{Q}_{t}-{Q}_{t}\|^{2}_{2})^{1/2}$, the second equality follows from the independence in Assumption \ref{Ass: DGP}(i) and the linearity of both expectation and trace operators, the first inequality follows since $\min_{t\leq T}\lambda_{\min}(Q_{t})\geq \min_{i\leq N,t\leq T}\lambda_{\min}(Q_{it})$, and the last inequality is due to the Cauchy-Schwartz inequality. Moreover, we have
\begin{align}\label{Eqn: Lem: TechB3: 7}
&\hspace{0.8cm}E\left[\sum_{t=1}^{T}\left(\sum_{s=1}^{T}\sum_{i=1}^{N}\sum_{j=1}^{N}\|\phi(z_{it})\|\|\phi(z_{js})\||E[\varepsilon_{it}\varepsilon_{js}]|\right)^{2}\right]\notag\\
&\leq \max_{i\leq N,t\leq T}E[\|\phi(z_{it})\|^4]\sum_{t=1}^{T}\left(\sum_{s=1}^{T}\sum_{i=1}^{N}\sum_{j=1}^{N}|E[\varepsilon_{it}\varepsilon_{js}]|\right)^{2}\notag\\
&\leq J^{2}M^{2}\max_{m\leq M, j\leq J, i\leq N, t\leq T} E[\phi^{4}_{j}(z_{it,m})]\sum_{t=1}^{T}\left(\sum_{s=1}^{T}\sum_{i=1}^{N}\sum_{j=1}^{N}|E[\varepsilon_{it}\varepsilon_{js}]|\right)^{2},
\end{align}
where the first inequality is due to the Cauchy-Schwartz inequality, the second one follows since $\max_{i\leq N,t\leq T}E[\|\phi(z_{it})\|^4]\leq J^{2}M^{2}\max_{m\leq M, j\leq J, i\leq N, t\leq T} E[\phi^{4}_{j}(z_{it,m})]$. By Assumptions \ref{Ass: Basis}(i), \ref{Ass: LoadingsFactors}(i), (ii), and \ref{Ass: Improvedrates}(ii), as well as Lemma \ref{Lem: TechB6}, we have $C^{\ast\ast}_{NT}=O_{p}(\sqrt{T}\xi_{J}\sqrt{\log J}/\sqrt{N})$. Combining this, \eqref{Eqn: Lem: TechB3: 6} and \eqref{Eqn: Lem: TechB3: 7} implies that $E_{\varepsilon}[\mathcal{T}_2]=O_{p}(\sqrt{N}TJ\xi_{J}\sqrt{\log J})$ by Assumptions \ref{Ass: Improvedrates}(i) and (iv). Thus, the latter\textemdash$\mathcal{T}_2=O_{p}(\sqrt{N}TJ\xi_{J}\sqrt{\log J})$\textemdash holds by the Markov's inequality and Lemma \ref{Lem: TechA5}. This proves the first result, and the proofs of other results are similar. \qed

\begin{lem}\label{Lem: TechB4}
\hspace{-0.1cm}Suppose Assumptions \ref{Ass: LoadingsFactors}(i), \ref{Ass: Intercept}, and \ref{Ass: Improvedrates}(ii) hold. Let ${Q}_{t}\equiv E[\hat{Q}_{t}]$. Then
\[\max_{i\leq N,t\leq T} E[\|B^{\prime}Q_{t}^{-1}\phi(z_{it})\|^2]<\infty \text{ and } \max_{i\leq N,t\leq T} E[|a^{\prime}Q_{t}^{-1}\phi(z_{it})|^2]<\infty.\]
\end{lem}
\noindent{\sc Proof:} Since $\|x\|^2 =\mathrm{tr}(xx^{\prime})$,
\begin{align}\label{Eqn: Lem: TechB4: 1}
E[\|B^{\prime}Q_{t}^{-1}\phi(z_{it})\|^2]&=E[\mathrm{tr}(B^{\prime}Q_{t}^{-1}\phi(z_{it})\phi(z_{it})^{\prime}Q_{t}^{-1}B)]=\mathrm{tr}(B^{\prime}Q_{t}^{-1}Q_{it}Q_{t}^{-1}B)\notag\\
&\leq \max_{i\leq N,t\leq T}\lambda_{\max}(Q_{it})\left(\min_{t\leq T}\lambda_{\min}(Q_{t})\right)^{-1}K\|B\|_{2}^{2}\notag\\
&\leq \max_{i\leq N,t\leq T}\lambda_{\max}(Q_{it})\left(\min_{i\leq N,t\leq T}\lambda_{\min}(Q_{it})\right)^{-1}K\|B\|_{2}^{2},
\end{align}
where the second equality follows from the linearity of both expectation and trace operators, the first inequality follows since $\mathrm{tr}(B^{\prime}B)=\|B\|_{F}^{2}\leq K\|B\|_{2}^{2}$, and the second inequality follows since $\min_{t\leq T}\lambda_{\min}(Q_{t})\geq \min_{i\leq N,t\leq T}\lambda_{\min}(Q_{it})$. Thus, the first result of the lemma follows from \eqref{Eqn: Lem: TechB4: 1}, along with Assumptions \ref{Ass: LoadingsFactors}(i) and \ref{Ass: Improvedrates}(ii). The proof of the second result is similar.\qed

\begin{lem}\label{Lem: TechB5}
Under Assumptions \ref{Ass: DGP}(i), \ref{Ass: Improvedrates}(i), and (iv),
\[\sum_{t=1}^{T}\|\Phi(Z_t)^{\prime}\varepsilon_{t}\|^{4}=O_{p}(N^{2}TJ^{2}).\]
\end{lem}
\noindent{\sc Proof:} The result follows by the Markov's inequality, since
\begin{align}\label{Eqn: Lem: TechB5: 1}
&\hspace{0.8cm}E\left[\sum_{t=1}^{T}\|\Phi(Z_t)^{\prime}\varepsilon_{t}\|^{4}\right]= E\left[\sum_{t=1}^{T}\left(\sum_{i=1}^{N}\sum_{j=1}^{N}\phi(z_{it})^{\prime}\phi(z_{jt})\varepsilon_{it}\varepsilon_{jt}\right)^2\right]\notag\\
&= E\left[\sum_{t=1}^{T}\sum_{i=1}^{N}\sum_{j=1}^{N}\sum_{k=1}^{N}\sum_{\ell=1}^{N}\phi(z_{it})^{\prime}\phi(z_{jt})\phi(z_{kt})^{\prime}\phi(z_{\ell t})\varepsilon_{it}\varepsilon_{jt}\varepsilon_{kt}\varepsilon_{\ell t}\right]\notag\\
&= \sum_{t=1}^{T}\sum_{i=1}^{N}\sum_{j=1}^{N}\sum_{k=1}^{N}\sum_{\ell=1}^{N}E[\phi(z_{it})^{\prime}\phi(z_{jt})\phi(z_{kt})^{\prime}\phi(z_{\ell t})]E[\varepsilon_{it}\varepsilon_{jt}\varepsilon_{kt}\varepsilon_{\ell t}]\notag\\
&\leq \max_{i\leq N,t\leq T}E[\|\phi(z_{it})\|^4]\sum_{t=1}^{T}\sum_{i=1}^{N}\sum_{j=1}^{N}\sum_{k=1}^{N}\sum_{\ell=1}^{N}|E[\varepsilon_{it}\varepsilon_{jt}\varepsilon_{kt}\varepsilon_{\ell t}]|\notag\\
&\leq J^{2}M^{2} \max_{m\leq M, j\leq J, i\leq N, t\leq T} E[\phi^{4}_{j}(z_{it,m})]\sum_{t=1}^{T}\sum_{i=1}^{N}\sum_{j=1}^{N}\sum_{k=1}^{N}\sum_{\ell=1}^{N}|E[\varepsilon_{it}\varepsilon_{jt}\varepsilon_{kt}\varepsilon_{\ell t}]|\notag\\
&= O(N^{2}TJ^{2}),
\end{align}
where the third equality follows by the independence in Assumption \ref{Ass: DGP}(i), the first inequality is due to the Cauchy Schwartz inequality, the second inequality follows since $\max_{i\leq N,t\leq T}E[\|\phi(z_{it})\|^4]\leq J^{2}M^{2}\max_{m\leq M, j\leq J, i\leq N, t\leq T} E[\phi^{4}_{j}(z_{it,m})]$, and the last equality follows from Assumptions \ref{Ass: Improvedrates}(i) and (iv).\qed

\begin{lem}\label{Lem: TechB6}
Suppose Assumptions \ref{Ass: Improvedrates}(ii) and (iii) hold. Let ${Q}_{t}\equiv E[\hat{Q}_{t}]$. Assume $J\geq 2$ and $\xi^{2}_{J}\log J=o(N)$. Then
\[\sum_{t=1}^{T}\|\hat{Q}_t-Q_t\|^{2}_{2}=O_{p}\left(\frac{T\xi^{2}_{J}\log J}{N}\right) \text{ and } \sum_{t=1}^{T}\|\hat{Q}_t-Q_t\|^{4}_{2}=O_{p}\left(\frac{T\xi^{4}_{J}\log^{2} J}{N^{2}}\right).\]
\end{lem}
\noindent{\sc Proof:} Recall that $\hat{Q}_{t}= \sum_{i=1}^{N}\phi(z_{it})\phi(z_{it})^{\prime}/N$. Let $\eta_1,\ldots, \eta_N$ be an i.i.d. sequence of Rademacher variables. It then follows that
\begin{align}\label{Eqn: Lem: TechB6: 1}
\mathcal{D}_{t}&\equiv E[\|\hat{Q}_t-Q_t\|^{4}_{2}]\notag\\
&\leq 16 E\left[\left\|\frac{1}{N}\sum_{i=1}^{N}\eta_i\phi(z_{it})\phi(z_{it})^{\prime}\right\|_2^{4}\right]\notag\\
&\leq16C\frac{\log^{2} JM}{N^{2}}\sup_{z}\|\phi(z)\|^{4}E\left[\left\|\frac{1}{N}\sum_{i=1}^{N}\phi(z_{it})\phi(z_{it})^{\prime}\right\|^{2}_2\right]\notag\\
&\leq 16 M^{2}C\frac{\xi_{J}^{4}\log^{2}JM}{N^{2}}E[\|\hat{Q}_{t}\|^{2}_2],
\end{align}
where the first inequality follows from the independence in Assumption \ref{Ass: Improvedrates}(iii) and the symmetrization lemma (e.g., Lemma 2.3.1 of \citet{VW_WeakConvergence_1996}), the second inequality follows by Lemma \ref{Lem: TechB7} and the fact that $\phi(z_{it})^{\prime}\phi(z_{it})\leq \sup_{z}\|\phi(z)\|^{2}$, the third inequality follows since $\sup_{z}\|\phi(z)\|^{2}\leq M\sup_{z}\|\bar{\phi}(z)\|^{2}=M\xi_{J}^{2}$. Let $A = 16M^2C{\xi_{J}^{4}\log^{2} JM}/{N^{2}}$. Combining $E[\|\hat{Q}_{t}\|_{2}^{2}]\leq 2\sqrt{\mathcal{D}_{t}} + 2\|Q_{t}\|^{2}_2$ and \eqref{Eqn: Lem: TechB6: 1} leads to the inequality: $\mathcal{D}_{t}\leq 2A(\sqrt{\mathcal{D}_{t}} + \|Q_{t}\|^{2}_2)$. Solving the inequality yields
\begin{align}\label{Eqn: Lem: TechB6: 2}
E[\|\hat{Q}_t-Q_t\|^{4}_{2}]\leq \left(A+\sqrt{A^{2}+2A\|Q_{t}\|_{2}^{2}}\right)^{2}.
\end{align}
Thus, by the fact that $\max_{t\leq T}\|Q_{t}\|_{2}\leq \max_{i\leq N,t\leq T}\lambda_{\max}(Q_{it})$ and the Markov's inequality, the second result of the lemma follows from \eqref{Eqn: Lem: TechB6: 2} and Assumption \ref{Ass: Improvedrates}(ii). The first result of the lemma follows similarly by noting that $E[\|\hat{Q}_t-Q_t\|^{2}_{2}]\leq (E[\|\hat{Q}_t-Q_t\|^{4}_{2}])^{1/2}$. This completes the proof of the lemma. \qed

\begin{lem}[Khinchin inequality]\label{Lem: TechB7}
Let $S_1,\ldots, S_N$ be a sequence of symmetric $k\times k$ matrices and $\eta_1,\ldots, \eta_N$ be an i.i.d. sequence of Rademacher variables. Then for $k\geq 2$,
\[E_{\eta}\left[\left\|\frac{1}{N}\sum_{i=1}^{N}\eta_iS_i\right\|_2^{4}\right]\leq C\frac{\log^{2} k}{N^{2}}\left\|\frac{1}{N}\sum_{i=1}^{N}S^{2}_i\right\|^{2}_2\]
for some constant $C$, where $E_{\eta}$ denotes the expectation with respect to $\{\eta_{i}\}_{i\leq N}$.
\end{lem}
\noindent{\sc Proof:} This is a modified version of Lemma 6.1 in \citet{BelloniChernozhukovChetverikovKato_SeriesEstimator_2015}. The result is trivial for $2\leq k\leq e^{6}$. For $k>e^6$, we have
\begin{align}\label{Eqn: Lem: TechB7: 1}
&\hspace{0.8cm}E_{\eta}\left[\left\|\frac{1}{N}\sum_{i=1}^{N}\eta_iS_i\right\|_2^{4}\right]\leq E_{\eta}\left[\left\|\frac{1}{N}\sum_{i=1}^{N}\eta_iS_i\right\|_{S_{\log k}}^{4}\right]\notag\\
&\leq \left(E_{\eta}\left[\left\|\frac{1}{N}\sum_{i=1}^{N}\eta_iS_i\right\|_{S_{\log k}}^{\log k}\right]\right)^{4/\log k}\notag\\
&\leq C_0^4\frac{\log^2 k}{N^{2}}\left\|\left(\frac{1}{N}\sum_{i=1}^{N}S^{2}_i\right)^{1/2}\right\|^{4}_{S_{\log k}}\leq C_0^4 e^{4}\frac{\log^2 k}{N^{2}}\left\|\frac{1}{N}\sum_{i=1}^{N}S^{2}_i\right\|^{2}_2,
\end{align}
where the first inequality follows by (6.44) in \citet{BelloniChernozhukovChetverikovKato_SeriesEstimator_2015} and the fact that $\|\cdot\|_{S_{\log k}}$ is the Schatten norm, the second inequality follows by the Jensen's inequality, the third inequality follows by (6.45) in \citet{BelloniChernozhukovChetverikovKato_SeriesEstimator_2015} and $C_0$ is some positive constant, and the fourth inequality follows by (6.44) in \citet{BelloniChernozhukovChetverikovKato_SeriesEstimator_2015} again. Thus, the result of the lemma follows by setting $C=C_0^4 e^{4}$.\qed

\subsection{Proof of Theorem \ref{Thm: AsymDis}}
\noindent{\sc Proof of Theorem \ref{Thm: AsymDis}:} Let us first look at \eqref{Eqn: Thm: ImprovedRates: 1}. The asymptotic distribution can be obtained by choosing large $J$ and assuming $T$ not too large such that the terms with $O_{p}(J^{-2\kappa})$ and $O_{p}(J/N^{2})$ are negligible relative to the term with $O_{p}(J/NT)$. Thus, the asymptotic distribution is determined by the term with $O_{p}(J/NT)$. Specifically, by the facts that $\|C+D\|_{F}\leq \|C\|_{F} + \|D\|_{F}$ and $\|CD\|_{F}\leq \|C\|_{2}\|D\|_{F}$, \eqref{Eqn: Thm: Rate: 3} implies
\begin{align}\label{Eqn: Thm: AsymDis: 1}
\|\sqrt{NT}(\hat{B} - B H) - \sqrt{NT}D_{4}\hat{B}V^{-1}\|_{F}&\leq  \sqrt{NT}\|V^{-1}\|_2\|D_{5}\hat{B}\|_F\notag\\
&\hspace{-4.5cm}+\sqrt{NT}\|\hat{B}\|_{2}\|V^{-1}\|_2\sum_{j\neq 4,5}^{6}\|D_{j}\|_F =O_{p}\left(\frac{\sqrt{NT}}{J^{\kappa}}+\frac{\sqrt{TJ}}{\sqrt{N}}\right),
\end{align}
where the equality follows by $J=o(\sqrt{N})$, Lemmas \ref{Lem: TechA1}(i)-(iii), \ref{Lem: TechA2}(i), and \ref{Lem: TechB1}(ii), along with the fact that $\|D_6\|_F = \|D_3\|_F$. Let $\mathcal{L}_{NT}\equiv \sum_{t=1}^{T}{Q}^{-1}_t\Phi(Z_t)^{\prime}\varepsilon_t (f_{t}-\bar{f})^{\prime}/\sqrt{NT}$. Since $J=o(\sqrt{N})$, $J^{(1/2-\kappa)}=o(\sqrt{NT}/J^{\kappa})$. By the fact that $\|C+D\|_{F}\leq \|C\|_{F} + \|D\|_{F}$, combining \eqref{Eqn: Thm: AsymDis: 1} and Lemma \ref{Lem: TechC1} implies
\begin{align}\label{Eqn: Thm: AsymDis: 2}
\hspace{-0.2cm}\|\sqrt{NT}(\hat{B} - B H) - \mathcal{L}_{NT}B^{\prime}B\mathcal{M}\|_{F}=O_{p}\left(\frac{\sqrt{NT}}{J^{\kappa}}+\frac{\sqrt{TJ}}{\sqrt{N}}+\frac{\sqrt{J\xi_{J}}\log^{1/4}J}{N^{1/4}}\right).
\end{align}
Note that $\mathbb{N}_2$ is a $JM\times K$ matrix from the last $K$ columns of $\mathbb{N}$. Thus, the second result of the theorem follows from \eqref{Eqn: Thm: AsymDis: 2} and Lemma \ref{Lem: TechC2}. We now look at \eqref{Eqn: Thm: Rate: 5}. By the fact that $\|x+y\|\leq \|x\| + \|y\|$, it implies
\begin{align}\label{Eqn: Thm: AsymDis: 3}
&\|\sqrt{NT}(\hat{a} - a) -(I_{JM} - \hat{B}\hat{B}^{\prime})[\sqrt{N/T}\tilde{E}1_T-\sqrt{NT}(\hat{B}-BH)H^{-1}\bar{f}]\notag\\
&\hspace{0.5cm}+ \hat{B}\sqrt{NT}(\hat{B}-BH)^{\prime}a\|\leq \|(I_{JM} - \hat{B}\hat{B}^{\prime})\sqrt{N/T}\tilde{\Delta}1_T\|=O_{p}\left(\frac{\sqrt{NT}}{J^{\kappa}}\right),
\end{align}
where the equality follows by Lemma \ref{Lem: TechA3}(i). Given the rate of $\|\hat{B}-BH\|_{F}$ in Theorem \ref{Thm: ImprovedRates} and the rate of $\|N\tilde{E}1_T\|$ in Lemma \ref{Lem: TechA4}(ii), we may replace all $\hat{B}$ except those in $\hat{B}-BH$ with $BH$ to obtain
\begin{align}\label{Eqn: Thm: AsymDis: 4}
&\|\sqrt{NT}(\hat{a} - a) -(I_{JM} - BHH^{\prime}B^{\prime})[\sqrt{N/T}\tilde{E}1_T-\sqrt{NT}(\hat{B}-BH)H^{-1}\bar{f}]\notag\\
&\hspace{0.5cm}+ BH\sqrt{NT}(\hat{B}-BH)^{\prime}a\|=O_{p}\left(\frac{\sqrt{NT}}{J^{\kappa}}+\frac{\sqrt{TJ}}{\sqrt{N}}+\frac{J}{\sqrt{NT}}\right)
\end{align}
by noting that $J=o(\sqrt{N})$ and $J^{(1/2-\kappa)}=o(\sqrt{NT}/J^{\kappa})$. Similarly, given the rate of $H-\mathcal{H}$ in Lemma \ref{Lem: TechC3}, we may replace all $H$ except those in $\hat{B}-BH$ with $\mathcal{H}$ to obtain
\begin{align}\label{Eqn: Thm: AsymDis: 5}
&\|\sqrt{NT}(\hat{a} - a) -(I_{JM} - B^{\prime}\mathcal{H}\mathcal{H}^{\prime}B^{\prime})[\sqrt{N/T}\tilde{E}1_T-\sqrt{NT}(\hat{B}-BH)\mathcal{H}^{-1}\bar{f}]\notag\\
&\hspace{0.5cm}+ B\mathcal{H}\sqrt{NT}(\hat{B}-BH)^{\prime}a\|=O_{p}\left(\frac{\sqrt{NT}}{J^{\kappa}}+\frac{\sqrt{TJ}}{\sqrt{N}}+\frac{J}{\sqrt{NT}}\right)
\end{align}
Let $\ell_{NT}\equiv \sum_{t=1}^{T}{Q}^{-1}_t\Phi(Z_t)^{\prime}\varepsilon_t/\sqrt{NT}$. Given the rate of $\|\sqrt{N/T}\tilde{E}1_{T}-\ell_{NT}\|$ in Lemma \ref{Lem: TechC1}, we may replace $\sqrt{N/T}\tilde{E}1_{T}$ with $\ell_{NT}$ to obtain
\begin{align}\label{Eqn: Thm: AsymDis: 6}
&\|\sqrt{NT}(\hat{a} - a) -(I_{JM} - B^{\prime}\mathcal{H}\mathcal{H}^{\prime}B^{\prime})[\ell_{NT}-\sqrt{NT}(\hat{B}-BH)\mathcal{H}^{-1}\bar{f}]\notag\\
&\hspace{0.5cm}+ B\mathcal{H}\sqrt{NT}(\hat{B}-BH)^{\prime}a\|=O_{p}\left(\frac{\sqrt{NT}}{J^{\kappa}}+\frac{\sqrt{TJ}}{\sqrt{N}}+\frac{\sqrt{J\xi_{J}}\log^{1/4}J}{N^{1/4}}\right)
\end{align}
by noting that $J/\sqrt{NT} = o({\sqrt{J\xi_{J}}\log^{1/4}J}/{N^{1/4}})$. The arguments in \eqref{Eqn: Thm: AsymDis: 4}-\eqref{Eqn: Thm: AsymDis: 6} are similar to those for the first result in Lemma \ref{Lem: TechC1}. Note that $\mathbb{N}_1$ is a $JM\times 1$ vector from the first column of $\mathbb{N}$. Thus, the first result of the theorem follows from \eqref{Eqn: Thm: AsymDis: 6}, Lemma \ref{Lem: TechC2}, and the second result of the theorem. \qed

\subsubsection{Technical Lemmas}
\begin{lem}\label{Lem: TechC1}
Suppose Assumptions \ref{Ass: Basis}-\ref{Ass: Improvedrates} and \ref{Ass: Asym}(i) and (ii) hold. Let $\tilde{E}$, $D_4$, and $V$ be given in the proof of Theorem \ref{Thm: Rate}, and $\ell_{NT}$ and $\mathcal{L}_{NT}$ be given in the proof of Theorem \ref{Thm: AsymDis}. Assume (i) $N\to\infty$; (ii) $T\geq K+1$; (iii) $J\to\infty$ with $\xi^{2}_{J}\log J=o(N)$. Then
\[\|\sqrt{NT}D_4\hat{B}V^{-1}-\mathcal{L}_{NT}B^{\prime}B\mathcal{M}\|_{F}=O_{p}\left(\frac{1}{J^{(\kappa-1/2)}}+\frac{\sqrt{J\xi_{J}}\log^{1/4}J}{N^{1/4}}\right)\]
and
\[\|\sqrt{N/T}\tilde{E}1_{T}-\ell_{NT}\|=O_{p}\left(\frac{\sqrt{J\xi_{J}}\log^{1/4}J}{N^{1/4}}\right),\]
where $\mathcal{M}$ is a nonrandom matrix given in Lemma \ref{Lem: TechC3}.
\end{lem}
\noindent{\sc Proof:} For the first result, we have the following decomposition
\begin{align}\label{Eqn: Lemma: TechC1: 1}
\sqrt{NT}D_4\hat{B}V^{-1} &= \sqrt{N/T}\tilde{E}M_TFB^{\prime}B\mathcal{M}+\sqrt{N/T}\tilde{E}M_TFB^{\prime}(\hat{B}-BH)V^{-1}\notag\\
&\hspace{0.5cm}+\sqrt{N/T}\tilde{E}M_TFB^{\prime}B(HV^{-1}-\mathcal{M})\equiv \mathcal{T}_1 +\mathcal{T}_2+ \mathcal{T}_3.
\end{align}
Therefore, it suffices to show that $\|\mathcal{T}_1-\mathcal{L}_{NT}B^{\prime}B\mathcal{M}\|_{F}=O_{p}(\sqrt{J\xi_{J}}\log^{1/4}J/N^{1/4})$, $\|\mathcal{T}_2\|_{F} = O_{p}(J^{(1/2-\kappa)}+J^{3/2}/N + J/\sqrt{NT})$ and $\|\mathcal{T}_3\|_{F} = O_{p}(J^{(1/2-\kappa)}+J^{3/2}/N + J/\sqrt{NT})$. The first one holds, since
\begin{align}\label{Eqn: Lemma: TechC1: 2}
\|\mathcal{T}_1-\mathcal{L}_{NT}B^{\prime}B\mathcal{M}\|_{F}&\leq \|B\|_{2}^{2}\|\mathcal{M}\|_{2}\left\|\frac{1}{\sqrt{NT}}\sum_{t=1}^{T}(\hat{Q}^{-1}_t-Q^{-1}_{t})\Phi(Z_t)^{\prime}\varepsilon_t f_{t}^{\prime}\right\|_{F}\notag\\
&\hspace{0.5cm}+\|B\|_{2}^{2}\|\mathcal{M}\|_{2}\|\bar{f}\|\left\|\frac{1}{\sqrt{NT}}\sum_{t=1}^{T}(\hat{Q}^{-1}_t-Q^{-1}_{t})\Phi(Z_t)^{\prime}\varepsilon_t \right\|\notag\\
&=O_{p}\left(\frac{\sqrt{J\xi_{J}}\log^{1/4}J}{N^{1/4}}\right),
\end{align}
where the equality follows from Assumptions \ref{Ass: LoadingsFactors}(i) and (ii), as well as Lemma \ref{Lem: TechC4}. The latter two follow by a similar argument. The second result also follows by a similar argument as in \eqref{Eqn: Lemma: TechC1: 2}. This completes the proof of the lemma.\qed

\begin{lem}\label{Lem: TechC2}
Suppose Assumptions \ref{Ass: LoadingsFactors}(ii), \ref{Ass: DGP}(i), (ii), \ref{Ass: Improvedrates}(i)-(iii), and \ref{Ass: Asym}(ii) and (iii) hold. Let $\ell_{NT}$ and $\mathcal{L}_{NT}$ be given in the proof of Theorem \ref{Thm: AsymDis}. Then there exists a $JM\times (K+1)$ random matrix $\mathbb{N}$ with $\mathrm{vec}(\mathbb{N})\sim N(0,\Omega)$ such that
\[\|(\ell_{NT},\mathcal{L}_{NT})-\mathbb{N}\|_{F} = O_{p}\left(\frac{J^{5/6}}{N^{1/6}}\right).\]
\end{lem}
\noindent{\sc Proof:} Let $\zeta_i\equiv \sum_{t=1}^{T}f_{t}^{\dag}\otimes Q^{-1}_{t}\phi(z_{it})\varepsilon_{it}/\sqrt{NT}$. Then $\mathrm{vec}((\ell_{NT},\mathcal{L}_{NT}))=\sum_{i=1}^{N}\zeta_i$. Note that $E[\zeta_{i}] = 0$ by Assumptions \ref{Ass: DGP}(i) and (ii), and $\zeta_{1},\ldots, \zeta_{N}$ are independent by Assumptions \ref{Ass: DGP}(i), \ref{Ass: Improvedrates}(iii), and \ref{Ass: Asym}(ii). Moreover,
\begin{align}\label{Eqn: Lem: TechC2: 1}
\sum_{i=1}^{N}E[\|\zeta_{i}\|^{3}]\leq \sum_{i=1}^{N}(E[\|\zeta_{i}\|^{4}])^{3/4} = O\left(\frac{J^{3/2}}{\sqrt{N}}\right),
\end{align}
where the inequality follows by the Liapounov’s inequality, and the equality follows from Assumptions \ref{Ass: LoadingsFactors}(ii), \ref{Ass: Improvedrates}(i), (ii), and \ref{Ass: Asym}(iii) because
\begin{align}\label{Eqn: Lem: TechC2: 2}
\hspace{-0.2cm}E[\|\zeta_{i}\|^{4}]&=\frac{1}{N^{2}T^{2}}E\left[\left(\sum_{t=1}^{T}\sum_{s=1}^{T} \phi(z_{it})^{\prime}Q^{-1}_tQ^{-1}_{s}\phi(z_{is})f_{t}^{\dag\prime}f_{s}^{\dag}\varepsilon_{it}\varepsilon_{is}\right)^2\right]\notag\\
&\hspace{-0.5cm}\leq C_{NT}\max_{i\leq N,t\leq T}E[\|\phi(z_{it})\|^{4}]\frac{1}{N^{2}T^{2}}\sum_{t=1}^{T}\sum_{s=1}^{T}\sum_{u=1}^{T}\sum_{v=1}^{T}|E[\varepsilon_{it}\varepsilon_{is}\varepsilon_{iu}\varepsilon_{iv}]|\notag\\
&\hspace{-0.5cm}\leq C_{NT}\hspace{-0.2cm}\max_{m\leq M, j\leq J, i\leq N, t\leq T} \hspace{-0.1cm}E[\phi^{4}_{j}(z_{it,m})]\frac{J^{2}M^{2}}{N^{2}T^{2}}\sum_{t=1}^{T}\sum_{s=1}^{T}\sum_{u=1}^{T}\sum_{v=1}^{T}|E[\varepsilon_{it}\varepsilon_{is}\varepsilon_{iu}\varepsilon_{iv}]|,
\end{align}
where $C_{NT}=\max_{t\leq T}\|f_{t}^{\dag}\|^{4}(\min_{i\leq N, t\leq T}\lambda_{\min}(Q_{it}))^{-4}$, the first inequality follows by the independence in Assumption \ref{Ass: DGP}(i), the Cauchy-Schwartz inequality, and the fact that $\min_{t\leq T}\lambda_{\min}(Q_{t})\geq \min_{i\leq N, t\leq T}\lambda_{\min}(Q_{it})$, and the second inequality follows since $\max_{i\leq N,t\leq T}E[\|\phi(z_{it})\|^4]\leq J^{2}M^{2}\max_{m\leq M, j\leq J, i\leq N, t\leq T} E[\phi^{4}_{j}(z_{it,m})]$. In addition, $\Omega=E[\mathrm{vec}((\ell_{NT},\mathcal{L}_{NT}))\mathrm{vec}((\ell_{NT},\mathcal{L}_{NT}))^{\prime}]$. Thus, Lemma \ref{Lem: TechC5} implies that there is a $JM\times (K+1)$ random matrix $\mathbb{N}$ with $\mathrm{vec}(\mathbb{N})\sim N(0,\Omega)$ such that
\begin{align}\label{Eqn: Lem: TechC2: 3}
\|(\mathcal{L}_{NT},\ell_{NT})-\mathbb{N}\|_{F} = \|\mathrm{vec}((\mathcal{L}_{NT},\ell_{NT}))-\mathrm{vec}(\mathbb{N})\| = O_{p}\left(\frac{J^{5/6}}{N^{1/6}}\right).
\end{align}
This completes the proof of the Lemma.\qed

\begin{lem}\label{Lem: TechC3}
Suppose Assumptions \ref{Ass: Basis}-\ref{Ass: Intercept} and \ref{Ass: Asym}(i) hold. Let $V$ be given in the proof of Theorem \ref{Thm: Rate}. Assume (i) $N\to\infty$; (ii) $T\geq K+1$; (iii) $J\to\infty$ with $J=o(\sqrt{N})$. Then
\[H=\mathcal{H} +O_{p}\left(\frac{1}{J^{\kappa}}+\frac{J}{N}+\frac{\sqrt{J}}{\sqrt{NT}}\right) \text{ and } HV^{-1}=\mathcal{M} +O_{p}\left(\frac{1}{J^{\kappa}}+\frac{J}{N}+\frac{\sqrt{J}}{\sqrt{NT}}\right),\]
where $\mathcal{H}=({F^{\prime}M_TF}/{T})^{1/2}\Upsilon\mathcal{V}^{-1/2}$, $\mathcal{M}=\mathcal{H}\mathcal{V}^{-1}$, $\mathcal{V}$ is a diagonal matrix of the eigenvalues of $({F^{\prime}M_TF}/{T})^{1/2}$ $B^{\prime}B({F^{\prime}M_TF}/{T})^{1/2}$ and $\Upsilon$ is the corresponding eigenvector matrix such that $\Upsilon^{\prime}\Upsilon=I_{K}$.
\end{lem}
\noindent{\sc Proof:} By the definition of $\hat{B}$, $(\tilde{Y}M_T\tilde{Y}^{\prime}/T)\hat{B} = \hat{B}V$. Pre-multiply it on both sides by $(F^{\prime}M_TF/T)^{1/2}B^{\prime}$ to obtain
\begin{align}\label{Eqn: Lemma: TechC3: 1}
({F^{\prime}M_TF}/{T})^{1/2}B^{\prime}({\tilde{Y}M_T\tilde{Y}^{\prime}}/{T})\hat{B} = ({F^{\prime}M_TF}/{T})^{1/2}B^{\prime}\hat{B}V.
\end{align}
To simplify notation, let $\delta_{NT}\equiv (F^{\prime}M_TF/T)^{1/2}B^{\prime}(\tilde{Y}M_T\tilde{Y}^{\prime}/T-B(F^{\prime}M_TF/T)B^{\prime})\hat{B}$ and $R_{NT}\equiv({F^{\prime}M_TF}/{T})^{1/2}$$B^{\prime}\hat{B}$. Then we can rewrite \eqref{Eqn: Lemma: TechC3: 1} as
\begin{align}\label{Eqn: Lemma: TechC3: 2}
[({F^{\prime}M_TF}/{T})^{1/2}B^{\prime}B({F^{\prime}M_TF}/{T})^{1/2} + \delta_{NT}R_{NT}^{-1}] R_{NT}= R_{NT}V.
\end{align}
Let $D_{NT}$ be a diagonal matrix consisting the diagonal elements of $R_{NT}^{\prime}R_{NT}$. Denote $\Upsilon_{NT}\equiv R_{NT}D^{-1/2}_{NT}$, which has a unit length. Then we can further rewrite \eqref{Eqn: Lemma: TechC3: 2} as
\begin{align}\label{Eqn: Lemma: TechC3: 3}
[({F^{\prime}M_TF}/{T})^{1/2}B^{\prime}B({F^{\prime}M_TF}/{T})^{1/2} + \delta_{NT}R_{NT}^{-1}] \Upsilon_{NT}= \Upsilon_{NT}V,
\end{align}
which implies that $({F^{\prime}M_TF}/{T})^{1/2}B^{\prime}B({F^{\prime}M_TF}/{T})^{1/2} + \delta_{NT}R_{NT}^{-1}$ has eigenvector matrix $\Upsilon_{NT}$ and eigenvalue matrix $V$. Since $R_{NT}= ({F^{\prime}M_TF}/{T})^{1/2}B^{\prime}BH+o_{p}(1)$ by simple algebra and Theorem \ref{Thm: Rate}, $R_{NT}^{-1}=O_{p}(1)$ by Assumptions \ref{Ass: LoadingsFactors}(i)-(iii) and Lemma \ref{Lem: TechA2}. This, along  with \eqref{Eqn: Lem: TechA2: 1} and Assumptions \ref{Ass: LoadingsFactors}(i) and (ii), implies
\begin{align}\label{Eqn: Lemma: TechC3: 4}
\delta_{NT}R_{NT}^{-1}=O_{p}\left(\frac{1}{J^{\kappa}}+\frac{J}{N}+\frac{\sqrt{J}}{\sqrt{NT}}\right).
\end{align}
Since the eigenvalues of $({F^{\prime}M_TF}/{T})B^{\prime}B$ are equal to those of $({F^{\prime}M_TF}/{T})^{1/2}B^{\prime}B$ $({F^{\prime}M_TF}/{T})^{1/2}$, the eigenvalues of $({F^{\prime}M_TF}/{T})^{1/2}B^{\prime}B({F^{\prime}M_TF}/{T})^{1/2}$ are distinct by Assumption \ref{Ass: Asym}(i). By the eigenvector perturbation theory, there exists a unique eigenvector matrix $\Upsilon$ of $({F^{\prime}M_TF}/{T})^{1/2}B^{\prime}B({F^{\prime}M_TF}/{T})^{1/2}$ such that
\begin{align}\label{Eqn: Lemma: TechC3: 5}
\Upsilon_{NT}=\Upsilon+O_{p}\left(\frac{1}{J^{\kappa}}+\frac{J}{N}+\frac{\sqrt{J}}{\sqrt{NT}}\right).
\end{align}
By \eqref{Eqn: Lem: TechA2: 1}, $R_{NT}^{\prime}R_{NT} = \hat{B}^{\prime}B({F^{\prime}M_TF}/{T})B^{\prime}\hat{B}= \hat{B}^{\prime}(\tilde{Y}M_T\tilde{Y}^{\prime}/T)\hat{B}+O_{p}(J^{-\kappa}+{J}/{N}+{\sqrt{J}}/{\sqrt{NT}}) = V+O_{p}(J^{-\kappa}+{J}/{N}+{\sqrt{J}}/{\sqrt{NT}})$. This implies that
\begin{align}\label{Eqn: Lemma: TechC3: 6}
D_{NT} = V + O_{p}\left(\frac{1}{J^{\kappa}}+\frac{J}{N}+\frac{\sqrt{J}}{\sqrt{NT}}\right).
\end{align}
Recall that $H^{\diamond}=(F^{\prime}M_TF/T)B^{\prime}\hat{B}V^{-1}$ as given in the proof of Lemma \ref{Lem: TechA2}(i). Thus, by \eqref{Eqn: Lemma: TechC3: 5} and \eqref{Eqn: Lemma: TechC3: 6}, $H^{\diamond}= ({F^{\prime}M_TF}/{T})^{1/2}R_{NT} V^{-1}  = ({F^{\prime}M_TF}/{T})^{1/2}\Upsilon_{NT} D^{1/2}_{NT}V^{-1} = \mathcal{H} + O_{p}({J^{-\kappa}}+J/N+\sqrt{J/NT})$, which together with \eqref{Eqn: Lem: TechA2: 2} and \eqref{Eqn: Lem: TechA2: 3} leads to the first result of the lemma. The second result of the lemma follows from \eqref{Eqn: Lem: TechA2: 2}, the first result of the lemma, and Lemma \ref{Lem: TechA2}(i). \qed

\begin{lem}\label{Lem: TechC4}
Suppose Assumptions \ref{Ass: Basis}(i), \ref{Ass: LoadingsFactors}(ii), \ref{Ass: DGP}(i), (ii), \ref{Ass: Improvedrates}, and \ref{Ass: Asym}(ii) hold. Assume $J\geq 2$ and $\xi^{2}_{J}\log J=o(N)$. Then
\[\left\|\frac{1}{\sqrt{NT}}\sum_{t=1}^{T}(\hat{Q}^{-1}_t-Q^{-1}_{t})\Phi(Z_t)^{\prime}\varepsilon_t f_{t}^{\prime}\right\|_{F} = O_{p}\left(\frac{\sqrt{J\xi_{J}}\log^{1/4}J}{N^{1/4}}\right)\]
 and
\[\left\|\frac{1}{\sqrt{NT}}\sum_{t=1}^{T}(\hat{Q}^{-1}_t-Q^{-1}_{t})\Phi(Z_t)^{\prime}\varepsilon_t\right\| = O_{p}\left(\frac{\sqrt{J\xi_{J}}\log^{1/4}J}{N^{1/4}}\right).\]
\end{lem}
\noindent{\sc Proof:} Let $\mathcal{T}\equiv \sum_{t=1}^{T}(\hat{Q}^{-1}_t-Q^{-1}_{t})\Phi(Z_t)^{\prime}\varepsilon_t f_{t}^{\prime}/\sqrt{NT}$ and $E_{\varepsilon}$ denote the expectation with respect to $\{\varepsilon_{t}\}_{t\leq T}$. Since $\|A\|_{F}^{2}=\mathrm{tr}(AA^{\prime})$,
\begin{align}\label{Eqn: Lem: TechC4: 1}
E_{\varepsilon}[\|\mathcal{T}\|^{2}_{F}]&=\frac{1}{NT}E_{\varepsilon}\left[\mathrm{tr}\left(\sum_{t=1}^{T}\sum_{s=1}^{T}(\hat{Q}^{-1}_t-Q^{-1}_{t})\Phi(Z_t)^{\prime}\varepsilon_tf_{t}^{\prime}
f_{s}\varepsilon_s^{\prime}\Phi(Z_s)(\hat{Q}^{-1}_s-Q^{-1}_{s})\right)\right]\notag\\
&=\frac{1}{NT}\sum_{i=1}^{N}\sum_{j=1}^{N}\sum_{t=1}^{T}\sum_{s=1}^{T}\phi(z_{it})^{\prime}(\hat{Q}^{-1}_t-Q^{-1}_{t})(\hat{Q}^{-1}_s-Q^{-1}_{s})\phi(z_{js})f_{t}^{\prime}f_{s}E[\varepsilon_{it}\varepsilon_{js}]\notag\\
&=\frac{1}{NT}\sum_{i=1}^{N}\sum_{t=1}^{T}\sum_{s=1}^{T}\phi(z_{it})^{\prime}(\hat{Q}^{-1}_t-Q^{-1}_{t})(\hat{Q}^{-1}_s-Q^{-1}_{s})\phi(z_{is})f_{t}^{\prime}f_{s}E[\varepsilon_{it}\varepsilon_{is}]\notag\\
&\leq C_{NT}^{\ast}\frac{1}{NT}\sum_{i=1}^{N}\sum_{t=1}^{T}\sum_{s=1}^{T}\|\hat{Q}_t-Q_{t}\|_{2}\|\phi(z_{it})\|\|\phi(z_{is})\||E[\varepsilon_{it}\varepsilon_{is}]|\notag\\
&\leq C_{NT}^{\ast\ast}\frac{1}{NT}\left(\sum_{t=1}^{T}\left(\sum_{i=1}^{N}\sum_{s=1}^{T}\|\phi(z_{it})\|\|\phi(z_{is})\||E[\varepsilon_{it}\varepsilon_{is}]|\right)^{2}\right)^{1/2},
\end{align}
where $C^{\ast}_{NT}= (\min_{t\leq T}\lambda_{\min}(\hat{Q}_{t}))^{-1}[(\min_{t\leq T}\lambda_{\min}(\hat{Q}_{t}))^{-1}+(\min_{i\leq N, t\leq T}\lambda_{\min}({Q}_{it}))^{-1}]$ $(\min_{i\leq N, t\leq T}\lambda_{\min}({Q}_{it}))^{-1}\max_{t\leq T}\|f_t\|^{2}$ and $C^{\ast\ast}_{NT}=C^{\ast}_{NT}(\sum_{t=1}^{T}\|\hat{Q}_{t}-{Q}_{t}\|^{2}_{2})^{1/2}$, the second equality follows from the independence in Assumption \ref{Ass: DGP}(i) and the linearity of both expectation and trace operators, the third equality follows by Assumption \ref{Ass: DGP}(ii) and the independence in Assumption \ref{Ass: Asym}(ii), the first inequality follows since $\min_{t\leq T}\lambda_{\min}(Q_{t})\geq \min_{i\leq N,t\leq T}\lambda_{\min}(Q_{it})$, and the last inequality is due to the Cauchy-Schwartz inequality. Moreover, we have
\begin{align}\label{Eqn: Lem: TechC4: 2}
&\hspace{0.8cm}E\left[\sum_{t=1}^{T}\left(\sum_{i=1}^{N}\sum_{s=1}^{T}\|\phi(z_{it})\|\|\phi(z_{is})\||E[\varepsilon_{it}\varepsilon_{is}]|\right)^{2}\right]\notag\\
&\leq \max_{i\leq N,t\leq T}E[\|\phi(z_{it})\|^4]\sum_{t=1}^{T}\left(\sum_{i=1}^{N}\sum_{s=1}^{T}|E[\varepsilon_{it}\varepsilon_{is}]|\right)^{2}\notag\\
&\leq J^{2}M^{2}\max_{m\leq M, j\leq J, i\leq N, t\leq T} E[\phi^{4}_{j}(z_{it,m})]\sum_{t=1}^{T}\left(\sum_{i=1}^{N}\sum_{s=1}^{T}|E[\varepsilon_{it}\varepsilon_{is}]|\right)^{2},
\end{align}
where the first inequality is due to the Cauchy-Schwartz inequality, the second one follows since $\max_{i\leq N,t\leq T}E[\|\phi(z_{it})\|^4]\leq J^{2}M^{2}\max_{m\leq M, j\leq J, i\leq N, t\leq T} E[\phi^{4}_{j}(z_{it,m})]$. By Assumptions \ref{Ass: Basis}(i), \ref{Ass: LoadingsFactors}(ii), and \ref{Ass: Improvedrates}(ii), along with Lemma \ref{Lem: TechB6}, we obtain that $C^{\ast\ast}_{NT} = O_{p}(\sqrt{T}\xi_{J}\sqrt{\log J}/\sqrt{N})$. Combining this, \eqref{Eqn: Lem: TechC4: 1} and \eqref{Eqn: Lem: TechC4: 2} implies that $E_{\varepsilon}[\|\mathcal{T}\|^{2}_{F}]=O_{p}(J\xi_{J}\sqrt{\log J}/\sqrt{N})$ by Assumptions \ref{Ass: Improvedrates}(i) and (iv). Thus, the first result of the lemma follows by the Markov's inequality and Lemma \ref{Lem: TechA5}. The proof of the second result is similar. \qed

\begin{lem}[Yurinskii’s coupling]\label{Lem: TechC5}
Let $\zeta_1, \ldots, \zeta_N$ be independent random $k-$vectors with $E[\zeta_i] = 0$ for each $i$ and $\beta = \sum_{i=1}^{N}E[\|\zeta_i\|^{3}]$ finite. Let $S=\sum_{i=1}^{N}\zeta_i$. For each $\delta>0$, there exists a random vector $\mathbb{S}$ in the same probability space with $S$ with a $N(0,E[SS^{\prime}])$ distribution such that
\begin{align*}
P\{\|S-\mathbb{S}\|>3\delta\}\leq C_0 D_0\left(1+\frac{|\log(1/D_0)|}{k}\right)
\end{align*}
for some universal constant $C_0$, where $D_0 = \beta k \delta^{-3}$.
\end{lem}
\noindent{\sc Proof:} This is the Yurinskii’s coupling, see Theorem 10 in \citet{Pollard_Probability_2002}.\qed

\subsection{Proof of Theorem \ref{Thm: Boot}}
\noindent{\sc Proof of Theorem \ref{Thm: Boot}:} Let us begin by defining some notation. For $A_t = \Delta_t\equiv R(Z_t) + \Delta(Z_{t})f_{t}$ and $\varepsilon_t$, let $\tilde{A}^{\ast}_{t} \equiv (\Phi(Z_t)^{\ast\prime}\Phi(Z_t))^{-1}\Phi(Z_t)^{\ast\prime}A_t$.  Let $\tilde{\Delta}^{\ast}\equiv (\tilde{\Delta}^\ast_1,\ldots, \tilde{\Delta}^{\ast}_T)$ and $\tilde{E}^{\ast}\equiv (\tilde{\varepsilon}^{\ast}_1,\ldots, \tilde{\varepsilon}^{\ast}_T)$. Then we have
\begin{align}\label{Eqn: Thm: Boot: 1}
\tilde{Y}^{\ast} = a1_{T}^{\prime} + BF^{\prime} + \tilde{\Delta}^{\ast} + \tilde{E}^{\ast},
\end{align}
where $1_{T}$ denotes a $T\times 1$ vector of ones. Recall $M_T= I_{T} - 1_T1_T^{\prime}/T$. Post-multiplying \eqref{Eqn: Thm: Boot: 1} by $M_T$ to remove $a$, we thus obtain
\begin{align}\label{Eqn: Thm: Boot: 2}
\tilde{Y}^{\ast}M_T = B(M_TF)^{\prime} + \tilde{\Delta}^{\ast}M_T + \tilde{E}^{\ast}M_T.
\end{align}
Recall that $V$ is a $K\times K$ diagonal matrix of the first $K$ largest eigenvalues of $\tilde{Y}M_T\tilde{Y}^{\prime}/T$ as defined in the proof of Theorem \ref{Thm: ImprovedRates}, $H=F^{\prime}M_T\hat{F}(\hat{F}^{\prime}M_T\hat{F})^{-1}$, and $\hat{F}^{\prime}M_T\hat{F}/T = V$ as showed in the proof of Theorem \ref{Thm: ImprovedRates}. By the definitions of $\hat{B}^{\ast}$, $\hat{B}^{\ast} = \tilde{Y}^{\ast}M_T\hat{F}(\hat{F}^{\prime}M_T\hat{F})^{-1}$. We may substitute \eqref{Eqn: Thm: Boot: 2} to it to obtain
\begin{align}\label{Eqn: Thm: Boot: 3}
\hat{B}^{\ast} - B H = [(\tilde{\Delta}^{\ast} + \tilde{E}^{\ast})M_T\tilde{Y}^{\prime}/T]\hat{B}V^{-1} = \sum_{j=1}^{6}D^{\ast}_{j}\hat{B}V^{-1},
\end{align}
where in the first equality we have used $\hat{F}^{\prime}M_T\hat{F}/T=V$ and $\hat{F} = \tilde{Y}^{\prime}\hat{B}$, in the second equality we have substituted \eqref{Eqn: Thm: Rate: 2} into the equation, and $D^{\ast}_1=\tilde{\Delta}^{\ast}M_TF B^{\prime}/T$, $D^{\ast}_2 = \tilde{\Delta}^{\ast}M_{T}\tilde{\Delta}^{\prime}/T$, $D^{\ast}_3=\tilde{\Delta}^{\ast}M_T\tilde{E}^{\prime}/T$, $D^{\ast}_4 =\tilde{E}^{\ast}M_TFB^{\prime}/T$, $D^{\ast}_5 = \tilde{E}^{\ast}M_T\tilde{E}^{\prime}/T$, and $D^{\ast}_6 = \tilde{E}^{\ast}M_T\tilde{\Delta}^{\prime}/T$. We can conduct the same exercise as in \eqref{Eqn: Thm: AsymDis: 1} to obtain
\begin{align}\label{Eqn: Thm: Boot: 4}
\|\sqrt{NT}(\hat{B}^{\ast} - B H) - \sqrt{NT}D^{\ast}_{4}\hat{B}V^{-1}\|_{F}&\leq  \sqrt{NT}\|V^{-1}\|_2\|D^{\ast}_{5}\hat{B}\|_F\notag\\
&\hspace{-4.5cm}+\sqrt{NT}\|\hat{B}\|_{2}\|V^{-1}\|_2\sum_{j\neq 4,5}^{6}\|D^{\ast}_{j}\|_F =O_{p}\left(\frac{\sqrt{NT}}{J^{\kappa}}+\frac{\sqrt{TJ}}{\sqrt{N}}\right),
\end{align}
where the equality follows by $J=o(\sqrt{N})$, Lemmas \ref{Lem: TechD1} and \ref{Lem: TechA2}(i). Let $\mathcal{L}^{\ast\ast}_{NT}\equiv \sum_{t=1}^{T}{Q}^{-1}_t\Phi(Z_t)^{\ast\prime}\varepsilon_t(f_{t}-\bar{f})^{\prime}/\sqrt{NT}$. Since $J=o(\sqrt{N})$, $J^{(1/2-\kappa)}=o(\sqrt{NT}/J^{\kappa})$. By the fact that $\|C+D\|_{F}\leq \|C\|_{F} + \|D\|_{F}$, combining \eqref{Eqn: Thm: Boot: 4} and Lemma \ref{Lem: TechD2} implies
\begin{align}\label{Eqn: Thm: Boot: 5}
\hspace{-0.4cm}\|\sqrt{NT}(\hat{B}^{\ast} - B H) - \mathcal{L}^{\ast\ast}_{NT}B^{\prime}B\mathcal{M}\|_{F}=O_{p}\left(\frac{\sqrt{NT}}{J^{\kappa}}+\frac{\sqrt{TJ}}{\sqrt{N}}+\frac{\sqrt{J\xi_{J}}\log^{1/4}J}{N^{1/4}}\right).
\end{align}
Let $\mathcal{L}^{\ast}_{NT}\equiv \sum_{t=1}^{T}{Q}^{-1}_t[\Phi(Z_t)^{\ast}-\Phi(Z_t)]^{\prime}\varepsilon_t(f_{t}-\bar{f})^{\prime}/\sqrt{NT}=\mathcal{L}^{\ast\ast}_{NT}-\mathcal{L}_{NT}$. Note that $\sqrt{NT}(\hat{B}^{\ast} - \hat{B})=\sqrt{NT}(\hat{B}^{\ast} - B H) - \sqrt{NT}(\hat{B} - B H)$. By the fact that $\|C+D\|_{F}\leq \|C\|_{F} + \|D\|_{F}$, we now may combine \eqref{Eqn: Thm: AsymDis: 2} and \eqref{Eqn: Thm: Boot: 5} to obtain
\begin{align}\label{Eqn: Thm: Boot: 6}
\|\sqrt{NT}(\hat{B}^{\ast} - \hat{B}) - \mathcal{L}^{\ast}_{NT}B^{\prime}B\mathcal{M}\|_{F}=O_{p}\left(\frac{\sqrt{NT}}{J^{\kappa}}+\frac{\sqrt{TJ}}{\sqrt{N}}+\frac{\sqrt{J\xi_{J}}\log^{1/4}J}{N^{1/4}}\right).
\end{align}
Note that $\mathbb{N}^{\ast}_2$ is a $JM\times K$ matrix from the last $K$ columns of $\mathbb{N}^{\ast}$. Thus, the second result of the theorem follows from \eqref{Eqn: Thm: Boot: 6} and Lemmas \ref{Lem: TechA5} and \ref{Lem: TechD3}. We now show the first result of the theorem. By the definition of $\hat{a}^{\ast}$,
\begin{align}\label{Eqn: Thm:  Boot: 7}
\hat{a}^{\ast}\hspace{-0.05cm}- \hspace{-0.05cm}a &= -\hspace{-0.05cm}\hat{B}^{\ast}(\hat{B}^{\ast\prime}\hat{B}^{\ast})^{-1}(\hat{B}^{\ast}\hspace{-0.05cm}-\hspace{-0.05cm}BH)^{\prime}a + (I_{JM} \hspace{-0.05cm}- \hat{B}^{\ast}(\hat{B}^{\ast\prime}\hat{B}^{\ast})^{-1}\hat{B}^{\ast\prime})(BH-\hat{B}^{\ast})H^{-1}\bar{f}\notag\\
&\hspace{-0.5cm}+ (I_{JM} - \hat{B}^{\ast}(\hat{B}^{\ast\prime}\hat{B}^{\ast})^{-1}\hat{B}^{\ast\prime})\tilde{\Delta}^{\ast}1_T/T + (I_{JM} - \hat{B}^{\ast}(\hat{B}^{\ast\prime}\hat{B}^{\ast})^{-1}\hat{B}^{\ast\prime})\tilde{E}^{\ast}1_T/T,
\end{align}
where $H^{-1}$ is well defined with probability approaching one by \eqref{Eqn: Thm: Rate: 4} and Lemma \ref{Lem: TechA2}(ii), and we have used $a^{\prime}B=0$ and $(I_{JM} - \hat{B}^{\ast}(\hat{B}^{\ast\prime}\hat{B}^{\ast})^{-1}\hat{B}^{\ast\prime})\hat{B}^{\ast}=0$. Let $\ell^{\ast\ast}_{NT}\equiv \sum_{t=1}^{T}{Q}^{-1}_t\Phi(Z_t)^{\ast\prime}\varepsilon_t/\sqrt{NT} $. By a similar argument as in \eqref{Eqn: Thm: AsymDis: 3}-\eqref{Eqn: Thm: AsymDis: 6}, we have
\begin{align}\label{Eqn: Thm:  Boot: 8}
&\|\sqrt{NT}(\hat{a}^{\ast} - a) -(I_{JM}- B\mathcal{H}\mathcal{H}^{\prime}B^{\prime})[\ell^{\ast\ast}_{NT}-\sqrt{NT}(\hat{B}^{\ast}-BH)\mathcal{H}^{-1}\bar{f}\notag\\
&\hspace{0.5cm}+ B\mathcal{H}\sqrt{NT}(\hat{B}^{\ast}-BH)^{\prime}a\|=O_{p}\left(\frac{\sqrt{NT}}{J^{\kappa}}+\frac{\sqrt{TJ}}{\sqrt{N}}+\frac{\sqrt{J\xi_{J}}\log^{1/4}J}{N^{1/4}}\right)
\end{align}
by noting that $\mathcal{H}^{\prime}B^{\prime}B\mathcal{H}=I_{K}$. Let $\ell^{\ast}_{NT}\equiv \sum_{t=1}^{T}{Q}^{-1}_t[\Phi(Z_t)^{\ast}-\Phi(Z_t)]^{\prime}\varepsilon_t/\sqrt{NT}=\ell^{\ast\ast}_{NT}-\ell_{NT}$. Note that $\sqrt{NT}(\hat{a}^{\ast} - \hat{a}) = \sqrt{NT}(\hat{a}^{\ast} - a) - \sqrt{NT}(\hat{a} - a)$ and $\sqrt{NT}(\hat{B}^{\ast} - \hat{B})=\sqrt{NT}(\hat{B}^{\ast} - B H) - \sqrt{NT}(\hat{B} - B H)$. By the fact that $\|x+y\|\leq \|x\| + \|y\|$, we now may combine \eqref{Eqn: Thm: AsymDis: 6} and \eqref{Eqn: Thm: Boot: 8} to obtain
\begin{align}\label{Eqn: Thm: Boot: 9}
&\|\sqrt{NT}(\hat{a}^{\ast} - \hat{a}) -(I_{JM} - B\mathcal{H}\mathcal{H}^{\prime}B^{\prime})[\ell^{\ast}_{NT}-\sqrt{NT}(\hat{B}^{\ast}-\hat{B})\mathcal{H}^{-1}\bar{f}\notag\\
&\hspace{0.5cm}+ B\mathcal{H}\sqrt{NT}(\hat{B}^{\ast}-\hat{B})^{\prime}a\|=O_{p}\left(\frac{\sqrt{NT}}{J^{\kappa}}+\frac{\sqrt{TJ}}{\sqrt{N}}+\frac{\sqrt{J\xi_{J}}\log^{1/4}J}{N^{1/4}}\right).
\end{align}
Note that $\mathbb{N}^{\ast}_1$ is a $JM\times 1$ vector from the first column of $\mathbb{N}^{\ast}$. Thus, the first result of the theorem follows from \eqref{Eqn: Thm: Boot: 9}, the second result of the theorem, and Lemmas \ref{Lem: TechA5} and \ref{Lem: TechD3}. This completes the proof of the theorem.\qed

\subsubsection{Technical Lemmas}
\begin{lem}\label{Lem: TechD1}
Let $D^{\ast}_1,D^{\ast}_2,D^{\ast}_3,D^{\ast}_5,D^{\ast}_6$ be given in the proof of Theorem \ref{Thm: Boot}.\\
(i) Under Assumptions \ref{Ass: LoadingsFactors}(i), (ii), (iv), \ref{Ass: Boot}(i), and (ii), $\|D^{\ast}_1\|^2_{F}=O_{p}(J^{-2\kappa})$.\\
(ii) Under Assumptions \ref{Ass: Basis}(i), \ref{Ass: LoadingsFactors}(ii), (iv), \ref{Ass: Boot}(i), and (ii), $\|D^{\ast}_2\|^{2}_{F}=O_{p}(J^{-4\kappa})$.\\
(iii) Under Assumptions \ref{Ass: Basis}, \ref{Ass: LoadingsFactors}(ii), (iv), \ref{Ass: DGP}, \ref{Ass: Boot}(i), (ii), $\|D^{\ast}_3\|^{2}_{F}\hspace{-0.05cm}=\hspace{-0.05cm}O_{p}(J^{-2\kappa}J/N)$.\\
(iv) Assume (i) $N\to\infty$; (ii) $T\geq K+1$; (iii) $J\to\infty$ with $J^{2}\xi^{2}_{J}\log J=o(N)$. Under Assumptions \ref{Ass: Basis}-\ref{Ass: Improvedrates}, \ref{Ass: Boot}(i), and (ii), $\|D^{\ast}_5\hat{B}\|^{2}_{F}=O_{p}(J/N^{2})$.\\
(v) Under Assumptions \ref{Ass: Basis}, \ref{Ass: LoadingsFactors}(ii), (iv), \ref{Ass: DGP}, \ref{Ass: Boot}(i), and (ii), $\|D^{\ast}_6\|^{2}_{F}=O_{p}(J^{-2\kappa}J/N)$.
\end{lem}
\noindent{\sc Proof:} (i) Since $\|M_T\|_2=1$, $\|D^{\ast}_1\|_{F}\leq \|B\|_2\|F\|_{2}\|\tilde{\Delta}^{\ast}\|_{F}/{T}$. The result then immediately follows from Assumptions \ref{Ass: LoadingsFactors}(i) and (ii), as well as Lemma \ref{Lem: TechD4}(i).

(ii) Since $\|M_T\|_2=1$, $\|D^{\ast}_2\|_{F}\leq \|\tilde{\Delta}\|_{F}\|\tilde{\Delta}^{\ast}\|_{F}/T$. The result then immediately follows from Lemmas \ref{Lem: TechA3}(i) and \ref{Lem: TechD4}(i).

(iii) Since $\|M_T\|_2=1$, $\|D^{\ast}_3\|_{F} \leq\|\tilde{\Delta}^{\ast}\|_{F}\|\tilde{E}\|_{F}/T$. The result then immediately follows from Lemmas \ref{Lem: TechA3}(ii) and \ref{Lem: TechD4}(i).

(iv) Since $\|M_T\|_2=1$, $\|D^{\ast}_5\hat{B}\|_{F}\leq \|\hat{B}^{\prime}\tilde{E}\|_{F}\|\tilde{E}^{\ast}\|_{F}/T$. The result then immediately follows from Lemmas \ref{Lem: TechB2}(i) and \ref{Lem: TechD4}(ii).

(v) Since $\|M_T\|_2=1$, $\|D^{\ast}_6\|_{F} \leq\|\tilde{\Delta}\|_{F}\|\tilde{E}^{\ast}\|_{F}/T$. The result then immediately follows from Lemmas \ref{Lem: TechA3}(i) and \ref{Lem: TechD4}(ii).\qed

\begin{lem}\label{Lem: TechD2}
Suppose Assumptions \ref{Ass: Basis}-\ref{Ass: Improvedrates}, \ref{Ass: Asym}(i), (ii), \ref{Ass: Boot}(i), and (ii) hold. Let $V$ be given in the proof of Theorem \ref{Thm: Rate}, and $\tilde{E}^{\ast}$, $D_4^{\ast}$, $\ell^{\ast\ast}_{NT}$, and $\mathcal{L}^{\ast\ast}_{NT}$ be given in the proof of Theorem \ref{Thm: Boot}. Assume (i) $N\to\infty$; (ii) $T\geq K+1$; (iii) $J\to\infty$ with $\xi^{2}_{J}\log J=o(N)$. Then
\[\|\sqrt{NT}D_4^{\ast}\hat{B}V^{-1}-\mathcal{L}^{\ast\ast}_{NT}B^{\prime}B\mathcal{M}\|_{F}=O_{p}\left(\frac{1}{J^{(\kappa-1/2)}}+\frac{\sqrt{J\xi_{J}}\log^{1/4}J}{N^{1/4}}\right)\]
and
\[\|\sqrt{N/T}\tilde{E}^{\ast}1_{T}-\ell^{\ast\ast}_{NT}\|=O_{p}\left(\frac{\sqrt{J\xi_{J}}\log^{1/4}J}{N^{1/4}}\right),\]
where $\mathcal{M}$ is a nonrandom matrix given in Lemma \ref{Lem: TechC3}.
\end{lem}
\noindent{\sc Proof:} For the first result, we have the following decomposition
\begin{align}\label{Eqn: Lemma: TechD2: 1}
\sqrt{NT}D_4^{\ast}\hat{B}V^{-1} &= \sqrt{N/T}\tilde{E}^{\ast}M_TFB^{\prime}B\mathcal{M}_2+\sqrt{N/T}\tilde{E}^{\ast}M_TFB^{\prime}(\hat{B}-BH)V^{-1}\notag\\
&\hspace{0.5cm}+\sqrt{N/T}\tilde{E}^{\ast}M_TFB^{\prime}B(HV^{-1}-\mathcal{M})\equiv \mathcal{T}_1 +\mathcal{T}_2+ \mathcal{T}_3.
\end{align}
Therefore, it suffices to show that $\|\mathcal{T}_1-\mathcal{L}^{\ast\ast}_{NT}B^{\prime}B\mathcal{M}_2\|_{F}=O_{p}(\sqrt{J\xi_{J}}\log^{1/4}J/N^{1/4})$, $\|\mathcal{T}_2\|_{F} = O_{p}(J^{(1/2-\kappa)}+J^{3/2}/N + J/\sqrt{NT})$, and $\|\mathcal{T}_3\|_{F} = O_{p}(J^{(1/2-\kappa)}+J^{3/2}/N + J/\sqrt{NT})$. The first one holds, since
\begin{align}\label{Eqn: Lemma: TechD2: 2}
\|\mathcal{T}_1-\mathcal{L}^{\ast\ast}_{NT}B^{\prime}B\mathcal{M}\|_{F}&\leq \|B\|_{2}^{2}\|\mathcal{M}\|_{2}\left\|\frac{1}{\sqrt{NT}}\sum_{t=1}^{T}(\hat{Q}^{\ast-1}_t-Q^{-1}_{t})\Phi(Z_t)^{\ast\prime}\varepsilon_t f_{t}^{\prime}\right\|_{F}\notag\\
&\hspace{0.5cm}+\|B\|_{2}^{2}\|\mathcal{M}\|_{2}\|\bar{f}\|\left\|\frac{1}{\sqrt{NT}}\sum_{t=1}^{T}(\hat{Q}^{\ast-1}_t-Q^{-1}_{t})\Phi(Z_t)^{\ast\prime}\varepsilon_t\right\|\notag\\
&=O_{p}\left(\frac{\sqrt{J\xi_{J}}\log^{1/4}J}{N^{1/4}}\right),
\end{align}
where the equality follows from Assumptions \ref{Ass: LoadingsFactors}(i) and (ii), along with Lemma \ref{Lem: TechD6}. The latter two follow by a similar argument. The second result also follows by a similar argument as in \eqref{Eqn: Lemma: TechD2: 2}. This completes the proof of the lemma.\qed

\begin{lem}\label{Lem: TechD3}
Suppose Assumptions \ref{Ass: LoadingsFactors}(ii), \ref{Ass: DGP}(i), (ii), \ref{Ass: Improvedrates}(i)-(iii), \ref{Ass: Asym}(ii), (iii), \ref{Ass: Boot}(i), and (iii) hold. Let $\ell^{\ast}_{NT}$ and $\mathcal{L}^{\ast}_{NT}$ be given in the proof of Theorem \ref{Thm: Boot}. Assume $J=o(\sqrt{N})$. Then there exists a $JM\times (K+1)$ random matrix $\mathbb{N}^{\ast}$ with $\mathrm{vec}(\mathbb{N}^{\ast})\sim N(0,\Omega)$ conditional on $\{Y_t, Z_t\}_{t\leq T}$ such that
\[\|(\ell^{\ast}_{NT},\mathcal{L}^{\ast}_{NT})-\sqrt{\omega_0}\mathbb{N}^{\ast}\|_{F} = O_{p}\left(\frac{J^{5/6}}{N^{1/6}}\right).\]
\end{lem}
\noindent{\sc Proof:} Let $\zeta_i\equiv (w_i-1)\sum_{t=1}^{T}f^{\dag}_{t}\otimes Q^{-1}_{t}\phi(z_{it})\varepsilon_{it}/\sqrt{NT}$. Then $\mathrm{vec}((\ell^{\ast}_{NT},\mathcal{L}^{\ast}_{NT}))=\sum_{i=1}^{N}\zeta_i$. Let $E_{w}$ denote the expectation with respect to $\{w_i\}_{i\leq N}$. Then conditional on $\{Y_t, Z_t\}_{t\leq T}$, $E_w[\zeta_{i}] = 0$ and $\zeta_{1},\ldots, \zeta_{N}$ are independent by Assumption \ref{Ass: Boot}(i). To proceed, let $\Omega_{NT}\equiv\sum_{i=1}^{N}$ $\sum_{t=1}^{T}\sum_{s=1}^{T}(f^{\dag}_{t}f_{s}^{\dag\prime})\otimes Q_{t}^{-1}\phi(z_{it})\phi(z_{is})^{\prime}Q_{s}^{-1}\varepsilon_{it}\varepsilon_{is}/NT$. Then $E_{w}[\mathrm{vec}((\ell^{\ast}_{NT},\mathcal{L}^{\ast}_{NT}))\mathrm{vec}((\ell^{\ast}_{NT},\mathcal{L}^{\ast}_{NT}))^{\prime}] = \omega_0\Omega_{NT}$. We now apply Lemma \ref{Lem: TechC5} to the independent random vectors $\zeta_1,\ldots, \zeta_N$ conditional on $\{Y_t, Z_t\}_{t\leq T}$. There exists a $JM\times (K+1)$ random matrix $\mathbb{N}^{\ast\ast}$ with $\mathrm{vec}(\mathbb{N}^{\ast\ast})\sim N(0,\Omega_{NT})$ conditional on $\{Y_t, Z_t\}_{t\leq T}$ such that the following holds:
\begin{align}\label{Eqn: Lem: TechD3: 1}
\|\mathrm{vec}((\ell^{\ast}_{NT},\mathcal{L}^{\ast}_{NT}))-\sqrt{\omega_0}\mathrm{vec}(\mathbb{N}^{\ast\ast})\| = O_{p^{\ast}}\left((J\beta)^{1/3}\right),
\end{align}
where $\beta = \sum_{i=1}^{N}E[\|\zeta_{i}\|^{3}]$. Next, we calculate $\beta$. To the end, we first calculate
\begin{align}\label{Eqn: Lem: TechD3: 2}
\hspace{-0.3cm}E[\|\zeta_{i}\|^{4}]&=E[(w_1-1)^{4}]\frac{1}{N^{2}T^{2}}E\left[\left(\sum_{t=1}^{T}\sum_{s=1}^{T} \phi(z_{it})^{\prime}Q^{-1}_tQ^{-1}_{s}\phi(z_{is})f^{\dag\prime}_tf_{s}^{\dag}\varepsilon_{it}\varepsilon_{is}\right)^2\right]\notag\\
&\hspace{-0.5cm}\leq C_{NT}\max_{i\leq N,t\leq T}E[\|\phi(z_{it})\|^{4}]\frac{1}{N^{2}T^{2}}\sum_{t=1}^{T}\sum_{s=1}^{T}\sum_{u=1}^{T}\sum_{v=1}^{T}|E[\varepsilon_{it}\varepsilon_{is}\varepsilon_{iu}\varepsilon_{iv}]|\notag\\
&\hspace{-0.5cm}\leq C_{NT}\hspace{-0.2cm}\max_{m\leq M, j\leq J, i\leq N, t\leq T} \hspace{-0.1cm}E[\phi^{4}_{j}(z_{it,m})]\frac{J^{2}M^{2}}{N^{2}T^{2}}\sum_{t=1}^{T}\sum_{s=1}^{T}\sum_{u=1}^{T}\sum_{v=1}^{T}|E[\varepsilon_{it}\varepsilon_{is}\varepsilon_{iu}\varepsilon_{iv}]|,
\end{align}
where $C_{NT}=E[(w_1-1)^{4}]\max_{t\leq T}\|f^{\dag}_{t}\|^{4}(\min_{i\leq N, t\leq T}\lambda_{\min}(Q_{it}))^{-4}$, the first inequality follows from the independence in Assumption \ref{Ass: DGP}(i) and the Cauchy-Schwartz inequality, as well as the fact that $\min_{t\leq T}\lambda_{\min}(Q_{t})\geq \min_{i\leq N, t\leq T}\lambda_{\min}(Q_{it})$, and the second one follows by $\max_{i\leq N,t\leq T}E[\|\phi(z_{it})\|^4]\leq J^{2}M^{2}\max_{m\leq M, j\leq J, i\leq N, t\leq T} E[\phi^{4}_{j}(z_{it,m})]$. Thus,
\begin{align}\label{Eqn: Lem: TechD3: 3}
\beta = \sum_{i=1}^{N}E[\|\zeta_{i}\|^{3}]\leq \sum_{i=1}^{N}(E[\|\zeta_{i}\|^{4}])^{3/4} = O\left(\frac{J^{3/2}}{\sqrt{N}}\right),
\end{align}
where the inequality follows by the Liapounov’s inequality, and the last equality follows from \eqref{Eqn: Lem: TechD3: 2} and Assumptions \ref{Ass: LoadingsFactors}(ii), \ref{Ass: Improvedrates}(i), (ii), \ref{Ass: Asym}(iii), and \ref{Ass: Boot}(i). We now may combine \eqref{Eqn: Lem: TechD3: 1}, \eqref{Eqn: Lem: TechD3: 3}, and Lemma \ref{Lem: TechA5} to obtain
\begin{align}\label{Eqn: Lem: TechD3: 4}
\|\mathrm{vec}((\ell^{\ast}_{NT},\mathcal{L}^{\ast}_{NT}))-\sqrt{\omega_0}\mathrm{vec}(\mathbb{N}^{\ast\ast})\| = O_{p}\left(\frac{J^{5/6}}{N^{1/6}}\right).
\end{align}
By Assumption \ref{Ass: Boot}(iii) and Lemma \ref{Lem: TechD8}, $\Omega^{-1/2}_{NT}$ is well defined with probability approaching one since $J=o(\sqrt{N})$. Define $\mathbb{N}^{\ast}$ such that $\mathrm{vec}(\mathbb{N}^{\ast})= \Omega^{1/2}\Omega^{-1/2}_{NT}\mathrm{vec}(\mathbb{N}^{\ast\ast})$. Then $\mathrm{vec}(\mathbb{N}^{\ast})\sim N(0,\Omega)$ conditional on $\{Y_t, Z_t\}_{t\leq T}$. It follows that
\begin{align}\label{Eqn: Lem: TechD3: 5}
&\|\mathrm{vec}((\ell^{\ast}_{NT},\mathcal{L}^{\ast}_{NT}))-\sqrt{\omega_0}\mathbb{N}^{\ast}\|_{F} \leq \|\mathrm{vec}((\ell^{\ast}_{NT},\mathcal{L}^{\ast}_{NT}))-\sqrt{\omega_0}\mathrm{vec}(\mathbb{N}^{\ast\ast})\| \notag\\
&\hspace{0.5cm}+  \sqrt{\omega_0}\|\mathrm{vec}(\mathbb{N}^{\ast})-\mathrm{vec}(\mathbb{N}^{\ast\ast})\|= O_{p}\left(\frac{J^{5/6}}{N^{1/6}}+\frac{J^{3/2}}{\sqrt{N}}\right) = O_{p}\left(\frac{J^{5/6}}{N^{1/6}}\right),
\end{align}
where the first equality follows by \eqref{Eqn: Lem: TechD3: 4} and the fact that $\|\mathrm{vec}(\mathbb{N}^{\ast})-\mathrm{vec}(\mathbb{N}^{\ast\ast})\|\leq \|\Omega^{1/2}_{NT}-\Omega^{1/2}\|_{2}\|\Omega^{-1/2}_{NT}\mathrm{vec}(\mathbb{N}^{\ast\ast})\| = O_{p}(J^{3/2}/\sqrt{N})$, which is due to Lemma \ref{Lem: TechD8}. This completes the proof of the lemma.
\qed

\begin{lem}\label{Lem: TechD4}
Let $\tilde{\Delta}^{\ast}$ and $\tilde{E}^{\ast}$ be given in the proof of Theorem \ref{Thm: Boot}. \\
(i) Under Assumptions \ref{Ass: LoadingsFactors}(ii), (iv), \ref{Ass: Boot}(i), and (ii), $\|\tilde{\Delta}^{\ast}\|_{F}^{2}/T = O_{p}(J^{-2\kappa})$.\\
(ii) Under Assumptions \ref{Ass: Basis}(ii), \ref{Ass: DGP}, \ref{Ass: Boot}(i), and (ii), $\|\tilde{E}^{\ast}\|_{F}^{2}/T = O_{p}(J/N)$.
\end{lem}
\noindent{\sc Proof:} (i) By the facts that $\|Ax\|\leq \|A\|_{2}\|x\|$ and $\|A\|_{2}\leq \|A\|_{F}$,
\begin{align}\label{Eqn: Lem: TechD4: 1}
\frac{1}{T}\|\tilde{\Delta}^{\ast}\|^{2}_{F}&= \frac{1}{T}\sum_{t=1}^{T}\|(\Phi(Z_t)^{\ast\prime}\Phi(Z_t))^{-1}\Phi(Z_t)^{\ast\prime}(R(Z_t)+\Delta(Z_t)f_{t})\|^{2}\notag\\
&\leq 2\max_{t\leq T}\|f_{t}\|^{2}\left(\min_{t\leq T}\lambda_{\min}(\hat{Q}^{\ast}_t)\right)^{-1}\hspace{-0.2cm}\frac{1}{NT}\sum_{t=1}^{T}\sum_{i=1}^{N}w_i\|\delta(z_{it})\|^2\notag\\
&\hspace{0.5cm}+2\left(\min_{t\leq T}\lambda_{\min}(\hat{Q}^{\ast}_t)\right)^{-1}\hspace{-0.2cm}\frac{1}{NT}\sum_{t=1}^{T}\sum_{i=1}^{N}w_i|r(z_{it})|^2=O_{p}\left(\frac{1}{J^{2\kappa}}\right),
\end{align}
where the last equality follows from Assumptions \ref{Ass: LoadingsFactors}(ii) and \ref{Ass: Boot}(ii), along with Lemma \ref{Lem: TechD5}(ii).

(ii) By the fact that $\|Ax\|\leq \|A\|_{2}\|x\|$,
\begin{align}\label{Eqn: Lem: TechD4: 2}
\frac{1}{T}\|\tilde{E}^{\ast}\|^{2}_{F}&= \frac{1}{T}\sum_{t=1}^{T}\|(\Phi(Z_t)^{\ast\prime}\Phi(Z_t))^{-1}\Phi(Z_t)^{\ast\prime}\varepsilon_{t}\|^{2}\notag\\
&\leq \left(\min_{t\leq T}\lambda_{\min}(\hat{Q}^{\ast}_t)\right)^{-2} \frac{1}{N^{2}T}\sum_{t=1}^{T}\|\Phi(Z_t)^{\ast\prime}\varepsilon_{t}\|^{2}= O_{p}\left(\frac{J}{N}\right),
\end{align}
where the last equality follows from Assumption \ref{Ass: Boot}(ii) and Lemma \ref{Lem: TechD5}(i). \qed

\begin{lem}\label{Lem: TechD5}
(i) Under Assumptions \ref{Ass: Basis}(ii), \ref{Ass: DGP}, and \ref{Ass: Boot}(i),
\[\sum_{t=1}^{T}\|\Phi(Z_t)^{\ast\prime}\varepsilon_{t}\|^{2}=O_{p}(NTJ).\]
(ii) Under Assumption \ref{Ass: LoadingsFactors}(iv) and \ref{Ass: Boot}(i),
\[\sum_{t=1}^{T}\sum_{i=1}^{N}w_i\|\delta(z_{it})\|^2=O_{p}(NTJ^{-2\kappa}) \text{ and } \sum_{t=1}^{T}\sum_{i=1}^{N}w_i|r(z_{it})|^2=O_{p}(NTJ^{-2\kappa}).\]
\end{lem}
\noindent{\sc Proof:} (i) The result follows by the Markov's inequality, since
\begin{align}\label{Eqn: Lem: TechD5: 1}
&\hspace{0.8cm}E\left[\sum_{t=1}^{T}\|\Phi(Z_t)^{\ast\prime}\varepsilon_{t}\|^{2}\right]= E\left[\sum_{t=1}^{T}\sum_{i=1}^{N}\sum_{j=1}^{N}\phi(z_{it})^{\prime}\phi(z_{jt})\varepsilon_{it}\varepsilon_{jt}w_{i}w_{j}\right]\notag\\
&\hspace{-0.5cm}= \sum_{t=1}^{T}\sum_{i=1}^{N}\sum_{j=1}^{N}E[\phi(z_{it})^{\prime}\phi(z_{jt})]E[\varepsilon_{it}\varepsilon_{jt}]E[w_iw_j]\notag\\
&\hspace{-0.5cm}\leq E[w_{1}^2]\max_{i\leq N,t\leq T}E[\|\phi(z_{it})\|^{2}]\sum_{t=1}^{T}\sum_{i=1}^{N}\sum_{j=1}^{N}|E[\varepsilon_{it}\varepsilon_{jt}]|\notag\\
&\hspace{-0.5cm}\leq TJM E[w_{1}^2]\max_{m\leq M, j\leq J, i\leq N, t\leq T} E[\phi^{2}_{j}(z_{it,m})]\max_{t\leq T}\sum_{i=1}^{N}\sum_{j=1}^{N}|E[\varepsilon_{it}\varepsilon_{jt}]|= O(NTJ),
\end{align}
where the second equality follows by the independence in Assumptions \ref{Ass: DGP}(i) and \ref{Ass: Boot}(i), the first inequality is due to the Cauchy Schwartz inequality, the second one follows by $\max_{i\leq N, t\leq T}E[\|\phi(z_{it})\|^{2}]\leq JM\max_{m\leq M, j\leq J, i\leq N, t\leq T} E[\phi^{2}_{j}(z_{it,m})]$, and the last equality follows from Assumptions \ref{Ass: Basis}(ii), \ref{Ass: DGP}(iii), and \ref{Ass: Boot}(i).

(iii) The first result follows since
\begin{align}\label{Eqn: Lem: TechD5: 2}
\hspace{-0.5cm}\sum_{t=1}^{T}\sum_{i=1}^{N}w_i\|\delta(z_{it})\|^2\leq T K M^{2}\max_{k\leq K,m\leq M}\sup_{z}|\delta_{km,J}(z)|^{2}\sum_{i=1}^{N}w_i=O_{p}(NTJ^{-2\kappa}),
\end{align}
where the inequality follows since $w_i$'s are positive and $\max_{i\leq N, t\leq T}\|\delta(z_{it})\|^2\leq M^{2}K$ $\sup_{k\leq K,m\leq M}\sup_{z}|\delta_{km,J}(z)|^{2}$, and the equality follows by the law of large numbers and Assumptions \ref{Ass: LoadingsFactors}(iv) and \ref{Ass: Boot}(i). The proof of the second result is similar. \qed

\begin{lem}\label{Lem: TechD6}
Suppose Assumptions \ref{Ass: LoadingsFactors}(ii), \ref{Ass: DGP}(i), (ii), \ref{Ass: Improvedrates}, \ref{Ass: Asym}(ii), \ref{Ass: Boot}(i), and (ii) hold. Assume $J\geq 2$ and $\xi^{2}_{J}\log J=o(N)$. Then
\[\left\|\frac{1}{\sqrt{NT}}\sum_{t=1}^{T}(\hat{Q}^{\ast-1}_t-Q^{-1}_{t})\Phi(Z_t)^{\ast\prime}\varepsilon_t f_{t}^{\prime}\right\|_{F} = O_{p}\left(\frac{\sqrt{J\xi_{J}}\log^{1/4}J}{N^{1/4}}\right)\]
and
\[\left\|\frac{1}{\sqrt{NT}}\sum_{t=1}^{T}(\hat{Q}^{\ast-1}_t-Q^{-1}_{t})\Phi(Z_t)^{\ast\prime}\varepsilon_t \right\| = O_{p}\left(\frac{\sqrt{J\xi_{J}}\log^{1/4}J}{N^{1/4}}\right).\]
\end{lem}
\noindent{\sc Proof:} Let $\mathcal{T}\equiv \sum_{t=1}^{T}(\hat{Q}^{\ast-1}_t-Q^{-1}_{t})\Phi(Z_t)^{\ast\prime}\varepsilon_t f_{t}^{\prime}/\sqrt{NT}$ and $E_{\varepsilon}$ denote the expectation with respect to $\{\varepsilon_{t}\}_{t\leq T}$. Since $\|A\|_{F}^{2}=\mathrm{tr}(AA^{\prime})$,
\begin{align}\label{Eqn: Lem: TechD6: 1}
E_{\varepsilon}[\|\mathcal{T}\|^{2}_{F}]&=\frac{1}{NT}E_{\varepsilon}\left[\mathrm{tr}\left(\sum_{t=1}^{T}\sum_{s=1}^{T}(\hat{Q}^{\ast-1}_t\hspace{-0.05cm}-\hspace{-0.05cm}Q^{-1}_{t})\Phi(Z_t)^{\ast\prime}\varepsilon_tf_{t}^{\prime}
f_{s}\varepsilon_s^{\prime}\Phi(Z_s)^{\ast}(\hat{Q}^{\ast-1}_s-Q^{-1}_{s})\right)\right]\notag\\
&\hspace{-1cm}=\frac{1}{NT}\sum_{i=1}^{N}\sum_{j=1}^{N}\sum_{t=1}^{T}\sum_{s=1}^{T}w_i\phi(z_{it})^{\prime}(\hat{Q}^{\ast-1}_t-Q^{-1}_{t})(\hat{Q}^{\ast-1}_s-Q^{-1}_{s})\phi(z_{js})w_jf_{t}^{\prime}f_{s}E[\varepsilon_{it}\varepsilon_{js}]\notag\\
&\hspace{-1cm}=\frac{1}{NT}\sum_{i=1}^{N}\sum_{t=1}^{T}\sum_{s=1}^{T}w_i^2\phi(z_{it})^{\prime}(\hat{Q}^{\ast-1}_t-Q^{-1}_{t})(\hat{Q}^{\ast-1}_s-Q^{-1}_{s})\phi(z_{is})f_{t}^{\prime}f_{s}E[\varepsilon_{it}\varepsilon_{is}]\notag\\
&\hspace{-1cm}\leq C_{NT}^{\ast}\frac{1}{NT}\sum_{i=1}^{N}\sum_{t=1}^{T}\sum_{s=1}^{T}\|\hat{Q}^{\ast}_t-Q_{t}\|_{2}w_i^2\|\phi(z_{it})\|\|\phi(z_{is})\||E[\varepsilon_{it}\varepsilon_{is}]|\notag\\
&\hspace{-1cm}\leq C_{NT}^{\ast\ast}\frac{1}{NT}\left(\sum_{t=1}^{T}\left(\sum_{i=1}^{N}\sum_{s=1}^{T}w_i^2\|\phi(z_{it})\|\|\phi(z_{is})\||E[\varepsilon_{it}\varepsilon_{is}]|\right)^{2}\right)^{1/2},
\end{align}
where $C^{\ast}_{NT}\hspace{-0.05cm}= \hspace{-0.05cm}(\min_{t\leq T}\lambda_{\min}(\hat{Q}_{t}^{\ast}))^{-1}[(\min_{t\leq T}\lambda_{\min}(\hat{Q}_{t}^{\ast}))^{-1}+(\min_{i\leq N, t\leq T}\lambda_{\min}({Q}_{it}))^{-1}]$ $\times (\min_{i\leq N, t\leq T}\lambda_{\min}({Q}_{it}))^{-1}\max_{t\leq T}\|f_t\|^{2}$ and $C^{\ast\ast}_{NT}=C^{\ast}_{NT}(\sum_{t=1}^{T}\|\hat{Q}^{\ast}_{t}-{Q}_{t}\|^{2}_{2})^{1/2}$, the second equality follows from the independence in Assumptions \ref{Ass: DGP}(i) and \ref{Ass: Boot}(i) and the linearity of both expectation and trace operators, the third equality follows by Assumption \ref{Ass: DGP}(ii) and the independence in Assumption \ref{Ass: Asym}(ii), the first inequality follows since $\min_{t\leq T}\lambda_{\min}(Q_{t})\geq \min_{i\leq N,t\leq T}\lambda_{\min}(Q_{it})$, and the last inequality is due to the Cauchy-Schwartz inequality. Moreover, we have
\begin{align}\label{Eqn: Lem: TechD6: 2}
&\hspace{0.8cm}E\left[\sum_{t=1}^{T}\left(\sum_{i=1}^{N}\sum_{s=1}^{T}w_i^2\|\phi(z_{it})\|\|\phi(z_{is})\||E[\varepsilon_{it}\varepsilon_{is}]|\right)^{2}\right]\notag\\
&\leq E[w_1^{4}]\max_{i\leq N,t\leq T}E[\|\phi(z_{it})\|^4]\sum_{t=1}^{T}\left(\sum_{i=1}^{N}\sum_{s=1}^{T}|E[\varepsilon_{it}\varepsilon_{is}]|\right)^{2}\notag\\
&\leq J^{2}M^{2}E[w_1^{4}]\max_{m\leq M, j\leq J, i\leq N, t\leq T} E[\phi^{4}_{j}(z_{it,m})]\sum_{t=1}^{T}\left(\sum_{i=1}^{N}\sum_{s=1}^{T}|E[\varepsilon_{it}\varepsilon_{is}]|\right)^{2},
\end{align}
where the first inequality follows by the Cauchy-Schwartz inequality and the independence in Assumption \ref{Ass: Boot}(i), the second one follows since $\max_{i\leq N,t\leq T}E[\|\phi(z_{it})\|^4]$ $\leq J^{2}M^{2}\max_{m\leq M, j\leq J, i\leq N, t\leq T} E[\phi^{4}_{j}(z_{it,m})]$. By Assumptions \ref{Ass: LoadingsFactors}(ii), \ref{Ass: Improvedrates}(ii), \ref{Ass: Boot}(ii), and Lemma \ref{Lem: TechD7}, $C^{\ast\ast}_{NT} = O_{p}(\sqrt{T}\xi_{J}\sqrt{\log J}/\sqrt{N})$. Combining this, \eqref{Eqn: Lem: TechD6: 1} and \eqref{Eqn: Lem: TechD6: 2} implies that $E_{\varepsilon}[\|\mathcal{T}\|^{2}_{F}]=O_{p}(J\xi_{J}\sqrt{\log J}/\sqrt{N})$ by Assumptions \ref{Ass: Improvedrates}(i), (iv), and \ref{Ass: Boot}(i). Thus, the result of the lemma follows by the Markov's inequality and Lemma \ref{Lem: TechA5}. The proof of the second result is similar.\qed

\begin{lem}\label{Lem: TechD7}
Suppose Assumptions \ref{Ass: Improvedrates}(ii), (iii), and \ref{Ass: Boot}(i) hold. Assume $J\geq 2$ and $\xi^{2}_{J}\log J=o(N)$. Then
\[\sum_{t=1}^{T}\|\hat{Q}^{\ast}_t-Q_t\|^{2}_{2}=O_{p}\left(\frac{T\xi^{2}_{J}\log J}{N}\right).\]
\end{lem}
\noindent{\sc Proof:} The proof is similar to the proof of Lemma \ref{Lem: TechB6}, thus omitted for brevity.\qed

\begin{lem}\label{Lem: TechD8}
Suppose Assumptions \ref{Ass: LoadingsFactors}(ii), \ref{Ass: DGP}(i), (ii), \ref{Ass: Improvedrates}(i)-(iii), \ref{Ass: Asym}(ii), (iii), and \ref{Ass: Boot}(iii) hold. Let $\Omega_{NT}$ be given in the proof of Lemma \ref{Lem: TechD3}. Then
\[\|\Omega_{NT}^{1/2}-\Omega^{1/2}\|_{2}=O_{p}\left(\frac{J}{\sqrt{N}}\right).\]
\end{lem}
\noindent{\sc Proof:} We first show $\|\Omega_{NT}-\Omega\|_{F}^{2}=O_{p}(J^{2}/N)$. Let $\zeta_i\equiv \sum_{t=1}^{T}f_{t}\otimes Q^{-1}_{t}\phi(z_{it})\varepsilon_{it}/\sqrt{NT}$. Then $\Omega_{NT}=\sum_{i=1}^{N}\zeta_i\zeta_i^{\prime}$ and $\Omega=\sum_{i=1}^{N}E[\zeta_i\zeta_i^{\prime}]$. Since $\|A\|_{F}^{2}=\mathrm{tr}(AA^{\prime})$,
\begin{align}\label{Eqn: Lem: TechD8: 1}
E[\|\Omega_{NT}-\Omega\|_{F}^{2}] &= E\left[\mathrm{tr}\left(\sum_{i=1}^{N}\sum_{j=1}^{N}(\zeta_i\zeta_i^{\prime}-E[\zeta_i\zeta_i^{\prime}])(\zeta_j\zeta_j^{\prime}-E[\zeta_j\zeta_j^{\prime}])^{\prime}\right)\right]\notag\\
& \hspace{-1cm}= \sum_{i=1}^{N}\left(E[(\zeta_i^{\prime}\zeta_i)^{2}]-\|E[\zeta_i\zeta_i^{\prime}]\|_{F}^{2}\right)\leq N\max_{i\leq N}E[\|\zeta_i\|^{4}] = O\left(\frac{J^2}{N}\right),
\end{align}
where the second equality follows because $\zeta_{1},\ldots, \zeta_{N}$ are independent by Assumptions \ref{Ass: DGP}(i), \ref{Ass: Improvedrates}(iii), and \ref{Ass: Asym}(ii), and because both expectation and trace operators are linear, the inequality follows by the Cauchy-Schwartz inequality since $\|E[\zeta_i\zeta_i^{\prime}]\|_{F}^{2}\geq 0$, and the last equality follows from \eqref{Eqn: Lem: TechC2: 2} and Assumptions \ref{Ass: LoadingsFactors}(ii), \ref{Ass: Improvedrates}(i), (ii), and \ref{Ass: Asym}(iii). Thus, $\|\Omega_{NT}-\Omega\|_{F}^{2}=O_{p}(J^{2}/N)$ follows from \eqref{Eqn: Lem: TechD8: 1} by the Markov's inequality. The result of the lemma follows from Assumption \ref{Ass: Boot}(iii) and Lemma A.2 of \citet{BelloniChernozhukovChetverikovKato_SeriesEstimator_2015}.\qed

\subsection{Proof of Theorem \ref{Thm: SpecTest}}
\noindent{\sc Proof of Theorem \ref{Thm: SpecTest}:} In order to show the first result, we assume that $\mathrm{H}_0$ is true. Since $\hat{\alpha}(z_{it}) = \hat{a}^{\prime}\phi(z_{it})$, $\hat{\beta}(z_{it}) = \hat{B}^{\prime}\phi(z_{it})$, $\alpha(z_{it})=a^{\prime}\phi(z_{it}) + r(z_{it}) = \gamma^{\prime}z_{it}$, and $\beta(z_{it})=B^{\prime}\phi(z_{it}) + \delta(z_{it}) = \Gamma^{\prime}z_{it}$, we have
\begin{align}\label{Eqn: Thm: SpecTest: 1}
\mathcal{S} &=\frac{1}{J}\sum_{i=1}^{N}\sum_{t=1}^{T}|(\hat{\gamma}-\gamma)^{\prime}z_{it} - (\hat{a}-a)^{\prime}\phi(z_{it})+r(z_{it})|^{2}\notag\\
&\hspace{0.5cm}+\frac{1}{J}\sum_{i=1}^{N}\sum_{t=1}^{T}\|(\hat{\Gamma} -\Gamma H)^{\prime} z_{it}- (\hat{B}-BH)^{\prime}\phi(z_{it}) + H^{\prime}\delta(z_{it})\|^{2}\notag\\
&=\frac{1}{J}\sum_{i=1}^{N}\sum_{t=1}^{T}|(\hat{\gamma}-\gamma)^{\prime}z_{it} - (\hat{a}-a)^{\prime}\phi(z_{it})|^{2}+ \mathcal{S}_{1}+2\mathcal{S}_{2}+2\mathcal{S}_{3}\notag\\
&\hspace{0.5cm}+\frac{1}{J}\sum_{i=1}^{N}\sum_{t=1}^{T}\|(\hat{\Gamma} -\Gamma H)^{\prime} z_{it} - (\hat{B}-BH)^{\prime}\phi(z_{it})\|^2 + \mathcal{S}_{4}+2\mathcal{S}_{5}+2\mathcal{S}_{6},
\end{align}
where $\mathcal{S}_1\hspace{-0.05cm}=\hspace{-0.05cm}\sum_{i=1}^{T}\sum_{t=1}^{N}|r(z_{it})|^2/J$, $\mathcal{S}_2\hspace{-0.05cm}=\hspace{-0.05cm} \sum_{i=1}^{N}\sum_{t=1}^{T}z_{it}^{\prime}(\hat{\gamma} -\gamma)r(z_{it})/J$, $\mathcal{S}_3 \hspace{-0.05cm}=\hspace{-0.05cm}\sum_{i=1}^{N}\sum_{t=1}^{T}$ $\phi(z_{it})^{\prime}(\hat{a}-a)r(z_{it})/J$, $\mathcal{S}_4 = \sum_{i=1}^{T}\sum_{t=1}^{N}\|H^{\prime}\delta(z_{it})\|^2/J$, $\mathcal{S}_5 = \sum_{i=1}^{N}\sum_{t=1}^{T}z_{it}^{\prime}(\hat{\Gamma} -\Gamma H)H^{\prime}$ $\delta(z_{it})/J$, and $\mathcal{S}_6 = \sum_{i=1}^{N}\sum_{t=1}^{T}\phi(z_{it})^{\prime}(\hat{B}-BH)H^{\prime}\delta(z_{it})/J$. Let $\mathcal{W}_{NT,a}\equiv (\sqrt{NT}(\hat{\gamma}-\gamma)^{\prime},-\sqrt{NT}(\hat{a}-a)^{\prime})^{\prime}$, $\mathcal{W}_{NT,B}\equiv (\sqrt{NT}(\hat{\Gamma}-\Gamma H)^{\prime},-\sqrt{NT}(\hat{B}-BH)^{\prime})^{\prime}$,  $\mathcal{W}_{NT}\equiv(\mathcal{W}_{NT,a},$ $\mathcal{W}_{NT,B})$, and $\hat{\mathcal{Q}}\equiv \sum_{i=1}^{N}\sum_{t=1}^{T}(z_{it}^{\prime},\phi(z_{it})^{\prime})^{\prime}(z_{it}^{\prime},\phi(z_{it})^{\prime})/NT$. By Lemma \ref{Lem: TechE1}, \eqref{Eqn: Thm: SpecTest: 1} implies
\begin{align}\label{Eqn: Thm: SpecTest: 2}
&\mathcal{S} -\frac{1}{J}\mathcal{W}_{NT,a}^{\prime}\hat{\mathcal{Q}}\mathcal{W}_{NT,a}-\frac{1}{J}\mathrm{tr}(\mathcal{W}_{NT,B}^{\prime}\hat{\mathcal{Q}}\mathcal{W}_{NT,B})\notag\\
&\hspace{-0.5cm}= \mathcal{S} -\frac{1}{J}\mathrm{tr}(\mathcal{W}_{NT}^{\prime}\hat{\mathcal{Q}}\mathcal{W}_{NT})= O_{p}\left(\frac{\sqrt{NT}}{J^{\kappa+1/2}}\right).
\end{align}
Let $\mathcal{Q}\equiv E[\hat{\mathcal{Q}}]$, $\mathbb{W}_a\equiv (\mathbb{G}_{\gamma}^{\prime},-\mathbb{G}_{a}^{\prime})^{\prime}$, $\mathbb{W}_{B}\equiv (\mathbb{G}_{\Gamma}^{\prime},-\mathbb{G}_{B}^{\prime})^{\prime}$, and $\mathbb{W}\equiv (\mathbb{W}_a,\mathbb{W}_B)$, where $\mathbb{G}_{\gamma}$ and $\mathbb{G}_{\Gamma}$ are given in Lemma \ref{Lem: TechE2}. By Lemmas \ref{Lem: TechE2} and \ref{Lem: TechE3} and Theorem \ref{Thm: AsymDis}, \eqref{Eqn: Thm: SpecTest: 2} implies
\begin{align}\label{Eqn: Thm: SpecTest: 3}
&\mathcal{S}- \frac{1}{J}\mathbb{W}_a^{\prime}\mathcal{Q}\mathbb{W}_a-\frac{1}{J}\mathrm{tr}(\mathbb{W}_B^{\prime}\mathcal{Q}\mathbb{W}_B)\notag\\
&\hspace{-0.5cm}=\mathcal{S}- \frac{1}{J}\mathrm{tr}(\mathbb{W}^{\prime}\mathcal{Q}\mathbb{W})=O_{p}\left(\frac{\sqrt{NT}}{J^{\kappa+1/2}}+\frac{J^{1/3}}{N^{1/6}} +\frac{\sqrt{\xi_{J}}\log^{1/4}J}{N^{1/4}}+ \sqrt{\frac{T}{N}}\right).
\end{align}
Let $\mathcal{W}_{NT,a}^{\ast}\equiv (\sqrt{NT/\omega_0}(\hat{\gamma}^{\ast}-\hat\gamma)^{\prime},-\sqrt{NT/\omega_0}(\hat{a}^{\ast}-\hat{a})^{\prime})^{\prime}$, $\mathcal{W}^{\ast}_{NT,B}\equiv (\sqrt{NT/\omega_0}(\hat{\Gamma}^{\ast}-\hat{\Gamma})^{\prime},$ $-\sqrt{NT/\omega_0}(\hat{B}^{\ast}-\hat{B})^{\prime})^{\prime}$, $\mathcal{W}_{NT}^{\ast}\equiv (\mathcal{W}_{NT,a}^{\ast},\mathcal{W}_{NT,B}^{\ast})$, $\mathbb{W}_{a}^{\ast}\equiv (\mathbb{G}_{\gamma}^{\ast\prime},-\mathbb{G}_{a}^{\ast\prime})^{\prime}$, $\mathbb{W}_{B}^{\ast}\equiv (\mathbb{G}_{\Gamma}^{\ast\prime},-\mathbb{G}_{B}^{\ast\prime})^{\prime}$, and $\mathbb{W}^{\ast}\equiv (\mathbb{W}_{a}^{\ast},\mathbb{W}_{B}^{\ast})$, where $\mathbb{G}^{\ast}_{\gamma}$ and $\mathbb{G}^{\ast}_{\Gamma}$ are given in Lemma \ref{Lem: TechE4}. Then \eqref{Eqn: BootStatistics} can be written as $\mathcal{S}^{\ast} = \mathcal{W}^{\ast\prime}_{NT,a}\hat{\mathcal{Q}}\mathcal{W}^{\ast}_{NT,a}/J+\mathrm{tr}(\mathcal{W}^{\ast\prime}_{NT,B}\hat{\mathcal{Q}}\mathcal{W}^{\ast}_{NT,B})/J = \mathrm{tr}(\mathcal{W}^{\ast\prime}_{NT}\hat{\mathcal{Q}}\mathcal{W}^{\ast}_{NT})/J$. By Lemmas \ref{Lem: TechA5}, \ref{Lem: TechE3}, and \ref{Lem: TechE4}, along with Theorem \ref{Thm: Boot},
\begin{align}\label{Eqn: Thm: SpecTest: 4}
&\mathcal{S}^{\ast}- \frac{1}{J}\mathbb{W}_a^{\ast\prime}\mathcal{Q}\mathbb{W}_a^{\ast}-\frac{1}{J}\mathrm{tr}(\mathbb{W}_B^{\ast\prime}\mathcal{Q}\mathbb{W}_B^{\ast})\notag\\
&\hspace{-0.5cm}=\mathcal{S}^{\ast}- \frac{1}{J}\mathrm{tr}(\mathbb{W}^{\ast\prime}\mathcal{Q}\mathbb{W}^{\ast})=O_{p}\left(\frac{\sqrt{NT}}{J^{\kappa+1/2}}+\frac{J^{1/3}}{N^{1/6}} +\frac{\sqrt{\xi_{J}}\log^{1/4}J}{N^{1/4}}+ \sqrt{\frac{T}{N}}\right).
\end{align}
Let $\gamma_{NT}\equiv (\sqrt{NT}J^{-\kappa}+{J^{5/6}}/{N^{1/6}}+ {\sqrt{J \xi_{J}}\log^{1/4}J}/{N^{1/4}}+\sqrt{{T}J/{N}})^{1/2}$, which is $o(1)$ by the assumption. Let $c_{0,1-\alpha}$ be the $1-\alpha$ quantile of $\mathrm{tr}(\mathbb{W}^{\ast\prime}\mathcal{Q}\mathbb{W}^{\ast})/J$, which is also the $1-\alpha$ quantile of $\mathrm{tr}(\mathbb{W}^{\prime}\mathcal{Q}\mathbb{W})/J$. Then in view of \eqref{Eqn: Thm: SpecTest: 4}, Lemma A.1 of \citet{BelloniChernozhukovChetverikovKato_SeriesEstimator_2015} implies that there exists a sequence $\{\nu_{NT}\}$ such that $\nu_{NT}=o(1)$ and
\begin{align}
P(c_{1-\alpha}<c_{0,1-\alpha-\nu_{NT}}-\gamma_{NT}/\sqrt{J}) &= o(1),\label{Eqn: Thm: SpecTest: 5}\\
P(c_{1-\alpha}>c_{0,1-\alpha+\nu_{NT}}+\gamma_{NT}/\sqrt{J}) &= o(1).\label{Eqn: Thm: SpecTest: 6}
\end{align}
Note that $\mathrm{tr}(\mathbb{W}^{\prime}\mathcal{Q}\mathbb{W}) = \mathrm{vec}(\mathbb{W})^{\prime}(I_{K}\otimes \mathcal{Q})\mathrm{vec}(\mathbb{W})$. Since $\mathcal{Q}$ has rank not smaller than $JM-M$ and the variance of $\mathrm{vec}(\mathbb{G}_{B})$ has full rank,  $\mathrm{tr}(\mathbb{W}^{\prime}\mathcal{Q}\mathbb{W})$ is bounded below by a random variable with a chi-squared distribution with degree of freedom $JM-M$ multiplied by a constant, and above by a random variable with a chi-squared distribution with degree of freedom $JM$ multiplied by a constant. Thus, it follows that
\begin{align}\label{Eqn: Thm: SpecTest: 7}
P(\mathcal{S}\leq c_{1-\alpha})&\leq P(\mathrm{tr}(\mathbb{W}^{\prime}\mathcal{Q}\mathbb{W})/J\leq c_{1-\alpha}+\gamma_{NT}/\sqrt{J})+o(1)\notag\\
&\leq P(\mathrm{tr}(\mathbb{W}^{\prime}\mathcal{Q}\mathbb{W})/J\leq c_{0,1-\alpha+\nu_{NT}}+2\gamma_{NT}/\sqrt{J})+o(1)\notag\\
&\leq P(\mathrm{tr}(\mathbb{W}^{\prime}\mathcal{Q}\mathbb{W})/\sqrt{J}\leq \sqrt{J}c_{0,1-\alpha+\nu_{NT}}+2\gamma_{NT})+o(1)\notag\\
&\leq P(\mathrm{tr}(\mathbb{W}^{\prime}\mathcal{Q}\mathbb{W})/\sqrt{J}\leq \sqrt{J}c_{0,1-\alpha+\nu_{NT}})+o(1)\notag\\
&\leq 1-\alpha+\nu_{NT}+o(1) = 1-\alpha+o(1),
\end{align}
where the first inequality follows since $P(|\mathcal{S}-\mathrm{tr}(\mathbb{G}^{\prime}Q\mathbb{G})/J|>\gamma_{NT}/\sqrt{J})=o(1)$ due to \eqref{Eqn: Thm: SpecTest: 3}, the second inequality follows from \eqref{Eqn: Thm: SpecTest: 6}, and the fourth inequality follows since $\gamma_{NT} = o(1)$ and $\mathrm{tr}(\mathbb{W}^{\prime}\mathcal{Q}\mathbb{W})$ is bounded by chi-squared random variables. By a similar argument, $P(\mathcal{S}> c_{1-\alpha})\leq 1-\alpha+o(1)$. Therefore, the first result of the theorem follows. To show the second result, we now assume that $\mathrm{H}_1$ is true. Since $(x+y)^{2}\geq x^{2}/2-y^{2}$,
\begin{align}\label{Eqn: Thm: SpecTest: 8}
\frac{2J}{NT}\mathcal{S} &\geq \frac{1}{NT}\sum_{i=1}^{N}\sum_{t=1}^{T}\|\hat{\Gamma}^{\prime}z_{it}-H^{\prime}\beta(z_{it})\|^2 - \frac{2}{NT}\sum_{i=1}^{N}\sum_{t=1}^{T}\|\hat{\beta}(z_{it})-H^{\prime}\beta(z_{it})\|^{2}\notag\\
&\hspace{0.5cm} + \frac{2}{NT}\sum_{i=1}^{N}\sum_{t=1}^{T}|\hat{\gamma}^{\prime}z_{it}-\hat{\alpha}(z_{it})|^2 \geq c_0 + o_{p}(1) \text{ for some } c_0>0,
\end{align}
where the second inequality follows from Lemmas \ref{Lem: TechE5} and \ref{Lem: TechE6}. We have
\begin{align}\label{Eqn: Thm: SpecTest: 9}
\frac{2J}{NT}\mathcal{S}^{\ast}&\leq \frac{4}{NT\omega_0}\sum_{i=1}^{N}\sum_{t=1}^{T}|(\hat{\gamma}^{\ast}-\hat{\gamma})^{\prime}z_{it}|^{2}+\frac{4}{NT\omega_0}\sum_{i=1}^{N}\sum_{t=1}^{T}|(\hat{a}^{\ast}-\hat{a})^{\prime}\phi(z_{it})|^2\notag\\
&\hspace{-1.5cm}+\frac{4}{NT\omega_0}\sum_{i=1}^{N}\sum_{t=1}^{T}\|(\hat{\Gamma}^{\ast}-\hat{\Gamma})^{\prime}z_{it}\|^{2}+\frac{4}{NT\omega_0}\sum_{i=1}^{N}\sum_{t=1}^{T}\|(\hat{B}^{\ast}-\hat{B})^{\prime}\phi(z_{it})\|^2=o_{p}(1),
\end{align}
where the equality follows from Lemma \ref{Lem: TechE7}. In view of \eqref{Eqn: Thm: SpecTest: 9}, Lemma A.1 of \citet{BelloniChernozhukovChetverikovKato_SeriesEstimator_2015} implies that $2c_{1-\alpha}J/(NT) = o_{p}(1)$. This together with \eqref{Eqn: Thm: SpecTest: 8} thus concludes the second result of the theorem.\qed

\subsubsection{Technical Lemmas}
\begin{lem}\label{Lem: TechE1}
Let $\mathcal{S}_1,\mathcal{S}_2,\mathcal{S}_3,\mathcal{S}_4,\mathcal{S}_5,\mathcal{S}_6$ be given in the proof of Theorem \ref{Thm: SpecTest}. Assume (i) $N\to\infty$; (ii) $T\geq K+1$ and $T=o(N)$; (iii) $J\to\infty$ with $J^{2}\xi^{2}_{J}\log J=o(N)$ and $NTJ^{-(2\kappa+1)}=o(1)$. Assume that $\mathrm{H}_0$ is true.\\
(i) Under Assumption \ref{Ass: LoadingsFactors}(iv), $\mathcal{S}_1 = O_{p}(NTJ^{-(2\kappa+1)})$.\\
(ii) Under Assumptions \ref{Ass: Basis}-\ref{Ass: Asym}, \ref{Ass: SpecTest}(i)-(iii), $\mathcal{S}_2 = O_{p}(\sqrt{NT}J^{-(\kappa+1)})$.\\
(iii) Under Assumptions \ref{Ass: Basis}-\ref{Ass: Improvedrates}, $\mathcal{S}_3 = O_{p}(\sqrt{NT}J^{-(\kappa+1/2)})$.\\
(iv) Under Assumptions \ref{Ass: Basis}-\ref{Ass: DGP}, $\mathcal{S}_4 = O_{p}(NTJ^{-(2\kappa+1)})$.\\
(v) Under Assumptions \ref{Ass: Basis}-\ref{Ass: Asym}, \ref{Ass: SpecTest}(i)-(iii), $\mathcal{S}_5 = O_{p}(\sqrt{NT}J^{-(\kappa+1)})$.\\
(vi) Under Assumptions \ref{Ass: Basis}-\ref{Ass: Improvedrates}, $\mathcal{S}_6 = O_{p}(\sqrt{NT}J^{-(\kappa+1/2)})$.
\end{lem}
\noindent{\sc Proof:} (i) The proof is similar to the proof of (iv).

(ii) The proof is similar to the proof of (v).

(iii) The proof is similar to the proof of (vi).

(iv) It follows that
\begin{align}\label{Eqn: Lem: TechE1: 1}
\hspace{-1cm}\mathcal{S}_4\leq \|H\|_{2}^{2}\sum_{i=1}^{T}\sum_{t=1}^{T}\|\delta(z_{it})\|^2/J\leq (NT/J)\|H\|_{2}^{2}M^{2}K\sup_{k\leq K,m\leq M}\sup_{z}|\delta_{km,J}(z)|^{2},
\end{align}
where the second inequality follows by $\max_{i\leq N, t\leq T}\|\delta(z_{it})\|^2\leq M^{2}K\sup_{k\leq K,m\leq M}$ $\sup_{z}|\delta_{km,J}(z)|^{2}$. Thus, the result of the lemma follows from \eqref{Eqn: Lem: TechE1: 1}, Assumption \ref{Ass: LoadingsFactors}(iv), and Lemma \ref{Lem: TechA2}(i).

(v) By Assumption \ref{Ass: SpecTest}(ii), $\sum_{i=1}^{N}\sum_{t=1}^{T}\|z_{it}\|^{2}/NT=O_{p}(1)$ by the Markov's inequality. It then follows that
\begin{align}\label{Eqn: Lem: TechE1: 2}
\frac{1}{J}\sum_{i=1}^{T}\sum_{t=1}^{T}\|(\hat{\Gamma}-\Gamma H)^{\prime}z_{it}\|^{2}&\leq \|\hat{\Gamma}-\Gamma H\|^{2}_{F}\frac{1}{J}\sum_{i=1}^{N}\sum_{t=1}^{T}\|z_{it}\|^{2}\notag\\
&=O_{p}\left(\frac{1}{J} + \frac{T}{J^{2\kappa+1}} + \frac{T}{NJ}\right) = O_{p}\left(\frac{1}{J}\right),
\end{align}
where the first equality follows from Lemma \ref{Lem: TechE2}, and the second equality follows since $T=o(N)$, $NTJ^{-(2\kappa+1)}=o(1)$, and $J=o(\sqrt{N})$. By the Cauchy-Schwartz inequality,
\begin{align}\label{Eqn: Lem: TechE1: 3}
|\mathcal{S}_5|\leq \mathcal{S}^{1/2}_4\left(\frac{1}{J}\sum_{i=1}^{T}\sum_{t=1}^{T}\|(\hat{\Gamma}-\Gamma H)^{\prime}z_{it}\|^{2}\right)^{1/2}.
\end{align}
Thus, the result of the lemma follows from \eqref{Eqn: Lem: TechE1: 2}, \eqref{Eqn: Lem: TechE1: 3}, and Lemma \ref{Lem: TechE1}(iv).

(vi) By the fact that $\|x\|^2 = \mathrm{tr}(xx^{\prime})$,
\begin{align}\label{Eqn: Lem: TechE1: 4}
&\hspace{0.8cm}\frac{1}{J}\sum_{i=1}^{T}\sum_{t=1}^{T}\|(\hat{B}-BH)^{\prime}\phi(z_{it})\|^{2}=\frac{N}{J}\sum_{t=1}^{T}\mathrm{tr}\left((\hat{B}-BH)^{\prime}\hat{Q}_t(\hat{B}-BH)\right)\notag\\
&\leq \frac{NT}{J}\max_{t\leq T}\lambda_{\max}(\hat{Q}_t)\|\hat{B}-BH\|_{F}^{2} = O_{p}\left(\frac{NT}{J^{2\kappa + 1}}+\frac{T}{N} +1\right)= O_{p}(1),
\end{align}
where the second equality follows from Assumption \ref{Ass: Basis}(i) and Theorem \ref{Thm: ImprovedRates}, and the last equality follows since $T=o(N)$ and $NTJ^{-(2\kappa+1)}=o(1)$. By the Cauchy-Schwartz inequality,
\begin{align}\label{Eqn: Lem: TechE1: 5}
|\mathcal{S}_6|\leq \mathcal{S}^{1/2}_4\left(\frac{1}{J}\sum_{i=1}^{T}\sum_{t=1}^{T}\|(\hat{B}-BH)^{\prime}\phi(z_{it})\|^{2}\right)^{1/2}.
\end{align}
Thus, the result follows from \eqref{Eqn: Lem: TechE1: 4}, \eqref{Eqn: Lem: TechE1: 5}, and Lemma \ref{Lem: TechE1}(iv). \qed

\begin{lem}\label{Lem: TechE2}
Suppose Assumptions \ref{Ass: Basis}-\ref{Ass: Asym} and \ref{Ass: SpecTest}(i)-(iii) hold. Let $\hat{\gamma}$ and $\hat{\Gamma}$ be given in Section \ref{Sec: 52}. Assume (i) $N\to\infty$; (ii) $T\geq K+1$; (iii) $J\to\infty$ with $J^{2}\xi^{2}_{J}\log J=o(N)$. Let $\Omega_z\equiv \sum_{i=1}^{N}\sum_{t=1}^{T}\sum_{s=1}^{T}f^{\dag}_{t}f_{s}^{\dag\prime}\otimes Q_{z,t}^{-1}E[z_{it}z_{is}^{\prime}]Q_{z,s}^{-1}E[\varepsilon_{it}\varepsilon_{is}]/NT$, where $Q_{z,t}= \sum_{i=1}^{N}E[z_{it}z_{it}^{\prime}]/N$. Assume that $\mathrm{H}_0$ is true. Then there exists an $M\times (K+1)$ random matrix $\mathbb{N}_z$ with $\mathrm{vec}(\mathbb{N}_z)\sim N(0,\Omega_z)$ such that
\begin{align*}
\|\sqrt{{NT}}(\hat{\gamma} - \gamma)-\mathbb{G}_\gamma\| = O_{p}\left(\frac{\sqrt{NT}}{J^{\kappa}}+\frac{\sqrt{TJ}}{\sqrt{N}}+\frac{J^{5/6}}{N^{1/6}}+\frac{\sqrt{J\xi_{J}}\log^{1/4}J}{N^{1/4}}\right)
\end{align*}
and
\begin{align*}
\|\sqrt{{NT}}(\hat{\Gamma} - \Gamma H)-\mathbb{G}_\Gamma\|_{F} = O_{p}\left(\frac{\sqrt{T}}{J^{\kappa}} + \frac{\sqrt{T}}{\sqrt{N}}+\frac{1}{N^{1/6}}\right),
\end{align*}
where  $\mathbb{G}_{\gamma} =\mathbb{N}_{z,1}-\mathbb{G}_{\Gamma}\mathcal{H}^{-1}\bar{f}-\Gamma\mathcal{H}\mathcal{H}^{\prime}B^{\prime}(\mathbb{N}_{1}-\mathbb{G}_{B}\mathcal{H}^{-1}\bar{f}) - \Gamma\mathcal{H}\mathbb{G}_{B}^{\prime}a$, $\mathbb{G}_{\Gamma} = \mathbb{N}_{z,2} B^{\prime}B\mathcal{M}$, $\mathcal{H}$, $\mathcal{M}$, $\mathbb{N}_1$ and $\mathbb{G}_B$ are given in Theorem \ref{Thm: AsymDis}, and $\mathbb{N}_{z,1}$ and $\mathbb{N}_{z,2}$ are the first column and the last $K$ columns of $\mathbb{N}_z$.
\end{lem}
\noindent{\sc Proof:} Let us begin by defining some notation. Let $\vec{\varepsilon}_{t} \equiv (Z_{t}^{\prime}Z_{t})^{-1}Z_{t}^{\prime}\varepsilon_t$ and $\vec{E}\equiv (\vec{\varepsilon}_{1},\ldots, \vec{\varepsilon}_{T})$. Then \eqref{Eqn: Model: Sieve: Vectors: Regressed} under $\mathrm{H}_0$ can be written as
\begin{align}\label{Eqn: Lem: TechE2: 1}
\vec{Y} = \gamma1_{T}^{\prime}+\Gamma F^{\prime} + \vec{E},
\end{align}
where $1_{T}$ denotes a $T\times 1$ vector of ones.  Recall $M_T= I_{T} - 1_T1_T^{\prime}/T$. Post-multiplying \eqref{Eqn: Lem: TechE2: 1} by $M_T$ to remove $\gamma$, we thus obtain
\begin{align}\label{Eqn: Lem: TechE2: 2}
\vec{Y}M_T = \Gamma(M_TF)^{\prime} + \vec{E}M_T.
\end{align}
Recall that $V$ is a $K\times K$ diagonal matrix of the first $K$ largest eigenvalues of $\tilde{Y}M_T\tilde{Y}^{\prime}/T$ as defined in the proof of Theorem \ref{Thm: Rate}, $H=F^{\prime}M_T\hat{F}(\hat{F}^{\prime}M_T\hat{F})^{-1}$ and $\hat{F}^{\prime}M_T\hat{F}/T = V$ as showed in the proof of Theorem \ref{Thm: Rate}. By the definition of $\hat{\Gamma}$, $\hat{\Gamma} = \vec{Y}M_T\hat{F}(\hat{F}^{\prime}M_T\hat{F})^{-1}$. We may substitute \eqref{Eqn: Lem: TechE2: 2} to $\hat{\Gamma} = \vec{Y}M_T\hat{F}(\hat{F}^{\prime}M_T\hat{F})^{-1}$ to obtain
\begin{align}\label{Eqn: Lem: TechE2: 3}
\hat{\Gamma}-\Gamma H = (\vec{E}M_T\tilde{Y}^{\prime}/T)\hat{B}V^{-1} = \sum_{j=1}^{3}\mathcal{D}_{j}\hat{B}V^{-1},
\end{align}
where in the first equality we have used $\hat{F}^{\prime}M_T\hat{F}/T=V$ and $\hat{F} = \tilde{Y}^{\prime}\hat{B}$, in the second equality we have substituted \eqref{Eqn: Thm: Rate: 2} into the equation, and $\mathcal{D}_1 =\vec{E}M_TFB^{\prime}/T$, $\mathcal{D}_2 = \vec{E}M_T\tilde{E}^{\prime}/T$, and $\mathcal{D}_3 = \vec{E}M_T\tilde{\Delta}^{\prime}/T$. We can conduct the same exercise as in \eqref{Eqn: Thm: AsymDis: 1} to obtain
\begin{align}\label{Eqn: Lem: TechE2: 4}
&\|\sqrt{NT}(\hat{\Gamma}-\Gamma H)-\sqrt{NT}\mathcal{D}_1\hat{B}V^{-1}\|_{F} \notag\\
&\hspace{-0.5cm}\leq \sqrt{NT}\|V^{-1}\|_{2}(\|\mathcal{D}_{2}\hat{B}\|_{F}+\|\mathcal{D}_{3}\|_{F}\|\hat{B}\|_2)=O_{p}\left(\frac{\sqrt{T}}{J^{\kappa}} + \frac{\sqrt{T}}{\sqrt{N}}\right),
\end{align}
where the equality follows by Lemmas \ref{Lem: TechA2}(i) and \ref{Lem: TechE8}. Thus, the second result of the lemma follows from \eqref{Eqn: Lem: TechE2: 4} and Lemma \ref{Lem: TechE9}. We now show the first result of the lemma. By the definition of $\hat{\gamma}$,
\begin{align}\label{Eqn: Lem: TechE2: 5}
\hat{\gamma}-\gamma &= \vec{E}1_{T}/T+(\Gamma H-\hat{\Gamma})H^{-1}\bar{f}-\hat{\Gamma}(\hat{B}-BH)^{\prime}a\notag\\
&\hspace{0.5cm}-\hat{\Gamma}\hat{B}^{\prime}(BH-\hat{B})H^{-1}\bar{f}-\hat{\Gamma}\hat{B}^{\prime}\tilde{E}1_{T}/T-\hat{\Gamma}\hat{B}^{\prime}\tilde{\Delta}1_{T}/T,
\end{align}
where $H^{-1}$ is well defined with probability approaching one by \eqref{Eqn: Thm: Rate: 4} and Lemma \ref{Lem: TechA2}(ii), and we have used $a^{\prime}B=0$ and $\hat{B}^{\prime}\hat{B}=I_{K}$. By a similar argument as in \eqref{Eqn: Thm: AsymDis: 3}-\eqref{Eqn: Thm: AsymDis: 5},
\begin{align}\label{Eqn: Lem: TechE2: 6}
&\|\sqrt{NT}(\hat{\gamma}-\gamma) - [\sqrt{N/T}\vec{E}1_{T}-\sqrt{NT}(\hat{\Gamma}-\Gamma H)\mathcal{H}^{-1}\bar{f}]\notag\\
&\hspace{0.5cm}+\Gamma\mathcal{H}\mathcal{H}^{\prime}B^{\prime}[\sqrt{N/T}\tilde{E}1_{T}-\sqrt{NT}(\hat{B}-BH)\mathcal{H}^{-1}\bar{f}]\notag\\
&\hspace{0.5cm}+\Gamma\mathcal{H}\sqrt{NT}(\hat{B}-BH)^{\prime}a\|=O_{p}\left(\frac{\sqrt{NT}}{J^{\kappa}}+\frac{\sqrt{TJ}}{\sqrt{N}}+\frac{J}{\sqrt{NT}}\right).
\end{align}
Thus, the first result of the lemma follows from \eqref{Eqn: Lem: TechE2: 6}, along with Lemmas \ref{Lem: TechE9}, \ref{Lem: TechC1}, \ref{Lem: TechC2}, and Theorem \ref{Thm: AsymDis}, as well as the second result of the lemma.\qed

\begin{lem}\label{Lem: TechE3}
Suppose Assumptions \ref{Ass: Improvedrates}(i), (iii), and \ref{Ass: SpecTest}(ii) hold. Let $\hat{\mathcal{Q}}$ and $\mathcal{Q}$ be given in the proof of Theorem \ref{Thm: SpecTest}. Then
\[\|\hat{\mathcal{Q}}-\mathcal{Q}\|_{F}^{2}= O_{p}\left(\frac{J^{2}}{N}\right).\]
\end{lem}
\noindent{\sc Proof:} Let $\hat{\mathcal{Q}}_{t}\equiv \sum_{i=1}^{N}(z_{it}^{\prime},\phi(z_{it})^{\prime})(z_{it}^{\prime},\phi(z_{it})^{\prime})^{\prime}/N$ and ${\mathcal{Q}}_{t}\equiv E[\hat{\mathcal{Q}}_{t}]$. Then $\hat{\mathcal{Q}} = \sum_{t=1}^{T}\hat{\mathcal{Q}}_{t}/T$ and ${\mathcal{Q}} = \sum_{t=1}^{T}{\mathcal{Q}}_{t}/T$. It follows that  $E[\|\hat{\mathcal{Q}}_{t}-{\mathcal{Q}}_{t}\|^{2}_{F}]\leq [((J+1)M)^{2}/N]$ $(\max_{m\leq M, j\leq J, i\leq N, t\leq T}E[\phi^{4}_{j}(z_{it,m})]+\max_{i\leq N, t\leq T}E[\|z_{it}\|^4])$ by the independence in Assumption \ref{Ass: Improvedrates}(iii). By the Cauchy-Schwartz inequality,
 \begin{align}\label{Eqn: Lem: TechE3: 1}
E[\|\hat{\mathcal{Q}}-\mathcal{Q}\|^{2}_{F}] \leq \frac{1}{T}\sum_{t=1}^{T}E[\|\hat{\mathcal{Q}}_{t}-{\mathcal{Q}}_{t}\|^{2}_{F}]=O\left(\frac{J^{2}}{N}\right),
\end{align}
where the equality follows from Assumptions \ref{Ass: Improvedrates}(i) and \ref{Ass: SpecTest}(ii). By the Markov's inequality, the result of the lemma thus follows from \eqref{Eqn: Lem: TechE3: 1}.\qed

\begin{lem}\label{Lem: TechE4}
Suppose Assumptions \ref{Ass: Basis}-\ref{Ass: Asym},  \ref{Ass: Boot}, and \ref{Ass: SpecTest}(ii)-(iv) hold. Let $\hat{\gamma}$, $\hat{\Gamma}$, $\hat{\gamma}^{\ast}$ and $\hat{\Gamma}^{\ast}$ be given in Section \ref{Sec: 52}. Assume (i) $N\to\infty$; (ii) $T\geq K+1$; (iii) $J\to\infty$ with $J^{2}\xi^{2}_{J}\log J=o(N)$. Assume that $\mathrm{H}_0$ is true. Then there exists an $M\times (K+1)$ random matrix $\mathbb{N}^{\ast}_z$ with $\mathrm{vec}(\mathbb{N}^{\ast}_z)\sim N(0,\Omega_z)$ conditional on $\{Y_{t},Z_{t}\}_{t\leq T}$ such that
\begin{align*}
\|\sqrt{{NT/\omega_0}}(\hat{\gamma}^{\ast} - \hat{\gamma})-\mathbb{G}^{\ast}_\gamma\| = O_{p^{\ast}}\left(\frac{\sqrt{NT}}{J^{\kappa}}+\frac{\sqrt{TJ}}{\sqrt{N}}+\frac{J^{5/6}}{N^{1/6}}+\frac{\sqrt{J\xi_{J}}\log^{1/4}J}{N^{1/4}}\right)
\end{align*}
and
\begin{align*}
\|\sqrt{{NT/\omega_0}}(\hat{\Gamma}^{\ast} - \hat{\Gamma})-\mathbb{G}^{\ast}_\Gamma\|_{F} = O_{p^{\ast}}\left(\frac{\sqrt{T}}{J^{\kappa}} + \frac{\sqrt{T}}{\sqrt{N}}+\frac{1}{N^{1/6}}\right),
\end{align*}
where $\Omega_z$ is given in Lemma \ref{Lem: TechE2}, $\mathbb{G}^{\ast}_{\gamma} =\mathbb{N}^{\ast}_{z,1}-\mathbb{G}^{\ast}_{\Gamma}\mathcal{H}^{-1}\bar{f}-\Gamma\mathcal{H}\mathcal{H}^{\prime}B^{\prime}(\mathbb{N}^{\ast}_{1}-\mathbb{G}^{\ast}_{B}\mathcal{H}^{-1}\bar{f}) - \Gamma\mathcal{H}\mathbb{G}_{B}^{\ast\prime}a$, $\mathbb{G}^{\ast}_{\Gamma} = \mathbb{N}^{\ast}_{z,2} B^{\prime}B\mathcal{M}$, $\mathcal{H}$, $\mathcal{M}$, $\mathbb{N}^{\ast}_1$ and $\mathbb{G}^{\ast}_B$ are given in Theorem \ref{Thm: Boot}, and $\mathbb{N}^{\ast}_{z,1}$ and $\mathbb{N}^{\ast}_{z,2}$ are the first column and the last $K$ columns of $\mathbb{N}^{\ast}_z$.
\end{lem}
\noindent{\sc Proof:} Let us begin by defining some notation. Let $\vec{\varepsilon}^{\ast}_{t} \equiv (Z_{t}^{\ast\prime}Z_{t})^{-1}Z_{t}^{\ast\prime}\varepsilon_t$ and $\vec{E}^{\ast}\equiv (\vec{\varepsilon}^{\ast}_{1},\ldots, \vec{\varepsilon}^{\ast}_{T})$. Then under $\mathrm{H}_0$, we have
\begin{align}\label{Eqn: Lem: TechE4: 1}
\vec{Y}^{\ast} = \gamma1_{T}^{\prime} +\Gamma F^{\prime} + \vec{E}^{\ast}.
\end{align}
where $1_{T}$ denotes a $T\times 1$ vector of ones.  Recall $M_T= I_{T} - 1_T1_T^{\prime}/T$. Post-multiplying \eqref{Eqn: Lem: TechE4: 1} by $M_T$ to remove $\gamma$, we thus obtain
\begin{align}\label{Eqn: Lem: TechE4: 2}
\vec{Y}^{\ast}M_T = \Gamma(M_TF)^{\prime} + \vec{E}^{\ast}M_T.
\end{align}
Recall that $V$ us a $K\times K$ diagonal matrix of the first $K$ largest eigenvalues of $\tilde{Y}M_T\tilde{Y}^{\prime}/T$ as defined in the proof of Theorem \ref{Thm: Rate}, $H=F^{\prime}M_T\hat{F}(\hat{F}^{\prime}M_T\hat{F})^{-1}$ and $\hat{F}^{\prime}M_T\hat{F}/T = V$ as showed in the proof of Theorem \ref{Thm: Rate}. By the definitions of $\hat{\Gamma}^{\ast}$, $\hat{\Gamma}^{\ast} = \vec{Y}^{\ast}M_T\hat{F}(\hat{F}^{\prime}M_T\hat{F})^{-1}$. We may substitute \eqref{Eqn: Lem: TechE4: 2} to $\hat{\Gamma}^{\ast} = \vec{Y}^{\ast}M_T\hat{F}(\hat{F}^{\prime}M_T\hat{F})^{-1}$ to obtain
\begin{align}\label{Eqn: Lem: TechE4: 3}
\hat{\Gamma}^{\ast}-\Gamma H =  (\vec{E}^{\ast}M_T\tilde{Y}^{\prime}/T)\hat{B}V^{-1} = \sum_{j=1}^{3}\mathcal{D}^{\ast}_{j}\hat{B}V^{-1},
\end{align}
where in the first equality we have used $\hat{F}^{\prime}M_T\hat{F}/T=V$ and $\hat{F} = \tilde{Y}^{\prime}\hat{B}$, in the second equality follows we have substituted \eqref{Eqn: Thm: Rate: 2} into the equation, and $\mathcal{D}^{\ast}_1 =\vec{E}^{\ast}M_TFB^{\prime}/T$, $\mathcal{D}^{\ast}_2 = \vec{E}^{\ast}M_T\tilde{E}^{\prime}/T$, and $\mathcal{D}^{\ast}_3 = \vec{E}^{\ast}M_T\tilde{\Delta}^{\prime}/T$. We can conduct the same exercise as in \eqref{Eqn: Thm: AsymDis: 1} to obtain
\begin{align}\label{Eqn: Lem: TechE4: 4}
&\|\sqrt{NT}(\hat{\Gamma}^{\ast}-\Gamma H)-\sqrt{NT}\mathcal{D}^{\ast}_1\hat{B}V^{-1}\|_{F}\notag\\
&\hspace{-0.5cm}\leq \sqrt{NT}\|V^{-1}\|_{2}(\|\mathcal{D}^{\ast}_{2}\hat{B}\|_{F}+\|\mathcal{D}^{\ast}_{3}\|_{F}\|\hat{B}\|_2)=O_{p}\left(\frac{\sqrt{T}}{J^{\kappa}} + \frac{\sqrt{T}}{\sqrt{N}}\right),
\end{align}
where the equality follows by Lemmas \ref{Lem: TechA2}(i) and \ref{Lem: TechE10}. By the fact that $\|C+D\|_{F}\leq \|C\|_{F} + \|D\|_{F}$, we may combine \eqref{Eqn: Lem: TechE2: 4} and \eqref{Eqn: Lem: TechE4: 4} to obtain
\begin{align}\label{Eqn: Lem: TechE4: 5}
\|\sqrt{NT}(\hat{\Gamma}^{\ast}-\hat{\Gamma})-\sqrt{NT}(\mathcal{D}^{\ast}_1-\mathcal{D}_1)\hat{B}V^{-1}\|_{F} =O_{p}\left(\frac{\sqrt{T}}{J^{\kappa}} + \frac{\sqrt{T}}{\sqrt{N}}\right).
\end{align}
Thus, the second result of the lemma follows from \eqref{Eqn: Lem: TechE4: 5} and Lemmas \ref{Lem: TechA5} and \ref{Lem: TechE11}. We now show the first result of the lemma. By the definition of $\hat{\gamma}^{\ast}$,
\begin{align}\label{Eqn: Lem: TechE4: 6}
\hat{\gamma}^{\ast}-\gamma &= \vec{E}^{\ast}1_{T}/T+(\Gamma H-\hat{\Gamma}^{\ast})H^{-1}\bar{f}-\hat{\Gamma}^{\ast}(\hat{B}^{\ast\prime}\hat{B}^{\ast})^{-1}(\hat{B}^{\ast}-BH)^{\prime}a\notag\\
&\hspace{0.5cm}-\hat{\Gamma}^{\ast}(\hat{B}^{\ast\prime}\hat{B}^{\ast})^{-1}\hat{B}^{\ast\prime}(BH-\hat{B}^{\ast})H^{-1}\bar{f}-\hat{\Gamma}^{\ast}(\hat{B}^{\ast\prime}\hat{B}^{\ast})^{-1}\hat{B}^{\ast\prime}\tilde{E}^{\ast}1_{T}/T\notag\\
&\hspace{0.5cm}-\hat{\Gamma}^{\ast}(\hat{B}^{\ast\prime}\hat{B}^{\ast})^{-1}\hat{B}^{\ast\prime}\tilde{\Delta}^{\ast}1_{T}/T,
\end{align}
where $H^{-1}$ is well defined with probability approaching one by \eqref{Eqn: Thm: Rate: 4} and Lemma \ref{Lem: TechA2}(ii), and we have used $a^{\prime}B=0$ and $(\hat{B}^{\ast\prime}\hat{B}^{\ast})^{-1}\hat{B}^{\ast\prime}\hat{B}^{\ast}=I_{K}$. By a similar argument as in \eqref{Eqn: Thm: AsymDis: 3}-\eqref{Eqn: Thm: AsymDis: 5},
\begin{align}\label{Eqn: Lem: TechE4: 7}
&\|\sqrt{NT}(\hat{\gamma}^{\ast}-\gamma) - [\sqrt{N/T}\vec{E}^{\ast}1_{T}-\sqrt{NT}(\hat{\Gamma}^{\ast}-\Gamma H)\mathcal{H}^{-1}\bar{f}]\notag\\
&\hspace{0.5cm}+\Gamma\mathcal{H}\mathcal{H}^{\prime}B^{\prime}[\sqrt{N/T}\tilde{E}^{\ast}1_{T}-\sqrt{NT}(\hat{B}^{\ast}-BH)\mathcal{H}^{-1}\bar{f}]\notag\\
&\hspace{0.5cm}+\Gamma\mathcal{H}\sqrt{NT}(\hat{B}^{\ast}-BH)^{\prime}a\|=O_{p}\left(\frac{\sqrt{NT}}{J^{\kappa}}+\frac{\sqrt{TJ}}{\sqrt{N}}+\frac{J}{\sqrt{NT}}\right).
\end{align}
By the fact that $\|x+y\|\leq \|x\|+\|y\|$, we may combine \eqref{Eqn: Lem: TechE2: 6} and \eqref{Eqn: Lem: TechE4: 7} to obtain
\begin{align}\label{Eqn: Lem: TechE4: 8}
&\|\sqrt{NT}(\hat{\gamma}^{\ast}-\hat{\gamma}) - [\sqrt{N/T}(\vec{E}^{\ast}1_{T}-\vec{E}1_{T})-\sqrt{NT}(\hat{\Gamma}^{\ast}-\hat\Gamma)\mathcal{H}^{-1}\bar{f}]\notag\\
&\hspace{0.5cm}+\Gamma\mathcal{H}\mathcal{H}^{\prime}B^{\prime}[\sqrt{N/T}(\tilde{E}^{\ast}1_{T}-\tilde{E}1_{T})-\sqrt{NT}(\hat{B}^{\ast}-\hat{B})\mathcal{H}^{-1}\bar{f}]\notag\\
&\hspace{0.5cm}+\Gamma\mathcal{H}\sqrt{NT}(\hat{B}^{\ast}-\hat{B})^{\prime}a\|=O_{p}\left(\frac{\sqrt{NT}}{J^{\kappa}}+\frac{\sqrt{TJ}}{\sqrt{N}}+\frac{J}{\sqrt{NT}}\right).\end{align}
Thus, the first result of the lemma follows from \eqref{Eqn: Lem: TechE4: 8}, along with Lemma \ref{Lem: TechE11}, \ref{Lem: TechC1}, \ref{Lem: TechD2}, \ref{Lem: TechD3}, and Theorem \ref{Thm: Boot}, as well as the second result of the lemma.\qed

\begin{lem}\label{Lem: TechE5}
Suppose Assumptions \ref{Ass: Basis}-\ref{Ass: Intercept} hold. Assume (i) $N\to\infty$; (ii) $T\geq K+1$; (iii) $J\to\infty$ with $J=o(\sqrt{N})$. Then
\[\frac{1}{NT}\sum_{i=1}^{N}\sum_{t=1}^{T}\|\hat{\beta}(z_{it})-H^{\prime}\beta(z_{it})\|^{2}=o_{p}(1). \]
\end{lem}
\noindent{\sc Proof:} Since $\hat{\beta}(z_{it}) = \hat{B}^{\prime}\phi(z_{it})$ and $\beta(z_{it})=B^{\prime}\phi(z_{it}) + \delta(z_{it})$,
\begin{align}\label{Eqn: Lem: TechE5: 1}
\frac{1}{J}\sum_{i=1}^{N}\sum_{t=1}^{T}\|\hat{\beta}(z_{it})-H^{\prime}\beta(z_{it})\|^{2}\leq \frac{2}{J}\sum_{i=1}^{T}\sum_{t=1}^{T}\|(\hat{B}-BH)^{\prime}\phi(z_{it})\|^{2} + 2\mathcal{S}_4,
\end{align}
where $\mathcal{S}_4 = \sum_{i=1}^{T}\sum_{t=1}^{N}\|H^{\prime}\delta(z_{it})\|^2/J$ as defined in the proof of Theorem \ref{Thm: SpecTest}. Note that \eqref{Eqn: Lem: TechE1: 4} and Lemma \ref{Lem: TechE1}(iv) continue to hold under $\mathrm{H}_1$. Thus, the result of the lemma follows from \eqref{Eqn: Lem: TechE1: 4} and Lemma \ref{Lem: TechE1}(iv). \qed

\begin{lem}\label{Lem: TechE6}
Suppose Assumptions \ref{Ass: Basis}-\ref{Ass: Intercept}, \ref{Ass: Improvedrates}(iii), \ref{Ass: SpecTest}(i), (ii), and (v) hold. Assume (i) $N\to\infty$; (ii) $T\geq K+1$; (iii) $J\to\infty$ with $J=o(\sqrt{N})$. Assume that $\mathrm{H}_1$ is true. Then there exists positive constant $c_0$ such that
\[\frac{1}{NT}\sum_{i=1}^{N}\sum_{t=1}^{T}\|\hat{\Gamma}^{\prime}z_{it}-H^{\prime}\beta(z_{it})\|^2\geq c_0 + o_{p}(1).\]
\end{lem}
\noindent{\sc Proof:} Let us begin by defining some notation. Let $\vec{A}_{t} \equiv (Z_{t}^{\prime}Z_{t})^{-1}Z_{t}^{\prime}A_t$ for $A_{t} = Y_{t}, \Psi_t, $ $\varepsilon_t$, where $\Psi_t = (\alpha(z_{1t}) + \beta(z_{1t})^{\prime}f_t,\ldots,\alpha(z_{Nt}) +\beta(z_{Nt})^{\prime}f_t)^{\prime}$. Let $\vec{Y}\equiv $
$(\vec{Y}_{1},\ldots, \vec{Y}_{T})$, $\vec{\Psi}\equiv (\vec{\Psi}_{1},\ldots, \vec{\Psi}_{T})$, and $\vec{E}\equiv (\vec{\varepsilon}_{1},\ldots, \vec{\varepsilon}_{T})$. Then $\hat{\Gamma}=(\vec{\Psi} + \vec{E})M_T\hat{F}(\hat{F}^{\prime}M_T\hat{F})^{-1}$. It is easy to show that $\hat{\Gamma} = (\vec{\Psi}M_TF/T)({F}^{\prime}M_T{F}/T)^{-1}H+o_{p}(1)$ by Theorem \ref{Thm: Rate}, and $\|(\vec{\Psi}M_TF/T)({F}^{\prime}M_T{F}/T)^{-1}\|_{F}\leq C^{\ast}$ for some $C^{\ast}$ with probability approaching one. This together with Lemma \ref{Lem: TechA2}(ii) implies that $P(\|\hat{\Gamma}H^{-1}\|_{F}>C)=o(1)$. Therefore, under $\mathrm{H}_1$,
\begin{align}\label{Eqn: Lem: TechE6: 1}
\frac{1}{NT}\sum_{i=1}^{N}\sum_{t=1}^{T}\|\hat{\Gamma}^{\prime}z_{it}-H^{\prime}\beta(z_{it})\|^2&\geq \lambda_{\min}(H^{\prime}H)\frac{1}{NT}\sum_{i=1}^{N}\sum_{t=1}^{T}\|(\hat{\Gamma}H^{-1})^{\prime}z_{it}-\beta(z_{it})\|^2\notag\\
& \hspace{-2.5cm}= \lambda_{\min}(H^{\prime}H) \frac{1}{NT}\sum_{i=1}^{N}\sum_{t=1}^{T}E\left[\|\beta(z_{it})- (\hat{\Gamma}H^{-1})^{\prime}z_{it}\|^2\right]+o_{p}(1)\notag\\
&\hspace{-2.5cm}\geq \lambda_{\min}(H^{\prime}H)\inf_{i\leq N,t\leq T}\inf_{\Pi}E[\|\beta(z_{it})-\Pi^{\prime} z_{it}\|^{2}]+o_{p}(1)\notag\\
&\hspace{-2.5cm}\geq c_0 +o_{p}(1) \text{ for some }c_0>0,
\end{align}
where the equality follows from Lemma \ref{Lem: TechE12} since $P(\|\hat{\Gamma}H^{-1}\|_{F}>C)=o(1)$, and the last inequality follows by Lemma \ref{Lem: TechA2}(ii).\qed

\begin{lem}\label{Lem: TechE7}
Suppose Assumptions \ref{Ass: Basis}-\ref{Ass: Intercept}, \ref{Ass: Improvedrates}(iii), \ref{Ass: SpecTest} hold. Assume (i) $N\to\infty$; (ii) $T\geq K+1$; (iii) $J\to\infty$ with $J=o(\sqrt{N})$. Then
\[\frac{1}{NT}\sum_{i=1}^{N}\sum_{t=1}^{T}|(\hat{\gamma}^{\ast}-\hat{\gamma})^{\prime}z_{it}|^{2}+\frac{1}{NT}\sum_{i=1}^{N}\sum_{t=1}^{T}|(\hat{a}^{\ast}-\hat{a})^{\prime}\phi(z_{it})|^2 = o_{p}(1)\]
and
\[\frac{1}{NT}\sum_{i=1}^{N}\sum_{t=1}^{T}\|(\hat{\Gamma}^{\ast}-\hat{\Gamma})^{\prime}z_{it}\|^{2}+\frac{1}{NT}\sum_{i=1}^{N}\sum_{t=1}^{T}\|(\hat{B}^{\ast}-\hat{B})^{\prime}\phi(z_{it})\|^2 = o_{p}(1).\]
\end{lem}
\noindent{\sc Proof:} We prove the second result, and the proof of the first result is similar. Note that \eqref{Eqn: Lem: TechE1: 4} continue to hold under $\mathrm{H}_1$, so the second term on the left-hand side of the second result is $o_{p}(1)$. For the first term, we have
\begin{align}\label{Eqn: Lem: TechE7: 1}
\frac{1}{NT}\sum_{i=1}^{N}\sum_{t=1}^{T}\|(\hat{\Gamma}^{\ast}-\hat{\Gamma})^{\prime}z_{it}\|^{2}\leq \|\hat{\Gamma}^{\ast}-\hat{\Gamma}\|_{F}^{2}\frac{1}{NT}\sum_{i=1}^{N}\sum_{t=1}^{T}\|z_{it}\|^{2}.
\end{align}
Let us define some notation. Let $\vec{A}^{\ast}_{t} \equiv (Z_{t}^{\ast\prime}Z_{t})^{-1}Z_{t}^{\ast\prime}A_t$ for $A_{t} = Y_{t}, \Psi_t, \varepsilon_t$, where $\Psi_t = (\alpha(z_{1t})+\beta(z_{1t})^{\prime}f_t,\ldots,\alpha(z_{Nt})+\beta(z_{Nt})^{\prime}f_t)^{\prime}$. Let $\vec{Y}^{\ast}\equiv (\vec{Y}^{\ast}_{1},\ldots, \vec{Y}^{\ast}_{T})$, $\vec{\Psi}^{\ast}\equiv (\vec{\Psi}^{\ast}_{1},\ldots, \vec{\Psi}^{\ast}_{T})$, and $\vec{E}^{\ast}\equiv (\vec{\varepsilon}^{\ast}_{1},\ldots, \vec{\varepsilon}^{\ast}_{T})$. Then $\hat{\Gamma}^{\ast}=(\vec{\Psi}^{\ast} + \vec{E}^{\ast})M_T\hat{F}(\hat{F}^{\prime}M_T\hat{F})^{-1}$. It is easy to show that $\hat{\Gamma}^{\ast} = (\vec{\Psi}^{\ast}M_TF/T)({F}^{\prime}M_T{F}/T)^{-1}H+o_{p}(1)$ by Theorem \ref{Thm: Rate}. From the proof of Lemma \ref{Lem: TechE6}, $\hat{\Gamma} = (\vec{\Psi}M_TF/T)({F}^{\prime}M_T{F}/T)^{-1}H+o_{p}(1)$. Moreover, it can be easily shown that $(\vec{\Psi}^{\ast}-\vec{\Psi})M_TF/T = o_{p}(1)$. Thus,
\begin{align}\label{Eqn: Lem: TechE7: 2}
\hat{\Gamma}^{\ast} - \hat{\Gamma} = (\vec{\Psi}^{\ast}-\vec{\Psi})F/T({F}^{\prime}{F}/T)^{-1}=o_{p}(1).
\end{align}
By Assumption \ref{Ass: SpecTest}(ii), $\sum_{i=1}^{N}\sum_{t=1}^{T}\|z_{it}\|^{2}/NT=O_{p}(1)$ by the Markov's inequality. This together with \eqref{Eqn: Lem: TechE7: 1} and \eqref{Eqn: Lem: TechE7: 2} implies that the first term is also $o_{p}(1)$. \qed

\begin{lem}\label{Lem: TechE8}
Let $\mathcal{D}_2$ and $\mathcal{D}_3$ be given in the proof of Lemma \ref{Lem: TechE2}.\\
(i) \hspace{-0.05cm}Assume (i) $N\to\infty$; (ii) $T\geq K+1$; (iii) $J\to\infty$ with $J^{2}\xi^{2}_{J}\log J=o(N)$. Under Assumptions \ref{Ass: Basis}-\ref{Ass: Improvedrates}, \ref{Ass: SpecTest}(i), and (ii), $\|\mathcal{D}_2\hat{B}\|^{2}_{F}=O_{p}(1/N^{2})$.\\
(ii) \hspace{-0.05cm}Under Assumptions \ref{Ass: Basis}(i), \ref{Ass: LoadingsFactors}(ii), (iv), \ref{Ass: DGP}, \ref{Ass: SpecTest}(i), (ii), $\|\mathcal{D}_3\|^{2}_{F}\hspace{-0.05cm}=\hspace{-0.05cm}O_{p}(J^{-2\kappa}/N)$.
\end{lem}
\noindent{\sc Proof:} (i) By Assumptions \ref{Ass: DGP}, \ref{Ass: SpecTest}(i), and (ii), we may follow a similar argument as in the proof of Lemma \ref{Lem: TechA3}(ii) to obtain $\|\vec{E}\|^{2}_{F}/T = O_{p}(1/N)$. Since $\|\mathcal{D}_2\hat{B}\|_{F}\leq \|\hat{B}^{\prime}\tilde{E}\|_{F}\|\vec{E}\|_{F}/T$, the result then follows from Lemmas \ref{Lem: TechB2}(i).

(ii) Note that $\|\vec{E}\|^{2}_{F}/T = O_{p}(1/N)$ from the proof of (i). Since $\|\mathcal{D}_3\|_{F}\leq \|\tilde{\Delta}\|_{F}\|\vec{E}\|_{F}/T$, the result then immediately follows from Lemmas \ref{Lem: TechA3}(i).\qed

\begin{lem}\label{Lem: TechE9}
Suppose Assumptions \ref{Ass: Basis}-\ref{Ass: DGP}, \ref{Ass: Improvedrates}(iii), (iv), \ref{Ass: Asym}, \ref{Ass: SpecTest}(i)-(iii) hold. Let $V$ be given in the proof of Theorem \ref{Thm: Rate}, $\mathcal{D}_1$ and $\vec{E}$ be given in the proof of Lemma \ref{Lem: TechE2}. Assume (i) $N\to\infty$; (ii) $T\geq K+1$; (iii) $J\to\infty$ with $J=o(\sqrt{N})$. Then there exists an $M\times (K+1)$ random matrix $\mathbb{N}_z$ with $\mathrm{vec}(\mathbb{N}_z)\sim N(0,\Omega_z)$ such that
\begin{align*}
\|\sqrt{NT}\mathcal{D}_1\hat{B}V^{-1}-\mathbb{G}_\Gamma\|_{F} = O_{p}\left(\frac{1}{J^{\kappa}}+\frac{1}{N^{1/6}}\right)
\end{align*}
and
\begin{align*}
\|\sqrt{N/T}\vec{E}1_{T}-\mathbb{N}_{z,1}\| = O_{p}\left(\frac{1}{N^{1/6}}\right),
\end{align*}
where $\Omega_z$ is given in Lemma \ref{Lem: TechE2}, $\mathbb{G}_\Gamma = \mathbb{N}_{z,2}B^{\prime}B\mathcal{M}$, $\mathcal{M}$ is a nonrandom matrix in Lemma \ref{Lem: TechC3}, and $\mathbb{N}_{z,1}$ and $\mathbb{N}_{z,2}$ are first column and the last $K$ columns of $\mathbb{N}_z$.
\end{lem}
\noindent{\sc Proof:} Let $\mathcal{L}_{NT,z}\equiv \sum_{t=1}^{T}Q_{t,z}^{-1}Z_{t}^{\prime}\varepsilon_t (f_{t}-\bar{f})^{\prime}/\sqrt{NT}$ and $\ell_{NT,z}\equiv \sum_{t=1}^{T}Q_{t,z}^{-1}Z_{t}^{\prime}\varepsilon_t/$ $\sqrt{NT}$. By a similar argument as in the proof of Lemma \ref{Lem: TechC1},
\begin{align}\label{Eqn: Lem: TechE9: 1}
\|\sqrt{NT}\mathcal{D}_1\hat{B}V^{-1}-\mathcal{L}_{NT,z}B^{\prime}B\mathcal{M}\|_{F} = O_{p}\left(\frac{1}{J^{\kappa}}+\frac{1}{N^{1/4}}\right)
\end{align}
and
\begin{align}\label{Eqn: Lem: TechE9: 2}
\|\sqrt{N/T}\vec{E}1_{T}-\ell_{NT,z}\| = O_{p}\left(\frac{1}{N^{1/4}}\right).
\end{align}
By a similar argument as in the proof of Lemma \ref{Lem: TechC2}, there exists an $M\times (K+1)$ random matrix $\mathbb{N}_z$ with $\mathrm{vec}(\mathbb{N}_z)\sim N(0,\Omega_z)$ such that
\begin{align}\label{Eqn: Lem: TechE9: 3}
\|(\ell_{NT,z},\mathcal{L}_{NT,z})-\mathbb{N}_z\|_{F} = O_{p}\left(\frac{1}{N^{1/6}}\right).
\end{align}
Thus the result of the lemma follows from \eqref{Eqn: Lem: TechE9: 1}-\eqref{Eqn: Lem: TechE9: 3}.\qed

\begin{lem}\label{Lem: TechE10}
Let $\mathcal{D}^{\ast}_2$ and $\mathcal{D}^{\ast}_3$ be given in the proof of Lemma \ref{Lem: TechE4}.\\
(i) Assume (i) $N\to\infty$; (ii) $T\geq K+1$; (iii) $J\to\infty$ with $J^{2}\xi^{2}_{J}\log J=o(N)$. Under Assumptions \ref{Ass: Basis}-\ref{Ass: Improvedrates}, \ref{Ass: Boot}(i), \ref{Ass: SpecTest}(ii), and (iv), $\|\mathcal{D}^{\ast}_2\hat{B}\|^{2}_{F}=O_{p}(1/N^{2})$.\\
(ii) Under Assumptions \ref{Ass: Basis}(i), \ref{Ass: LoadingsFactors}(ii), (iv), \ref{Ass: DGP}, \ref{Ass: Boot}(i), \ref{Ass: SpecTest}(ii), and (iv), $\|\mathcal{D}^{\ast}_3\|^{2}_{F}=O_{p}(J^{-2\kappa}/N)$.
\end{lem}
\noindent{\sc Proof:} (i) By Assumptions \ref{Ass: DGP}, \ref{Ass: Boot}(i), \ref{Ass: SpecTest} (ii), and (iv), we may follow a similar argument as in the proof of Lemma \ref{Lem: TechD4}(ii) to obtain $\|\vec{E}^{\ast}\|^{2}_{F}/T = O_{p}(1/N)$. Since $\|\mathcal{D}^{\ast}_2\hat{B}\|_{F}\leq \|\hat{B}^{\prime}\tilde{E}\|_{F}\|\vec{E}^{\ast}\|_{F}/T$, the result then follows from Lemmas \ref{Lem: TechB2}(i).

(ii) Note that $\|\vec{E}^{\ast}\|^{2}_{F}/T = O_{p}(1/N)$ from the proof of (i). Since $\|\mathcal{D^{\ast}}_3\|_{F}\leq \|\tilde{\Delta}\|_{F}\|\vec{E}^{\ast}\|_{F}/T$, the result then immediately follows from Lemmas \ref{Lem: TechA3}(i).\qed

\begin{lem}\label{Lem: TechE11}
Suppose Assumptions \ref{Ass: Basis}-\ref{Ass: DGP}, \ref{Ass: Improvedrates}(iii), (iv), \ref{Ass: Asym}, \ref{Ass: Boot}(i), and \ref{Ass: SpecTest}(ii)-(iv) hold. Let $V$ be given in the proof of Theorem \ref{Thm: Rate}, $\mathcal{D}_1$ and $\vec{E}$ be given in the proof of Lemma \ref{Lem: TechE2}, and $\mathcal{D}^{\ast}_1$ and $\vec{E}^{\ast}$ be given in the proof of Lemma \ref{Lem: TechE4}. Assume (i) $N\to\infty$; (ii) $T\geq K+1$; (iii) $J\to\infty$ with $J=o(\sqrt{N})$. Then there exists an $M\times (K+1)$ random matrix $\mathbb{N}^{\ast}_z$ with $\mathrm{vec}(\mathbb{N}^{\ast}_z)\sim N(0,\Omega_z)$ conditional on $\{Y_{t},Z_{t}\}_{t\leq T}$ such that
\begin{align*}
\|\sqrt{NT}(\mathcal{D}^{\ast}_1-\mathcal{D}_1)\hat{B}V^{-1}-\sqrt{\omega_0}\mathbb{G}^{\ast}_\Gamma\|_{F} = O_{p}\left(\frac{1}{J^{\kappa}}+\frac{1}{N^{1/6}}\right)
\end{align*}
and
\begin{align*}
\|\sqrt{N/T}(\vec{E}^{\ast}1_{T}-\vec{E}1_{T})-\sqrt{\omega_0}\mathbb{N}^{\ast}_{z,1}\| = O_{p}\left(\frac{1}{N^{1/6}}\right),
\end{align*}
where $\Omega_z$ is given in Lemma \ref{Lem: TechE2}, $\mathbb{G}^{\ast}_\Gamma = \mathbb{N}^{\ast}_{z,2}B^{\prime}B\mathcal{M}$, $\mathcal{M}$ is a nonrandom matrix in Lemma \ref{Lem: TechC3}, and $\mathbb{N}^{\ast}_{z,1}$ and $\mathbb{N}^{\ast}_{z,2}$ are first column and the last $K$ columns of $\mathbb{N}^{\ast}_z$.
\end{lem}
\noindent{\sc Proof:} Let $\mathcal{L}^{\ast\ast}_{NT,z}\equiv \sum_{t=1}^{T}Q_{t,z}^{-1}Z_{t}^{\ast\prime}\varepsilon_t(f_{t}-\bar{f})^{\prime}/\sqrt{NT}$ and $\ell^{\ast\ast}_{NT,z}\equiv \sum_{t=1}^{T}Q_{t,z}^{-1}Z_{t}^{\ast\prime}\varepsilon_t/$ $\sqrt{NT}$. By a similar argument as in the proof of Lemma
\ref{Lem: TechD2},
\begin{align}\label{Eqn: Lem: TechE11: 1}
\|\sqrt{NT}\mathcal{D}^{\ast}_1\hat{B}V^{-1}-\mathcal{L}^{\ast\ast}_{NT,z}B^{\prime}B\mathcal{M}\|_{F} = O_{p}\left(\frac{1}{J^{\kappa}}+\frac{1}{N^{1/4}}\right)
\end{align}
and
\begin{align}\label{Eqn: Lem: TechE11: 2}
\|\sqrt{N/T}\vec{E}^{\ast}1_{T}-\ell^{\ast\ast}_{NT,z}\| = O_{p}\left(\frac{1}{N^{1/4}}\right).
\end{align}
Let $\mathcal{L}^{\ast}_{NT,z}\equiv \sum_{t=1}^{T}Q_{t,z}^{-1}(Z_{t}^{\ast}-Z_{t})^{\prime}\varepsilon_t (f_{t}-\bar{f})^{\prime}/\sqrt{NT} = \mathcal{L}^{\ast\ast}_{NT,z}-\mathcal{L}_{NT,z}$ and $\ell^{\ast}_{NT,z}\equiv \sum_{t=1}^{T}Q_{t,z}^{-1}(Z_{t}^{\ast}-Z_{t})^{\prime}\varepsilon_t/\sqrt{NT}=\ell^{\ast\ast}_{NT,z}-\ell_{NT,z}$. By a similar argument as in the proof of Lemma \ref{Lem: TechD3}, there exists an $M\times (K+1)$ random matrix $\mathbb{N}^{\ast}_z$ with $\mathrm{vec}(\mathbb{N}^{\ast}_z)\sim N(0,\Omega_z)$ conditional on $\{Y_{t},Z_{t}\}_{t\leq T}$ such that
\begin{align}\label{Eqn: Lem: TechE11: 3}
\|(\ell^{\ast}_{NT,z},\mathcal{L}^{\ast}_{NT,z})-\sqrt{\omega_0}\mathbb{N}^{\ast}_z\|_{F} = O_{p}\left(\frac{1}{N^{1/6}}\right).
\end{align}
Thus, the result of the lemma follows from \eqref{Eqn: Lem: TechE9: 1},\eqref{Eqn: Lem: TechE9: 2}, and \eqref{Eqn: Lem: TechE11: 1}-\eqref{Eqn: Lem: TechE11: 3}.\qed

\begin{lem}\label{Lem: TechE12}
Suppose Assumptions \ref{Ass: Improvedrates}(iii), \ref{Ass: SpecTest}(ii), and (v) hold. For any given positive constant $C$,
\[\sup_{\|\Pi\|_{F}\leq C}\left|\frac{1}{NT}\sum_{i=1}^{N}\sum_{t=1}^{T}\|\beta(z_{it})- \Pi^{\prime}z_{it}\|^2 - \frac{1}{NT}\sum_{i=1}^{N}\sum_{t=1}^{T}E\left[\|\beta(z_{it})- \Pi^{\prime}z_{it}\|^2\right]\right|=o_{p}(1).\]
\end{lem}
\noindent{\sc Proof:} Let $\mathcal{A}_{C}\equiv \{\Pi\in\mathbf{R}^{M\times K}, \|\Pi\|_{F}\leq C\}$ for $C>0$, and $\mathcal{F}_{C}\equiv\{\zeta(\cdot,\Pi): \zeta(z_1,\cdots,z_T,\Pi)=\sum_{t=1}^{T}\|\beta(z_{t})- \Pi^{\prime}z_{t}\|^2/T \text{ for } \Pi\in\mathcal{A}_{C}\}$ be a class of functions $\zeta(\cdot,\Pi)$ indexed by $\Pi\in\mathcal{A}_{C}$. We aim to show $\sup_{\Pi\in\mathcal{A}_{C}}|\frac{1}{N}\sum_{i=1}^{N}\zeta(z_{i1},\cdots,z_{iT},\Pi) - \frac{1}{N}\sum_{i=1}^{N}E[\zeta(z_{i1},\cdots,z_{iT},\Pi)]|=o_{p}(1)$. It follows that for any $\Pi_1,\Pi_2\in\mathcal{A}_{C}$,
\begin{align}\label{Eqn: Lem: TechE12: 1}
&\hspace{0.8cm}|\zeta(z_1,\cdots,z_T,\Pi_1)-\zeta(z_1,\cdots,z_T,\Pi_2)|\notag\\
&\leq \|\Pi_1 - \Pi_2\|_{F}\frac{1}{T}\sum_{t=1}^{T}\|z_{t}\|(\|\beta(z_{t})- \Pi_1^{\prime}z_{t}\| + \|\beta(z_{t})- \Pi_2^{\prime}z_{t}\|)\notag\\
&\leq \|\Pi_1 - \Pi_2\|_{F}\frac{2}{T}\sum_{t=1}^{T}(\|z_{t}\|\|\beta(z_{t})\| + C\|z_{t}\|^{2}) \equiv \|\Pi_1 - \Pi_2\|_{F}G(z_1,\cdots,z_T).
\end{align}
By Assumptions \ref{Ass: SpecTest}(ii) and (v), $\max_{i\leq N}E[G(z_{i1},\cdots,z_{iT})]<\infty$. This together with \eqref{Eqn: Lem: TechE12: 1} implies that  and $\mathcal{F}_{C}$ is a class of functions that are Lipschitz in the index $\Pi\in\mathcal{A}_{C}$ with envelop function $G$. Since $\mathcal{A}_{C}$ is compact, for every $\epsilon>0$, the covering number $N(\epsilon,\mathcal{A}_{C},\|\cdot\|_{F})$ of $\mathcal{A}_{C}$ with respect to $\|\cdot\|_{F}$ is bounded. By Theorem 2.7.11 of \citet{VW_WeakConvergence_1996}, for every $\epsilon>0$, the bracketing number $N_{[]}(\epsilon,\mathcal{F}_{C},L_1(P))$ of $\mathcal{F}_{C}$ with respect to $L_1(P)$ is bounded. Thus, the result of the lemma follows by the Glivenko-Cantelli theorem (e.g., Theorem 2.4.1 of \citet{VW_WeakConvergence_1996}). \qed

\subsection{Proof of Theorem \ref{Thm: Numberfactors}}
\noindent{\sc Proof of Theorem \ref{Thm: Numberfactors}:} (A) Let $\theta_{k}\equiv {\lambda_{k}(\tilde{Y}M_T\tilde{Y}^{\prime}/T)}/{\lambda_{k+1}(\tilde{Y}M_T\tilde{Y}^{\prime}/T)}$. If $\hat{K}\neq K$, then there exists some $1\leq k\leq K-1$ or $K+1\leq k\leq JM/2$ such that $\theta_{k}\geq \theta_{K}$. Let $\underline{JM/2}$ be the integer part of $JM/2$. Since ${\lambda_{1}(\tilde{Y}M_T\tilde{Y}^{\prime}/T)}/{\lambda_{K}(\tilde{Y}M_T\tilde{Y}^{\prime}/T)}\geq \theta_{k}$ for all $1\leq k\leq K-1$ and ${\lambda_{K+1}(\tilde{Y}M_T\tilde{Y}^{\prime}/T)}/{\lambda_{\underline{JM/2}}(\tilde{Y}M_T\tilde{Y}^{\prime}/T)}\geq \theta_{k}$ for all $K+1\leq k\leq JM/2$, the event of $\hat{K}\hspace{-0.05cm}\neq\hspace{-0.05cm}K$ implies the event of ${\lambda_{1}(\tilde{Y}M_T\tilde{Y}^{\prime}/T)}/{\lambda_{K}(\tilde{Y}M_T\tilde{Y}^{\prime}/T)}$ $\geq \theta_{K}$ or the event of ${\lambda_{K+1}(\tilde{Y}M_T\tilde{Y}^{\prime}/T)}/{\lambda_{\underline{JM/2}}(\tilde{Y}M_T\tilde{Y}^{\prime}/T)}\geq \theta_{K}$. Thus,
\begin{align}\label{Eqn: Thm: Numberfactors: 1}
P(\hat{K}\neq K)\leq P\left(\frac{\lambda_{1}(\tilde{Y}M_T\tilde{Y}^{\prime}/T)}{\lambda_{K}(\tilde{Y}M_T\tilde{Y}^{\prime}/T)}\geq \theta_{K}\right) + P\left(\frac{\lambda_{K+1}(\tilde{Y}M_T\tilde{Y}^{\prime}/T)}{\lambda_{\underline{JM/2}}(\tilde{Y}M_T\tilde{Y}^{\prime}/T)}\geq \theta_{K}\right).
\end{align}
By Lemmas \ref{Lem: TechF1} and \ref{Lem: TechF2}, ${\lambda_{1}(\tilde{Y}M_T\tilde{Y}^{\prime}/T)}/{\lambda_{K}(\tilde{Y}M_T\tilde{Y}^{\prime}/T)}=O_{p}(1)$, $\theta_{K}/N=C+o_{p}(1)$ for some positive constant $C$, and ${\lambda_{K+1}(\tilde{Y}M_T\tilde{Y}^{\prime}/T)}/{\lambda_{\underline{JM/2}}(\tilde{Y}M_T\tilde{Y}^{\prime}/T)}=O_{p}(1)$, since $\underline{JM/2}+1<JM-K-1$ for large $J$. Thus, $P(\hat{K}\neq K)\to 0$.

(B) If $\tilde{K}\hspace{-0.05cm}\neq\hspace{-0.05cm}K$, then $\lambda_{K-1}(\tilde{Y}M_T\tilde{Y}^{\prime}/T)\hspace{-0.05cm}<\hspace{-0.05cm}\lambda_{NT}$ or $\lambda_{K+1}(\tilde{Y}M_T\tilde{Y}^{\prime}/T)\hspace{-0.05cm}\geq\hspace{-0.05cm}\lambda_{NT}$. Thus,
\begin{align}\label{Eqn: Thm: NumberfactorsAlt: 1}
\hspace{-0.4cm}P(\tilde{K}\neq K)\leq P\left(\lambda_{K-1}(\tilde{Y}M_T\tilde{Y}^{\prime}/T)< \lambda_{NT}\right) + P\left(\lambda_{K+1}(\tilde{Y}M_T\tilde{Y}^{\prime}/T)\geq \lambda_{NT}\right).
\end{align}
By Lemma \ref{Lem: TechF1} and $\lambda_{NT}\to0$, $P(\lambda_{K-1}(\tilde{Y}M_T\tilde{Y}^{\prime}/T)< \lambda_{NT})\to 0$. For a matrix $A$, let $\sigma_{k}(A)$ denote the $k$th largest singular value of $A$. Since $\lambda_{k}(AA^{\prime})=\sigma^{2}_{k}(A)$,
\begin{align}\label{Eqn: Thm: NumberfactorsAlt: 2}
\lambda_{K+1}(\tilde{Y}M_T\tilde{Y}^{\prime}/T)&=\sigma^{2}_{K+1}(\tilde{Y}M_T/\sqrt{T})\hspace{-0.05cm}=\hspace{-0.05cm}|\sigma_{K+1}(\tilde{Y}M_T/\sqrt{T})\hspace{-0.05cm}-\hspace{-0.05cm}\sigma_{K+1}(BF^{\prime}M_T/\sqrt{T})|^2\notag\\
&\hspace{-2cm}\leq\frac{1}{T}\|\tilde{Y}M_T-B(M_TF)^{\prime}\|_{F}^{2}\leq \frac{2}{T}\|\tilde{\Delta}\|_{F}^{2} + \frac{2}{T}\|\tilde{E}\|_{F}^{2} = O_{p}\left(\frac{1}{J^{2\kappa}} + \frac{J}{N}\right),
\end{align}
where the second equality follows since the rank of $B(M_TF)^{\prime}$ is not greater than $K$, the first inequality follows by the Weyl's inequality, the second inequality by \eqref{Eqn: Thm: Rate: 2} and the Cauchy-Schwartz inequality, and the last equality follows from Lemmas \ref{Lem: TechA3}(i) and (ii). Since $\lambda_{NT}\min\{N/J,\hspace{-0.05cm}J^{2\kappa}\}\hspace{-0.05cm}\to \hspace{-0.05cm}\infty$,  \eqref{Eqn: Thm: NumberfactorsAlt: 2} implies $P(\lambda_{K+1}(\tilde{Y}M_T\tilde{Y}^{\prime}/T)\hspace{-0.05cm}\geq \lambda_{NT})\to 0$. This completes the proof of the theorem. \qed

\subsubsection{Technical Lemmas}
\begin{lem}\label{Lem: TechF1}
Suppose Assumptions \ref{Ass: Basis}-\ref{Ass: DGP} hold.  Assume (i) $N\to\infty$; (ii) $T\geq K+1$; (iii) $J\to\infty$ with $J=o(\sqrt{N})$. Then there exist positive constants $c_1$ and $c_2$ such that
\[c_1+o_{p}(1)\leq\lambda_{K}({\tilde{Y}M_T\tilde{Y}^{\prime}}/{T})\leq \lambda_{1}({\tilde{Y}M_T\tilde{Y}^{\prime}}/{T})\leq c_2+o_{p}(1). \]
\end{lem}
\noindent{\sc Proof:} By \eqref{Eqn: Lem: TechA2: 2}, $\lambda_{k}({\tilde{Y}M_T\tilde{Y}^{\prime}}/{T})=\lambda_{k}(({F^{\prime}M_TF}/{T})B^{\prime}B)+o_{p}(1)$ for $k=1,\ldots,K$. Thus, the result immediately follows from Assumptions \ref{Ass: LoadingsFactors}(i)-(iii). \qed

\begin{lem}\label{Lem: TechF2}
Suppose  Assumptions \ref{Ass: Basis}(i), \ref{Ass: LoadingsFactors}(ii), (iv), \ref{Ass: DGP}(i), \ref{Ass: Improvedrates}(i), and \ref{Ass: Numberfactors} hold. Assume (i) $N\to\infty$; (ii) $T\to\infty$; (iii) $J\to\infty$ with $J=o(\min\{\sqrt{N},\sqrt{T}\})$ and $J^{-2\kappa}N=o(1)$. Then there exist positive constants $c_3$ and $c_4$ such that
\[c_3+o_{p}(1)\leq N \lambda_{JM-K-1}({\tilde{Y}M_T\tilde{Y}^{\prime}}/{T})\leq N \lambda_{K+1}({\tilde{Y}M_T\tilde{Y}^{\prime}}/{T}) \leq c_4+o_{p}(1).\]
\end{lem}
\noindent{\sc Proof:} For a matrix $A$, let $\sigma_{k}(A)$ denote the $k$th largest singular value of $A$. Noting that $\lambda_{k}(AA^{\prime})=\sigma^{2}_{k}(A)$, it follows that for $k=1,\ldots,JM-K$,
\begin{align}\label{Eqn: TechF2: 1}
&\hspace{0.5cm}|\lambda_{K+k}(\tilde{Y}M_T\tilde{Y}^{\prime})-\lambda_{K+k}((BF^{\prime}+\tilde{E})M_T(BF^{\prime}+\tilde{E})^{\prime})|\notag\\
&\leq|\sigma_{K+k}(\tilde{Y}M_T)-\sigma_{K+k}((BF^{\prime}+\tilde{E})M_T)|^2+2|\sigma_{K+k}(\tilde{Y}M_T)\notag\\
&\hspace{0.5cm}-\sigma_{K+k}((BF^{\prime}+\tilde{E})M_T)|\sigma_{K+k}((BF^{\prime}+\tilde{E})M_T)\notag\\
&\leq\|\tilde{Y}M_T-(BF^{\prime}+\tilde{E})M_T\|^{2}_{F}+2\|\tilde{Y}M_T-(BF^{\prime}+\tilde{E})M_T\|_{F}\notag\\
&\hspace{0.5cm} \times \lambda^{1/2}_{K+k}((BF^{\prime}+\tilde{E})M_T(BF^{\prime}+\tilde{E})^{\prime})\notag\\
& \leq\|\tilde{\Delta}\|^{2}_{F}+2\|\tilde{\Delta}\|_{F}\lambda^{1/2}_{K+1}((BF^{\prime}+\tilde{E})M_{T}(BF^{\prime}+\tilde{E})^{\prime}),
\end{align}
where the first inequality is due to the triangle inequality, the second inequality follows by the Weyl's inequality, and the third inequality follows from \eqref{Eqn: Thm: Rate: 2} and the fact that $\lambda_{K+k}((BF^{\prime}+\tilde{E})M_T(BF^{\prime}+\tilde{E})^{\prime})\leq \lambda_{K+1}((BF^{\prime}+\tilde{E})M_T(BF^{\prime}+\tilde{E})^{\prime})$ for $k\geq 1$. We next show that the right-hand side of \eqref{Eqn: TechF2: 1} is asymptotically negligible and study the behavior of $\lambda_{K+k}((BF^{\prime}+\tilde{E})M_T(BF^{\prime}+\tilde{E})^{\prime})$. Let $\tilde{B}=B+\tilde{E}M_TF(F^{\prime}M_TF)^{-1}$ and $M_{F} = I_{T}-M_TF(F^{\prime}M_TF)^{-1}(M_TF)^{\prime}$. We may decompose $(BF^{\prime}+\tilde{E})M_T(BF^{\prime}+\tilde{E})^{\prime}$ by
\begin{align}\label{Eqn: TechF2: 2}
(BF^{\prime}+\tilde{E})M_T(BF^{\prime}+\tilde{E})^{\prime} = \tilde{B}F^{\prime}M_TF\tilde{B}^{\prime}+\tilde{E}M_TM_{F}M_T\tilde{E}^{\prime}.
\end{align}
Then, \eqref{Eqn: TechF2: 2} implies that for $k=1,\ldots,JM-K$,
\begin{align}\label{Eqn: TechF2: 3}
&\lambda_{K+k}((BF^{\prime}+\tilde{E})M_T(BF^{\prime}+\tilde{E})^{\prime})\leq \lambda_{K+1}(\tilde{B}F^{\prime}M_TF\tilde{B}^{\prime})\notag\\
&\hspace{2cm}+\lambda_{k}(\tilde{E}M_TM_{F}M_T\tilde{E}^{\prime})\leq \lambda_{k}(\tilde{E}M_T\tilde{E}^{\prime})\leq \lambda_{k}(\tilde{E}\tilde{E}^{\prime}),
\end{align}
where the first inequality follows by Lemma \ref{Lem: TechF3}(i), the second inequality follows by Lemma \ref{Lem: TechF3}(ii) as well as the fact that the rank of $\tilde{B}F^{\prime}M_TF\tilde{B}^{\prime}$ is not greater than $K$ and $I-M_F$ is positive semi-definite, and the third inequality follows since $I-M_T$ is positive semi-definite. Moreover, \eqref{Eqn: TechF2: 2} also implies that for $k=1,\ldots,JM-2K-1$,
\begin{align}\label{Eqn: TechF2: 4}
&\lambda_{K+k}((BF^{\prime}+\tilde{E})M_T(BF^{\prime}+\tilde{E})^{\prime})\geq\lambda_{K+k}(\tilde{E}M_TM_{F}M_T\tilde{E}^{\prime})\notag\\
&\hspace{0.5cm}=\lambda_{K+k}(\tilde{EM_T}M_{F}M_T\tilde{E}^{\prime})+\lambda_{K+1}(\tilde{E}M_T(I-M_{F})M_T\tilde{E}^{\prime})\geq\lambda_{2K+k}(\tilde{E}M_T\tilde{E}^{\prime})\notag\\
&\hspace{0.5cm}=\lambda_{2K+k}(\tilde{E}M_T\tilde{E}^{\prime})+\lambda_{2}(\tilde{E}(I_T-M_T)\tilde{E}^{\prime})\geq \lambda_{2K+k+1}(\tilde{E}\tilde{E}^{\prime}),
\end{align}
where the first inequality follows by Lemma \ref{Lem: TechF3}(ii), the first equality follows since the rank of $\tilde{E}M_T(I-M_{F})M_T\tilde{E}^{\prime}$ is not greater than $K$, the second inequality follows by Lemma \ref{Lem: TechF3}(i), and the second equality and the third inequality follow similarly. Putting \eqref{Eqn: TechF2: 3} and \eqref{Eqn: TechF2: 4} together implies that eigenvalues of $(BF^{\prime}+\tilde{E})M_{T}(BF^{\prime}+\tilde{E})^{\prime}$ are bounded by those of $\tilde{E}\tilde{E}^{\prime}$. Thus, we  may study the behavior of the eigenvalues of $\tilde{E}\tilde{E}^{\prime}$. Recall that $\mathcal{A}_{NT}= \sum_{t=1}^{T}\hat{Q}_{t}^{-1}\Phi(Z_{t})^{\prime}E[\varepsilon_{t}\varepsilon_{t}^{\prime}]\Phi(Z_{t})\hat{Q}_{t}^{-1}/NT$ in Lemma \ref{Lem: TechF4}. By the Weyl's inequality and Lemma \ref{Lem: TechF4},
\begin{align}\label{Eqn: TechF2: 5}
\sup_{k\leq JM}|\lambda_{k}({N}\tilde{E}\tilde{E}^{\prime}/T)-\lambda_{k}(\mathcal{A}_{NT})|\leq \|{N}\tilde{E}\tilde{E}^{\prime}/{T}-\mathcal{A}_{NT}\|_{F}=o_{p}(1).
\end{align}
This implies that the eigenvalues of $N\tilde{E}\tilde{E}^{\prime}/T$ and $\mathcal{A}_{NT}$ are asymptotically equivalent. Then, it follows from \eqref{Eqn: TechF2: 3} and \eqref{Eqn: TechF2: 5} that
\begin{align}\label{Eqn: TechF2: 6}
&\lambda_{K+1}({N}(BF^{\prime}+\tilde{E})M_{T}(BF^{\prime}+\tilde{E})^{\prime}/{T})\notag\\
&\hspace{2cm}\leq \lambda_{1}({N}\tilde{E}\tilde{E}^{\prime}/{T})\leq \lambda_{1}\left(\mathcal{A}_{NT}\right)+o_{p}(1)=O_{p}(1),
\end{align}
because $\lambda_{1}(\mathcal{A}_{NT})\leq (\min_{t\leq T} \lambda_{\min}(\hat{Q}_t))^{-1}\max_{t\leq T}\lambda_{\max}(E[\varepsilon_{t}\varepsilon_{t}^{\prime}])=O_{p}(1)$ by Assumptions \ref{Ass: Basis}(i) and \ref{Ass: Numberfactors} (i). Combining \eqref{Eqn: TechF2: 1}, \eqref{Eqn: TechF2: 6}, and Lemma \ref{Lem: TechA3}(i) yields
\begin{align}\label{Eqn: TechF2: 7}
\hspace{-0.2cm}\sup_{k\leq JM-K}|N\lambda_{K+k}({\tilde{Y}M_T\tilde{Y}^{\prime}}/{T})-N\lambda_{K+k}({(BF^{\prime}+\tilde{E})M_T(BF^{\prime}+\tilde{E})^{\prime}}/{T})|=o_p(1).
\end{align}
This means that $N\lambda_{K+k}({\tilde{Y}M_T\tilde{Y}^{\prime}}/{T})$ and $ N\lambda_{K+k}({(BF^{\prime}+\tilde{E})M_T(BF^{\prime}+\tilde{E})^{\prime}}/{T})$ are asymptotically equivalent. By the triangle inequality, it follows from \eqref{Eqn: TechF2: 3}-\eqref{Eqn: TechF2: 5} and \eqref{Eqn: TechF2: 7} that
\begin{align}\label{Eqn: TechF2: 8}
&\lambda_{JM}(\mathcal{A}_{NT}) + o_{p}(1)\leq N\lambda_{JM-K-1}({\tilde{Y}M_T\tilde{Y}^{\prime}}/{T})\notag\\
&\hspace{2cm}\leq N\lambda_{K+1}({\tilde{Y}M_T\tilde{Y}^{\prime}}/{T})\leq \lambda_{1}(\mathcal{A}_{NT}) + o_{p}(1).
\end{align}
Because $\lambda_{1}(\mathcal{A}_{NT})\leq (\min_{t\leq T} \lambda_{\min}(\hat{Q}_t))^{-1}\max_{t\leq T}\lambda_{\max}(E[\varepsilon_{t}\varepsilon_{t}^{\prime}])$ and $\lambda_{JM}(\mathcal{A}_{NT})\geq $ $(\max_{t\leq T} \lambda_{\max}(\hat{Q}_t))^{-1}\min_{t\leq T}\lambda_{\min}(E[\varepsilon_{t}\varepsilon_{t}^{\prime}])$, the result of the lemma then follows from \eqref{Eqn: TechF2: 8} along with Assumptions \ref{Ass: Basis}(i) and \ref{Ass: Numberfactors}(i).\qed

\begin{lem}[Weyl's inequalities]\label{Lem: TechF3}
Let $C$ and $D$ be $k\times k$ symmetric matrices.\\
(i) For every $i,j\geq 1$ and $i+j-1\leq k$,
\[\lambda_{i+j-1}(C+D)\leq \lambda_{i}(C)+\lambda_{j}(D).\]
(ii) If $D$ is positive semi-definite, for all $1\leq i\leq k$,
\[\lambda_{i}(C+D)\geq \lambda_{i}(C).\]
\end{lem}
\noindent{\sc Proof:} The results can be found in Section III.2 of \citet{Bhatia_MatrixAnalysis_1997}. Also, see the appendices of \citet{AhnHorenstein_EigenvalueRatio_2013} and \citet{Fanetal_ProjectedPCASupp_2016}.\qed

\begin{lem}\label{Lem: TechF4}
Let $\mathcal{A}_{NT}\hspace{-0.1cm}\equiv \hspace{-0.1cm}\sum_{t=1}^{T}\hat{Q}_{t}^{-1}\Phi(Z_{t})^{\prime}E[\varepsilon_{t}\varepsilon_{t}^{\prime}]\Phi(Z_{t})\hat{Q}_{t}^{-1}/NT$ and $\tilde{E}$ be given in the proof of Theorem \ref{Thm: Rate}. Under Assumptions \ref{Ass: Basis}(i), \ref{Ass: DGP}(i), \ref{Ass: Improvedrates}(i), and \ref{Ass: Numberfactors}(ii),
\[\|N\tilde{E}\tilde{E}^{\prime}/{T}-\mathcal{A}_{NT}\|^{2}_{F}=O_{p}\left(\frac{J^{2}}{N}+\frac{J^{2}}{T}\right).\]
\end{lem}
\noindent{\sc Proof:} Let $E_{\varepsilon}$ denote the expectation with respect to $\{\varepsilon_{t}\}_{t\leq T}$. To simplify the notation, let $\hat{\psi}_{it}\equiv \phi(z_{it})\hat{Q}_{t}^{-1}$ and $\nu_{ijt}\equiv\varepsilon_{it}\varepsilon_{jt}-E[\varepsilon_{it}\varepsilon_{jt}]$. Since $\|A\|_{F}^2=\mathrm{tr}(AA^{\prime})$,
\begin{align}\label{Eqn: Lem: TechF4: 1}
&E_{\varepsilon}[\|\bar{E}\bar{E}^{\prime}/NT-\mathcal{A}_{NT}\|^{2}_{F}]\hspace{-0.05cm}=\hspace{-0.05cm} \frac{1}{N^{2}T^{2}}E_{\varepsilon}\hspace{-0.05cm}\left[\mathrm{tr}\hspace{-0.1cm}\left(\sum_{t=1}^{T}\sum_{s=1}^{T}\sum_{i=1}^{N}\sum_{j= 1}^{N}\sum_{k=1}^{N}\sum_{\ell= 1}^{N}\hat{\psi}_{it}\hat{\psi}_{jt}^{\prime}\nu_{ijt}\nu_{k\ell s}\hat{\psi}_{\ell s}\hat{\psi}_{ks}^{\prime}\right)\right]\notag\\
&=\frac{1}{N^{2}T^{2}}\sum_{t=1}^{T}\sum_{s=1}^{T}\sum_{i=1}^{N}\sum_{j= 1}^{N}\sum_{k=1}^{N}\sum_{\ell= 1}^{N}\hat{\psi}_{it}^{\prime}\hat{\psi}_{ks}\hat{\psi}_{jt}^{\prime}\hat{\psi}_{\ell s}cov(\varepsilon_{it}\varepsilon_{jt},\varepsilon_{ks}\varepsilon_{\ell s})\notag\\
&=(\min_{t\leq T}\lambda_{\min}(\hat{Q}_{t}))^{-4}\frac{1}{N^{2}T^{2}}\hspace{-0.05cm}\sum_{t=1}^{T}\sum_{s=1}^{T}\sum_{i=1}^{N}\sum_{j= 1}^{N}\sum_{k=1}^{N}\sum_{\ell= 1}^{N}\hspace{-0.05cm}\|\phi(z_{it})\|\|\phi(z_{jt})\|\|\phi(z_{ks})\|\|\phi(z_{\ell s})\|\notag\\
&\hspace{0.5cm} \times |cov(\varepsilon_{it}\varepsilon_{jt},\varepsilon_{ks}\varepsilon_{\ell s})|,
\end{align}
where the second equality follows from the independence in Assumption \ref{Ass: DGP} (i) and the linearity of both expectation and trace operators, and the inequality follows since $\|\hat{\psi}_{it}\|\leq (\lambda_{\min}(\hat{Q}_{t}))^{-1}\|\phi(z_{it})\|$. Moreover,
\begin{align}\label{Eqn: Lem: TechF4: 2}
&\hspace{0.8cm}E\left[\sum_{t=1}^{T}\sum_{s=1}^{T}\sum_{i=1}^{N}\sum_{j= 1}^{N}\sum_{k=1}^{N}\sum_{\ell= 1}^{N}\hspace{-0.05cm}\|\phi(z_{it})\|\|\phi(z_{jt})\|\|\phi(z_{ks})\|\|\phi(z_{\ell s})\||cov(\varepsilon_{it}\varepsilon_{jt},\varepsilon_{ks}\varepsilon_{\ell s})|\right]\notag\\
&\leq \max_{i\leq N,t\leq T}E[\|\phi(z_{it})\|^{4}]\sum_{t=1}^{T}\sum_{s=1}^{T}\sum_{i=1}^{N}\sum_{j= 1}^{N}\sum_{k=1}^{N}\sum_{\ell= 1}^{N}|cov(\varepsilon_{it}\varepsilon_{jt},\varepsilon_{ks}\varepsilon_{\ell s})|\notag\\
&\leq J^{2}M^{2}\max_{\ell\leq JM, i\leq N, t\leq T} E[\phi^{4}(z_{it,m})]\sum_{t=1}^{T}\sum_{s=1}^{T}\sum_{i=1}^{N}\sum_{j= 1}^{N}\sum_{k=1}^{N}\sum_{\ell= 1}^{N}|cov(\varepsilon_{it}\varepsilon_{jt},\varepsilon_{ks}\varepsilon_{\ell s})|,
\end{align}
where the first inequality is due to the Cauchy-Schwartz inequality, and the second one follows since $\max_{i\leq N,t\leq T}E[\|\phi(z_{it})\|^{4}]=J^{2}M^{2}\max_{\ell\leq JM, i\leq N, t\leq T} E[\phi^{4}(z_{it,m})]$.
Combining \eqref{Eqn: Lem: TechF4: 1} and \eqref{Eqn: Lem: TechF4: 2} implies that $E_{\varepsilon}[\|\bar{E}\bar{E}^{\prime}/NT-\mathcal{A}_{NT}\|^{2}_{F}]=O_{p}({J^{2}}/{N}+{J^{2}}/{T})$ by Assumptions \ref{Ass: Basis}(i), \ref{Ass: Improvedrates}(i), and \ref{Ass: Numberfactors}(ii). Thus, the result of the lemma follows by the Markov's inequality and Lemma \ref{Lem: TechA5}. \qed

\section{Additional Discussions}\label{App: Sec: C}

\subsection{Regressed-PCA: A Special Case}\label{App: Sec: C1}
We consider the regressed-PCA approach when $N>JM$ and$\Phi(Z_t)'\Phi(Z_t)/N=I_{JM}$. There exists an $N\times (N-JM)$ matrix $\Psi_{t}$ such that $(\Phi(Z_t)/\sqrt{N}, \Psi_{t})$ is orthonormal, i.e., $(\Phi(Z_t)/\sqrt{N}, \Psi_{t})(\Phi(Z_t)/\sqrt{N}, \Psi_{t})^{\prime} = I_{N}$. It follows that
\begin{align}
&\hspace{0.5cm}({Y}_t - \Phi(Z_t)a -\Phi(Z_t)Bf_{t})^{\prime}  ({Y}_t - \Phi(Z_t)a -\Phi(Z_t) Bf_{t})\notag\\
&=({Y}_t \hspace{-0.05cm}-\hspace{-0.05cm} \Phi(Z_t)a \hspace{-0.05cm}-\hspace{-0.05cm}\Phi(Z_t)Bf_{t})^{\prime} (\Phi(Z_t)/\sqrt{N}, \Psi_{t})(\Phi(Z_t)/\sqrt{N}, \Psi_{t})^{\prime} ({Y}_t\hspace{-0.05cm} -\hspace{-0.05cm} \Phi(Z_t)a \hspace{-0.05cm}-\hspace{-0.05cm}\Phi(Z_t) Bf_{t})\notag\\
&= N[({Y}_t\hspace{-0.05cm} - \hspace{-0.05cm}\Phi(Z_t)a \hspace{-0.05cm}-\hspace{-0.05cm}\Phi(Z_t)Bf_{t})^{\prime} \Phi(Z_t)\Phi(Z_t)^{\prime} ({Y}_t \hspace{-0.05cm}-\hspace{-0.05cm} \Phi(Z_t)a\hspace{-0.05cm} -\hspace{-0.05cm}\Phi(Z_t) Bf_{t})/N^2] + {Y}_t^{\prime}\Psi_{t}\Psi_{t}^{\prime}{Y}_t\notag\\
& =N[({Y}_t - \Phi(Z_t)a -\Phi(Z_t)Bf_{t})^{\prime} S_t ({Y}_t - \Phi(Z_t)a -\Phi(Z_t) Bf_{t})] +{Y}_t^{\prime}\Psi_{t}\Psi_{t}^{\prime}{Y}_t.
\end{align}
Thus, the objective function in \eqref{Eqn: objls} is equivalent to the objective function in \eqref{Eqn: objregpca}, scaled by a factor of $N$ and plus a constant term that does not depend on $a$, $B$, or $f_t$. Therefore, when the condition $\Phi(Z_t)'\Phi(Z_t)/N=I_{JM}$ holds, the regressed-PCA method reduces to the least squares approach.

\subsection{Optimality of Fama-MacBeth Managed Portfolios}\label{App: Sec: C11}
\cite{Fama_76} demonstrates that the $j$th column of $\Phi(Z_t)(\Phi(Z_t)^{\prime}\Phi(Z_t))^{-1}$ solves the following dynamic programming problem:
\begin{align}
\min_{\omega} \omega^{\prime}\omega \text{ such that } \Phi(Z_t)^{\prime}\omega = e_{j},
\end{align}
 where $e_j$ is the $j$th column of $I_{JM}$. The proof is straightforward, relying on the use of a lagrangian function. The first order conditions are:
\begin{align}
2\omega + \Phi(Z_t)\lambda &= 0,\\
\Phi(Z_t)^{\prime}\omega &= e_{j},
\end{align}
where $\lambda$ is a lagrangian multiplier. Solving these yields $\omega=\Phi(Z_t)(\Phi(Z_t)^{\prime}\Phi(Z_t))^{-1}e_{j} $, which corresponds to the $j$th column of $\Phi(Z_t)(\Phi(Z_t)^{\prime}\Phi(Z_t))^{-1}$.
If individual asset returns are i.i.d. (conditional on $Z_{t}$), the objective function presents the variance of the portfolio $Y_{t}^{\prime}\omega$ multiplied by a constant. Hence, under the i.i.d. assumption, each portfolio in $\tilde{Y}_{t} = (\Phi(Z_t)^{\prime}\Phi(Z_t))^{-1}\Phi(Z_t)^{\prime}Y_{t}$ is minimum-variance portfolio. Moreover, the covariance matrix of $\tilde{Y}_{t}$ is proportional to $(\Phi(Z_t)^{\prime}\Phi(Z_t))^{-1}$, suggesting low correlations among the portfolios in $\tilde{Y}_{t}$ when the columns of $\Phi(Z_t)$ have low correlations.

\subsection{Sorting: Regression on Dummies}\label{App: Sec: C13}
At each time $t$, sorting divides the space of $z_{it}$ into $q$ distinct, non-overlapping regions $R_{t,1}, R_{t,2},\ldots, R_{q,t}$, which may vary over time. Let $D_{it,1}, D_{it,2}, \ldots, D_{it,q}$ represent group dummy variables for these regions, where $D_{it,j} = 1\{z_{it}\in R_{j,t}\}$. The equally weighted returns of the sorted portfolios are then the coefficient estimates from a cross-sectional regression of $y_{it}$ on $D_{it,1}, D_{it,2}, \ldots, D_{it,q}$ without an intercept. These are calculated as $\tilde{Y}_t=(\sum_{i=1}^{N}\phi(z_{it})\phi(z_{it})^{\prime})^{-1}\sum_{i=1}^{N}\phi(z_{it})y_{it}$, where $\phi(z_{it}) = (D_{it,1}, D_{it,2}, \ldots, D_{it,q})^{\prime}$. Specially, the return of the portfolio corresponding to $R_{j,t}$ is equal to the coefficient on $D_{it,j}$. For value weighted returns, book-to-market ratios can be used as weights in the regression. In the example of constructing high-minus-low and small-minus-big factors, $z_{it}$ consists of capitalization ($c_{it}$) and book-to-market ratio ($bm_{it}$) with $q=6$:
\begin{align}
D_{it,1} &= 1\{c_{it}\leq Q_{0.5,t}(c_{it})\}1\{bm_{it}\leq Q_{0.3,t}(bm_{it})\},\notag\\
D_{it,2} &= 1\{c_{it}\leq Q_{0.5,t}(c_{it})\}1\{ Q_{0.3,t}(bm_{it})<bm_{it}\leq Q_{0.7,t}(bm_{it})\}, \notag\\
D_{it,3} &= 1\{c_{it}\leq Q_{0.5,t}(c_{it})\}1\{ bm_{it}> Q_{0.7,t}(bm_{it})\}\notag\\
D_{it,4} &= 1\{c_{it}> Q_{0.5,t}(c_{it})\}1\{bm_{it}\leq Q_{0.3,t}(bm_{it})\},\notag\\
D_{it,5} &= 1\{c_{it}> Q_{0.5,t}(c_{it})\}1\{ Q_{0.3,t}(bm_{it})<bm_{it}\leq Q_{0.7,t}(bm_{it})\}, \notag\\
D_{it,6} &= 1\{c_{it}> Q_{0.5,t}(c_{it})\}1\{ bm_{it}> Q_{0.7,t}(bm_{it})\},
\end{align}
where $Q_{0.5,t}(c_{it})$ is the 50\% quantile of $\{c_{1t}, c_{2t}, \ldots, c_{nt}\}$ at each time $t$.

\subsection{Misspecifications of Alpha and Beta Functions}\label{App: Sec: C12}
We use simple examples to illustrate how misspefications of $\alpha(\cdot)$ and $\beta(\cdot)$ may result in inconsistent estimation of $F$. Consider the following two models:
\begin{align}
Y_t &=  W_t\Pi+ Z_{t}\Gamma f_t +\varepsilon_{t},\label{Eqn: Simplelienar1}\\
Y_t &= (Z_{t}\Gamma + W_t\Pi) f_t +\varepsilon_{t},\label{Eqn: Simplelienar2}
\end{align}
where $Z_t$ and $W_t$ are $N\times 1$ vectors, $f_t$ is a scalar factor, and $\varepsilon_{t}$ is independent of $Z_t$ and $W_t$. Further assume $\Pi = \Gamma$ and $W_t=Z_tg_t + v_t$, where $g_t$ is a scalar coefficient and $v_t$ is independent of $Z_t$. In this case, model \eqref{Eqn: Simplelienar1} and model \eqref{Eqn: Simplelienar2} can be rewritten as:
\begin{align}
Y_t &= Z_{t}\Gamma f_t^{\star} + \varepsilon^{\star}_{t},\label{Eqn: Simplelienarequiv1}\\
Y_t &= Z_{t}\Gamma f_t^{\star\star} + \varepsilon^{\star\star}_{t},\label{Eqn: Simplelienarequiv2}
\end{align}
where $f_t^{\star} = f_t+g_t$, $\varepsilon^{\star}_{t}=v_t\Gamma + \varepsilon_{t}$, $f_t^{\star\star} = f_t(1+g_t)$, and $\varepsilon^{\star\star}_{t}=v_t\Gamma f_t + \varepsilon_{t}$. If only $Z_t$ is used for estimating model \eqref{Eqn: Simplelienar1} (i.e., $\alpha(\cdot)$ is misspecified), then $\hat{F}$ consistently estimates $F^{\star} = (f_1^{\star},\ldots,f_T^{\star})^{\prime}$ up to a scalar. Similarly, if only $Z_t$ is used for estimating model \eqref{Eqn: Simplelienar2} (i.e., $\beta(\cdot)$ is misspecified), then $\hat{F}$ consistently estimates $F^{\star\star} = (f_1^{\star\star},\ldots,f_T^{\star\star})^{\prime}$ up to a scalar. In both cases, $\hat{F}$ fails to consistently estimate the space spanned by $F$ unless $g_t$ is proportional to $f_t$ in the first case or remains constant over time in the second case.

\subsection{On Bootstrap Failure}\label{App: Sec: C2}
A more natural bootstrap estimator for $B$ is given by $\hat{B}^{\ast\ast}$, whose columns are the eigenvectors of $\tilde{Y}^{\ast}M_T\tilde{Y}^{\ast\prime}/T$ corresponding to its first $K$ largest eigenvalues. We notice that the distribution of $\sqrt{NT/\omega_0}(\hat{B}^{\ast\ast} - \hat{B})$ conditional on the data may fail to estimate the distribution of $\mathbb{G}_{B}$. The key part of the proof for Theorem \ref{Thm: Boot} is to show that $\sqrt{{NT}}(\hat{B}^{\ast} - B H)$ and $\sqrt{{NT}}(\hat{B} - B H)$ share a similar asymptotic expansion. Specifically, we show
\begin{align}
\left\|\sqrt{NT}(\hat{B} - B H) - \frac{1}{\sqrt{NT}}\sum_{t=1}^{T}{Q}^{-1}_t\Phi(Z_t)^{\prime}\varepsilon_t(f_{t}-\bar{f})^{\prime}B^{\prime}B\mathcal{M}\right\|_{F}=O_{p}(\delta_{NT})
\end{align}
and
\begin{align}
\left\|\sqrt{NT}(\hat{B}^{\ast} - B H) - \frac{1}{\sqrt{NT}}\sum_{t=1}^{T}{Q}^{-1}_t\Phi(Z_t)^{\ast\prime}\varepsilon_t (f_{t}-\bar{f})^{\prime}B^{\prime}B\mathcal{M}\right\|_{F}=O_{p}(\delta_{NT}),
\end{align}
where $\delta_{NT}={\sqrt{NT}}{J^{-\kappa}}+{\sqrt{TJ/N}}+\sqrt{J\xi_{J}}(\log J /N)^{1/4}$. Let $\hat{F}^{\ast}\equiv \tilde{Y}^{\ast\prime}\hat{B}^{\ast\ast}$ and $H^{\ast}\equiv(F^{\prime}M_T\hat{F}^{\ast})(\hat{F}^{^{\ast}\prime}M_T\hat{F}^{\ast})^{-1}$. Similarly, we can also show
\begin{align}
\left\|\sqrt{NT}(\hat{B}^{\ast\ast} - B H^{\ast}) - \frac{1}{\sqrt{NT}}\sum_{t=1}^{T}{Q}^{-1}_t\Phi(Z_t)^{\ast\prime}\varepsilon_t(f_{t}-\bar{f})^{\prime}B^{\prime}B\mathcal{M}\right\|_{F}=O_{p}(\delta_{NT}).
\end{align}
Thus, the distribution of $\sqrt{NT/\omega_0}(\hat{B}^{\ast\ast} - \hat{B})$ conditional on the data may fail to estimate the distribution of $\mathbb{G}_{B}$, since $\sqrt{NT/\omega_0}(H^{\ast}-H)$ is not asymptotically negligible due to the relatively slow convergence rate of $\hat{F}$ and $\hat{F}^{\ast}$. Since $\hat{B}^{\ast\ast} = \tilde{Y}^{\ast}M_T\hat{F}^{\ast}(\hat{F}^{\ast\prime}M_T\hat{F}^{\ast})^{-1}$, it is crucial to use $\hat{F}$ rather than $\hat{F}^{\ast}$ in \eqref{Eqn: BootstrappedEstimators} to ensure that $\hat{B}^{\ast}$ and $\hat{B}$ share a common rotational transformation matrix and are centered around the same quantity $BH$, rendering the validity of the bootstrap.

\subsection{Three Versions of \texorpdfstring{$R^2$}{R Square}}\label{App: Sec: D6}
From a mathematical perspective, $R_{T,N}^2$ and $R_{N,T}^2$ can be expressed as weighted versions of $R^2$. However, the variation in these weights is not due to an unbalanced panel. Let $\hat{\varepsilon}_{it} \equiv y_{it}- \hat{\alpha}(z_{i,t-1}) - \hat{\beta}(z_{i,t-1})^{\prime}\hat{f}_t$. It follows that
\begin{align}
R_{T,N}^2 = 1-\frac{1}{N}\sum_{i}\frac{\sum_{t}\hat{\varepsilon}_{it}^2}{\sum_{t}{y}_{it}^2} = 1-\frac{\sum_{i} \frac{\sum_{i,t}{y}_{it}^2}{N\sum_{t}{y}_{it}^2}\sum_{t}\hat{\varepsilon}_{it}^2}{\sum_{i,t}{y}_{it}^2} =  1-\frac{\sum_{i} \omega_i\sum_{t}\hat{\varepsilon}_{it}^2}{\sum_{i,t}{y}_{it}^2},
\end{align}
where $w_i = {\sum_{i,t}{y}_{it}^2}/{N\sum_{t}{y}_{it}^2}$. The variation in $\omega_i$ arises from differences in the total time variation across individual stocks. The same reasoning applies to $R_{N,T}^2$.

\subsection{Sufficient Conditions for Assumptions}
We provide sufficient conditions for Assumptions \ref{Ass: Basis}(i) and \ref{Ass: Numberfactors}(ii) in the following two propositions, justifying that the two assumptions are not restrictive.
\begin{pro}[Assumption \ref{Ass: Basis}(i)]\label{Pro: JustNum1}
Suppose Assumptions \ref{Ass: Improvedrates}(ii) and (iii) hold.  Assume $J\geq 2$ and $\sqrt{T}\xi^{2}_{J}\log J=o(N)$, where $\xi_{J}$ is given above Theorem \ref{Thm: ImprovedRates}. Then Assumption \ref{Ass: Basis}(i) holds.
\end{pro}
\noindent{\sc Proof:} Let ${Q}_{t}\equiv E[\hat{Q}_{t}]$ Since $\sqrt{T}\xi^{2}_{J}\log J=o(N)$, by Lemma \ref{Lem: TechB6},
\begin{align}\label{Eqn: Pro: JustNum1: 1}
\max_{t\leq T}\|\hat{Q}_t - Q_{t}\|_{2}\leq\left(\sum_{t=1}^{T}\|\hat{Q}_t - Q_{t}\|^{4}_{2}\right)^{1/4}=O_{p}\left(\frac{T^{1/4}\xi_{J}\log^{1/2}J}{\sqrt{N}}\right) = o_{p}(1).
\end{align}
By \eqref{Eqn: Pro: JustNum1: 1} and the Weyl's inequality,
\begin{align}\label{Eqn: Pro: JustNum1: 2}
\left|\min_{t\leq T}\lambda_{\min}(\hat{Q}_t) - \min_{t\leq T}\lambda_{\min}({Q}_t)\right|\leq \max_{t\leq T}\|\hat{Q}_t - Q_{t}\|_{2} = o_{p}(1)
\end{align}
and
\begin{align}\label{Eqn: Pro: JustNum1: 3}
\left|\max_{t\leq T}\lambda_{\max}(\hat{Q}_t) - \max_{t\leq T}\lambda_{\max}({Q}_t)\right|\leq \max_{t\leq T}\|\hat{Q}_t - Q_{t}\|_{2} = o_{p}(1).
\end{align}
The result of the lemma thus follows from \eqref{Eqn: Pro: JustNum1: 2} and \eqref{Eqn: Pro: JustNum1: 3} as well as Assumption \ref{Ass: Improvedrates}(ii) by noting that $\min_{t\leq T}\lambda_{\min}(Q_{t})\geq \min_{i\leq N,t\leq T}\lambda_{\min}(Q_{it})$ and $\max_{t\leq T}\lambda_{\max}(Q_{t})\leq \max_{i\leq N,t\leq T}\lambda_{\max}(Q_{it})$. \qed

\begin{pro}[Assumption \ref{Ass: Numberfactors}(ii)]\label{Pro: JustNum2}
Suppose Assumptions \ref{Ass: DGP}(ii) and \ref{Ass: Asym}(ii) hold. Assume $\max_{i\leq N,t\leq T}E[\varepsilon_{it}^{4}]<\infty$ and there is $0<C_5<\infty$ such that
\[\max_{i\leq N}\frac{1}{T}\sum_{t=1}^{T}\sum_{s=1}^{T}|E[\varepsilon_{it}\varepsilon_{is}]|^2<C_5.\]
Then Assumption \ref{Ass: Numberfactors}(ii) holds.
\end{pro}
\noindent{\sc Proof:} By the independence condition and Assumption \ref{Ass: DGP}(ii), $E[\varepsilon_{it}\varepsilon_{jt}]=0$ for $i\neq j$. Thus, we may have the following decomposition
\begin{align}\label{Eqn: Pro: JustNum2: 1}
&\hspace{0.8cm}\sum_{t=1}^{T}\sum_{s=1}^{T}\sum_{i=1}^{N}\sum_{j= 1}^{N}\sum_{k=1}^{N}\sum_{\ell= 1}^{N}|cov(\varepsilon_{it}\varepsilon_{jt},\varepsilon_{ks}\varepsilon_{\ell s})|\notag\\
&=\sum_{t=1}^{T}\sum_{s=1}^{T}\sum_{i=1}^{N}\sum_{k=1}^{N}|cov(\varepsilon^{2}_{it},\varepsilon^{2}_{ks})|+\sum_{t=1}^{T}\sum_{s=1}^{T}\sum_{i=1}^{N}\sum_{j\neq i}^{N}\sum_{k=1}^{N}\sum_{\ell\neq k}^{N}|E[\varepsilon_{it}\varepsilon_{jt}\varepsilon_{ks}\varepsilon_{\ell s}]|\notag\\
&\hspace{0.5cm}+2\sum_{t=1}^{T}\sum_{s=1}^{T}\sum_{i=1}^{N}\sum_{k=1}^{N}\sum_{\ell\neq k}^{N}|E[(\varepsilon^{2}_{it}-E[\varepsilon^{2}_{it}])\varepsilon_{ks}\varepsilon_{\ell s}]|\notag\\
&\equiv \mathcal{T}_1 +\mathcal{T}_2+ \mathcal{T}_3.
\end{align}
We next establish bound for $\mathcal{T}_1, \mathcal{T}_2$, and $\mathcal{T}_3$. By the independence condition,
\begin{align}\label{Eqn: Pro: JustNum2: 2}
\mathcal{T}_1 = \sum_{t=1}^{T}\sum_{s=1}^{T}\sum_{i=1}^{N}var(\varepsilon^{2}_{it})\leq \sum_{t=1}^{T}\sum_{s=1}^{T}\sum_{i=1}^{N}E[\varepsilon^{4}_{it}]\leq NT^{2}\max_{i\leq N,t\leq T}E[\varepsilon_{it}^{4}],
\end{align}
where the first inequality follows from $var(\varepsilon^{2}_{it})\leq E[\varepsilon^{4}_{it}]$.  By the independence condition and Assumption \ref{Ass: DGP}(ii), $E[\varepsilon_{it}\varepsilon_{jt}\varepsilon_{ks}\varepsilon_{\ell s}]=0$ unless $i=k$ and $j=\ell$ or $i=\ell$ and $j=k$ given $i\neq j$. It then follows that
\begin{align}\label{Eqn: Pro: JustNum2: 3}
\mathcal{T}_2&=2\sum_{t=1}^{T}\sum_{s=1}^{T}\sum_{i=1}^{N}\sum_{j\neq i}^{N}|E[\varepsilon_{it}\varepsilon_{is}\varepsilon_{jt}\varepsilon_{j s}]|= 2\sum_{t=1}^{T}\sum_{s=1}^{T}\sum_{i=1}^{N}\sum_{j\neq 1}^{N}|E[\varepsilon_{it}\varepsilon_{is}]||E[\varepsilon_{jt}\varepsilon_{j s}]|\notag\\
&\leq 2\sum_{t=1}^{T}\sum_{s=1}^{T}\sum_{i=1}^{N}\sum_{j= 1}^{N}|E[\varepsilon_{it}\varepsilon_{is}]||E[\varepsilon_{jt}\varepsilon_{j s}]|=2\sum_{t=1}^{T}\sum_{s=1}^{T}\left(\sum_{i=1}^{N}|E[\varepsilon_{it}\varepsilon_{is}]|\right)^{2}\notag\\
&\leq 2N\sum_{i=1}^{N}\sum_{t=1}^{T}\sum_{s=1}^{T}|E[\varepsilon_{it}\varepsilon_{is}]|^{2} \leq 2N^{2}\max_{i\leq N}\sum_{t=1}^{T}\sum_{s=1}^{T}|E[\varepsilon_{it}\varepsilon_{is}]|^{2},
\end{align}
where the second equality follows by the independence condition, the first inequality follows since $|E[\varepsilon_{it}\varepsilon_{is}]|^{2}\geq 0$, the second inequality is due to the Cauchy-Schwartz inequality. Again by the independence condition and Assumption \ref{Ass: DGP}(ii), $E[(\varepsilon^{2}_{it}-E[\varepsilon^{2}_{it}])\varepsilon_{ks}\varepsilon_{\ell s}]=0$ for $k\neq \ell$, so $\mathcal{T}_{3}=0$. This together with \eqref{Eqn: Pro: JustNum2: 1}-\eqref{Eqn: Pro: JustNum2: 3} and the assumptions thus concludes the result of the proposition.\qed
\section{Monte Carlo Simulations}\label{App: Sec: E}
In this appendix, we conduct small-scale Monte Carlo simulations to evaluate the finite sample performance of our estimators and tests.

We consider the following data-generating process, assuming
\begin{align}\label{Eqn: alphabeta}
\alpha(z_{it}) = \theta z_{it,1} + \delta z_{it,1}^{2} \text{ and } \beta(z_{it}) = (z_{it,2} + \delta z_{it,2}^{2}, 2z_{it,3} + 2\delta z_{it,3}^{2})^{\prime}
\end{align}
for $\theta\geq 0$ and $\delta\geq 0$, where $K=2$ and $M=3$. Note that  $\alpha(\cdot)=0$ when $\theta = \delta=0$, and both $\alpha(z_{it})$ and $\beta(z_{it})$ become nonlinear functions of $z_{it}$ when $\delta>0$. We define:
\begin{align}\label{Eqn: Covariates}
z_{it,1}=\sigma_{t}*u_{it,1}, z_{it,2} = 0.3z_{i(t-1),2} + u_{it,2}, \text{ and } z_{it,3}=u_{it,3},
\end{align}
where $u_{it} = (u_{it,1},u_{it,2},u_{it,3})^{\prime}$ are i.i.d. $N(0,I_3)$ across both $i$ and $t$, $\sigma_{t}$'s are i.i.d. $U(1,2)$ over $t$, and $z_{i0,2}$'s are i.i.d. $N(0,1)$. All components of $z_{it}$ vary over $t$, but in different ways. We also define $f_{t} = 0.3 f_{t-1} + \eta_t$, where $\eta_t$'s are i.i.d. $N(0,I_K)$ and $f_0\sim N(0,I_K/0.91)$. For $0\leq \rho<1$, we specify:
\begin{align}\label{Eqn: WeakDep}
\varepsilon_t = \rho \varepsilon_{t-1} + e_{t},
\end{align}
where $e_t$'s are i.i.d. $N(0,I_N)$ and $\varepsilon_0\sim N(0,I_N/(1-\rho^2))$. The parameter $\rho$ measures the weak dependence of $\varepsilon_{it}$ over $t$. Here, $u_{it}$'s, $\sigma_{t}$'s, $z_{i0}$'s $\eta_t$'s, $f_0$, $e_t$'s, and $\varepsilon_0$ are mutually independent. We generate $y_{it}$ based on the model \eqref{Eqn: Model}.

\begin{figure}[!htbp]
\centering
\caption{Histograms of the 2nd entry in $\sqrt{NT}(\hat{a} - a)$ (\textit{blue}) and $\sqrt{NT}(\hat{a}^{\ast} - \hat{a})$ (\textit{yellow}, based on the first simulation replication) when $\theta = 1$, $\delta = 0.5$, and $\rho = 0.3$}\label{Fig: Sim1}
\includegraphics[width=14.5cm, height=18cm]{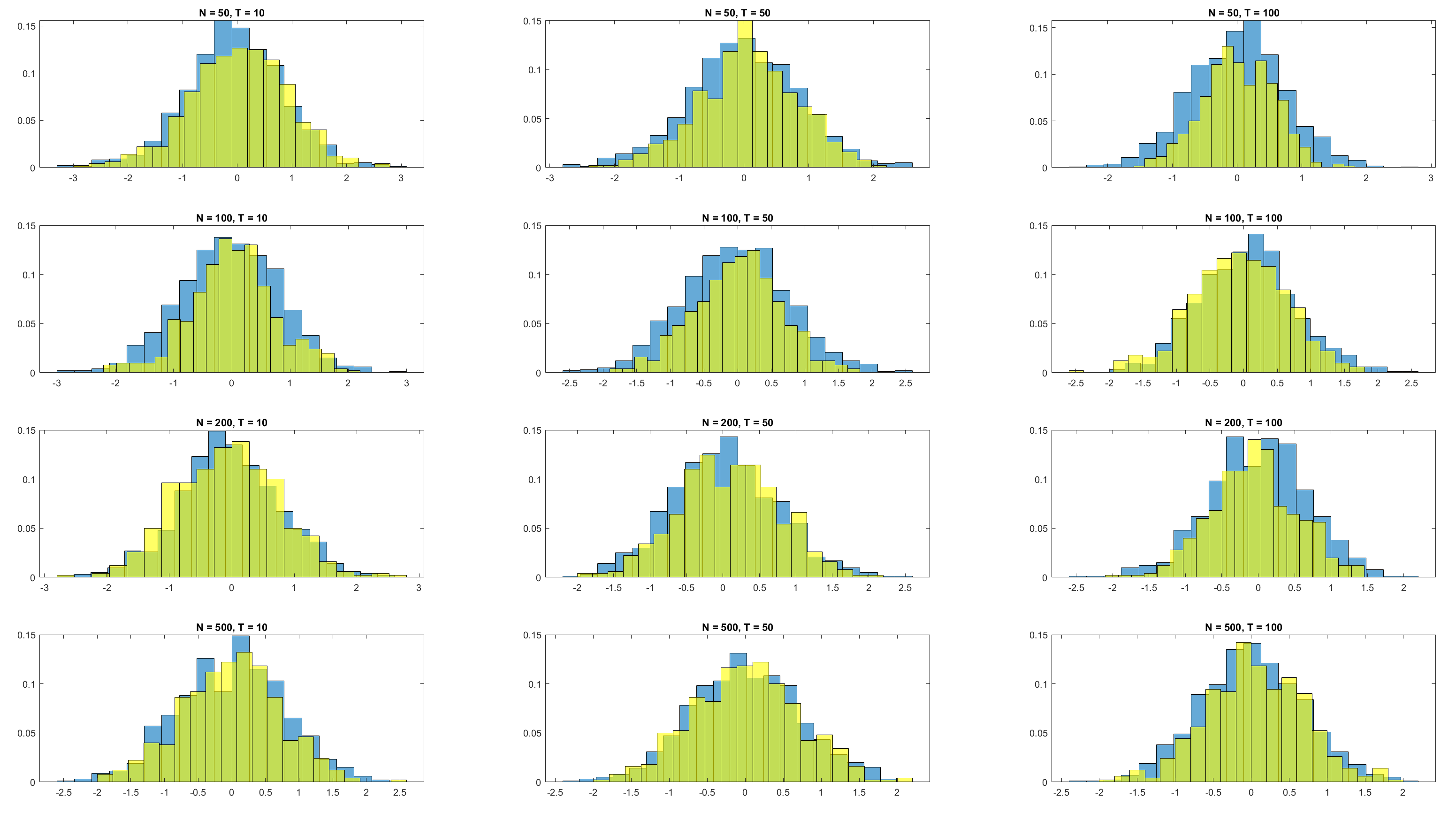}
\end{figure}

\begin{figure}[!htbp]
\centering
\caption{Histograms of the $(1,2)$th entry in $\sqrt{NT}(\hat{B} - BH)$ (\textit{blue}) and $\sqrt{NT}(\hat{B}^{\ast} - \hat{B})$ (\textit{yellow}, based on the first simulation replication) when $\theta = 1$, $\delta = 0.5$, and $\rho = 0.3$}\label{Fig: Sim2}
\includegraphics[width=14.5cm, height=18cm]{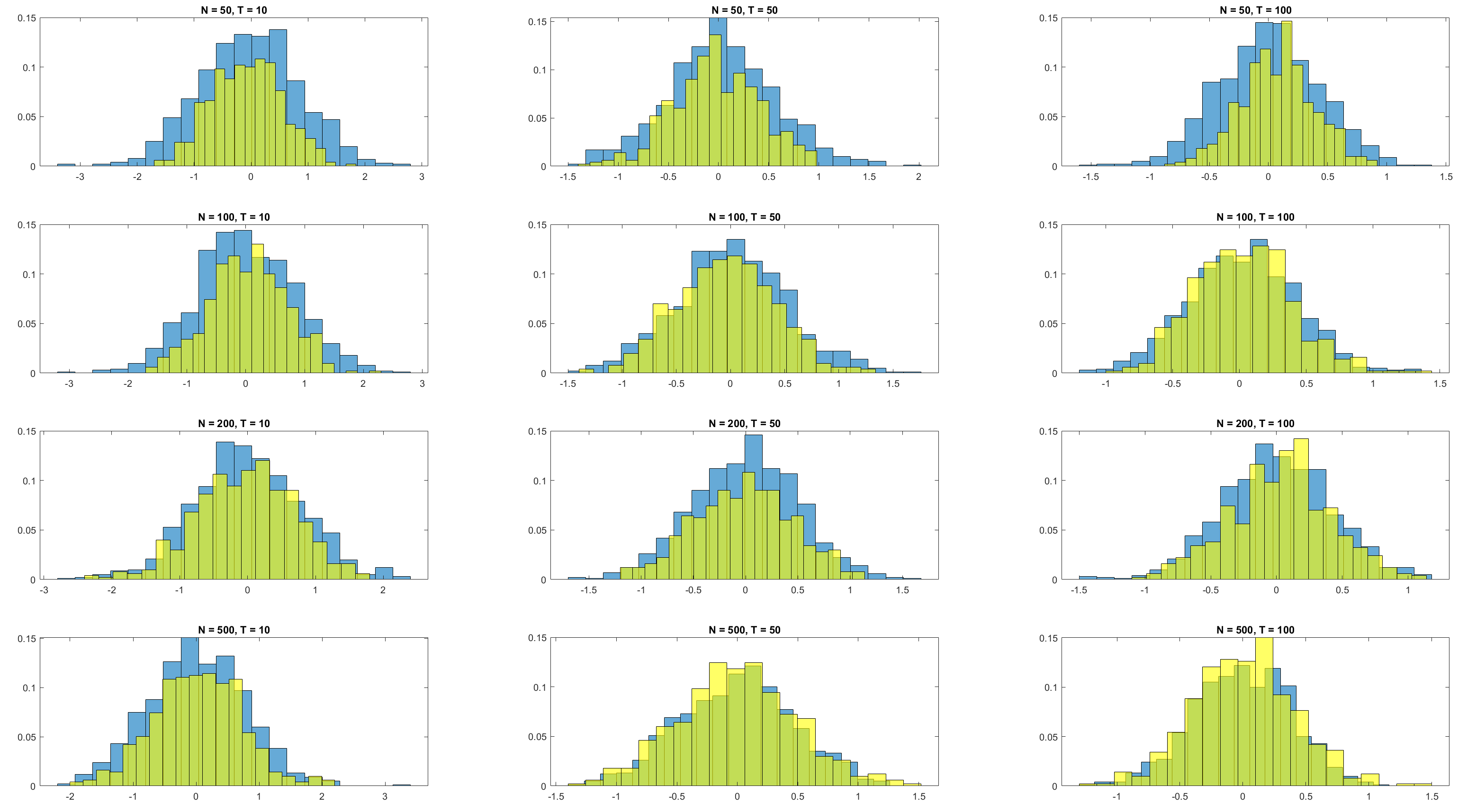}
\end{figure}

To implement regressed-PCA, we select $\phi(z_{it}) = (z_{it,1},z_{it,1}^{2},z_{it,2},z_{it,2}^{2},$ $z_{it,3},z_{it,3}^{2})^{\prime}$, so $J=2$, and the sieve approximation error is zero. We let $\lambda_{NT} = 1/\log(N)$ when implementing $\tilde{K}$ in \eqref{Eqn: Kestiamtoralt}. For the weighted bootstrap, we let $w_{i}$'s be i.i.d. standard exponential random variables. We first analyse the performance of $\hat{a}$, $\hat{B}$, $\hat{F}$, $\hat{K}$ in \eqref{Eqn: Kestiamtor}, and $\tilde{K}$ under varying $(N,T)$ values, with $\theta=1$, $\delta=0.5$, and $\rho=0,0.3,0.7$. We report the mean square errors of $\hat{a}$, $\hat{B}$, and $\hat{F}$ in Table \ref{Tab: MSE} and the correct rates of $\hat{K}$ and $\tilde{K}$ in Table \ref{Tab: CorrectRate}. Figures \ref{Fig: Sim1} and \ref{Fig: Sim2} present histograms of $\sqrt{NT}(\hat{a} - a)$ and $\sqrt{NT}(\hat{B} - BH)$ and their bootstrap estimates (i.e., $\sqrt{NT}(\hat{a}^{\ast} - \hat{a})$ and $\sqrt{NT}(\hat{B}^{\ast} - \hat{B})$) for $\rho = 0.3$, with similar results for $\rho = 0$ and $0.7$ available upon request. Due to space limitations, we only display one entry of $\sqrt{NT}(\hat{a} - a)$ and $\sqrt{NT}(\hat{B} - BH)$, with similar results for other entries available on request.

Next, we assess the performance of tests for $\alpha(\cdot) = 0$ and the linearity of $\alpha(\cdot)$ and $\beta(\cdot)$. To test $\alpha(\cdot) = 0$, we fix $\delta = 0$. Hence, $\alpha(\cdot) = 0$ if and only if $\theta = 0$. We report the rejection rates for $\theta = 0, 0.01, 0.02, \ldots, 0.1$ under $\rho = 0.3$, with similar results for $\rho = 0$ and $0.7$ available on request. For the linearity test, fixing $\theta = 1$, $\alpha(\cdot)$ and $\beta(\cdot)$ are linear if and only if $\delta = 0$. We report the rejection rates for $\delta = 0, 0.01, 0.02, \ldots, 0.1$ under $\rho = 0.3$, with similar results for other values of $\rho$ available on request. We set the number of simulation replications to 1,000 and the number of bootstrap draws to 499 for each replication.

The main findings are as follows. First, as shown in Table \ref{Tab: MSE}, the mean square errors of $\hat{a}$, $\hat{B}$, and $\hat{F}$ decrease as $N$ increases, even for $T = 10$, indicating consistency of the estimators as $N \to \infty$, even for small $T$. Increasing $T$ further reduces the mean square errors of $\hat{a}$ and $\hat{B}$, but does not affect that of $\hat{F}$. Both results hold regardless of $\rho$, confirming that the estimators remain valid under weak dependence of $\varepsilon_{it}$. These findings align with Theorem \ref{Thm: ImprovedRates}. Second, as shown in Table \ref{Tab: CorrectRate}, $\hat{K}$ and $\tilde{K}$ correctly estimate $K$ in all cases, except for small $N$ and $T$, consistent with Theorem \ref{Thm: Numberfactors}. Third, Figures \ref{Fig: Sim1} and \ref{Fig: Sim2} show that both $\sqrt{NT}(\hat{a} - a)$ and $\sqrt{NT}(\hat{B} - BH)$, as well as their bootstrap estimates, follow bell-shaped distributions, even for $T = 10$, suggesting asymptotic normality as per Theorems \ref{Thm: AsymDis} and \ref{Thm: Boot}. The two distributions converge as $N$ increases, though the approximation may be unsatisfactory for $N = 50$.

Finally, both tests perform well. Table \ref{Tab: RejectionRateAlpha} shows that the first test may slightly overreject $\alpha(\cdot) = 0$ (which holds when $\theta = 0$) when $N = 50$, but this corrects as $N$ increases, even for $T = 10$. The test is consistent as $N \to \infty$ for small $T$, and increasing $T$ can improve power, though it may slightly affect the size (e.g., for $\theta = 0$, the rejection rate increases as $T$ grows from 10 to 100 for $N = 200$). This is in line with the $T = o(N)$ requirement in Theorem \ref{Thm: SpecTest} or underlying in Theorem \ref{Thm: Boot}. The second test exhibits similar performance, as shown in Table \ref{Tab: RejectionRateLinear}, and the details are omitted for brevity. The findings of the second test are consistent with Theorem \ref{Thm: SpecTest}.

In conclusion, our estimators and bootstrap inference methods show strong performance for large $N$, even when $T$ is small.

\setlength{\tabcolsep}{3pt}
\begin{table}[!htbp]
\centering
\begin{threeparttable}
\renewcommand{\arraystretch}{1.2}
\caption{Mean square errors of $\hat{a}$, $\hat{B}$ and $\hat{F}$ when $\theta = 1$ and $\delta = 0.5$\tnote{\dag}}\label{Tab: MSE}
\begin{tabular}{ccccccccccccccc}
\hline\hline
&\multirow{2}{*}{(N,T)}&&\multicolumn{3}{c}{$\rho = 0$}&&\multicolumn{3}{c}{$\rho = 0.3$}&&\multicolumn{3}{c}{$\rho = 0.7$}&\\
\cline{4-6}\cline{8-10}\cline{12-14}
&&&$\hat{a}$&$\hat{B}$&$\hat{F}$ &&$\hat{a}$&$\hat{B}$&$\hat{F}$ &&$\hat{a}$&$\hat{B}$&$\hat{F}$&\\
\cline{2-14}
&$(50,10)$    &&0.0077&0.0154&0.0394&&0.0088&0.0170&0.0435&&0.0171&0.0295&0.0799&\\
&$(100,10)$   &&0.0034&0.0064&0.0168&&0.0039&0.0071&0.0186&&0.0075&0.0127&0.0336&\\
&$(200,10)$   &&0.0016&0.0030&0.0079&&0.0018&0.0034&0.0087&&0.0033&0.0058&0.0155&\\
&$(500,10)$   &&0.0006&0.0012&0.0030&&0.0007&0.0013&0.0033&&0.0013&0.0022&0.0060&\\
\cline{2-14}
&$(50,50)$    &&0.0012&0.0022&0.0423&&0.0014&0.0025&0.0466&&0.0028&0.0049&0.0842&\\
&$(100,50)$   &&0.0005&0.0009&0.0184&&0.0006&0.0010&0.0203&&0.0012&0.0019&0.0365&\\
&$(200,50)$   &&0.0002&0.0004&0.0086&&0.0003&0.0004&0.0095&&0.0006&0.0008&0.0170&\\
&$(500,50)$   &&0.0000&0.0001&0.0033&&0.0001&0.0002&0.0037&&0.0002&0.0003&0.0065&\\
\cline{2-14}
&$(50,100)$   &&0.0005&0.0010&0.0431&&0.0006&0.0011&0.0473&&0.0013&0.0024&0.0850&\\
&$(100,100)$  &&0.0002&0.0004&0.0187&&0.0003&0.0004&0.0206&&0.0006&0.0008&0.0370&\\
&$(200,100)$  &&0.0001&0.0002&0.0087&&0.0001&0.0002&0.0096&&0.0003&0.0003&0.0172&\\
&$(500,100)$  &&0.0000&0.0001&0.0034&&0.0000&0.0001&0.0037&&0.0001&0.0001&0.0066&\\
\hline\hline
\end{tabular}
\begin{tablenotes}
      \small
      \item[\dag] The mean square errors of $\hat{a}$ , $\hat{B}$ and $\hat{F}$ are given by $\sum_{\ell=1}^{1000}\|\hat{a}^{(\ell)}-a\|^2/1000$, $\sum_{\ell=1}^{1000}\|\hat{B}^{(\ell)}-BH^{(\ell)}\|_{F}^2/1000$ and $\sum_{\ell=1}^{1000}\|\hat{F}^{(\ell)}- F(H^{{(\ell)}\prime})^{-1}\|_{F}^2/{1000T}$, where $\hat{a}^{(\ell)}$, $\hat{B}^{(\ell)}$ and $\hat{F}^{(\ell)}$ are estimators in the $\ell$th simulation replication, and $H^{(\ell)}\equiv (F^{\prime}M_T\hat{F}^{(\ell)})(\hat{F}^{{(\ell)}\prime}M_T\hat{F}^{(\ell)})^{-1}$ is a rotational transformation matrix.
    \end{tablenotes}
\end{threeparttable}
\end{table}%

\setlength{\tabcolsep}{7.3pt}
\begin{table}[!htbp]
\centering
\begin{threeparttable}
\renewcommand{\arraystretch}{1.2}
\caption{Correct rates of $\hat{K}$ and $\tilde{K}$ when $\theta = 1$ and $\delta = 0.5$}\label{Tab: CorrectRate}
\begin{tabular}{cccccccccccc}
\hline\hline
&\multirow{2}{*}{(N,T)}&&\multicolumn{2}{c}{$\rho = 0$}&&\multicolumn{2}{c}{$\rho = 0.3$}&&\multicolumn{2}{c}{$\rho = 0.7$}&\\
\cline{4-5}\cline{7-8}\cline{10-11}
&&&$\hat{K}$&$\tilde{K}$ &&$\hat{K}$&$\tilde{K}$ &&$\hat{K}$&$\tilde{K}$&\\
\cline{2-11}
&$(50,10)$   &&0.999&1.000&&0.999&1.000&&0.994&1.000&\\
&$(100,10)$  &&1.000&1.000&&1.000&1.000&&0.999&1.000&\\
&$(200,10)$  &&1.000&1.000&&1.000&1.000&&1.000&1.000&\\
&$(500,10)$  &&1.000&1.000&&1.000&1.000&&1.000&1.000&\\
&$(50,50)$   &&1.000&1.000&&1.000&1.000&&1.000&1.000&\\
&$(100,50)$  &&1.000&1.000&&1.000&1.000&&1.000&1.000&\\
&$(200,50)$  &&1.000&1.000&&1.000&1.000&&1.000&1.000&\\
&$(500,50)$  &&1.000&1.000&&1.000&1.000&&1.000&1.000&\\
&$(50,100)$  &&1.000&1.000&&1.000&1.000&&1.000&1.000&\\
&$(100,100)$ &&1.000&1.000&&1.000&1.000&&1.000&1.000&\\
&$(200,100)$ &&1.000&1.000&&1.000&1.000&&1.000&1.000&\\
&$(500,100)$ &&1.000&1.000&&1.000&1.000&&1.000&1.000&\\
\hline\hline
\end{tabular}
\end{threeparttable}
\end{table}%

\setlength{\tabcolsep}{3pt}
\begin{table}[!htbp]
\centering
\begin{threeparttable}
\renewcommand{\arraystretch}{1.2}
\caption{Rejection rates of testing $\alpha(\cdot) = 0 $ when $\delta = 0$ and $\rho = 0.3$\tnote{\dag}}\label{Tab: RejectionRateAlpha}
\begin{tabular}{ccccccccccccccc}
\hline\hline
&\multirow{2}{*}{(N,T)}&&\multicolumn{11}{c}{$\theta$}&\\
\cline{4-14}
&&&$0$&$0.01$&$0.02$&$0.03$&$0.04$&$0.05$&$0.06$&$0.07$&$0.08$&$0.09$&$0.1$&\\
\cline{2-14}
&$(50,10)$   &&0.089&0.096&0.117&0.150&0.186&0.222&0.283&0.349&0.435&0.512&0.593&\\
&$(100,10)$  &&0.096&0.113&0.133&0.184&0.274&0.383&0.502&0.616&0.727&0.827&0.904&\\
&$(200,10)$  &&0.057&0.080&0.162&0.270&0.442&0.628&0.790&0.901&0.970&0.990&0.999&\\
&$(500,10)$  &&0.048&0.099&0.297&0.573&0.822&0.951&0.994&1.000&1.000&1.000&1.000&\\
\cline{2-14}
&$(50,50)$   &&0.094&0.129&0.232&0.415&0.615&0.784&0.915&0.978&0.997&0.998&1.000&\\
&$(100,50)$  &&0.085&0.165&0.391&0.691&0.913&0.989&0.998&1.000&1.000&1.000&1.000&\\
&$(200,50)$  &&0.073&0.235&0.643&0.941&0.996&1.000&1.000&1.000&1.000&1.000&1.000&\\
&$(500,50)$  &&0.052&0.451&0.960&1.000&1.000&1.000&1.000&1.000&1.000&1.000&1.000&\\
\cline{2-14}
&$(50,100)$  &&0.089&0.151&0.360&0.693&0.901&0.985&0.999&1.000&1.000&1.000&1.000&\\
&$(100,100)$ &&0.076&0.256&0.685&0.956&0.997&1.000&1.000&1.000&1.000&1.000&1.000&\\
&$(200,100)$ &&0.073&0.381&0.925&1.000&1.000&1.000&1.000&1.000&1.000&1.000&1.000&\\
&$(500,100)$ &&0.059&0.737&1.000&1.000&1.000&1.000&1.000&1.000&1.000&1.000&1.000&\\
\hline\hline
\end{tabular}
\begin{tablenotes}
      \small
      \item[\dag] The significance level $\alpha = 5\%$.
    \end{tablenotes}
\end{threeparttable}
\end{table}%

\setlength{\tabcolsep}{3.2pt}
\begin{table}[!htbp]
\centering
\begin{threeparttable}
\renewcommand{\arraystretch}{1.2}
\caption{Rejection rates of testing linearity of $\alpha(\cdot)$ and $\beta(\cdot)$ when $\theta = 1$ and $\rho = 0.3$\tnote{\dag}}\label{Tab: RejectionRateLinear}
\begin{tabular}{ccccccccccccccc}
\hline\hline
&\multirow{2}{*}{(N,T)}&&\multicolumn{11}{c}{$\delta$}&\\
\cline{4-14}
&&&$0$&$0.01$&$0.02$&$0.03$&$0.04$&$0.05$&$0.06$&$0.07$&$0.08$&$0.09$&$0.1$&\\
\cline{2-14}
&$(50,10)$   &&0.086&0.097&0.158&0.288&0.464&0.641&0.801&0.910&0.963&0.990&0.998&\\
&$(100,10)$  &&0.080&0.130&0.309&0.565&0.839&0.962&0.993&1.000&1.000&1.000&1.000&\\
&$(200,10)$  &&0.058&0.181&0.555&0.932&0.995&1.000&1.000&1.000&1.000&1.000&1.000&\\
&$(500,10)$  &&0.038&0.397&0.963&1.000&1.000&1.000&1.000&1.000&1.000&1.000&1.000&\\
\cline{2-14}
&$(50,50)$   &&0.093&0.248&0.669&0.965&0.999&1.000&1.000&1.000&1.000&1.000&1.000&\\
&$(100,50)$  &&0.100&0.443&0.966&1.000&1.000&1.000&1.000&1.000&1.000&1.000&1.000&\\
&$(200,50)$  &&0.070&0.771&1.000&1.000&1.000&1.000&1.000&1.000&1.000&1.000&1.000&\\
&$(500,50)$  &&0.047&0.998&1.000&1.000&1.000&1.000&1.000&1.000&1.000&1.000&1.000&\\
\cline{2-14}
&$(50,100)$  &&0.096&0.459&0.971&1.000&1.000&1.000&1.000&1.000&1.000&1.000&1.000&\\
&$(100,100)$ &&0.085&0.846&1.000&1.000&1.000&1.000&1.000&1.000&1.000&1.000&1.000&\\
&$(200,100)$ &&0.066&0.994&1.000&1.000&1.000&1.000&1.000&1.000&1.000&1.000&1.000&\\
&$(500,100)$ &&0.057&1.000&1.000&1.000&1.000&1.000&1.000&1.000&1.000&1.000&1.000&\\
\hline\hline
\end{tabular}
\begin{tablenotes}
      \small
      \item[\dag] The significance level $\alpha = 5\%$.
    \end{tablenotes}
\end{threeparttable}
\end{table}%
\newpage
\section{Additional Empirical Results}\label{App: Sec: F}
In this appendix, we provide additional results for Section \ref{Sec: 8}. First, we provide formulas for various $R^2$'s. The formulas for the in-sample fits by excluding $\hat{\alpha}(z_{i,t-1})$ are as follows:
\begin{align}
	R^2_{f}& = 1-\frac{\sum_{i, t}[y_{it}- \hat{\beta}(z_{i,t-1})^{\prime}\hat{f}_t]^2}{\sum_{i, t} y_{it}^{2}}, \label{Eqn: R24}\\
	R^2_{f, T,N} & = 1 - \frac{1}{N} \sum_{i} \frac{\sum_{t}[y_{it}- \hat{\beta}(z_{i,t-1})^{\prime}\hat{f}_t]^2}{\sum_{t}y_{it}^{2}},\label{Eqn: R52}\\
	R^2_{f, N,T} & = 1 - \frac{1}{T} \sum_{t} \frac{\sum_{i}[y_{it}- \hat{\beta}(z_{i,t-1})^{\prime}\hat{f}_t]^2}{\sum_{i}y_{it}^{2}}.\label{Eqn: R26}
\end{align}
Three versions of out-of-sample predictive $R^2$'s are as follows:
\begin{align}
	R^2_O & = 1-\frac{\sum_{i, t\geq 120}[y_{it}- \hat{\alpha}_{t-1}(z_{i,t-1}) - \hat{\beta}_{t-1}(z_{i,t-1})^{\prime}\hat{\lambda}_t]^2}{\sum_{i, t\geq 120} y_{it}^{2}}, \label{Eqn: R21Predictive}\\
	R^2_{T,N,O} & = 1 - \frac{1}{N} \sum_{i} \frac{\sum_{t\geq 120}[y_{it}- \hat{\alpha}_{t-1}(z_{i,t-1}) - \hat{\beta}_{t-1}(z_{i,t-1})^{\prime}\hat{\lambda}_t]^2}{\sum_{t\geq 120}y_{it}^{2}},\label{Eqn: R22Predictive}\\
	R^2_{N,T,O} & = 1 - \frac{1}{T-120} \sum_{t\geq 120} \frac{\sum_{i}[y_{it}- \hat{\alpha}_{t-1}(z_{i,t-1})- \hat{\beta}_{t-1}(z_{i,t-1})^{\prime}\hat{\lambda}_t]^2}{\sum_{i}y_{it}^{2}}.\label{Eqn: R23Predictive}
\end{align}
Three versions of out-of-sample fit $R^2$'s are as follows:
\begin{align}
	R^2_{f,O} & = 1-\frac{\sum_{i, t\geq 120}[y_{it}- \hat{\beta}_{t-1}(z_{i,t-1})^{\prime}{\hat f_{t-1,t}}]^2}{\sum_{i, t\geq 120} y_{it}^{2}}, \label{Eqn: R21OOS}\\
	R^2_{f,T,N,O} & = 1 - \frac{1}{N} \sum_{i} \frac{\sum_{t\geq 120}[y_{it}-  \hat{\beta}_{t-1}(z_{i,t-1})^{\prime}\hat f_{t-1,t}]^2}{\sum_{t\geq 120}y_{it}^{2}},\label{Eqn: R22OOS}\\
	R^2_{f, N,T,O} & = 1 - \frac{1}{T-120} \sum_{t\geq 120} \frac{\sum_{i}[y_{it}- \hat{\beta}_{t-1}(z_{i,t-1})^{\prime}\hat f_{t-1,t}]^2}{\sum_{i}y_{it}^{2}}.\label{Eqn: R23OOS}
\end{align}

\subsection{Model Estimation}\label{App: Sec: F1}
Table \ref{Tab: factor_compare_app} show the first two moments of the long-short factors of \citet{FamaFrench_FiveFactor_2015} constructed based on the dataset used in this paper and the corresponding factors from Kenneth R. French’s Website as well as their correlations.

Table \ref{Tab: chars} collects characteristics used in different specifications.

Tables \ref{Tab: Emprical_apd1}-\ref{Tab: Emprical_apd3} present the model estimation results under the restriction $\alpha(\cdot) = 0$. Similar to the unrestricted cases, the nonlinear specifications demonstrate both better in-sample and out-of-sample performance than the linear specification in most cases. The linearity of $\beta(\cdot)$ is also rejected at the $1\%$ level in all cases. However, $R^2, R^2_{T,N}$, and $R^2_{N,T}$ are slightly smaller than those in the unrestricted cases. Unlike the unrestricted cases, increasing the number of factors may worsen the out-of-sample predictive $R^2$'s. Specifically, $R^{2}_{N,T,O}$ can become negative for $K\geq 4$ in the linear specification. In the nonlinear specifications, $R_{O}^{2}$ and $R_{O, T, N}^{2}$ exhibit a hump-shaped relationship with the number factors, peaking at three or four factors depending on the specification of $\beta(\cdot)$.



Table \ref{Tab: Emprical_apd10} shows the correlations between our factors and the long-short factors of \citet{FamaFrench_FiveFactor_2015}, and Tables \ref{Tab: Emprical_apd14}-\ref{Tab: Emprical_apd13} present the projection regression results. Our findings reveal substantial correlations between these factors and our factors, with both sets explaining significant variations in each other. However, the long-short factors fail to price most of our factors derived from our nonlinear specifications, while the pricing errors for most of the factors from our linear specification are not statistically significant. This suggests that the nonlinear specifications capture additional common variations in stock returns not accounted by the long-short factors. It is also observed that our factors are unable to fully explain the cross-sectional variations of some long-short factors. This observation arises because long-short factors are mixed with pricing errors, as they do not effectively distinguish between the risk and mispricing explanations of the role of characteristics in predicting stock returns, as discussed in Section \ref{SubSec:Comparefactor}.

\begin{figure}[!htbp]
	\centering
	\includegraphics[width=14.5cm, height=16cm]{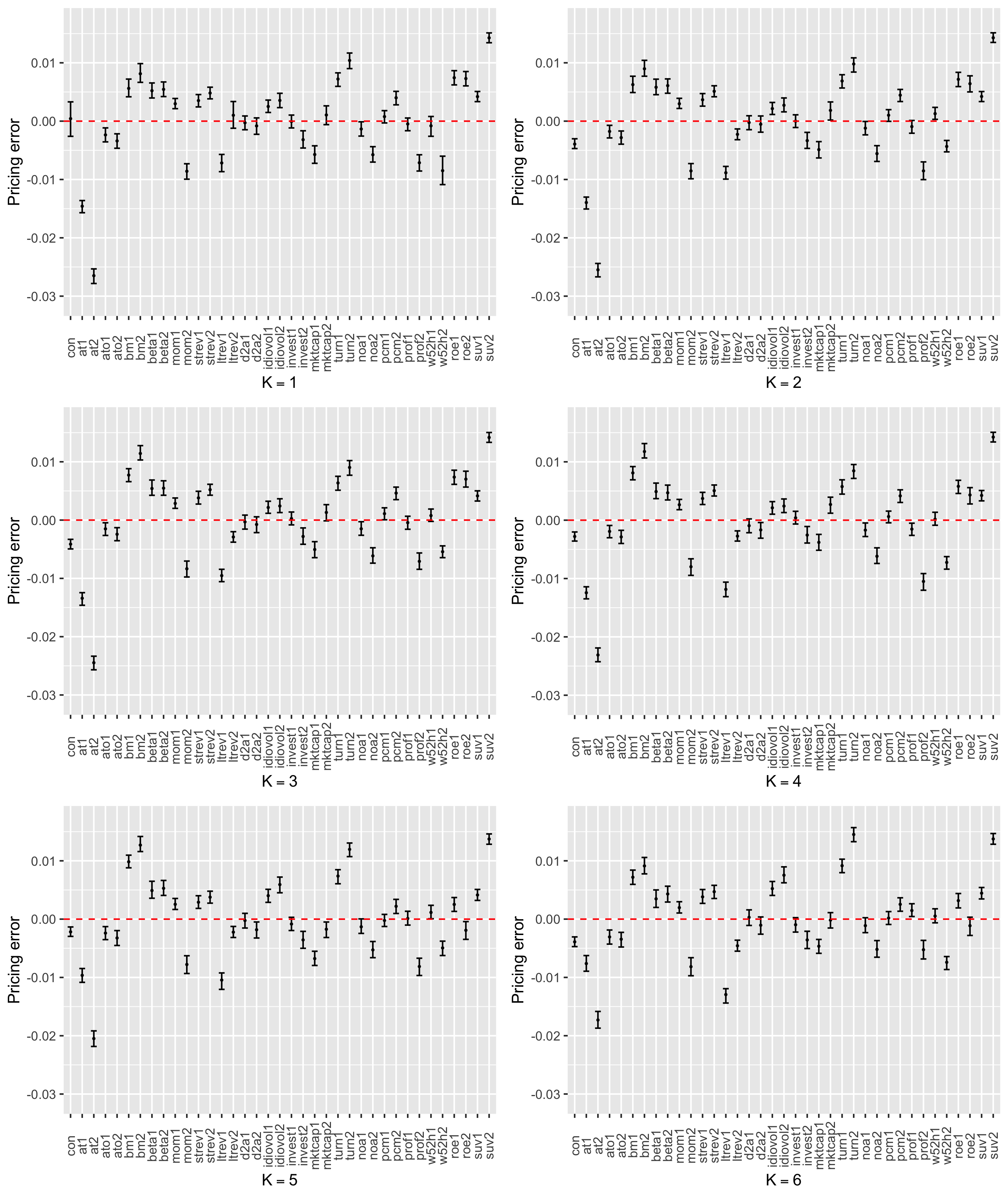}
\caption{$95\%$ confidence intervals for coefficients in $\alpha(\cdot)$ under nonlinear specifications of $\alpha(\cdot)$ and $\beta(\cdot)$ with 18 characteristics}\label{Fig: Emprical_apd17}
\end{figure}

\begin{figure}[!htbp]
	\centering
	\includegraphics[width=14.5cm, height=16cm]{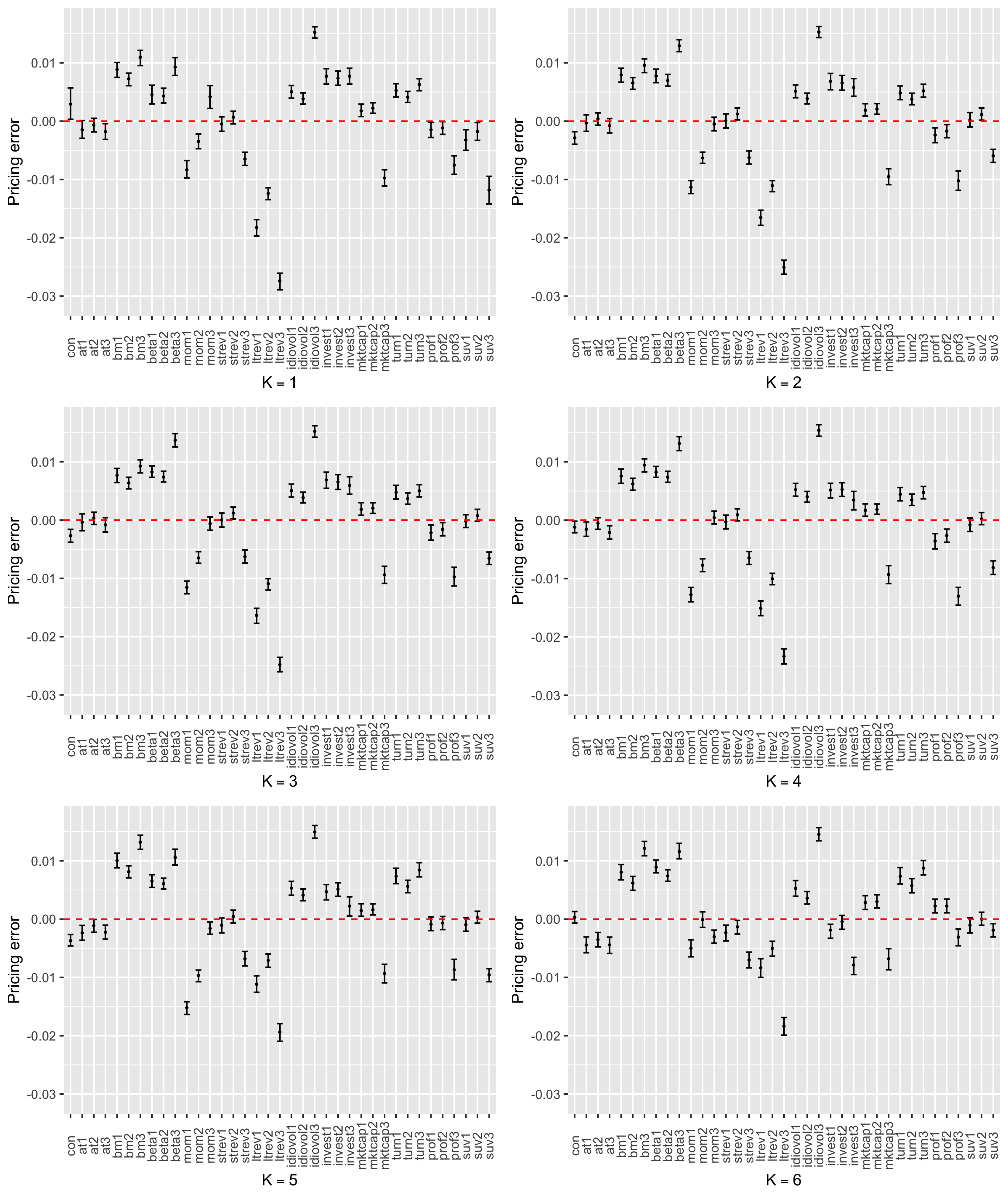}
\caption{$95\%$ confidence intervals for coefficients in $\alpha(\cdot)$ under nonlinear specifications of $\alpha(\cdot)$ and $\beta(\cdot)$ with 12 characteristics}\label{Fig: Emprical_apd18}
\end{figure}

\begin{figure}[!htbp]
	\centering
	\includegraphics[width=14.5cm, height=16cm]{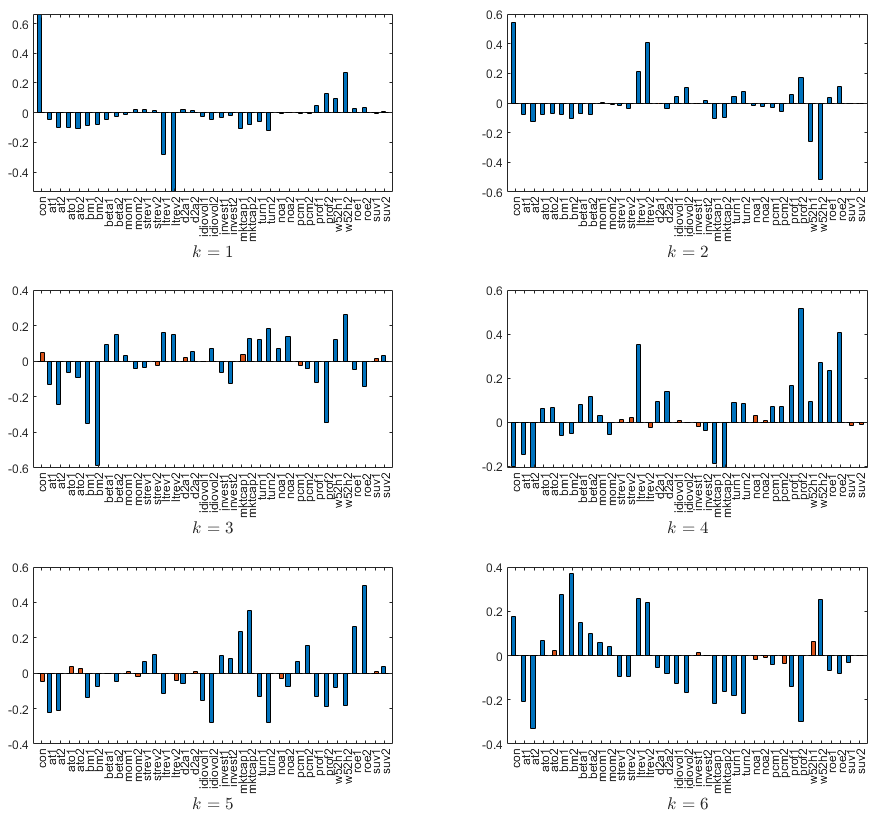}
	\caption{Estimates of coefficients in $\beta(\cdot)$ under nonlinear specifications of $\alpha(\cdot)$ and $\beta(\cdot)$ with 18 characteristics (blue: significant at the $5\%$ level; red: insignificant)}\label{Fig: Emprical_apd19}
\end{figure}

\begin{figure}[!htbp]
	\centering
	\includegraphics[width=14.5cm, height=16cm]{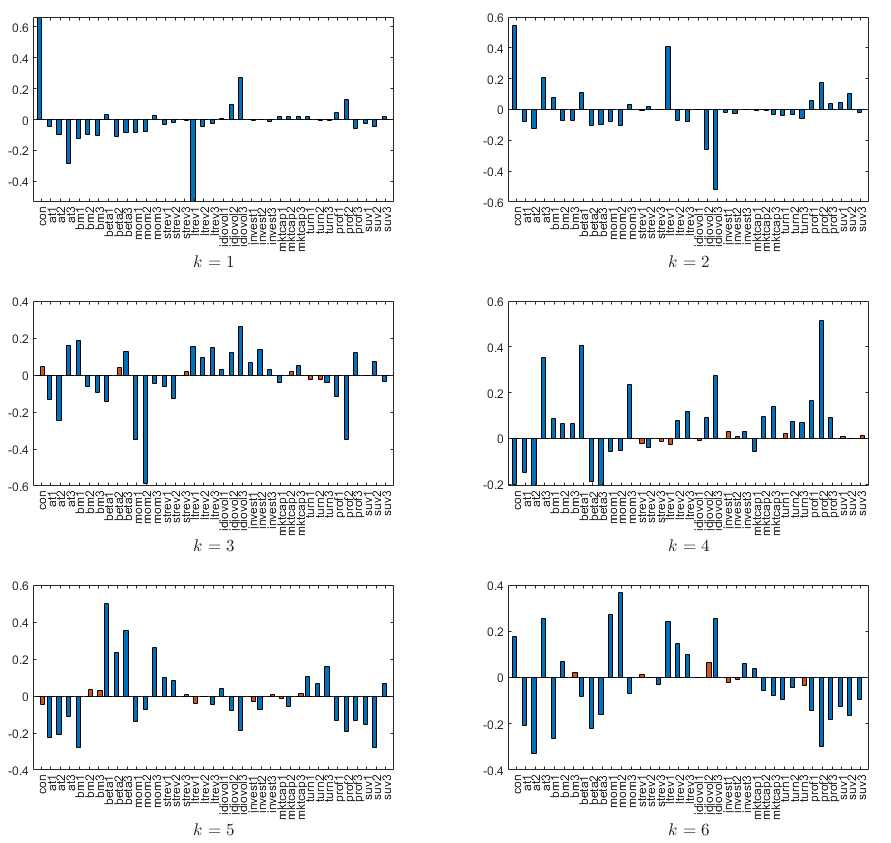}
	\caption{Estimates of coefficients in $\beta(\cdot)$ under nonlinear specifications of $\alpha(\cdot)$ and $\beta(\cdot)$ with 12 characteristics (blue: significant at the $5\%$ level; red: insignificant)}\label{Fig: Emprical_apd20}
\end{figure}

Figures \ref{Fig: Emprical_apd17}-\ref{Fig: Emprical_apd20} illustrate the contribution of each characteristic to pricing errors and risk exposures under the two nonlinear specifications. Almost all sieve coefficients are significant, which aligns with the strong evidence of nonlinearity found in Tables \ref{Tab: Emprical1}-\ref{Tab: Emprical3}.

\begin{figure}[!htbp]
	\centering
	\includegraphics[width=14cm, height=19cm]{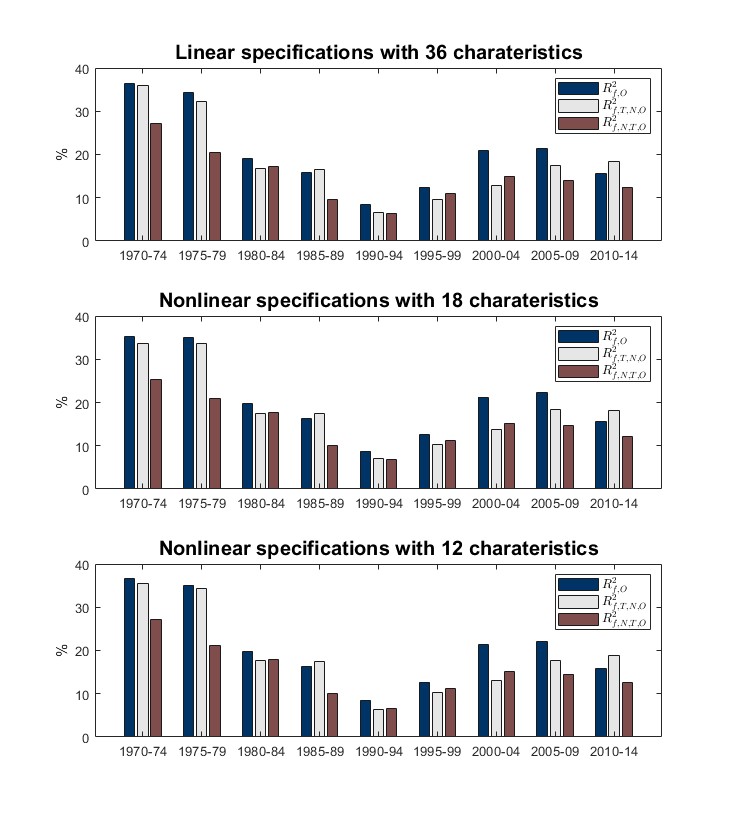}
	\caption{$R^2_{f,O}$, $R^2_{f,T,N,O}$, and $R^2_{f,N,T,O}$ (\eqref{Eqn: R21OOS}-\eqref{Eqn: R23OOS}) with $K=10$: subsample analysis}\label{Fig: Emprical_apd22}
\end{figure}

Figure \ref{Fig: Emprical_apd22} displays the patterns of $R^{2}_{f,O}$, $R^{2}_{f,T,N,O}$, and $R^{2}_{f,N,T,O}$ for ten factor models in subsample analysis. It shares the similar findings with Figure \ref{Fig: Empirical6}.

\subsection{Asset Pricing Tests}\label{App: Sec: F2}

The following characteristics are used in double sorted portfolios in Table \ref{Tab: factor_compare1_main} and this section. Size: market capitalization; BM: book-to-market equity ratio; OP: operating profitability; INV: growth rate of assets; MOM: momentum; Beta: market beta; AC: accruals; NI: net stock issues; Var: variance of daily total returns.

Table \ref{Tab: corr_testasset_app} shows the bilateral correlations and standard deviations of the testing portfolios in our asset pricing tests.

Tables \ref{Tab: factor_compare3} and \ref{Tab: factor_compare3_continue} present the results for five groups of testing portfolios with $K=6$ (corresponding to Tables \ref{Tab: factor_compare1_main} and \ref{Tab: factor_compare1_main_continue}), where ``MOM'' represent the momentum factor
Tables \ref{Tab: factor_compare_add_K5_1}-\ref{Tab: factor_compare_add_K6_5} collect results for remaining testing portfolios, where Tables \ref{Tab: factor_compare_add_K5_1}-\ref{Tab: factor_compare_add_K5_5} correspond to $K=5$ and Tables \ref{Tab: factor_compare_add_K6_1}-\ref{Tab: factor_compare_add_K6_5} correspond to $K=6$.
In explaining the Fama-MacBeth managed portfolios, our factors and IPCA's factors continue to outperform others for $K=6$ in terms of average absolute intercepts. In explaining the sorted portfolios, our factors continue to outperform IPCA's factors for $K=6$, as evidenced by smaller pricing errors, $t$-statistics, and $GRS$ statistics. The finding of larger regression $R^2$'s persists. In explaining IPCA's managed portfolios, the inferior performance of IPCA's factors is also observed for $K=6$. The findings are robust for additional testing portfolios, as evidenced in Tables \ref{Tab: factor_compare_add_K5_1}-\ref{Tab: factor_compare_add_K6_5}.

\setlength{\tabcolsep}{8pt}
	\begin{table}[!htbp]
		\centering
		\begin{threeparttable}
			\renewcommand{\arraystretch}{1.6}
			\caption{Summary statistics of factors\tnote{\dag}}\label{Tab: factor_compare_app}
			\begin{tabular}{clccccccc}
				\hline\hline
				&Factors  & MKT & SMB & HML & RMW & CMA & MOM &\\
				\cline{2-8}
				&\multicolumn{8}{l}{Based on the dataset in \cite{Kellyetal_Characteristics_2019} }\\
				\cline{2-8}
				& Mean & 0.50 & 0.16 & 0.63 & 0.27 & 0.45 & 0.85 &  \\
				& Standard deviation & 4.64 & 4.04 & 2.74 & 2.06 & 2.04 & 4.59 &  \\
				& $t$ & 2.53 & 0.91 & 5.44 & 3.05 & 5.11 & 4.35 &  \\
				
				\cline{2-8}
				&\multicolumn{8}{l}{Kenneth R. French's Website}\\
				\cline{2-8}
				& Mean & 0.48 & 0.18 & 0.38 & 0.30 & 0.37 & 0.67 &  \\
				& Standard deviation & 4.61 & 3.06 & 2.91 & 2.32 & 2.00 & 4.38 &  \\
				& $t$ & 2.46 & 1.36 & 3.04 & 3.03 & 4.35 & 3.58 &  \\
				
				\cline{2-8}
				& Correlation & 0.99 & 0.84 & 0.70 & 0.24 & 0.85 & 0.95 &  \\
				
				\hline\hline
			\end{tabular}
			\begin{tablenotes}
				\small
			    \item[\dag] MKT: market excess return; SMB: small-minus-big factor; HML: high-minus-low factor; RMW: robust-minus-weak factor; CMA: conservative-minus-aggressive factor; MOM: momentum factor.
			\end{tablenotes}
		\end{threeparttable}
	\end{table}%

\setlength{\tabcolsep}{3pt}
\begin{table}[!htbp]
	\centering
	\begin{threeparttable}
		\renewcommand{\arraystretch}{1.0}
		\caption{Characteristics used in different specifications\tnote{\dag}}\label{Tab: chars}
		\begin{tabular}{clcccc}
			\hline\hline
			& & Linear specification & Nonlinear specification 1& Nonlinear specification 2 & \\
			\cline{2-5}
			& constant & $\checkmark$ &  $\checkmark$  &  $\checkmark$  &  \\
			& a2me & $\checkmark$ &   &   &  \\
			& assets & $\checkmark$ & $\checkmark$ & $\checkmark$ &  \\
			& ato & $\checkmark$ & $\checkmark$ &   &  \\
			& beta & $\checkmark$ & $\checkmark$ & $\checkmark$ &  \\
			& bidask & $\checkmark$ &   &   &  \\
			& bm & $\checkmark$ & $\checkmark$ & $\checkmark$ &  \\
			& c & $\checkmark$ &   &   &  \\
			& cto & $\checkmark$ &   &   &  \\
			& d2a & $\checkmark$ & $\checkmark$ &   &  \\
			& dpi2a & $\checkmark$ &   &   &  \\
			& e2p & $\checkmark$ &   &   &  \\
			& fc2y & $\checkmark$ &   &   &  \\
			& freecf & $\checkmark$ &   &   &  \\
			& idiovol & $\checkmark$ & $\checkmark$ & $\checkmark$ &  \\
			& intmom & $\checkmark$ &   &   &  \\
			& invest & $\checkmark$ & $\checkmark$ & $\checkmark$ &  \\
			& lev & $\checkmark$ &   &   &  \\
			& ltrev  & $\checkmark$ & $\checkmark$ & $\checkmark$ &  \\
			& mktcap & $\checkmark$ & $\checkmark$ & $\checkmark$ &  \\
			& mom & $\checkmark$ & $\checkmark$ & $\checkmark$ &  \\
			& noa & $\checkmark$ & $\checkmark$ &   &  \\
			& oa & $\checkmark$ &   &   &  \\
			& ol & $\checkmark$ &   &   &  \\
			& pcm & $\checkmark$ & $\checkmark$ &   &  \\
			& pm & $\checkmark$ &   &   &  \\
			& prof & $\checkmark$ & $\checkmark$ & $\checkmark$ &  \\
			& q & $\checkmark$ &   &   &  \\
			& rna & $\checkmark$ &   &   &  \\
			& roa & $\checkmark$ &   &   &  \\
			& roe & $\checkmark$ & $\checkmark$ &   &  \\
			& s2p & $\checkmark$ &   &   &  \\
			& sga2s & $\checkmark$ &   &   &  \\
			& strev  & $\checkmark$ & $\checkmark$ & $\checkmark$ &  \\
			& suv & $\checkmark$ & $\checkmark$ & $\checkmark$ &  \\
			& turn & $\checkmark$ & $\checkmark$ & $\checkmark$ &  \\
			& w52h & $\checkmark$ & $\checkmark$ &   &  \\
			\hline\hline
		\end{tabular}
	\end{threeparttable}
\end{table}%

\setlength{\tabcolsep}{3.3pt}
\begin{table}[!htbp]
	\centering
	\begin{threeparttable}
		\renewcommand{\arraystretch}{1.6}
		\caption{Results under linear specification of $\beta(\cdot)$ with 36 characteristics\tnote{\dag}}\label{Tab: Emprical_apd1}
		\begin{tabular}{clccccccccccc}
			\hline\hline
			&\multicolumn{11}{c}{Restricted ($\alpha(\cdot) = 0$)}&\\
			\cline{2-12}
			&$K$&$R^{2}_{K}$&$R^{2}$&$R^{2}_{T,N}$&$R^{2}_{N,T}$&$R^{2}_{f,O}$&$R^{2}_{f,T,N,O}$&$R^{2}_{f,N,T,O}$&$R^{2}_{O}$&$R^{2}_{T,N,O}$&$R^{2}_{N,T,O}$\\
			\cline{2-12}
			&1&26.62&2.14&0.58&0.06 &6.79&	4.10&	5.98&0.20&0.09&0.07\\
			&2&36.48&4.18&1.72&1.37&13.66	&10.55&	11.33&0.28&0.34&0.02 \\
			&3&45.10&5.32&2.98&2.30 &14.20	&11.17&	11.77&0.26&0.31&0.01\\
			&4&52.62&11.45&8.03&8.86&14.74	&12.16&	12.16&0.31&0.39&-0.01\\
			&5&58.72&11.69&8.18&9.10&15.13	&12.70&	12.48&0.36&0.47&-0.04\\
			&6&64.28&13.85&10.06&11.58&15.32&12.96& 12.69&0.38&0.47&-0.11\\
			&7&69.26&15.20&11.71&13.17 &15.58&13.18& 12.96&0.40&0.50&-0.13\\
			&8&72.98&15.53&11.99&13.44&15.90&13.46& 13.24&0.41& 0.53&-0.13\\
			&9&76.40&15.73&12.15&13.68&16.25&13.96& 13.50&0.40& 0.53&-0.08\\
			&10&79.29&15.90&12.37&13.85&16.42&14.21& 13.70& 0.41&0.51&-0.06\\
			\cline{2-12}
			&$K$&$R^{2}_{\tilde Y}$& $R^{2}_{f}$&$R^{2}_{f,T,N}$&$R^{2}_{f,N,T}$&$p_{\alpha}$&$p_{\text{lin}}$\\
			\cline{2-12}
			&1-10&20.89&$R^{2}$&$R^{2}_{T,N}$&$R^{2}_{N,T}$ &NA&$<1\%$&\\
			\hline\hline
		\end{tabular}
		\begin{tablenotes}
			\small
            \item[\dag] $K$: the number of factors specified; $R^{2}_{\tilde Y}$: Fama-MacBeth cross-sectional regression $R^2$ ($\%$); $R^2_{K}$: the variation of the Fama-MacBeth managed portfolios $\tilde{Y}_t$ captured by the extracted factors $\hat{f}_t$ ($\%$);  $R^{2}$, $R^{2}_{T,N}$, $R^{2}_{N,T}$: various in-sample $R^2$'s ($\%$), see \eqref{Eqn: R21}-\eqref{Eqn: R23} with $\hat{\alpha}(\cdot) = 0$; $R^{2}_{f}$, $R^{2}_{f,T,N}$, $R^{2}_{f,N,T}$: various in-sample $R^2$'s without $\alpha(\cdot)$ ($\%$), see \eqref{Eqn: R24}-\eqref{Eqn: R26}; $R^{2}_{f,O}$, $R^{2}_{f,T,N,O}$, $R^{2}_{f,N,T,O}$: various out-of-sample fit $R^2$'s ($\%$), see \eqref{Eqn: R21OOS}-\eqref{Eqn: R23OOS};  $R^{2}_O$, $R^{2}_{T,N,O}$, $R^{2}_{N,T,O}$: various out-of-sample predictive $R^2$'s ($\%$), see \eqref{Eqn: R21Predictive}-\eqref{Eqn: R23Predictive} with $\hat{\alpha}_{t-1}(\cdot) = 0$; $p_{\alpha}$ and $p_{\text{lin}}$: the $p$-values of \textit{alpha} test ($\alpha(\cdot)=0$) and model specification test (linearity of $\beta(\cdot)$), respectively.
		\end{tablenotes}
	\end{threeparttable}
\end{table}%

\setlength{\tabcolsep}{3.3pt}
\begin{table}[!htbp]
	\centering
	\begin{threeparttable}
		\renewcommand{\arraystretch}{1.6}
		\caption{Results under nonlinear specification of $\beta(\cdot)$ with 18 characteristics\tnote{\dag}}\label{Tab: Emprical_apd2}
		\begin{tabular}{clccccccccccc}
			\hline\hline
			&\multicolumn{11}{c}{Restricted ($\alpha(\cdot) = 0$)}&\\
			\cline{2-12}
			&$K$&$R^{2}_{K}$&$R^{2}$&$R^{2}_{T,N}$&$R^{2}_{N,T}$&$R^{2}_{f,O}$&$R^{2}_{f,T,N,O}$&$R^{2}_{f,N,T,O}$&$R^{2}_{O}$&$R^{2}_{T,N,O}$&$R^{2}_{N,T,O}$\\
			\cline{2-12}
			&1  & 41.75         & 5.61  & 3.00  & 3.14  &11.32&	7.84&	8.99        & 0.30  & 0.34  & -0.12   \\
			&2 & 59.20 & 9.14  & 5.56  & 6.26  &14.01	&11.40	&11.34      & 0.34  & 0.30  & -0.38   \\
			&3   & 65.00        & 9.80  & 6.24  & 7.12  &14.70&	12.15&	12.08      & 0.60  & 0.76  & 0.29     \\
			&4   & 70.17        & 10.79 & 7.23  & 8.37  & 15.24&	13.00	&12.66     & 0.60  & 0.80  & 0.19  \\
			&5   & 74.44        & 14.28 & 10.57 & 11.98 &16.12&	13.86&	13.31      & 0.52  & 0.66  & 0.29      \\
			&6  & 77.39         & 14.58 & 10.88 & 12.18 &16.38&	14.16&	13.62      & 0.52  & 0.63  & 0.22       \\
			&7   & 80.12        & 14.91 & 11.07 & 12.61 &16.82&	14.79&	13.86      & 0.53  & 0.58  & 0.22       \\
			&8  & 82.36         & 15.43 & 11.93 & 13.17 &17.01&	14.87&	14.06      & 0.54  & 0.57  & 0.27       \\
			&9  & 84.34         & 15.80 & 12.28 & 13.45 & 17.18&	15.08&	14.22     & 0.53  & 0.54  & 0.27    \\
			&10    & 86.23      & 15.94 & 12.37 & 13.59 & 17.35&	15.22&	14.37     & 0.53  & 0.54  & 0.27    \\
			\cline{2-12}
			&$K$&$R^{2}_{\tilde Y}$& $R^{2}_{f}$&$R^{2}_{f,T,N}$&$R^{2}_{f,N,T}$&$p_{\alpha}$&$p_{\text{lin}}$\\
			\cline{2-12}
			&1-10&21.11&$R^{2}$&$R^{2}_{T,N}$&$R^{2}_{N,T}$ &NA&$<1\%$&\\
			\hline\hline
		\end{tabular}
		\begin{tablenotes}
			\small
            \item[\dag] $K$: the number of factors specified; $R^{2}_{\tilde Y}$: Fama-MacBeth cross-sectional regression $R^2$ ($\%$); $R^2_{K}$: the variation of the Fama-MacBeth managed portfolios $\tilde{Y}_t$ captured by the extracted factors $\hat{f}_t$ ($\%$);  $R^{2}$, $R^{2}_{T,N}$, $R^{2}_{N,T}$: various in-sample $R^2$'s ($\%$), see \eqref{Eqn: R21}-\eqref{Eqn: R23} with $\hat{\alpha}(\cdot) = 0$; $R^{2}_{f}$, $R^{2}_{f,T,N}$, $R^{2}_{f,N,T}$: various in-sample $R^2$'s without $\alpha(\cdot)$ ($\%$), see \eqref{Eqn: R24}-\eqref{Eqn: R26}; $R^{2}_{f,O}$, $R^{2}_{f,T,N,O}$, $R^{2}_{f,N,T,O}$: various out-of-sample fit $R^2$'s ($\%$), see \eqref{Eqn: R21OOS}-\eqref{Eqn: R23OOS};  $R^{2}_O$, $R^{2}_{T,N,O}$, $R^{2}_{N,T,O}$: various out-of-sample predictive $R^2$'s ($\%$), see \eqref{Eqn: R21Predictive}-\eqref{Eqn: R23Predictive} with $\hat{\alpha}_{t-1}(\cdot) = 0$; $p_{\alpha}$ and $p_{\text{lin}}$: the $p$-values of \textit{alpha} test ($\alpha(\cdot)=0$) and model specification test (linearity of $\beta(\cdot)$), respectively.
		\end{tablenotes}
	\end{threeparttable}
\end{table}%

\setlength{\tabcolsep}{3.3pt}
\begin{table}[!htbp]
	\centering
	\begin{threeparttable}
		\renewcommand{\arraystretch}{1.6}
		\caption{Results under nonlinear specification of $\beta(\cdot)$ with 12 characteristics\tnote{\dag}}\label{Tab: Emprical_apd3}
		\begin{tabular}{clccccccccccc}
			\hline\hline
			&\multicolumn{11}{c}{Restricted ($\alpha(\cdot) = 0$)}&\\
			\cline{2-12}
			&$K$&$R^{2}_{K}$&$R^{2}$&$R^{2}_{T,N}$&$R^{2}_{N,T}$&$R^{2}_{f,O}$&$R^{2}_{f,T,N,O}$&$R^{2}_{f,N,T,O}$&$R^{2}_{O}$&$R^{2}_{T,N,O}$&$R^{2}_{N,T,O}$\\
			\cline{2-12}
			&1   & 42.95  & 5.34  & 2.59  & 2.90  & 11.15&	7.66&	8.84& 0.32  & 0.34  & -0.10\\
			&2  & 61.58   & 9.12  & 5.45  & 6.15  & 13.79&	11.11&	11.04& 0.33  & 0.21  & -0.56\\
			&3   & 68.02   & 10.15 & 6.08  & 7.25  & 14.47&	11.86&	11.74& 0.62  & 0.68  & 0.16\\
			&4  & 74.08      & 10.77 & 6.99  & 7.98  & 15.38&	13.20&	12.75& 0.57  & 0.65  & 0.24 \\
			&5  & 78.98   & 14.15 & 10.49 & 11.93 &16.04	&14.23	&13.34& 0.55  & 0.57  & 0.23 \\
			&6  & 82.66     & 14.43 & 10.68 & 12.32 & 16.48&	14.71&	13.67& 0.56  & 0.53  & 0.23\\
			&7     & 85.44   & 14.93 & 11.31 & 12.76 & 16.81&	14.97&	13.93& 0.55  & 0.55  & 0.25\\
			&8   & 87.85    & 15.37 & 11.78 & 13.13 & 17.11&	15.10	&14.14& 0.56  & 0.54  & 0.27\\
			&9  & 89.53    & 16.28 & 12.57 & 13.85  & 17.30&	15.34	&14.33 & 0.56  & 0.52  & 0.27\\
			&10    & 91.13   & 16.49 & 12.78 & 14.08 &17.45&	15.52	&14.48& 0.57  & 0.55  & 0.27\\
			\cline{2-12}
			&$K$&$R^{2}_{\tilde Y}$& $R^{2}_{f}$&$R^{2}_{f,T,N}$&$R^{2}_{f,N,T}$&$p_{\alpha}$&$p_{\text{lin}}$\\
			\cline{2-12}
			&1-10&20.72&$R^{2}$&$R^{2}_{T,N}$&$R^{2}_{N,T}$ &NA&$<1\%$&\\
			\hline\hline
		\end{tabular}
		\begin{tablenotes}
			\small
            \item[\dag] $K$: the number of factors specified; $R^{2}_{\tilde Y}$: Fama-MacBeth cross-sectional regression $R^2$ ($\%$); $R^2_{K}$: the variation of the Fama-MacBeth managed portfolios $\tilde{Y}_t$ captured by the extracted factors $\hat{f}_t$ ($\%$);  $R^{2}$, $R^{2}_{T,N}$, $R^{2}_{N,T}$: various in-sample $R^2$'s ($\%$), see \eqref{Eqn: R21}-\eqref{Eqn: R23} with $\hat{\alpha}(\cdot) = 0$; $R^{2}_{f}$, $R^{2}_{f,T,N}$, $R^{2}_{f,N,T}$: various in-sample $R^2$'s without $\alpha(\cdot)$ ($\%$), see \eqref{Eqn: R24}-\eqref{Eqn: R26}; $R^{2}_{f,O}$, $R^{2}_{f,T,N,O}$, $R^{2}_{f,N,T,O}$: various out-of-sample fit $R^2$'s ($\%$), see \eqref{Eqn: R21OOS}-\eqref{Eqn: R23OOS};  $R^{2}_O$, $R^{2}_{T,N,O}$, $R^{2}_{N,T,O}$: various out-of-sample predictive $R^2$'s ($\%$), see \eqref{Eqn: R21Predictive}-\eqref{Eqn: R23Predictive} with $\hat{\alpha}_{t-1}(\cdot) = 0$; $p_{\alpha}$ and $p_{\text{lin}}$: the $p$-values of \textit{alpha} test ($\alpha(\cdot)=0$) and model specification test (linearity of $\beta(\cdot)$),              respectively.
		\end{tablenotes}
	\end{threeparttable}
\end{table}%

\setlength{\tabcolsep}{11pt}
\begin{table}[!htbp]
	\centering
	\begin{threeparttable}
		\renewcommand{\arraystretch}{1.1}
		\caption{Factor correlations\tnote{\dag}}\label{Tab: Emprical_apd10}
		\begin{tabular}{clccccccc}
			\hline\hline
			&&MKT&SMB&HML&RMW&CMA&MOM&\\
			\cline{2-8}
			\multicolumn{9}{c}{Linear specifications of $\alpha(\cdot)$ and $\beta(\cdot)$ with 36 characteristics}\\
			\cline{2-8}
			&Factor  1 & 0.02  & 0.11  & 0.10  & -0.11 & 0.02  & -0.36  &\\
			&Factor  2 & 0.26  & 0.26  & 0.05  & -0.20 & -0.02 & -0.16  & \\
			&Factor  3 & -0.26 & -0.22 & -0.05 & 0.11  & 0.07  & 0.18   &\\
			&Factor  4 & 0.58  & 0.46  & -0.33 & -0.30 & -0.30 & -0.06  &\\
			&Factor  5 & 0.10  & 0.05  & -0.15 & 0.01  & -0.16 & 0.05   &\\
			&Factor  6 & 0.32  & 0.24  & -0.13 & -0.20 & -0.13 & -0.04  &\\
			&Factor  7 & 0.30  & 0.19  & 0.02  & -0.09 & -0.10 & -0.16  & \\
			&Factor  8 & -0.04 & -0.05 & 0.21  & 0.14  & 0.16  & -0.41  &\\
			&Factor  9 & 0.02  & 0.03  & -0.08 & -0.06 & -0.03 & 0.29   & \\
			&Factor  10 & -0.03 & 0.01  & 0.13  & 0.09  & 0.08  & 0.01 &\\
			\cline{2-8}
			\multicolumn{9}{c}{Nonlinear specifications of $\alpha(\cdot)$ and $\beta(\cdot)$ with 18 characteristics}\\
			\cline{2-8}
			&Factor  1 & 0.24  & 0.37  & 0.07  & -0.30 & -0.02 & -0.37 & \\
			&Factor  2 & 0.41  & 0.31  & -0.37 & -0.30 & -0.29 & -0.09 &\\
			&Factor  3 & -0.22 & -0.11 & 0.45  & 0.23  & 0.33  & -0.50 & \\
			&Factor  4 & 0.48  & 0.21  & -0.04 & -0.15 & -0.18 & -0.34 &\\
			&Factor  5 & -0.12 & -0.01 & -0.25 & -0.06 & -0.10 & 0.14  &\\
			&Factor  6 & 0.07  & -0.21 & -0.04 & 0.08  & -0.08 & 0.09  & \\
			&Factor  7 & -0.15 & -0.22 & -0.09 & -0.06 & -0.03 & -0.12 & \\
			&Factor  8 & -0.01 & -0.08 & 0.14  & 0.09  & 0.05  & 0.05  & \\
			&Factor  9 & -0.17 & 0.03  & 0.08  & 0.10  & 0.09  & 0.10  &\\
			&Factor  10 & -0.21 & -0.04 & 0.11  & 0.06  & 0.06  & 0.07 & \\
			\cline{2-8}
			\multicolumn{9}{c}{Nonlinear specifications of $\alpha(\cdot)$ and $\beta(\cdot)$ with 12 characteristics}\\
			\cline{2-8}
			&Factor  1 & 0.22  & 0.36  & 0.05  & -0.32 & -0.01 & -0.32 &\\
			&Factor  2 & 0.39  & 0.31  & -0.27 & -0.26 & -0.26 & -0.34 &\\
			&Factor  3 & -0.19 & -0.19 & 0.49  & 0.25  & 0.30  & -0.57 & \\
			&Factor  4 & 0.50  & 0.21  & -0.12 & -0.16 & -0.24 & -0.13 &\\
			&Factor  5 & -0.04 & -0.15 & -0.11 & -0.02 & -0.07 & 0.12  &\\
			&Factor  6 & -0.20 & -0.20 & -0.14 & 0.06  & -0.07 & 0.08  &\\
			&Factor  7 & -0.05 & -0.04 & 0.16  & 0.12  & 0.04  & 0.00  &\\
			&Factor  8 & -0.04 & 0.19  & 0.28  & 0.06  & 0.22  & 0.18  &\\
			&Factor  9 & 0.27  & 0.06  & 0.18  & -0.14 & 0.16  & 0.11  & \\
			&Factor  10 & 0.17  & -0.27 & -0.22 & 0.21  & -0.20 & -0.06 &\\
			\hline\hline
		\end{tabular}
		\begin{tablenotes}
			\small
				\item[\dag]  MKT: market excess return; SMB: small-minus-big factor; HML: high-minus-low factor; RMW: robust-minus-weak factor; CMA: conservative-minus-aggressive factor; MOM: momentum factor.
		\end{tablenotes}
	\end{threeparttable}
\end{table}%

\begin{landscape}
	\setlength{\tabcolsep}{4.5pt}
	\begin{table}[!htbp]
		\centering
		\begin{threeparttable}
			\renewcommand{\arraystretch}{1.6}
			\caption{Factor projections: linear specifications of $\alpha(\cdot)$ and $\beta(\cdot)$  with 36 characteristics\tnote{\dag}}\label{Tab: Emprical_apd14}
			\begin{tabular}{ccccccccccccc}
				\hline\hline
				& & Factor 1 & Factor 2 & Factor 3 & Factor 4 & Factor 5 & Factor 6 & Factor 7 & Factor 8 & Factor 9 & Factor 10 &\\
				\cline{2-12}	
				& Constant & 3.71*** & 0.14 & 0.39 & -0.05 & 0.25 & 0.28 & 0.46 & 0.62*** & 0.04 & -0.10 &  \\
				& $t$ & [5.83] & [0.36] & [1.03] & [-0.17] & [0.75] & [0.94] & [1.63] & [2.71] & [0.17] & [-0.42] &  \\
				& MKT & -0.20 & 0.43*** & -0.42*** & 0.79*** & 0.09 & 0.40*** & 0.37*** & 0.02 & 0.04 & 0.02 &  \\
				& $t$ & [-1.31] & [4.61] & [-4.76] & [12.30] & [1.12] & [5.80] & [5.64] & [0.40] & [0.73] & [0.32] &  \\
				& SMB & 0.42** & 0.54*** & -0.46*** & 0.80*** & 0.07 & 0.31*** & 0.27*** & 0.02 & -0.00 & 0.09 &  \\
				& $t$ & [1.98] & [4.07] & [-3.68] & [8.73] & [0.64] & [3.11] & [2.87] & [0.24] & [-0.01] & [1.14] &  \\
				& HML & 0.31 & 0.38** & -0.46** & -0.61*** & -0.15 & -0.13 & 0.37*** & -0.04 & 0.03 & 0.25** &  \\
				& $t$ & [1.02] & [2.03] & [-2.57] & [-4.69] & [-0.93] & [-0.89] & [2.72] & [-0.41] & [0.24] & [2.30] &  \\
				& RMW & -0.39 & -0.33* & -0.02 & -0.16 & 0.10 & -0.24* & 0.03 & 0.47*** & -0.20* & 0.19* &  \\
				& $t$ & [-1.34] & [-1.83] & [-0.11] & [-1.31] & [0.68] & [-1.81] & [0.27] & [4.59] & [-1.88] & [1.83] &  \\
				& CMA & -0.17 & 0.01 & 0.28 & 0.26 & -0.33 & 0.11 & -0.31 & 0.57*** & -0.11 & -0.03 &  \\
				& $t$ & [-0.37] & [0.04] & [1.06] & [1.33] & [-1.41] & [0.51] & [-1.53] & [3.51] & [-0.62] & [-0.19] &  \\
				& MOM & -1.20*** & -0.19** & 0.22*** & -0.02 & 0.08 & 0.02 & -0.13** & -0.57*** & 0.38*** & 0.04 &  \\
				& $t$ & [-8.43] & [-2.14] & [2.64] & [-0.28] & [1.13] & [0.24] & [-1.98] & [-11.10] & [7.00] & [0.78] &  \\
                \cline{2-12}
				& $R^2$ & 14.82 & 13.77 & 12.47 & 46.37 & 3.23 & 13.87 & 12.74 & 23.66 & 9.23 & 2.56 &  \\
				
				\hline\hline
			\end{tabular}
			\begin{tablenotes}
				\small
				\item[\dag]  MKT: market excess return; SMB: small-minus-big factor; HML: high-minus-low factor; RMW: robust-minus-weak factor; CMA: conservative-minus-aggressive factor; MOM: momentum factor; $^{\ast\ast\ast}$: $p$-value $<1\%$; $^{\ast\ast}$: $p$-value $<5\%$; $^{\ast}$: $p$-value $<10\%$.
			\end{tablenotes}
		\end{threeparttable}
	\end{table}%
\end{landscape}

\begin{landscape}
	\setlength{\tabcolsep}{4.5pt}
	\begin{table}[!htbp]
		\centering
		\begin{threeparttable}
			\renewcommand{\arraystretch}{1.6}
			\caption{Factor projections: nonlinear specifications of $\alpha(\cdot)$ and $\beta(\cdot)$ with 18 characteristics\tnote{\dag}}
			\label{Tab: Emprical_apd15}
			\begin{tabular}{ccccccccccccc}
				\hline\hline
				& & Factor 1 & Factor 2 & Factor 3 & Factor 4 & Factor 5 & Factor 6 & Factor 7 & Factor 8 & Factor 9 & Factor 10 &\\
				\cline{2-12}
				
				& Constant & 2.96*** & 1.03*** & 0.43** & 0.61*** & 1.49*** & 0.95*** & 0.82*** & 2.25*** & 0.81*** & 2.26*** &  \\
				& $t$ & [5.21] & [2.79] & [2.42] & [3.31] & [8.57] & [5.51] & [5.53] & [14.77] & [5.70] & [16.25] &  \\
				& MKT & 0.36*** & 0.55*** & -0.14*** & 0.44*** & -0.18*** & 0.10** & -0.15*** & 0.04 & -0.10*** & -0.15*** &  \\
				& $t$ & [2.68] & [6.36] & [-3.26] & [10.14] & [-4.38] & [2.54] & [-4.26] & [1.14] & [-3.02] & [-4.55] &  \\
				& SMB & 1.33*** & 0.51*** & 0.03 & 0.15** & -0.02 & -0.31*** & -0.30*** & -0.07 & 0.12*** & 0.02 &  \\
				& $t$ & [6.98] & [4.11] & [0.52] & [2.33] & [-0.34] & [-5.28] & [-5.98] & [-1.34] & [2.60] & [0.47] &  \\
				& HML & 0.45* & -1.11*** & 0.35*** & 0.18** & -0.41*** & 0.08 & -0.24*** & 0.25*** & 0.05 & 0.17** &  \\
				& $t$ & [1.67] & [-6.33] & [4.19] & [2.02] & [-4.98] & [0.96] & [-3.38] & [3.45] & [0.73] & [2.58] &  \\
				& RMW & -0.95*** & -0.51*** & 0.57*** & -0.00 & -0.16** & 0.01 & -0.26*** & 0.06 & 0.13** & -0.01 &  \\
				& $t$ & [-3.69] & [-3.08] & [7.04] & [-0.03] & [-2.01] & [0.17] & [-3.85] & [0.93] & [2.06] & [-0.12] &  \\
				& CMA & 0.06 & 0.34 & 0.45*** & -0.18 & 0.02 & -0.19 & 0.04 & -0.14 & 0.02 & -0.21** &  \\
				& $t$ & [0.15] & [1.30] & [3.53] & [-1.37] & [0.14] & [-1.58] & [0.37] & [-1.30] & [0.19] & [-2.13] &  \\
				& MOM & -1.03*** & -0.20** & -0.63*** & -0.28*** & 0.07* & 0.09** & -0.14*** & 0.07** & 0.06* & 0.05 &  \\
				& $t$ & [-8.06] & [-2.44] & [-15.77] & [-6.75] & [1.68] & [2.35] & [-4.18] & [2.03] & [1.87] & [1.62] &  \\
                \cline{2-12}
				& $R^2$ & 29.29 & 30.86 & 48.20 & 31.51 & 11.52 & 7.45 & 13.83 & 3.81 & 5.15 & 5.93 &  \\
				\hline\hline
			\end{tabular}
			\begin{tablenotes}
				\small
				\item[\dag]  MKT: market excess return; SMB: small-minus-big factor; HML: high-minus-low factor; RMW: robust-minus-weak factor; CMA: conservative-minus-aggressive factor; MOM: momentum factor; $^{\ast\ast\ast}$: $p$-value $<1\%$; $^{\ast\ast}$: $p$-value $<5\%$; $^{\ast}$: $p$-value $<10\%$.
			\end{tablenotes}
		\end{threeparttable}
	\end{table}%
\end{landscape}

\begin{landscape}
	\setlength{\tabcolsep}{4.5pt}
	\begin{table}[!htbp]
		\centering
		\begin{threeparttable}
			\renewcommand{\arraystretch}{1.6}
			\caption{Factor projections: nonlinear specifications of $\alpha(\cdot)$ and $\beta(\cdot)$ with 12 characteristics\tnote{\dag}}\label{Tab: Emprical_apd16}
			\begin{tabular}{ccccccccccccc}
				\hline\hline
				& & Factor 1 & Factor 2 & Factor 3 & Factor 4 & Factor 5 & Factor 6 & Factor 7 & Factor 8 & Factor 9 & Factor 10 &\\
				\cline{2-12}				
				& Constant & 3.57*** & 1.99*** & 0.27 & 0.47** & 1.43*** & 2.67*** & 2.43*** & 1.93*** & -0.16 & 0.36*** &  \\
				& $t$ & [5.47] & [4.83] & [1.37] & [2.08] & [6.36] & [14.11] & [12.86] & [12.03] & [-1.19] & [2.70] &  \\
				& MKT & 0.36** & 0.51*** & -0.11** & 0.56*** & -0.05 & -0.22*** & -0.01 & 0.04 & 0.30*** & 0.16*** &  \\
				& $t$ & [2.38] & [5.23] & [-2.41] & [10.38] & [-1.03] & [-4.90] & [-0.16] & [1.17] & [9.24] & [5.17] &  \\
				& SMB & 1.39*** & 0.68*** & -0.17*** & 0.14* & -0.30*** & -0.25*** & 0.02 & 0.33*** & -0.05 & -0.30*** &  \\
				& $t$ & [6.35] & [4.93] & [-2.62] & [1.89] & [-3.94] & [-3.88] & [0.30] & [6.22] & [-1.07] & [-6.72] &  \\
				& HML & 0.39 & -1.03*** & 0.64*** & 0.17 & -0.12 & -0.27*** & 0.34*** & 0.45*** & 0.27*** & -0.28*** &  \\
				& $t$ & [1.25] & [-5.26] & [6.83] & [1.60] & [-1.10] & [-2.99] & [3.75] & [5.90] & [4.16] & [-4.46] &  \\
				& RMW & -1.28*** & -0.32* & 0.64*** & -0.06 & -0.21** & -0.08 & 0.19** & 0.18** & -0.16*** & 0.28*** &  \\
				& $t$ & [-4.32] & [-1.72] & [7.16] & [-0.59] & [-2.02] & [-0.90] & [2.25] & [2.43] & [-2.60] & [4.66] &  \\
				& CMA & 0.19 & 0.24 & 0.27* & -0.31* & -0.18 & -0.11 & -0.25* & 0.03 & 0.24** & 0.07 &  \\
				& $t$ & [0.40] & [0.80] & [1.91] & [-1.93] & [-1.09] & [-0.82] & [-1.86] & [0.25] & [2.43] & [0.71] &  \\
				& MOM & -1.01*** & -0.86*** & -0.83*** & -0.06 & 0.12** & 0.01 & 0.03 & 0.22*** & 0.16*** & -0.07** &  \\
				& $t$ & [-6.89] & [-9.29] & [-18.86] & [-1.17] & [2.36] & [0.28] & [0.79] & [6.01] & [5.34] & [-2.46] &  \\
                \cline{2-12}
				& $R^2$ & 26.23 & 33.78 & 57.02 & 26.17 & 5.62 & 10.69 & 4.49 & 19.40 & 21.20 & 20.97 &  \\
				\hline\hline
			\end{tabular}
			\begin{tablenotes}
				\small
				\item[\dag]  MKT: market excess return; SMB: small-minus-big factor; HML: high-minus-low factor; RMW: robust-minus-weak factor; CMA: conservative-minus-aggressive factor; MOM: momentum factor; $^{\ast\ast\ast}$: $p$-value $<1\%$; $^{\ast\ast}$: $p$-value $<5\%$; $^{\ast}$: $p$-value $<10\%$.
			\end{tablenotes}
		\end{threeparttable}
	\end{table}%
\end{landscape}

	\setlength{\tabcolsep}{7pt}
	\begin{table}[!htbp]
		\centering
		\begin{threeparttable}
			\renewcommand{\arraystretch}{1.5}
			\caption{Factor projections: \quad linear specifications of $\alpha(\cdot)$ and $\beta(\cdot)$  with 36 characteristics\tnote{\dag}}\label{Tab: Emprical_apd11}
			\begin{tabular}{ccccccccccccc}
				\hline\hline
				& & MKT & SMB & HML & RMW & CMA & MOM &\\
				\cline{2-8}
				& Constant & 0.12 & -0.05 & 0.32*** & 0.40*** & 0.39*** & 1.20*** &  \\
				& $t$ & [1.05] & [-0.46] & [2.83] & [4.45] & [4.84] & [8.46] &  \\
				& Factor 1 & 0.01 & 0.02*** & 0.02*** & -0.02*** & 0.00 & -0.11*** &  \\
				& $t$ & [0.82] & [3.30] & [2.77] & [-2.79] & [0.74] & [-11.65] &  \\
				& Factor 2 & 0.13*** & 0.08*** & 0.02 & -0.05*** & -0.00 & -0.07*** &  \\
				& $t$ & [10.58] & [7.92] & [1.35] & [-5.21] & [-0.43] & [-5.06] &  \\
				& Factor 3 & -0.14*** & -0.08*** & -0.02 & 0.03*** & 0.01 & 0.09*** &  \\
				& $t$ & [-10.75] & [-6.63] & [-1.30] & [2.77] & [1.63] & [5.72] &  \\
				& Factor 4 & 0.33*** & 0.17*** & -0.12*** & -0.08*** & -0.07*** & -0.03* &  \\
				& $t$ & [23.52] & [13.95] & [-8.79] & [-7.83] & [-7.64] & [-1.79] &  \\
				& Factor 5 & 0.06*** & 0.02 & -0.06*** & 0.00 & -0.04*** & 0.03* &  \\
				& $t$ & [4.11] & [1.34] & [-3.85] & [0.15] & [-3.99] & [1.66] &  \\
				& Factor 6 & 0.22*** & 0.11*** & -0.05*** & -0.07*** & -0.04*** & -0.02 &  \\
				& $t$ & [13.25] & [7.35] & [-3.43] & [-5.38] & [-3.23] & [-1.09] &  \\
				& Factor 7 & 0.21*** & 0.09*** & 0.01 & -0.03** & -0.03** & -0.10*** &  \\
				& $t$ & [12.14] & [5.78] & [0.69] & [-2.40] & [-2.53] & [-4.98] &  \\
				& Factor 8 & -0.03 & -0.03 & 0.11*** & 0.06*** & 0.06*** & -0.32*** &  \\
				& $t$ & [-1.63] & [-1.55] & [5.51] & [3.66] & [4.31] & [-13.25] &  \\
				& Factor 9 & 0.02 & 0.02 & -0.04** & -0.03* & -0.01 & 0.23*** &  \\
				& $t$ & [1.00] & [0.83] & [-2.19] & [-1.66] & [-0.93] & [9.10] &  \\
				& Factor 10 & -0.03 & 0.01 & 0.08*** & 0.04** & 0.03** & 0.01 &  \\
				& $t$ & [-1.31] & [0.29] & [3.45] & [2.20] & [2.03] & [0.45] &  \\
                \cline{2-8}
				& $R^2$ & 67.68 & 42.93 & 23.21 & 22.88 & 18.01 & 47.39 &  \\		
				\hline\hline
			\end{tabular}
			\begin{tablenotes}
				\small
				\item[\dag]  MKT: market excess return; SMB: small-minus-big factor; HML: high-minus-low factor; RMW: robust-minus-weak factor; CMA: conservative-minus-aggressive factor; MOM: momentum factor; $^{\ast\ast\ast}$: $p$-value $<1\%$; $^{\ast\ast}$: $p$-value $<5\%$; $^{\ast}$: $p$-value $<10\%$.
			\end{tablenotes}
		\end{threeparttable}
	\end{table}%

	\setlength{\tabcolsep}{7pt}
	\begin{table}[!htbp]
		\centering
		\begin{threeparttable}
			\renewcommand{\arraystretch}{1.5}
			\caption{Factor projections: nonlinear specifications of $\alpha(\cdot)$ and $\beta(\cdot)$ with 18 characteristics\tnote{\dag}}\label{Tab: Emprical_apd12}
			\begin{tabular}{ccccccccccccc}
				\hline\hline
				& & MKT & SMB & HML & RMW & CMA & MOM &\\
				\cline{2-8}
				& Constant & 1.04*** & 0.33** & 0.06 & 0.19 & 0.32*** & 0.63*** &  \\
				& $t$ & [5.58] & [2.14] & [0.44] & [1.46] & [2.83] & [3.36] &  \\
				& Factor 1 & 0.07*** & 0.08*** & 0.01** & -0.05*** & -0.00 & -0.11*** &  \\
				& $t$ & [8.98] & [11.00] & [2.24] & [-8.34] & [-0.16] & [-12.95] &  \\
				& Factor 2 & 0.20*** & 0.10*** & -0.11*** & -0.07*** & -0.06*** & -0.04*** &  \\
				& $t$ & [15.60] & [9.20] & [-11.81] & [-8.19] & [-7.78] & [-3.27] &  \\
				& Factor 3 & -0.19*** & -0.06*** & 0.25*** & 0.10*** & 0.12*** & -0.41*** &  \\
				& $t$ & [-8.27] & [-3.32] & [14.35] & [6.54] & [9.10] & [-17.69] &  \\
				& Factor 4 & 0.45*** & 0.13*** & -0.02 & -0.07*** & -0.07*** & -0.30*** &  \\
				& $t$ & [18.03] & [6.31] & [-0.90] & [-4.22] & [-4.90] & [-11.88] &  \\
				& Factor 5 & -0.13*** & -0.01 & -0.18*** & -0.03* & -0.06*** & 0.15*** &  \\
				& $t$ & [-4.44] & [-0.31] & [-8.09] & [-1.68] & [-3.12] & [5.03] &  \\
				& Factor 6 & 0.08** & -0.17*** & -0.03 & 0.05** & -0.04** & 0.10*** &  \\
				& $t$ & [2.47] & [-6.33] & [-1.18] & [2.33] & [-2.25] & [3.06] &  \\
				& Factor 7 & -0.20*** & -0.20*** & -0.08*** & -0.04* & -0.01 & -0.15*** &  \\
				& $t$ & [-5.72] & [-6.62] & [-2.97] & [-1.81] & [-0.51] & [-4.12] &  \\
				& Factor 8 & -0.01 & -0.07** & 0.12*** & 0.06** & 0.03 & 0.07* &  \\
				& $t$ & [-0.39] & [-2.45] & [4.35] & [2.39] & [1.44] & [1.83] &  \\
				& Factor 9 & -0.24*** & 0.03 & 0.07** & 0.07*** & 0.06** & 0.14*** &  \\
				& $t$ & [-6.32] & [1.03] & [2.53] & [2.81] & [2.54] & [3.51] &  \\
				& Factor 10 & -0.31*** & -0.04 & 0.10*** & 0.05* & 0.04* & 0.10** &  \\
				& $t$ & [-8.00] & [-1.24] & [3.49] & [1.82] & [1.74] & [2.51] &  \\
                \cline{2-8}
				& $R^2$ & 62.06 & 39.35 & 46.33 & 29.54 & 26.49 & 56.75 &  \\
				\hline\hline
			\end{tabular}
			\begin{tablenotes}
				\small
				\item[\dag]  MKT: market excess return; SMB: small-minus-big factor; HML: high-minus-low factor; RMW: robust-minus-weak factor; CMA: conservative-minus-aggressive factor; MOM: momentum factor; $^{\ast\ast\ast}$: $p$-value $<1\%$; $^{\ast\ast}$: $p$-value $<5\%$; $^{\ast}$: $p$-value $<10\%$.
			\end{tablenotes}
		\end{threeparttable}
	\end{table}%

	\setlength{\tabcolsep}{7pt}
	\begin{table}[!htbp]
		\centering
		\begin{threeparttable}
			\renewcommand{\arraystretch}{1.5}
			\caption{Factor projections: nonlinear specifications of $\alpha(\cdot)$ and $\beta(\cdot)$ with 12 characteristics\tnote{\dag}} \label{Tab: Emprical_apd13}
			\begin{tabular}{ccccccccccccc}
				\hline\hline
				& & MKT & SMB & HML & RMW & CMA & MOM &\\
				\cline{2-8}
				& Constant & 0.45** & 0.03 & 0.01 & 0.18 & 0.29*** & 0.44** &  \\
				& $t$ & [2.53] & [0.20] & [0.11] & [1.50] & [2.83] & [2.58] &  \\
				& Factor 1 & 0.06*** & 0.07*** & 0.01* & -0.04*** & 0.00 & -0.09*** &  \\
				& $t$ & [8.55] & [11.46] & [1.89] & [-9.17] & [0.01] & [-12.37] &  \\
				& Factor 2 & 0.16*** & 0.09*** & -0.07*** & -0.05*** & -0.05*** & -0.14*** &  \\
				& $t$ & [14.71] & [10.03] & [-9.30] & [-7.35] & [-7.39] & [-13.05] &  \\
				& Factor 3 & -0.13*** & -0.09*** & 0.22*** & 0.09*** & 0.10*** & -0.38*** &  \\
				& $t$ & [-7.18] & [-6.19] & [16.92] & [7.23] & [8.85] & [-21.59] &  \\
				& Factor 4 & 0.40*** & 0.11*** & -0.06*** & -0.06*** & -0.08*** & -0.10*** &  \\
				& $t$ & [18.92] & [6.65] & [-3.94] & [-4.54] & [-6.67] & [-5.05] &  \\
				& Factor 5 & -0.04 & -0.09*** & -0.07*** & -0.01 & -0.03** & 0.11*** &  \\
				& $t$ & [-1.60] & [-4.85] & [-3.91] & [-0.34] & [-2.26] & [4.60] &  \\
				& Factor 6 & -0.21*** & -0.14*** & -0.09*** & 0.03* & -0.03* & 0.08*** &  \\
				& $t$ & [-7.53] & [-6.33] & [-4.86] & [1.78] & [-1.96] & [3.01] &  \\
				& Factor 7 & -0.06** & -0.03 & 0.11*** & 0.07*** & 0.02 & 0.00 &  \\
				& $t$ & [-1.98] & [-1.35] & [5.39] & [3.80] & [1.36] & [0.17] &  \\
				& Factor 8 & -0.05 & 0.15*** & 0.21*** & 0.03 & 0.11*** & 0.20*** &  \\
				& $t$ & [-1.45] & [6.11] & [9.56] & [1.61] & [6.05] & [6.75] &  \\
				& Factor 9 & 0.38*** & 0.05* & 0.15*** & -0.10*** & 0.09*** & 0.15*** &  \\
				& $t$ & [10.35] & [1.84] & [6.12] & [-4.24] & [4.32] & [4.22] &  \\
				& Factor 10 & 0.25*** & -0.25*** & -0.20*** & 0.15*** & -0.13*** & -0.07** &  \\
				& $t$ & [6.57] & [-8.67] & [-7.56] & [6.02] & [-5.84] & [-2.03] &  \\
                \cline{2-8}
				& $R^2$ & 62.97 & 47.96 & 54.55 & 34.67 & 34.05 & 62.92 &  \\
				\hline\hline
			\end{tabular}
			\begin{tablenotes}
				\small
				\item[\dag]  MKT: market excess return; SMB: small-minus-big factor; HML: high-minus-low factor; RMW: robust-minus-weak factor; CMA: conservative-minus-aggressive factor; MOM: momentum factor; $^{\ast\ast\ast}$: $p$-value $<1\%$; $^{\ast\ast}$: $p$-value $<5\%$; $^{\ast}$: $p$-value $<10\%$.
			\end{tablenotes}
		\end{threeparttable}
	\end{table}%

\begin{landscape}
	\setlength{\tabcolsep}{1.5pt}
	\begin{table}[!htbp]
		\centering
        \adjustbox{max width=1.4\textwidth, max height=1.4\textheight}{
		\begin{threeparttable}
			\renewcommand{\arraystretch}{1.6}
			\caption{Bilateral correlations and standard deviations of testing portfolios\tnote{\dag}}\label{Tab: corr_testasset_app}
			\begin{tabular}{clccccccc|cccc}
				\hline\hline
				& &\multicolumn{7}{c}{Correlation} &\multicolumn{3}{c}{Standard deviation}&\\
				\cline{2-12}
				& Testing portfolios & Min & Median & Max & AMean & AMin & AMedian & AMax & Min & Median &  Max & \\
				\cline{2-12}
				& Regressed-PCA's 36 managed portfolios  & -0.87 & -0.01 & 0.63 & 0.09 & 0.00 & 0.06 & 0.87 & 0.01 & 0.04 & 0.12 &  \\
				& 100 double sorted portfolios on Size and BM, OP, INV, and MOM & 0.53 & 0.85 & 0.99 & 0.84 & 0.53 & 0.85 & 0.99 & 0.04 & 0.06 & 0.08 &  \\
				& 110 double sorted portfolios on Size and Beta, AC, NI, and Var& 0.42 & 0.85 & 0.98 & 0.83 & 0.42 & 0.85 & 0.98 & 0.04 & 0.06 & 0.09 &  \\
				& IPCA's 36 managed portfolios & -0.97 & -0.00 & 0.98 & 0.36 & 0.00 & 0.33 & 0.98 & 0.00 & 0.00 & 0.01 &   \\
				& 110 single sorted portfolios on 55 characteristics (P1\&10)  & 0.34 & 0.80 & 1.00 & 0.79 & 0.34 & 0.80 & 1.00 & 0.04 & 0.06 & 0.09 &  \\
				& 72 single sorted portfolios on 36 characteristics (P1\&10)  & 0.39 & 0.75 & 1.00 & 0.74 & 0.39 & 0.75 & 1.00 & 0.04 & 0.06 & 0.10 &  \\				
				& 25 double sorted portfolios on Size and BM & 0.53 & 0.84 & 0.97 & 0.82 & 0.53 & 0.84 & 0.97 & 0.04 & 0.06 & 0.08 &  \\
				& 25 double sorted portfolios on Size and OP & 0.64 & 0.86 & 0.97 & 0.86 & 0.64 & 0.86 & 0.97 & 0.04 & 0.06 & 0.07 &  \\
				& 25 double sorted portfolios on Size and INV & 0.63 & 0.87 & 0.97 & 0.85 & 0.63 & 0.87 & 0.97 & 0.04 & 0.06 & 0.07 &  \\
				& 25 double sorted portfolios on Size and MOM & 0.54 & 0.82 & 0.97 & 0.81 & 0.54 & 0.82 & 0.97 & 0.04 & 0.06 & 0.08 &  \\
				& 25 double sorted portfolios on Size and Beta & 0.65 & 0.89 & 0.97 & 0.86 & 0.65 & 0.89 & 0.97 & 0.04 & 0.06 & 0.07 &  \\
				& 25 double sorted portfolios on Size and AC & 0.44 & 0.83 & 0.97 & 0.81 & 0.44 & 0.83 & 0.97 & 0.04 & 0.06 & 0.08 &  \\
				& 35 double sorted portfolios on Size and NI & 0.47 & 0.84 & 0.97 & 0.83 & 0.47 & 0.84 & 0.97 & 0.04 & 0.05 & 0.08 &  \\
				& 25 double sorted portfolios on Size and Var & 0.46 & 0.84 & 0.97 & 0.83 & 0.46 & 0.84 & 0.97 & 0.04 & 0.06 & 0.09 & \\
				& Regressed-PCA S1's 36 managed portfolios & -0.77 & 0.00 & 0.83 & 0.12 & 0.00 & 0.08 & 0.83 & 0.01 & 0.02 & 0.09&  \\
				& Regressed-PCA S2's 36 managed portfolios &-0.74 & 0.01 & 0.96 & 0.16 & 0.00 & 0.10 & 0.96 & 0.01 & 0.03 & 0.09 &  \\
				\hline\hline
			\end{tabular}
			\begin{tablenotes}
				\small
				\item[\dag]  Min: minimum value; Median: median value; Max: maximal value; AMean: average absolute value; AMin: minimum absolute value; AMedian: median absolute value; AMax: maximal absolute value.
			\end{tablenotes}
		\end{threeparttable}
}
	\end{table}%
\end{landscape}

\begin{landscape}
	\setlength{\tabcolsep}{4.2pt}
	\begin{table}[!htbp]
		\centering
		\begin{threeparttable}
			\renewcommand{\arraystretch}{0.95}
			\caption{Comparing asset pricing tests: $K=6$\tnote{\dag}}\label{Tab: factor_compare3}
			\begin{tabular}{clcccccccccccccc}
				\hline\hline
				&Testing portfolios/Factors & $A|a|$ & $A|t(a)|$ & $Aa^2/V\overline{r}$ & $A\lambda^2/V\overline{r}$ & $As(a)$ & $As(e)$ & $Sh^2(a)$ & $Sh^2(f)$ & $AR^2$ & $GRS$ & $p(GRS)$&\\
				\cline{2-13}
				&\multicolumn{12}{l}{Group I: Regressed-PCA's 36 managed portfolios}\\
				\cline{2-13}			
				& Regressed-PCA & 0.39 & 3.51 & 0.64 & 0.61 & 0.12 & 2.75 & 1.97 & 0.04 & 33.27 & 26.64 & 0.00 \\
				& Regressed-PCA S1 & 0.41 & 3.08 & 0.53 & 0.47 & 0.17 & 3.60 & 2.47 & 0.22 & 19.43 & 28.49 & 0.00 \\
				& Regressed-PCA S2 & 0.40 & 2.76 & 0.50 & 0.43 & 0.19 & 3.68 & 2.48 & 0.42 & 17.68 & 24.52 & 0.00 \\
				& IPCA & 0.43 & 3.07 & 0.48 & 0.41 & 0.18 & 3.67 & 3.13 & 0.33 & 20.17 & 33.15 & 0.00 \\
				& IPCA$\setminus$Regressed-PCA & 0.53 & 3.22 & 1.02 & 0.94 & 0.19 & 3.92 & 3.14 & 0.28 & 14.33 & 34.41 & 0.00 \\
				& FF5+MOM & 0.52 & 3.18 & 1.28 & 1.21 & 0.18 & 3.95 & 2.59 & 0.17 & 10.83 & 31.28 & 0.00 \\
				& KNS & 0.50 & 3.35 & 1.12 & 1.05 & 0.17 & 3.95 & 2.55 & 0.05 & 11.25 & 34.33 & 0.00 \\
				
				\cline{2-13}
				&\multicolumn{12}{l}{Group II: 100 double sorted portfolios on Size and BM, OP, INV, and MOM}\\
				\cline{2-13}
				 & Regressed-PCA & 0.74 & 4.95 & 11.01 & 10.58 & 0.15 & 3.52 & 0.99 & 0.04 & 60.32 & 4.22 & 0.00 \\
				& Regressed-PCA S1 & 0.55 & 3.01 & 6.27 & 5.66 & 0.18 & 3.88 & 1.22 & 0.22 & 52.70 & 4.43 & 0.00 \\
				& Regressed-PCA S2 & 0.99 & 5.19 & 19.77 & 19.09 & 0.19 & 3.78 & 1.37 & 0.42 & 54.71 & 4.28 & 0.00 \\
				& IPCA & 0.86 & 9.66 & 14.83 & 14.65 & 0.10 & 1.96 & 3.92 & 0.33 & 86.71 & 13.05 & 0.00 \\
				& IPCA$\setminus$Regressed-PCA & 1.29 & 5.54 & 30.77 & 29.71 & 0.24 & 4.93 & 1.57 & 0.28 & 27.02 & 5.43 & 0.00  \\
				& FF5+MOM & 0.38 & 4.53 & 2.75 & 2.61 & 0.09 & 1.90 & 1.71 & 0.17 & 88.67 & 6.48 & 0.00 \\
				& KNS & 0.87 & 5.81 & 13.95 & 13.54 & 0.15 & 3.42 & 0.94 & 0.05 & 62.92 & 3.96 & 0.00 \\
				
				\cline{2-13}
				&\multicolumn{12}{l}{Group III: 110 double sorted portfolios on Size and Beta, AC, NI, and Var}\\
				\cline{2-13}
				& Regressed-PCA & 0.75 & 4.96 & 12.51 & 12.02 & 0.15 & 3.51 & 1.25 & 0.04 & 59.88 & 4.70 & 0.00 \\
				& Regressed-PCA S1 & 0.53 & 2.91 & 6.75 & 6.04 & 0.19 & 3.92 & 1.52 & 0.22 & 51.27 & 4.90 & 0.00 \\
				& Regressed-PCA S2 & 0.98 & 5.10 & 21.83 & 21.03 & 0.20 & 3.82 & 1.66 & 0.42 & 53.13 & 4.59 & 0.00 \\
				& IPCA & 0.88 & 9.52 & 16.84 & 16.62 & 0.10 & 2.00 & 3.94 & 0.33 & 85.66 & 11.65 & 0.00 \\
				& IPCA$\setminus$Regressed-PCA & 1.30 & 5.59 & 35.65 & 34.38 & 0.24 & 4.99 & 1.83 & 0.28 & 26.38 & 5.60 & 0.00  \\
				& FF5+MOM & 0.39 & 4.38 & 3.42 & 3.23 & 0.10 & 2.05 & 2.08 & 0.17 & 86.41 & 7.00 & 0.00 \\
				& KNS & 0.86 & 5.72 & 15.90 & 15.42 & 0.15 & 3.47 & 1.10 & 0.05 & 61.06 & 4.14 & 0.00 \\
				\hline\hline
			\end{tabular}
			\begin{tablenotes}
				\small
				\item[\dag]  $A|a|$: average absolute intercept;  $A|t(a)|$: average absolute $t$-statistic for the intercepts;  $Aa^2/V\overline{r}$: average squared intercept over the cross-section variance of  $\overline{r}$, average returns of the testing portfolios; $A\lambda^2/V\overline{r}$: average difference between each squared intercept and its squared standard error divided by the variance of $\overline{r}$; $As(a)$: average standard error of the intercepts; $As(e)$: average residual standard deviation; $Sh^2(a)$: maximized squared Sharpe ratio for the intercepts; $Sh^2(f)$: maximized squared Sharpe ratio for the factors; $AR^2$: average regression $R^2$ (\%);  $GRS$: $GRS$ statistic of \citet{Gibbonsetal_efficiency_1989}; $p(GRS)$: $p$-value of $GRS$.
			\end{tablenotes}
		\end{threeparttable}
	\end{table}%
\end{landscape}

\begin{landscape}
	\setlength{\tabcolsep}{4.2pt}
	\begin{table}[!htbp]
		\centering
		\begin{threeparttable}
			\renewcommand{\arraystretch}{0.95}
			\caption{Comparing asset pricing tests: $K=6$ (continued)\tnote{\dag}}\label{Tab: factor_compare3_continue}
			\begin{tabular}{clcccccccccccccc}
				\hline\hline
				&Testing portfolios/Factors & $A|a|$ & $A|t(a)|$ & $Aa^2/V\overline{r}$ & $A\lambda^2/V\overline{r}$ & $As(a)$ & $As(e)$ & $Sh^2(a)$ & $Sh^2(f)$ & $AR^2$ & $GRS$ &$p(GRS)$&\\
				\cline{2-13}
				&\multicolumn{12}{l}{Group IV: IPCA's 36 managed portfolios}\\
				\cline{2-13}
				  & Regressed-PCA & 0.05 & 3.60 & 0.91 & 0.86 & 0.01 & 0.32 & 1.55 & 0.04 & 28.06 & 20.98 & 0.00 \\
				 & Regressed-PCA S1 & 0.06 & 5.36 & 1.47 & 1.44 & 0.01 & 0.25 & 1.86 & 0.22 & 51.86 & 21.46 & 0.00 \\
				 & Regressed-PCA S2 & 0.07 & 4.92 & 1.45 & 1.41 & 0.01 & 0.25 & 1.74 & 0.42 & 52.31 & 17.24 & 0.00 \\
				 & IPCA & 0.06 & 5.77 & 1.24 & 1.22 & 0.01 & 0.21 & 2.29 & 0.33 & 66.51 & 24.27 & 0.00 \\
				 & IPCA$\setminus$Regressed-PCA & 0.06 & 5.06 & 1.38 & 1.33 & 0.01 & 0.31 & 1.92 & 0.28 & 40.25 & 21.06 & 0.00 \\
				 & FF5+MOM & 0.04 & 3.39 & 0.79 & 0.76 & 0.01 & 0.25 & 1.39 & 0.17 & 54.97 & 16.76 & 0.00 \\
				 & KNS & 0.04 & 3.83 & 0.76 & 0.73 & 0.01 & 0.25 & 1.46 & 0.05 & 55.27 & 19.69 & 0.00 \\
				
				\cline{2-13}
				&\multicolumn{12}{l}{Group V: 110 single sorted portfolios on 55 characteristics (P1\&10)}\\
				\cline{2-13}
				 & Regressed-PCA & 0.63 & 3.81 & 5.61 & 5.27 & 0.17 & 3.87 & 1.36 & 0.04 & 52.76 & 5.12 & 0.00 \\
				& Regressed-PCA S1 & 0.42 & 2.27 & 3.16 & 2.73 & 0.19 & 3.99 & 1.36 & 0.22 & 49.50 & 4.37 & 0.00 \\
				& Regressed-PCA S2 & 0.73 & 3.67 & 8.01 & 7.51 & 0.20 & 4.00 & 1.44 & 0.42 & 49.28 & 3.99 & 0.00 \\
				& IPCA & 0.73 & 5.87 & 7.55 & 7.34 & 0.13 & 2.65 & 2.83 & 0.33 & 77.17 & 8.37 & 0.00 \\
				& IPCA$\setminus$Regressed-PCA & 1.10 & 4.78 & 15.34 & 14.63 & 0.24 & 4.99 & 1.62 & 0.28 & 25.12 & 4.96 & 0.00 \\
				& FF5+MOM & 0.39 & 4.36 & 2.26 & 2.12 & 0.10 & 2.24 & 2.54 & 0.17 & 83.76 & 8.58 & 0.00 \\
				& KNS & 0.82 & 5.34 & 8.60 & 8.32 & 0.15 & 3.52 & 1.30 & 0.05 & 59.97 & 4.90 & 0.00 \\
				
				\cline{2-13}
				&\multicolumn{12}{l}{Group VI: 72 single sorted portfolios on 36 characteristics (P1\&10)}\\
				\cline{2-13}
			 & Regressed-PCA & 0.59 & 3.32 & 4.33 & 3.96 & 0.19 & 4.28 & 0.81 & 0.04 & 51.08 & 5.07 & 0.00 \\
			& Regressed-PCA S1 & 0.43 & 2.17 & 2.73 & 2.29 & 0.21 & 4.33 & 1.08 & 0.22 & 49.42 & 5.81 & 0.00 \\
			& Regressed-PCA S2 & 0.71 & 3.32 & 6.75 & 6.24 & 0.22 & 4.34 & 0.89 & 0.42 & 49.14 & 4.12 & 0.00 \\
			& IPCA & 0.67 & 4.77 & 5.76 & 5.51 & 0.15 & 3.06 & 1.74 & 0.33 & 74.73 & 8.57 & 0.00 \\
			& IPCA$\setminus$Regressed-PCA & 1.08 & 4.26 & 13.28 & 12.51 & 0.27 & 5.46 & 1.27 & 0.28 & 24.47 & 6.49 & 0.00 \\
			& FF5+MOM & 0.41 & 4.27 & 2.44 & 2.26 & 0.12 & 2.65 & 2.48 & 0.17 & 81.54 & 13.91 & 0.00 \\
			& KNS & 0.82 & 5.24 & 7.69 & 7.43 & 0.16 & 3.63 & 0.73 & 0.05 & 63.15 & 4.57 & 0.00 \\
				\hline\hline
			\end{tabular}
			\begin{tablenotes}
				\small
				\item[\dag]  $A|a|$: average absolute intercept;  $A|t(a)|$: average absolute $t$-statistic for the intercepts;  $Aa^2/V\overline{r}$: average squared intercept over the cross-section variance of  $\overline{r}$, average returns of the testing portfolios; $A\lambda^2/V\overline{r}$: average difference between each squared intercept and its squared standard error divided by the variance of $\overline{r}$; $As(a)$: average standard error of the intercepts; $As(e)$: average residual standard deviation; $Sh^2(a)$: maximized squared Sharpe ratio for the intercepts; $Sh^2(f)$: maximized squared Sharpe ratio for the factors; $AR^2$: average regression $R^2$ (\%);  $GRS$: $GRS$ statistic of \citet{Gibbonsetal_efficiency_1989}; $p(GRS)$: $p$-value of $GRS$.
			\end{tablenotes}
		\end{threeparttable}
	\end{table}%
\end{landscape}

\begin{landscape}
	\setlength{\tabcolsep}{4.2pt}
	\begin{table}[!htbp]
		\centering
		\begin{threeparttable}
			\renewcommand{\arraystretch}{0.95}
			\caption{Additional asset pricing tests: $K=5$\tnote{\dag}}\label{Tab: factor_compare_add_K5_1}
			\begin{tabular}{clcccccccccccccc}
				\hline\hline
				&Testing portfolios/Factors & $A|a|$ & $A|t(a)|$ & $Aa^2/V\overline{r}$ & $A\lambda^2/V\overline{r}$ & $As(a)$ & $As(e)$ & $Sh^2(a)$ & $Sh^2(f)$ & $AR^2$ & $GRS$ & $p(GRS)$&\\
				\cline{2-13}
				&\multicolumn{12}{l}{Group I: 25 double sorted portfolios on Size and BM}\\
				\cline{2-13}
				 & Regressed-PCA & 0.86 & 5.03 & 17.75 & 17.07 & 0.17 & 3.99 & 0.32 & 0.04 & 49.64 & 6.42 & 0.00 \\
				& Regressed-PCA S1 & 0.58 & 3.23 & 8.72 & 7.97 & 0.18 & 3.94 & 0.45 & 0.16 & 51.07 & 8.05 & 0.00 \\
				& Regressed-PCA S2 & 0.44 & 2.40 & 5.20 & 4.44 & 0.18 & 4.05 & 0.44 & 0.13 & 47.97 & 8.14 & 0.00 \\
				& IPCA & 0.90 & 10.41 & 19.45 & 19.25 & 0.09 & 2.04 & 1.83 & 0.10 & 85.43 & 34.55 & 0.00 \\
				& IPCA$\setminus$Regressed-PCA & 1.12 & 5.76 & 29.70 & 28.74 & 0.20 & 4.61 & 0.67 & 0.06 & 36.10 & 13.16 & 0.00 \\
				& FF5 & 0.39 & 4.57 & 3.67 & 3.47 & 0.09 & 2.04 & 0.69 & 0.12 & 86.80 & 12.88 & 0.00 \\
				& KNS & 0.97 & 6.19 & 22.03 & 21.48 & 0.16 & 3.61 & 0.30 & 0.04 & 58.51 & 5.99 & 0.00 \\
				
				\cline{2-13}
				&\multicolumn{12}{l}{Group II: 25 double sorted portfolios on Size and OP}\\
				\cline{2-13}
				 & Regressed-PCA & 0.84 & 5.04 & 27.72 & 26.64 & 0.17 & 3.89 & 0.22 & 0.04 & 51.55 & 4.32 & 0.00 \\
				& Regressed-PCA S1 & 0.50 & 2.86 & 10.83 & 9.63 & 0.18 & 3.87 & 0.28 & 0.16 & 52.03 & 4.93 & 0.00 \\
				& Regressed-PCA S2 & 0.37 & 2.08 & 5.86 & 4.63 & 0.18 & 3.98 & 0.28 & 0.13 & 49.34 & 5.07 & 0.00 \\
				& IPCA & 0.86 & 11.19 & 30.45 & 30.16 & 0.09 & 1.90 & 2.23 & 0.10 & 87.16 & 42.02 & 0.00 \\
				& IPCA$\setminus$Regressed-PCA & 1.09 & 5.59 & 46.29 & 44.74 & 0.20 & 4.55 & 0.54 & 0.06 & 35.88 & 10.64 & 0.00 \\
				& FF5 & 0.40 & 4.96 & 6.47 & 6.17 & 0.09 & 1.93 & 0.73 & 0.12 & 88.20 & 13.65 & 0.00 \\
				& KNS & 0.94 & 6.16 & 33.83 & 32.94 & 0.15 & 3.53 & 0.22 & 0.04 & 59.64 & 4.41 & 0.00 \\

                \cline{2-13}
				&\multicolumn{12}{l}{Group III: 25 double sorted portfolios on Size and INV}\\
				\cline{2-13}
				 & Regressed-PCA & 0.88 & 5.35 & 21.28 & 20.54 & 0.17 & 3.83 & 0.36 & 0.04 & 51.90 & 7.24 & 0.00 \\
				& Regressed-PCA S1 & 0.56 & 3.24 & 9.25 & 8.44 & 0.18 & 3.82 & 0.51 & 0.16 & 52.11 & 9.06 & 0.00 \\
				& Regressed-PCA S2 & 0.42 & 2.42 & 5.47 & 4.63 & 0.18 & 3.92 & 0.49 & 0.13 & 49.49 & 9.09 & 0.00 \\
				& IPCA & 0.90 & 11.22 & 23.08 & 22.89 & 0.08 & 1.88 & 2.10 & 0.10 & 87.18 & 39.53 & 0.00 \\
				& IPCA$\setminus$Regressed-PCA & 1.13 & 5.89 & 34.84 & 33.76 & 0.20 & 4.53 & 0.83 & 0.06 & 35.53 & 16.28 & 0.00 \\
				& FF5 & 0.43 & 5.24 & 5.05 & 4.86 & 0.08 & 1.84 & 0.75 & 0.12 & 89.09 & 13.92 & 0.00 \\
				& KNS & 0.97 & 6.40 & 25.75 & 25.14 & 0.15 & 3.50 & 0.37 & 0.04 & 59.02 & 7.44 & 0.00 \\
				
				\hline\hline
			\end{tabular}
			\begin{tablenotes}
				\small
				\item[\dag]  $A|a|$: average absolute intercept;  $A|t(a)|$: average absolute $t$-statistic for the intercepts;  $Aa^2/V\overline{r}$: average squared intercept over the cross-section variance of  $\overline{r}$, average returns of the testing portfolios; $A\lambda^2/V\overline{r}$: average difference between each squared intercept and its squared standard error divided by the variance of $\overline{r}$; $As(a)$: average standard error of the intercepts; $As(e)$: average residual standard deviation; $Sh^2(a)$: maximized squared Sharpe ratio for the intercepts; $Sh^2(f)$: maximized squared Sharpe ratio for the factors; $AR^2$: average regression $R^2$ (\%);  $GRS$: $GRS$ statistic of \citet{Gibbonsetal_efficiency_1989}; $p(GRS)$: $p$-value of $GRS$.
			\end{tablenotes}
		\end{threeparttable}
	\end{table}%
\end{landscape}

\begin{landscape}
	\setlength{\tabcolsep}{4.2pt}
	\begin{table}[!htbp]
		\centering
		\begin{threeparttable}
			\renewcommand{\arraystretch}{0.95}
			\caption{Additional asset pricing tests: $K=5$ (continued)\tnote{\dag}}\label{Tab: factor_compare_add_K5_3}
			\begin{tabular}{clcccccccccccccc}
				\hline\hline
				&Testing portfolios/Factors & $A|a|$ & $A|t(a)|$ & $Aa^2/V\overline{r}$ & $A\lambda^2/V\overline{r}$ & $As(a)$ & $As(e)$ & $Sh^2(a)$ & $Sh^2(f)$ & $AR^2$ & $GRS$ & $p(GRS)$&\\
				\cline{2-13}
				&\multicolumn{12}{l}{Group IV: 25 double sorted portfolios on Size and MOM}\\
				\cline{2-13}
			                \cline{2-13}
			& Regressed-PCA & 0.82 & 4.73 & 7.27 & 6.97 & 0.18 & 4.16 & 0.41 & 0.04 & 50.16 & 8.13 & 0.00 \\
			& Regressed-PCA S1 & 0.64 & 3.53 & 4.17 & 3.87 & 0.18 & 3.94 & 0.58 & 0.16 & 54.13 & 10.40 & 0.00 \\
			& Regressed-PCA S2 & 0.52 & 2.78 & 3.01 & 2.71 & 0.18 & 4.04 & 0.67 & 0.13 & 51.89 & 12.34 & 0.00 \\
			& IPCA & 0.94 & 11.45 & 9.20 & 9.12 & 0.09 & 2.03 & 2.29 & 0.10 & 87.24 & 43.22 & 0.00 \\
			& IPCA$\setminus$Regressed-PCA &  1.01 & 5.24 & 11.77 & 11.36 & 0.21 & 4.75 & 0.81 & 0.06 & 37.26 & 15.76 & 0.00\\
			& FF5 & 0.44 & 4.47 & 2.36 & 2.25 & 0.11 & 2.40 & 0.47 & 0.12 & 83.53 & 8.68 & 0.00 \\
			& KNS & 0.97 & 6.27 & 8.79 & 8.57 & 0.16 & 3.57 & 0.26 & 0.04 & 62.15 & 5.15 & 0.00 \\

				\cline{2-13}
				&\multicolumn{12}{l}{Group V: 25 double sorted portfolios on Size and Beta}\\
				\cline{2-13}
				& Regressed-PCA & 0.87 & 5.31 & 41.95 & 40.35 & 0.17 & 3.92 & 0.23 & 0.04 & 49.54 & 4.51 & 0.00 \\
				& Regressed-PCA S1 & 0.53 & 3.12 & 18.66 & 16.91 & 0.18 & 3.91 & 0.28 & 0.16 & 48.81 & 4.97 & 0.00 \\
				& Regressed-PCA S2 & 0.40 & 2.31 & 10.28 & 8.48 & 0.18 & 4.01 & 0.27 & 0.13 & 46.17 & 5.04 & 0.00 \\
				& IPCA & 0.90 & 11.17 & 47.69 & 47.25 & 0.09 & 1.97 & 1.92 & 0.10 & 85.50 & 36.20 & 0.00 \\
				& IPCA$\setminus$Regressed-PCA & 1.13 & 6.12 & 70.09 & 67.78 & 0.20 & 4.55 & 0.63 & 0.06 & 36.26 & 12.40 & 0.00 \\
				& FF5 & 0.41 & 4.70 & 9.73 & 9.24 & 0.09 & 2.07 & 0.37 & 0.12 & 85.48 & 6.94 & 0.00 \\
				& KNS & 0.96 & 6.27 & 49.66 & 48.36 & 0.16 & 3.56 & 0.19 & 0.04 & 56.91 & 3.72 & 0.00 \\

				\cline{2-13}
				&\multicolumn{12}{l}{Group VI: 25 double sorted portfolios on Size and AC}\\
				\cline{2-13}
			 & Regressed-PCA & 0.83 & 4.90 & 36.79 & 35.26 & 0.17 & 3.92 & 0.30 & 0.04 & 53.86 & 5.93 & 0.00 \\
			& Regressed-PCA S1 & 0.42 & 2.40 & 10.30 & 8.62 & 0.18 & 3.89 & 0.28 & 0.16 & 54.78 & 5.00 & 0.00 \\
			& Regressed-PCA S2 & 0.28 & 1.61 & 4.99 & 3.30 & 0.18 & 3.96 & 0.28 & 0.13 & 52.94 & 5.14 & 0.00 \\
			& IPCA & 0.81 & 9.39 & 34.66 & 34.25 & 0.09 & 1.95 & 0.96 & 0.10 & 87.20 & 18.04 & 0.00 \\
			& IPCA$\setminus$Regressed-PCA & 1.05 & 5.04 & 57.71 & 55.31 & 0.21 & 4.80 & 0.35 & 0.06 & 33.49 & 6.78 & 0.00 \\
			& FF5 & 0.46 & 5.09 & 11.74 & 11.29 & 0.09 & 2.04 & 0.83 & 0.12 & 87.79 & 15.34 & 0.00 \\
			& KNS & 0.94 & 6.10 & 46.85 & 45.60 & 0.16 & 3.56 & 0.31 & 0.04 & 61.58 & 6.30 & 0.00 \\
				\hline\hline
			\end{tabular}
			\begin{tablenotes}
				\small
				\item[\dag]  $A|a|$: average absolute intercept;  $A|t(a)|$: average absolute $t$-statistic for the intercepts;  $Aa^2/V\overline{r}$: average squared intercept over the cross-section variance of  $\overline{r}$, average returns of the testing portfolios; $A\lambda^2/V\overline{r}$: average difference between each squared intercept and its squared standard error divided by the variance of $\overline{r}$; $As(a)$: average standard error of the intercepts; $As(e)$: average residual standard deviation; $Sh^2(a)$: maximized squared Sharpe ratio for the intercepts; $Sh^2(f)$: maximized squared Sharpe ratio for the factors; $AR^2$: average regression $R^2$ (\%);  $GRS$: $GRS$ statistic of \citet{Gibbonsetal_efficiency_1989}; $p(GRS)$: $p$-value of $GRS$.
			\end{tablenotes}
		\end{threeparttable}
	\end{table}%
\end{landscape}

\begin{landscape}
	\setlength{\tabcolsep}{4.2pt}
	\begin{table}[!htbp]
		\centering
		\begin{threeparttable}
			\renewcommand{\arraystretch}{1}
			\caption{Additional asset pricing tests: $K=5$ (continued)\tnote{\dag}}\label{Tab: factor_compare_add_K5_4}
			\begin{tabular}{clcccccccccccccc}
				\hline\hline
				&Testing portfolios/Factors & $A|a|$ & $A|t(a)|$ & $Aa^2/V\overline{r}$ & $A\lambda^2/V\overline{r}$ & $As(a)$ & $As(e)$ & $Sh^2(a)$ & $Sh^2(f)$ & $AR^2$ & $GRS$ & $p(GRS)$&\\
				\cline{2-13}
				&\multicolumn{12}{l}{Group VII: 35 double sorted portfolios on Size and NI}\\
				\cline{2-13}
				 & Regressed-PCA & 0.84 & 4.85 & 16.08 & 15.43 & 0.18 & 4.02 & 0.40 & 0.04 & 49.84 & 5.63 & 0.00 \\
				& Regressed-PCA S1 & 0.54 & 2.94 & 7.14 & 6.41 & 0.18 & 4.00 & 0.56 & 0.16 & 50.20 & 7.07 & 0.00 \\
				& Regressed-PCA S2 & 0.41 & 2.18 & 4.11 & 3.37 & 0.19 & 4.10 & 0.53 & 0.13 & 47.68 & 6.84 & 0.00 \\
				& IPCA & 0.87 & 9.73 & 17.60 & 17.39 & 0.10 & 2.15 & 2.69 & 0.10 & 83.75 & 35.50 & 0.00 \\
				& IPCA$\setminus$Regressed-PCA & 1.09 & 5.43 & 26.60 & 25.67 & 0.21 & 4.71 & 0.95 & 0.06 & 34.16 & 13.02 & 0.00  \\
				& FF5 & 0.42 & 4.44 & 4.06 & 3.84 & 0.10 & 2.21 & 0.83 & 0.12 & 84.30 & 10.85 & 0.00 \\
				& KNS & 0.94 & 5.86 & 19.28 & 18.73 & 0.16 & 3.68 & 0.39 & 0.04 & 57.44 & 5.45 & 0.00 \\
				
				\cline{2-13}
				&\multicolumn{12}{l}{Group VIII: 25 double sorted portfolios on Size and Var}\\
				\cline{2-13}
			 & Regressed-PCA & 0.89 & 5.34 & 8.00 & 7.71 & 0.18 & 4.04 & 0.45 & 0.04 & 48.41 & 8.98 & 0.00 \\
			& Regressed-PCA S1 & 0.70 & 3.99 & 5.19 & 4.88 & 0.18 & 3.92 & 0.70 & 0.16 & 49.80 & 12.47 & 0.00 \\
			& Regressed-PCA S2 & 0.58 & 3.24 & 3.90 & 3.59 & 0.18 & 4.04 & 0.72 & 0.13 & 46.55 & 13.29 & 0.00 \\
			& IPCA & 0.95 & 12.60 & 9.63 & 9.57 & 0.08 & 1.86 & 2.21 & 0.10 & 87.24 & 41.58 & 0.00 \\
			& IPCA$\setminus$Regressed-PCA & 1.11 & 6.32 & 12.81 & 12.40 & 0.20 & 4.57 & 1.15 & 0.06 & 39.10 & 22.45 & 0.00 \\
			& FF5 & 0.45 & 5.46 & 2.07 & 1.99 & 0.09 & 1.96 & 0.71 & 0.12 & 87.38 & 13.12 & 0.00 \\
			& KNS & 0.99 & 6.48 & 9.51 & 9.30 & 0.15 & 3.53 & 0.43 & 0.04 & 57.72 & 8.57 & 0.00 \\
				\hline\hline
			\end{tabular}
			\begin{tablenotes}
				\small
				\item[\dag]  $A|a|$: average absolute intercept;  $A|t(a)|$: average absolute $t$-statistic for the intercepts;  $Aa^2/V\overline{r}$: average squared intercept over the cross-section variance of  $\overline{r}$, average returns of the testing portfolios; $A\lambda^2/V\overline{r}$: average difference between each squared intercept and its squared standard error divided by the variance of $\overline{r}$; $As(a)$: average standard error of the intercepts; $As(e)$: average residual standard deviation; $Sh^2(a)$: maximized squared Sharpe ratio for the intercepts; $Sh^2(f)$: maximized squared Sharpe ratio for the factors; $AR^2$: average regression $R^2$ (\%);  $GRS$: $GRS$ statistic of \citet{Gibbonsetal_efficiency_1989}; $p(GRS)$: $p$-value of $GRS$.
			\end{tablenotes}
		\end{threeparttable}
	\end{table}%
\end{landscape}

\begin{landscape}
	\setlength{\tabcolsep}{4.2pt}
	\begin{table}[!htbp]
		\centering
		\begin{threeparttable}
			\renewcommand{\arraystretch}{1}
			\caption{Additional asset pricing tests: $K=5$ (continued)\tnote{\dag}}\label{Tab: factor_compare_add_K5_5}
			\begin{tabular}{clcccccccccccccc}
				\hline\hline
				&Testing portfolios/Factors & $A|a|$ & $A|t(a)|$ & $Aa^2/V\overline{r}$ & $A\lambda^2/V\overline{r}$ & $As(a)$ & $As(e)$ & $Sh^2(a)$ & $Sh^2(f)$ & $AR^2$ & $GRS$ & $p(GRS)$&\\
				\cline{2-13}
				&\multicolumn{12}{l}{Group IX: Regressed-PCA S1's 36 managed portfolios}\\
				\cline{2-13}
				                 \cline{2-13}
				& Regressed-PCA & 0.54 & 5.20 & 0.84 & 0.82 & 0.11 & 2.44 & 3.17 & 0.04 & 18.11 & 43.05 & 0.00 \\
				& Regressed-PCA S1 & 0.52 & 5.92 & 0.73 & 0.72 & 0.09 & 1.88 & 3.13 & 0.16 & 38.14 & 38.17 & 0.00 \\
				& Regressed-PCA S2 & 0.53 & 5.80 & 0.71 & 0.69 & 0.09 & 2.07 & 3.18 & 0.13 & 33.15 & 39.73 & 0.00 \\
				& IPCA & 0.53 & 5.39 & 0.80 & 0.78 & 0.11 & 2.33 & 4.75 & 0.10 & 27.85 & 60.84 & 0.00 \\
				& IPCA$\setminus$Regressed-PCA &0.59 & 5.38 & 0.99 & 0.96 & 0.12 & 2.61 & 3.80 & 0.06 & 16.89 & 50.50 & 0.00\\
				& FF5 & 0.56 & 4.93 & 0.99 & 0.96 & 0.12 & 2.59 & 3.18 & 0.12 & 16.02 & 40.13 & 0.00 \\
				& KNS & 0.57 & 5.37 & 1.00 & 0.98 & 0.11 & 2.56 & 3.25 & 0.04 & 16.28 & 44.31 & 0.00 \\
				
				\cline{2-13}
				&\multicolumn{12}{l}{Group X: Regressed-PCA S2's 36 managed portfolios}\\
				\cline{2-13}
			 & Regressed-PCA & 0.63 & 5.42 & 0.83 & 0.81 & 0.12 & 2.82 & 2.88 & 0.04 & 20.45 & 39.12 & 0.00 \\
			& Regressed-PCA S1 & 0.60 & 5.84 & 0.72 & 0.71 & 0.10 & 2.21 & 2.89 & 0.16 & 42.04 & 35.15 & 0.00 \\
			& Regressed-PCA S2 & 0.62 & 7.27 & 0.73 & 0.72 & 0.08 & 1.86 & 1.46 & 0.13 & 50.54 & 18.20 & 0.00 \\
			& IPCA & 0.64 & 5.88 & 0.82 & 0.80 & 0.12 & 2.56 & 4.59 & 0.10 & 32.44 & 58.83 & 0.00 \\
		& IPCA$\setminus$Regressed-PCA &0.69 & 5.53 & 1.00 & 0.97 & 0.13 & 2.96 & 3.68 & 0.06 & 17.67 & 48.85 & 0.00 \\
			& FF5 & 0.62 & 4.76 & 0.96 & 0.94 & 0.13 & 2.90 & 2.88 & 0.12 & 21.41 & 36.42 & 0.00 \\
			& KNS & 0.65 & 5.53 & 1.00 & 0.98 & 0.13 & 2.86 & 2.97 & 0.04 & 20.33 & 40.52 & 0.00 \\

				\hline\hline
			\end{tabular}
			\begin{tablenotes}
				\small
				\item[\dag]  $A|a|$: average absolute intercept;  $A|t(a)|$: average absolute $t$-statistic for the intercepts;  $Aa^2/V\overline{r}$: average squared intercept over the cross-section variance of  $\overline{r}$, average returns of the testing portfolios; $A\lambda^2/V\overline{r}$: average difference between each squared intercept and its squared standard error divided by the variance of $\overline{r}$; $As(a)$: average standard error of the intercepts; $As(e)$: average residual standard deviation; $Sh^2(a)$: maximized squared Sharpe ratio for the intercepts; $Sh^2(f)$: maximized squared Sharpe ratio for the factors; $AR^2$: average regression $R^2$ (\%);  $GRS$: $GRS$ statistic of \citet{Gibbonsetal_efficiency_1989}; $p(GRS)$: $p$-value of $GRS$.
			\end{tablenotes}
		\end{threeparttable}
	\end{table}%
\end{landscape}

\begin{landscape}
	\setlength{\tabcolsep}{4.2pt}
	\begin{table}[!htbp]
		\centering
		\begin{threeparttable}
			\renewcommand{\arraystretch}{0.95}
			\caption{Additional asset pricing tests: $K=6$\tnote{\dag}}\label{Tab: factor_compare_add_K6_1}
			\begin{tabular}{clcccccccccccccc}
				\hline\hline
				&Testing portfolios/Factors & $A|a|$ & $A|t(a)|$ & $Aa^2/V\overline{r}$ & $A\lambda^2/V\overline{r}$ & $As(a)$ & $As(e)$ & $Sh^2(a)$ & $Sh^2(f)$ & $AR^2$ & $GRS$ & $p(GRS)$&\\
				\cline{2-13}
				&\multicolumn{12}{l}{Group I: 25 double sorted portfolios on Size and BM}\\
				\cline{2-13}
			 & Regressed-PCA & 0.76 & 4.93 & 13.84 & 13.28 & 0.16 & 3.56 & 0.32 & 0.04 & 58.73 & 6.37 & 0.00 \\
			& Regressed-PCA S1 & 0.57 & 3.10 & 8.41 & 7.63 & 0.19 & 3.92 & 0.46 & 0.22 & 51.46 & 7.87 & 0.00 \\
			& Regressed-PCA S2 & 1.04 & 5.33 & 27.04 & 26.18 & 0.20 & 3.83 & 0.52 & 0.42 & 53.23 & 7.64 & 0.00 \\
			& IPCA & 0.89 & 9.35 & 18.91 & 18.66 & 0.10 & 2.04 & 1.83 & 0.33 & 85.34 & 28.54 & 0.00 \\
			& IPCA$\setminus$Regressed-PCA &1.31 & 5.64 & 39.81 & 38.48 & 0.24 & 4.92 & 0.66 & 0.28 & 27.28 & 10.60 & 0.00  \\
			& FF5+MOM & 0.36 & 4.22 & 3.06 & 2.86 & 0.09 & 2.00 & 0.65 & 0.17 & 87.25 & 11.61 & 0.00 \\
			& KNS & 0.87 & 5.75 & 17.68 & 17.16 & 0.15 & 3.47 & 0.30 & 0.05 & 61.83 & 5.92 & 0.00 \\
				
				\cline{2-13}
				&\multicolumn{12}{l}{Group II: 25 double sorted portfolios on Size and OP}\\
				\cline{2-13}
			  & Regressed-PCA & 0.72 & 4.88 & 21.02 & 20.17 & 0.15 & 3.43 & 0.21 & 0.04 & 61.43 & 4.21 & 0.00 \\
			& Regressed-PCA S1 & 0.47 & 2.61 & 10.10 & 8.84 & 0.18 & 3.86 & 0.24 & 0.22 & 52.34 & 4.10 & 0.00 \\
			& Regressed-PCA S2 & 0.98 & 5.12 & 38.54 & 37.15 & 0.19 & 3.76 & 0.39 & 0.42 & 54.34 & 5.69 & 0.00 \\
			& IPCA & 0.85 & 9.90 & 28.88 & 28.52 & 0.09 & 1.91 & 1.87 & 0.33 & 87.08 & 29.12 & 0.00 \\
		& IPCA$\setminus$Regressed-PCA& 1.27 & 5.49 & 61.94 & 59.78 & 0.24 & 4.88 & 0.42 & 0.28 & 26.23 & 6.83 & 0.00 \\
			& FF5+MOM & 0.35 & 4.38 & 5.13 & 4.82 & 0.09 & 1.90 & 0.72 & 0.17 & 88.59 & 12.71 & 0.00 \\
			& KNS & 0.84 & 5.69 & 27.20 & 26.37 & 0.15 & 3.39 & 0.21 & 0.05 & 62.68 & 4.17 & 0.00 \\

				\cline{2-13}
				&\multicolumn{12}{l}{Group III: 25 double sorted portfolios on Size and INV}\\
				\cline{2-13}
			 & Regressed-PCA & 0.76 & 5.24 & 16.41 & 15.84 & 0.15 & 3.36 & 0.36 & 0.04 & 61.87 & 7.15 & 0.00 \\
			& Regressed-PCA S1 & 0.55 & 3.11 & 9.11 & 8.26 & 0.18 & 3.80 & 0.49 & 0.22 & 52.50 & 8.33 & 0.00 \\
			& Regressed-PCA S2 & 1.03 & 5.50 & 29.94 & 29.00 & 0.19 & 3.70 & 0.61 & 0.42 & 54.61 & 8.92 & 0.00 \\
			& IPCA & 0.91 & 10.23 & 23.31 & 23.08 & 0.09 & 1.88 & 1.85 & 0.33 & 87.11 & 28.89 & 0.00 \\
			& IPCA$\setminus$Regressed-PCA & 1.34 & 5.82 & 48.05 & 46.54 & 0.24 & 4.86 & 0.73 & 0.28 & 25.88 & 11.76 & 0.00 \\
			& FF5+MOM & 0.37 & 4.66 & 3.83 & 3.64 & 0.08 & 1.80 & 0.72 & 0.17 & 89.56 & 12.86 & 0.00 \\
			& KNS & 0.87 & 5.95 & 20.66 & 20.10 & 0.15 & 3.35 & 0.39 & 0.05 & 62.54 & 7.66 & 0.00 \\

				\hline\hline
			\end{tabular}
			\begin{tablenotes}
				\small
				\item[\dag]  $A|a|$: average absolute intercept;  $A|t(a)|$: average absolute $t$-statistic for the intercepts;  $Aa^2/V\overline{r}$: average squared intercept over the cross-section variance of  $\overline{r}$, average returns of the testing portfolios; $A\lambda^2/V\overline{r}$: average difference between each squared intercept and its squared standard error divided by the variance of $\overline{r}$; $As(a)$: average standard error of the intercepts; $As(e)$: average residual standard deviation; $Sh^2(a)$: maximized squared Sharpe ratio for the intercepts; $Sh^2(f)$: maximized squared Sharpe ratio for the factors; $AR^2$: average regression $R^2$ (\%);  $GRS$: $GRS$ statistic of \citet{Gibbonsetal_efficiency_1989}; $p(GRS)$: $p$-value of $GRS$.
			\end{tablenotes}
		\end{threeparttable}
	\end{table}%
\end{landscape}

\begin{landscape}
	\setlength{\tabcolsep}{4.2pt}
	\begin{table}[!htbp]
		\centering
		\begin{threeparttable}
			\renewcommand{\arraystretch}{0.95}
			\caption{Additional asset pricing tests: $K=6$ (continued)\tnote{\dag}}\label{Tab: factor_compare_add_K6_3}
			\begin{tabular}{clcccccccccccccc}
				\hline\hline
				&Testing portfolios/Factors & $A|a|$ & $A|t(a)|$ & $Aa^2/V\overline{r}$ & $A\lambda^2/V\overline{r}$ & $As(a)$ & $As(e)$ & $Sh^2(a)$ & $Sh^2(f)$ & $AR^2$ & $GRS$ & $p(GRS)$&\\
                \cline{2-13}
				&\multicolumn{12}{l}{Group IV: 25 double sorted portfolios on Size and MOM}\\
				\cline{2-13}
			 & Regressed-PCA & 0.73 & 4.72 & 5.84 & 5.59 & 0.16 & 3.73 & 0.40 & 0.04 & 59.23 & 8.00 & 0.00 \\
			& Regressed-PCA S1 & 0.60 & 3.23 & 3.64 & 3.33 & 0.19 & 3.93 & 0.52 & 0.22 & 54.48 & 8.87 & 0.00 \\
			& Regressed-PCA S2 & 0.91 & 4.80 & 9.25 & 8.90 & 0.20 & 3.82 & 0.62 & 0.42 & 56.68 & 9.03 & 0.00 \\
			& IPCA & 0.80 & 9.14 & 7.22 & 7.12 & 0.10 & 2.02 & 1.82 & 0.33 & 87.32 & 28.34 & 0.00 \\
			& IPCA$\setminus$Regressed-PCA  & 1.22 & 5.21 & 14.36 & 13.79 & 0.25 & 5.07 & 0.50 & 0.28 & 28.70 & 8.04 & 0.00 \\
			& FF5+MOM & 0.42 & 4.85 & 1.74 & 1.67 & 0.09 & 1.92 & 0.42 & 0.17 & 89.28 & 7.42 & 0.00 \\
			& KNS & 0.88 & 5.84 & 7.24 & 7.03 & 0.15 & 3.45 & 0.25 & 0.05 & 64.65 & 4.90 & 0.00 \\

				\cline{2-13}
				&\multicolumn{12}{l}{Group V: 25 double sorted portfolios on Size and Beta}\\
				\cline{2-13}
			                \cline{2-13}
			& Regressed-PCA & 0.75 & 5.15 & 32.54 & 31.29 & 0.15 & 3.46 & 0.22 & 0.04 & 58.93 & 4.41 & 0.00 \\
			& Regressed-PCA S1 & 0.51 & 2.91 & 16.81 & 14.97 & 0.18 & 3.89 & 0.23 & 0.22 & 49.29 & 3.94 & 0.00 \\
			& Regressed-PCA S2 & 1.02 & 5.47 & 59.59 & 57.55 & 0.19 & 3.79 & 0.37 & 0.42 & 51.55 & 5.36 & 0.00 \\
			& IPCA & 0.89 & 9.83 & 43.36 & 42.82 & 0.10 & 1.97 & 1.60 & 0.33 & 85.49 & 24.98 & 0.00 \\
			& IPCA$\setminus$Regressed-PCA  & 1.31 & 5.79 & 92.63 & 89.41 & 0.24 & 4.88 & 0.41 & 0.28 & 27.17 & 6.69 & 0.00 \\
			& FF5+MOM & 0.36 & 4.10 & 7.36 & 6.87 & 0.09 & 2.03 & 0.34 & 0.17 & 85.92 & 6.03 & 0.00 \\
			& KNS & 0.86 & 5.81 & 39.84 & 38.62 & 0.15 & 3.42 & 0.17 & 0.05 & 60.06 & 3.46 & 0.00 \\

				\cline{2-13}
				&\multicolumn{12}{l}{Group VI: 25 double sorted portfolios on Size and AC}\\
				\cline{2-13}
				 & Regressed-PCA & 0.71 & 4.77 & 26.92 & 25.77 & 0.15 & 3.40 & 0.29 & 0.04 & 64.24 & 5.80 & 0.00 \\
				& Regressed-PCA S1 & 0.43 & 2.40 & 10.71 & 8.96 & 0.18 & 3.87 & 0.25 & 0.22 & 55.22 & 4.18 & 0.00 \\
				& Regressed-PCA S2 & 0.90 & 4.69 & 43.07 & 41.14 & 0.19 & 3.77 & 0.36 & 0.42 & 56.97 & 5.18 & 0.00 \\
				& IPCA & 0.88 & 9.35 & 41.19 & 40.69 & 0.10 & 1.95 & 1.05 & 0.33 & 87.13 & 16.36 & 0.00 \\
				& IPCA$\setminus$Regressed-PCA  & 1.33 & 5.38 & 93.52 & 90.18 & 0.25 & 5.15 & 0.34 & 0.28 & 23.54 & 5.48 & 0.00 \\
				& FF5+MOM & 0.40 & 4.43 & 8.91 & 8.45 & 0.09 & 2.00 & 0.84 & 0.17 & 88.20 & 14.93 & 0.00 \\
				& KNS & 0.86 & 5.67 & 39.04 & 37.84 & 0.15 & 3.47 & 0.30 & 0.05 & 63.58 & 6.00 & 0.00 \\
				\hline\hline
			\end{tabular}
			\begin{tablenotes}
				\small
				\item[\dag]  $A|a|$: average absolute intercept;  $A|t(a)|$: average absolute $t$-statistic for the intercepts;  $Aa^2/V\overline{r}$: average squared intercept over the cross-section variance of  $\overline{r}$, average returns of the testing portfolios; $A\lambda^2/V\overline{r}$: average difference between each squared intercept and its squared standard error divided by the variance of $\overline{r}$; $As(a)$: average standard error of the intercepts; $As(e)$: average residual standard deviation; $Sh^2(a)$: maximized squared Sharpe ratio for the intercepts; $Sh^2(f)$: maximized squared Sharpe ratio for the factors; $AR^2$: average regression $R^2$ (\%);  $GRS$: $GRS$ statistic of \citet{Gibbonsetal_efficiency_1989}; $p(GRS)$: $p$-value of $GRS$.
			\end{tablenotes}
		\end{threeparttable}
	\end{table}%
\end{landscape}

\begin{landscape}
	\setlength{\tabcolsep}{4.2pt}
	\begin{table}[!htbp]
		\centering
		\begin{threeparttable}
			\renewcommand{\arraystretch}{1}
			\caption{Additional asset pricing tests: $K=6$ (continued)\tnote{\dag}}\label{Tab: factor_compare_add_K6_4}
			\begin{tabular}{clcccccccccccccc}
				\hline\hline
				&Testing portfolios/Factors & $A|a|$ & $A|t(a)|$ & $Aa^2/V\overline{r}$ & $A\lambda^2/V\overline{r}$ & $As(a)$ & $As(e)$ & $Sh^2(a)$ & $Sh^2(f)$ & $AR^2$ & $GRS$ & $p(GRS)$&\\
				\cline{2-13}
				&\multicolumn{12}{l}{Group VII: 35 double sorted portfolios on Size and NI}\\
				\cline{2-13}
			 & Regressed-PCA & 0.73 & 4.73 & 12.35 & 11.84 & 0.16 & 3.57 & 0.40 & 0.04 & 59.32 & 5.56 & 0.00 \\
			& Regressed-PCA S1 & 0.53 & 2.80 & 6.71 & 5.96 & 0.19 & 3.99 & 0.53 & 0.22 & 50.60 & 6.36 & 0.00 \\
			& Regressed-PCA S2 & 0.97 & 4.92 & 22.54 & 21.70 & 0.20 & 3.89 & 0.69 & 0.42 & 52.43 & 7.01 & 0.00 \\
			& IPCA & 0.87 & 8.83 & 17.37 & 17.11 & 0.11 & 2.15 & 2.35 & 0.33 & 83.69 & 25.70 & 0.00 \\
			& IPCA$\setminus$Regressed-PCA  & 1.30 & 5.42 & 36.67 & 35.37 & 0.24 & 5.03 & 0.76 & 0.28 & 25.00 & 8.65 & 0.00 \\
			& FF5+MOM & 0.37 & 3.90 & 3.13 & 2.90 & 0.10 & 2.17 & 0.80 & 0.17 & 84.75 & 9.96 & 0.00 \\
			& KNS & 0.84 & 5.43 & 15.68 & 15.16 & 0.16 & 3.55 & 0.38 & 0.05 & 60.17 & 5.33 & 0.00 \\
				
				\cline{2-13}
				&\multicolumn{12}{l}{Group VIII: 25 double sorted portfolios on Size and Var}\\
				\cline{2-13}
			 & Regressed-PCA & 0.80 & 5.31 & 6.68 & 6.45 & 0.16 & 3.60 & 0.45 & 0.04 & 57.27 & 8.99 & 0.00 \\
			& Regressed-PCA S1 & 0.64 & 3.59 & 4.38 & 4.06 & 0.19 & 3.90 & 0.64 & 0.22 & 50.26 & 10.91 & 0.00 \\
			& Regressed-PCA S2 & 1.01 & 5.41 & 11.24 & 10.89 & 0.20 & 3.82 & 0.73 & 0.42 & 51.84 & 10.58 & 0.00 \\
			& IPCA & 0.87 & 10.34 & 7.72 & 7.64 & 0.09 & 1.87 & 1.74 & 0.33 & 87.10 & 27.04 & 0.00 \\
			& IPCA$\setminus$Regressed-PCA  & 1.26 & 5.84 & 15.16 & 14.59 & 0.24 & 4.90 & 0.79 & 0.28 & 30.37 & 12.67 & 0.00 \\
			& FF5+MOM & 0.44 & 5.28 & 1.97 & 1.89 & 0.09 & 1.96 & 0.68 & 0.17 & 87.45 & 12.00 & 0.00 \\
			& KNS & 0.90 & 6.07 & 7.90 & 7.69 & 0.15 & 3.41 & 0.42 & 0.05 & 60.78 & 8.31 & 0.00 \\
				
				\hline\hline
			\end{tabular}
			\begin{tablenotes}
				\small
				\item[\dag]  $A|a|$: average absolute intercept;  $A|t(a)|$: average absolute $t$-statistic for the intercepts;  $Aa^2/V\overline{r}$: average squared intercept over the cross-section variance of  $\overline{r}$, average returns of the testing portfolios; $A\lambda^2/V\overline{r}$: average difference between each squared intercept and its squared standard error divided by the variance of $\overline{r}$; $As(a)$: average standard error of the intercepts; $As(e)$: average residual standard deviation; $Sh^2(a)$: maximized squared Sharpe ratio for the intercepts; $Sh^2(f)$: maximized squared Sharpe ratio for the factors; $AR^2$: average regression $R^2$ (\%);  $GRS$: $GRS$ statistic of \citet{Gibbonsetal_efficiency_1989}; $p(GRS)$: $p$-value of $GRS$.
			\end{tablenotes}
		\end{threeparttable}
	\end{table}%
\end{landscape}

\begin{landscape}
	\setlength{\tabcolsep}{4.2pt}
	\begin{table}[!htbp]
		\centering
		\begin{threeparttable}
			\renewcommand{\arraystretch}{1}
			\caption{Additional asset pricing tests: $K=6$ (continued)\tnote{\dag}}\label{Tab: factor_compare_add_K6_5}
			\begin{tabular}{clcccccccccccccc}
				\hline\hline
				&Testing portfolios/Factors & $A|a|$ & $A|t(a)|$ & $Aa^2/V\overline{r}$ & $A\lambda^2/V\overline{r}$ & $As(a)$ & $As(e)$ & $Sh^2(a)$ & $Sh^2(f)$ & $AR^2$ & $GRS$ & $p(GRS)$&\\
				\cline{2-13}
				&\multicolumn{12}{l}{Group IX: Regressed-PCA S1's 36 managed portfolios}\\
				\cline{2-13}
				 & Regressed-PCA & 0.54 & 5.27 & 0.84 & 0.82 & 0.11 & 2.41 & 3.19 & 0.04 & 19.82 & 43.08 & 0.00 \\
				& Regressed-PCA S1 & 0.51 & 6.27 & 0.69 & 0.68 & 0.08 & 1.77 & 5.39 & 0.22 & 41.99 & 62.20 & 0.00 \\
				& Regressed-PCA S2 & 0.47 & 4.85 & 0.56 & 0.54 & 0.10 & 2.00 & 2.89 & 0.42 & 35.74 & 28.58 & 0.00 \\
				& IPCA & 0.50 & 4.86 & 0.60 & 0.57 & 0.11 & 2.26 & 3.85 & 0.33 & 31.41 & 40.80 & 0.00 \\
				& IPCA$\setminus$Regressed-PCA  & 0.57 & 5.05 & 0.77 & 0.74 & 0.12 & 2.56 & 3.50 & 0.28 & 19.43 & 38.43 & 0.00 \\
				& FF5+MOM & 0.56 & 4.92 & 1.09 & 1.07 & 0.12 & 2.51 & 3.13 & 0.17 & 19.79 & 37.82 & 0.00 \\
				& KNS & 0.57 & 5.36 & 1.03 & 1.00 & 0.11 & 2.53 & 3.25 & 0.05 & 18.03 & 43.79 & 0.00 \\
				\cline{2-13}
				&\multicolumn{12}{l}{Group X: Regressed-PCA S2's 36 managed portfolios}\\
				\cline{2-13}
			& Regressed-PCA & 0.63 & 5.46 & 0.83 & 0.81 & 0.12 & 2.77 & 2.88 & 0.04 & 22.56 & 38.82 & 0.00 \\
			& Regressed-PCA S1 & 0.62 & 6.29 & 0.72 & 0.70 & 0.10 & 2.09 & 2.85 & 0.22 & 46.56 & 32.87 & 0.00 \\
			& Regressed-PCA S2 & 0.54 & 6.03 & 0.55 & 0.54 & 0.09 & 1.72 & 3.40 & 0.42 & 54.10 & 33.71 & 0.00 \\
			& IPCA & 0.64 & 5.61 & 0.69 & 0.67 & 0.12 & 2.43 & 3.63 & 0.33 & 37.52 & 38.48 & 0.00 \\
			& IPCA$\setminus$Regressed-PCA  & 0.67 & 5.35 & 0.79 & 0.76 & 0.14 & 2.89 & 3.26 & 0.28 & 20.86 & 35.72 & 0.00 \\
			& FF5+MOM & 0.63 & 4.97 & 1.07 & 1.05 & 0.13 & 2.78 & 2.85 & 0.17 & 25.91 & 34.35 & 0.00 \\
			& KNS & 0.65 & 5.51 & 1.02 & 1.00 & 0.12 & 2.84 & 2.96 & 0.05 & 21.88 & 39.89 & 0.00 \\
				\hline\hline
			\end{tabular}
			\begin{tablenotes}
				\small
				\item[\dag]  $A|a|$: average absolute intercept;  $A|t(a)|$: average absolute $t$-statistic for the intercepts;  $Aa^2/V\overline{r}$: average squared intercept over the cross-section variance of  $\overline{r}$, average returns of the testing portfolios; $A\lambda^2/V\overline{r}$: average difference between each squared intercept and its squared standard error divided by the variance of $\overline{r}$; $As(a)$: average standard error of the intercepts; $As(e)$: average residual standard deviation; $Sh^2(a)$: maximized squared Sharpe ratio for the intercepts; $Sh^2(f)$: maximized squared Sharpe ratio for the factors; $AR^2$: average regression $R^2$ (\%);  $GRS$: $GRS$ statistic of \citet{Gibbonsetal_efficiency_1989}; $p(GRS)$: $p$-value of $GRS$.
			\end{tablenotes}
		\end{threeparttable}
	\end{table}%
\end{landscape}

\clearpage
%
%

%
%
%
%


\end{appendices}
\addcontentsline{toc}{section}{References}
\putbib
\end{bibunit}

\bibliography{Mybibliography_Infinity}

\end{document}